\documentclass[a4paper,12pt]{article}
\pdfoutput=1
\usepackage[nokeyprefix]{refstyle}
\usepackage{varioref}
\usepackage{xr}
\usepackage{epsfig,amsmath,amsfonts,amsthm}
\usepackage[figuresright]{rotating}
\usepackage[left=28mm,right=20mm,top=30mm,bottom=20mm]{geometry}
\usepackage{subfig}
\usepackage{graphicx}
\graphicspath{{C:/Local Storage/New SMD/plots1/}}
\usepackage{rotating}
\usepackage{booktabs}
\usepackage{moreverb}

\DeclareGraphicsExtensions{.pdf,.png,.jpg}
\usepackage{natbib}
\numberwithin{equation}{section}

\def\e{\hbox{E}}
\def\cov{\hbox{Cov}}
\def\diag{\hbox{diag}}
\def\var{\hbox{Var}}

\def\max{\hbox{max}}

\providecommand{\keywords}[1]{\textbf{{Keywords}} #1}

\begin{document}

\title{Simulations for estimation of random effects and overall effect in three-level meta-analysis of standardized mean differences using constant and inverse-variance weights}

\author{Elena Kulinskaya and  David C. Hoaglin }

\date{\today}

\maketitle

\begin{abstract}
We consider a three-level meta-analysis of standardized mean differences.
The standard method of estimation uses inverse-variance weights and REML/PL estimation of variance components for the random effects.

We introduce new moment-based point and interval estimators for the two variance components and related estimators of the overall mean. Similar to traditional analysis of variance, our method is based on two conditional $Q$ statistics with effective-sample-size weights. We study, by simulation, bias and coverage of these new estimators.  For comparison,  we also study bias and coverage of the REML/PL-based approach as implemented in  {\it  rma.mv} in {\it metafor}.
Our results demonstrate that the new methods are often considerably better and do not have convergence problems, which plague the standard analysis.

\end{abstract}


\section{Introduction}
When viewed as a multilevel model (\cite{Raudenbush1985}, \cite{Goldstein2000}), the standard random-effects model of meta-analysis is a two-level hierarchical model. Typically, the level-1 units comprise the sample within each study, and the level-2 units are the studies. A three-level model involves a further level of structure; for example, the studies may be nested within clusters. \cite{Konstantopoulos2011} discusses the traditional analysis of a three-level model in detail. \cite{Cheung2014} formulated a three-level model in the framework of structural-equation modelling.  Multilevel meta-analysis is implemented in the {\it rma.mv} procedure in {\it metafor} \citep{metafor}. Traditional analysis uses REML-based estimation of the variance components for the random effects and profile-likelihood estimation of their confidence intervals.

In a meta-analysis involving three levels, one can apply moment estimation similar to that in analysis of variance, separating the total variation into within and between components. Our approach uses constant weights based on effective sample sizes.

In a related development, in multivariate meta-analysis, \cite{Jackson2024} proposed a method of  moment-based estimation of the between-study covariance matrix (analogous to the heterogeneity variance in the usual random-effects model). In that method, the weights for estimating the overall effects involve an estimate of the covariance matrix, which is then updated to reflect the current estimate of the overall effects. By using constant weights, we avoid that complication.

In a systematic review of multilevel meta-analyses, \cite{FernandezCastilla2020} listed existing simulation studies (their Table 1) and the parameters from 178 actual meta-analyses in behavioral, social, biological, and medical sciences (their Tables 3--5). Our simulation scenarios were largely guided by those tables. Of those 178 meta-analyses, 162 fitted the three-level model, and 80 used the standardized mean difference (SMD) as the effect measure.


For SMD as the effect measure, our simulation study  evaluated our novel  moment-based method of estimation in a three-level meta-analysis and compared it with the standard REML/PL-based methods implemented in {\it rma.mv} in {\it metafor}.

\section{A three-level hierarchical model for meta-analysis}


We consider a 3-level  hierarchical model involving $M$ clusters of  studies: $K_g$ studies in cluster $g$ ($g = 1, \ldots, M$) and sample size $n_{gi}$ for study $i$ in cluster $g$ ($ i = 1,\ldots, K_g$). The model includes random variation at each level: level-1, within-study random errors $\epsilon_g \sim MVN(0, \diag(\sigma_{gi}^2))$; level-2, within-cluster between-studies random effects $\eta_g \sim MVN(0, \tau^2 I_{K_g})$; and level-3, between-clusters random effects $\zeta \sim MVN(0, \omega^2 I_M)$; where $I_{K_g}$ and $I_M$ are unit matrices of dimension $K_g$ and $M$, respectively. In this model, the level-2 parameter, $\tau^2$, is not cluster-specific, so estimation can use the full data. Cluster-specific variances, $\tau^2_g$, would require within-cluster estimation. Our methods can handle the latter scenario, but we have not done any simulations for it.

Similar to a two-level meta-regression model with random effects, our model for the $T_g$, the summary statistics in cluster $g$, involves $\eta_g$ and $\epsilon_g$:
\begin{equation} \label{3lM_1}
T_g =  a_{0g} 1_g + Y_g \beta_0 + \eta_g + \epsilon_g,
\end{equation}
where   $a_{0g}$ is an effect of cluster $g$, $1_{K_g}$ is a vector of 1's of dimension $K_g$, $Y_g$ is a $K_g \times p$ design matrix, $\beta_0$ is a $p$-vector of fixed effects (without an intercept), and $\eta_g$ and $\epsilon_g$ represent independent between- and within-study variation.  
It follows that, conditionally, $\cov(T_g | a_{0g}) = \Sigma_g = \diag(\tau^2 + \sigma_{gi}^2)$.

To write the model in Equation~(\ref{3lM_1}) in terms of the stacked vector $T = (T_1^T, \ldots, T_M^T)^T$ of length ${\mathcal K}=\sum K_g$, we define the parameter vector $\beta=(a_{01}, a_{02}, \ldots, a_{0M}, \beta_0^T)^T$ of dimension $(M + p)$ and the design matrices $X_g = (\delta_{g1} 1_{K_g}, \ldots, \delta_{gM} 1_{K_g}, Y_g)$, where $\delta_{gt}$ is the Kronecker delta ($\delta_{gt} = 1$ if $g = t$ and 0 otherwise). We form the ${\mathcal K} \times (M+p)$ design matrix $X$ by stacking the $X_g$. Thus, also using stacked error vectors $\eta$ and $\epsilon$, Equation~(\ref{3lM_1}) becomes
\begin{equation} \label{3lM_00}
T = X \beta + \eta + \epsilon.
\end{equation}
This model differs from the standard meta-regression model  only by including in $\beta$ the $M$-vector of random intercepts $a_0 = (a_{01}, \ldots, a_{0M})^T$; the $K_g$ studies in cluster $g$ have the same $a_{0g}$.

We model the intercepts $a_{0g}$ via a mixed-effects meta-regression and a level-3 random effect:
\begin{equation} \label{3lM_2a}
a_{0g} = W_g \gamma + Z_g \xi + \zeta_g,
\end{equation}
where  $W_g$ is a  $1 \times q$ matrix of fixed effects, $\gamma$ is a $q$-vector of unknown coefficients, $Z_g$ is a $1 \times r$ random-effects design matrix, $\xi \sim MVN(0, \Omega)$ is an $r$-vector of random effects, and  $\zeta = (\zeta_1, \ldots, \zeta_M)^T \sim MVN(0, \omega^2 I_M)$.  
The variance is $\var(a_{0g}) = Z_g \Omega Z_g^T + \omega^2$. 
Thus, the unconditional covariance matrix for cluster $g$ is
\begin{equation} \label{3lM_cov}
V_g = \cov(T_g) = (Z_g \Omega Z_g^T + \omega^2) J_{K_g} + \Sigma_g = (Z_g \Omega Z_g^T + \omega^2) J_g + \diag(\tau^2 + \sigma_{gi}^2),
\end{equation}
where $J_{K_g}$ is the $K_g \times K_g$ matrix of 1's. The full covariance matrix of the stacked vector $T$, denoted by $V$, is the block-diagonal matrix of the $M$ matrices $V_g$.

\section{Estimation} \label{sec:3level_estimation}
As in ANOVA.we approach estimation by subdividing the total variation into within and between components and using moments. Thus, we work with Equations~(\ref{3lM_00}) and (\ref{3lM_2a}) separately.

\subsection{Estimating 2nd-level parameters (within-cluster analysis)} \label{sec:L2}

At level 2, conditionally on the values of $(a_{01}, \ldots, a_{0M})$, we can estimate $\beta$ and $\tau^2$ as in a standard meta-regression.

To accommodate various choices of weights, we introduce $A$, a symmetric  ${\mathcal K} \times {\mathcal K}$ matrix of full rank. Then from Equation~(\ref{3lM_00}), $\e(X^T A T) = X^T A X \beta$ and
\begin{equation} \label{BhatA}
\hat\beta_A = (X^T A X)^{-1} X^T A T
\end{equation}
is an unbiased estimator of $\beta$ with conditional covariance
\begin{equation} \label{covBhatA}
\cov(\hat\beta_A | a_0) = (X^T A X)^{-1} X^T A \Sigma  A  X (X^T A X)^{-1},
\end{equation}
where $\Sigma = \diag(\Sigma_g)$. The maximum-likelihood-based estimator corresponds to $A = \Sigma^{-1}$ (i.e., to inverse-variance weights).
For ML/IV, $\hat\beta_A$ given by Equation~(\ref{BhatA}) needs to be calculated simultaneously with $\hat{\tau}^2$. For any fixed $A$, however, $\hat{\beta}$ and $\hat{\tau}^2$ can be estimated separately.

The residuals based on $\hat\beta_A$ are
$R_A = T - X \hat\beta_A = T - X (X^T A X)^{-1} X^T A T = (I - H_A)T$, where $H_A =X (X^T A X)^{-1} X^T A$ is a hat matrix, similar to that in least-squares regression. As projections, $H_A$ and $(I - H_A)$ are idempotent. That is, $H_A^2 = H_A$.  Also,  $H_A^T A = A H_A$.

Moment estimation of the variance component $\tau^2$ is based on the generalized $Q$ statistic, a quadratic form in the residuals:
\begin{equation} \label{Q_A}
Q_A = R_A^T A R_A = T^T (I - H_A^T) A (I - H_A) T = T^T A (I - H_A) T
\end{equation}
(discussed by \cite{dersimonian2007random}).
Given unbiased estimators $v_{gi}^2 = \hat\sigma_{gi}^2$ and a nonrandom $A$ matrix, and
denoting the weight matrix by $C = A (I - H_A)$, $Q_A = T^T C T$ and $\e(Q_A) = \sum_g \sum_i c_{(g)ii} (\tau^2 + \e(v_{gi}^2))$, where the weight $c_{(g)ii}$ is applied to $T_{gi}^2$ in Equation~(\ref{Q_A}). Hence, a moment estimator of $\tau^2$ is
\begin{equation} \label{tau_A}
\hat{\tau}^2 = \max((\sum \sum c_{(g)ii})^{-1} (Q_A - \sum \sum c_{(g)ii} \e(v_{gi}^2)), 0).
\end{equation}
The distribution of $Q_A$ is that of a quadratic form in normal random variables, and we can obtain a confidence interval for $\tau^2$ from percentage points of an approximation to this distribution under the hypothesis of no heterogeneity.

In the simplest case $p = 0$ (i.e., no meta-regression), and the within-cluster model in Equation~(\ref{3lM_1}) is the standard random-effects model of meta-analysis.  Then the only unknown parameters are $\beta = a_0 = (a_{01}, \ldots, a_{0M})^T$. For a block-diagonal matrix $A = \diag(A^1, \ldots, A^M)$, in which $A^g$ is $K_g \times K_g$, Equation~(\ref{BhatA}) results in $\hat a_{0g} = \sum_i A^g_{+i} T_{gi} / A^g_{++}$,
where $A^g_{+i} = \sum_j A^g_{ji}$ and $A^g_{++} = \sum_i A^g_{+i}$. The conditional variances are $\var(\hat a_{0g} | a_{0g}) = \sum_i (A^g_{+i})^2 (\tau^2 + \sigma_{gi}^2) / (A^g_{++})^2 = \sum_i b_{gi}^2(\tau^2+\sigma_{gi}^2)$ for $b_{gi} = A^g_{+i} / A^g_{++}$.

\subsection{Estimating 3rd-level parameters (between-cluster analysis)} \label{sec:L3}

At level 3 the models for $a_{01}, \ldots, a_{0M}$ (Equation~(\ref{3lM_2a})) share the fixed effects ($\gamma$), the covariance matrix $\Omega$, and the variance $\omega^2$. Since $\e(\hat a_{0g} | a_{0g}) = a_{0g}$, the unconditional expected value $\e(\hat a_{0g}) = \e(a_{0g}) =W_g \gamma$. In matrix form, $\e(\hat a_{0}) = W \gamma$ for the $(M \times q)$ matrix $W$ with rows $W_g$. Hence, for a symmetric matrix $F$ of a full rank,
\begin{equation} \label{Ghat_F}
\hat\gamma_{F} = (W^T F W)^{-1} W^T F \hat a_0
\end{equation}
is an unbiased estimator of $\gamma$ with covariance
\begin{equation} \label{covGhatA_g}
\cov(\hat\gamma_{F}) = (W^T F W)^{-1} W^T F \cov(\hat a_0) F^T W (W^T F W)^{-1}.
\end{equation}
Next, $\cov(\hat a_0)=\e(\cov(\hat a_0|a_0))+\cov (\e(\hat a_0|a_0))$.
Since the $a_{0g}$ are independent, $\cov(\hat a_0 | a_0) = \diag(\var(\hat a_{0g} | a_{0g}))$, and $\cov (\e(\hat a_0 | a_0)) = \diag(Z_g^T \Omega  Z_g + \omega^2)$.

Now the \lq\lq residuals'' $R_F = \hat a_0 - W \hat\gamma_F = (I - W (W^T F W)^{-1} W^T F) \hat a_0 = (I - H_F) \hat a_0$, where $H_F = W (W^T F W)^{-1} W^T F$ is an idempotent matrix.

Similar to the 2nd-level case, we consider a generalized $Q$ statistic
\begin{equation} \label{Q_F}
Q_F = R_F^T F R_F = \hat{a}_0^T (I - H_F^T) F (I - H_F) \hat a_0 = \hat a_0^T F (I - H_F) \hat a_0.
\end{equation}
Given unbiased estimators $v_i^2 = \hat\sigma_i^2$ and a nonrandom $F$ matrix, and
denoting the weight matrix by $P = F (I - H_F)$, $\e(Q_F )= \sum p_{gg} \var(\hat a_{0g}) = \sum p_{gg} (\sum_i b_{gi}^2 (\tau^2 + \sigma_{gi}^2) + Z_g \Omega Z_g^T + \omega^2).$  We have already estimated $\sigma^2_{gi}$ and $\tau^2$, but equating $Q_F$ to $\e(Q_F)$ provides only one equation with $||\Omega|| + 1$ parameters. Only when $\Omega$ is known or zero can we find a moment estimator of $w^2$  {\`a} la \cite{dersimonian2007random}:
\begin{equation} \label{omega_F}
\hat\omega^2 = \max \left( \left(\sum p_{gg} \right)^{-1} \left(Q_F - \sum p_{gg} \left(\sum_i b_{gi}^2 (\hat\tau^2 + v_{gi}^2) + Z_g \Omega Z_g^T \right) \right), 0 \right).
\end{equation}
The distribution of $Q_F$ is that of a quadratic form in normal random variables, and a confidence interval for $\omega^2$ can be obtained from percentage points of an approximation to this distribution under the hypothesis of no heterogeneity.

\section{Simulation design  and results} \label{simul}

Our main objective was to evaluate  our moment-based method of estimation in a 3-level meta-analysis and to compare it with standard methods of hierarchical
meta-analysis ({\it rma.mv} in {\it metafor}) for standardized mean difference as an effect measure.

\subsection{Simulation design}
At the end of Section~\ref{sec:L2} we introduced, for generality, the block-diagonal matrix $A$. In our simulations we considered comparative studies with two arms, C and T. The $A^g$ were diagonal: $A^g = \diag(\tilde{n}_{gi}, i = 1, \ldots, K_g$). In Study $i$, with $n_{gi}^C$ and $n_{gi}^T$ observations in the arms, $n_{gi} = n_{gi}^C + n_{gi}^T$ is the total sample size, and $\tilde{n}_{gi} = n_{gi}^C n_{gi}^T / n_{gi}$ is the within-study effective sample size. Then, with $\tilde{n}_g = \sum_i \tilde{n}_{gi}$, $\hat{a}_{0g} = \sum_i \tilde{n}_{gi} T_{gi} / \tilde{n}_g$ are the sample-size-weighted means (SSW) of the elements of $T_g$. The matrix $F$ in Equation~(\ref{Ghat_F}) was also diagonal: $F = \diag(\tilde{n}_{g}, g=1, \ldots, M)$.

In our simulations we took $p=0$ in Equation~(\ref{3lM_1}), and $q=1$ and $r=0$ in Equation~(\ref{3lM_2a}). Thus, the 3-level model of meta-analysis had one fixed effect (the overall mean, $\gamma$) and two variance components ($\tau^2$ at level 2, and $\omega^2$ at level 3).

To emphasize that we are dealing with meta-analysis of SMD, we change the notation for the effects in what follows. We denote the overall true effect ($\gamma$ previously) by $\delta$ and the cluster-specific means $a_{0g}$ by $\delta_{0g}$.

Our simulations varied five parameters:   the overall true effect $\delta$, the between-studies variance $\tau^2$ and the between-clusters variance ($\omega^2 = \tau^2$ or $\omega^2 = 4 \tau^2$),  the number of clusters ($M$), the (equal) number of studies in each cluster ($K_g \equiv  K$), and the studies' (equal) total sample size ($n_{gi} \equiv n$). The proportion of observations in the control arm was fixed at $1/2$.  Table~\ref{tab:design} lists the values of each parameter.

Given a true effect $\delta$, we generated the $M$ cluster effects $\delta_{0g}$ from a normal distribution with mean $\delta$ and variance $\omega^2$.
Next, we generated the means for the $K$ studies in each cluster by adding shifts distributed as $N(0, \tau^2)$.    
Observed SMD values were then generated from scaled noncentral  $t$-distributions with noncentrality parameters based on the  study means.  
 Finally, Hedges's correction \citep{hedges1983random}  was used to obtain values of Hedges's $g$.

In the main series of simulations (1600 combinations of parameters), we chose $\omega^2 = \tau^2$, which is typical for medical meta-analyses \citep{FernandezCastilla2020}. In the additional series of simulations (160 combinations of parameters),  for selected values of $\tau^2$, $K$ and $M$, we chose $\omega^2 = 4 \tau^2$, typical for behavioral and social sciences \citep{FernandezCastilla2020}. We used \emph{R} statistical software \citep{rrr}.

In addition to the effective-sample-size-based methods for the 3-level model introduced in Section~\ref{sec:3level_estimation}, we included, for comparison, REML and profile-likelihood-based methods \citep{Hardy_1996_StatMed_619} (implemented in the {\it rma.mv} procedure in {\it metafor} \citep{metafor}).

For assessing heterogeneity in a 3-level model, we used the matrices $Q_A$ and $Q_F$.  We used approximations to their distributions (as quadratic forms in approximately normal variables) to obtain empirical level and power for heterogeneity tests based on $Q_A$ and $Q_F$, and confidence intervals for $\tau^2$ and $\omega^2$. For these distributions, we used the R package CompQuadForm \citep{ComQuadForm}. The Farebrother approximation \citep{Farebrother1984} did not work well for the 3-level model because of sparseness of the resulting covariance matrix, so we used the Davies algorithm \citep{Davies1980} instead.

User-friendly R programs implementing our methods are available at osf.io/zgtpn .

Initially, we aimed at a total of $10,000$ repetitions for each combination of parameters. However, {\it rma.mv}, which we used with its default optimizer {\it nlminb}, was rather slow for $M \geq 10$. Therefore, we decided on $1000$ repetitions for all combinations of parameters. Even then, {\it rma.mv} using {\it nlminb} struggled to converge for $M=50$ when $K=20$. For these values, time constraints prevented us from obtaining results beyond $n = 20$.  Appendix A reports results on non-convergence rates of {\it rma.mv}.

We compared the performance of the two classes of methods on
\begin{itemize}
\item Bias in estimation of the overall effect $\delta$;
\item Coverage of the overall effect $\delta$ by 95\% confidence intervals, centered on the particular estimate of $\delta$ and  based on quantiles from the normal distribution and the $t$-distribution with $M - 1$ df;
\item Bias in estimation of the variance components $\tau^2$ and $\omega^2$;
\item  Coverage of the 95\% confidence  intervals for $\tau^2$ and $\omega^2$.
\end{itemize}

We also studied empirical levels of the heterogeneity tests based on $Q_A$ and $Q_F$ using the  Davies approximation \citep{Davies1980}.

\begin{table}[ht]
	\caption{ \label{tab:design} \emph{Values of parameters in the simulations}}
	\begin{footnotesize}
		\begin{center}
			\begin{tabular}
				{|l|l|l|}
				\hline
				Parameter & Values  \\
				\hline
                $M$ (number of clusters) & 5*, 10, 25, 50*\\
				$K$ (number of studies in a cluster) & 2*, 5, 10, 20*  \\
				\hline
				$n$ (study size, total of the two arms) & 20, 40, 100, 200, 1000  \\
								                			\hline
				$f$ (fraction of observations in the control arm) & 1/2   \\
							\hline
				$\delta$ (true value of SMD) & $0$, $0.2$, $0.5$, $1$ \\
				\hline
                $\tau^2$ (variance of random effects at level 2) & 0*, 0.02, 0.1, 0.2*, 0.5* \\
                			\hline
		$\omega^2$ (variance of random effects at level 3) & $\tau^2$ or $4 \tau^2$ \\
				\hline			
			\end{tabular}
		\end{center}
* Only in simulations for $\omega^2 = \tau^2$.
	\end{footnotesize}
\end{table}

\subsection{Summary of simulation results}

\subsubsection{  Non-convergence rates in three-level analysis as implemented in {\it  metafor} (Appendix A)}

Our simulations used the default {\it nlminb} optimizer  for {\it rma.mv}. Here we report on the non-convergence rates, defined as output of zero length.
Additionally,  even when there was some output, the search for confidence limits for either or both of the variance components may not have converged. Typically, the non-convergence rates of those confidence limits were somewhat higher, up to 9.5 percentage points higher for $\tau^2$ and up to 8.3 percentage points higher for $\omega^2$, but we do not report on them.

The non-convergence rates were considerable (7 to 11\%) when the variance components were close to zero, but they decreased and then became negligible further on.  They did not depend on the overall effect $\delta$, but they improved greatly with increases in the number of studies $K$, in the study size $n$, and, to some extent, in the number of clusters $M$ for $M \leq 25$.

They increased dramatically for $M = 50$ and $K = 20$, reaching up to 45\% at $\tau^2 = 0$ when $n = 20$.

 The new moment-based methods encountered no instances of non-convergence.

\subsubsection{ Empirical level of the two $Q$ tests, at $\alpha = .05$, for heterogeneity (Appendix B)}

Empirical levels of $Q_A$  were typically above the nominal $.05$ level (range .039 to .125), and those of $Q_F$ were below the nominal level (range .019 to .058).

The results do not improve uniformly with $n$ but appear rather heterogeneous. This behavior may be due to the relatively small number of repetitions. Typically, empirical levels were closer to nominal for larger values of $M$, $K$, and $n$. The levels seem not to depend on $\delta$.

\subsubsection{ Bias  of point estimators of $\tau^2$ (Appendices C and I)}

In the main series of simulations, the REML and $Q_A$-based estimators of $\tau^2$ were both positively biased at $\tau^2 = 0$, the $Q_A$-based estimator somewhat more. However, the REML estimator had increasingly negative bias for $\tau^2 \geq 0.02$, whereas the $Q_A$-based estimator had a  positive bias at $n=20$ but negligible bias for $n \geq 40$.

The bias of REML was greatest for $n = 20$ and $K \leq 5$, reaching $-0.1$ at $\tau^2 = 0.5$ when $M = 5$ and $K = 2$. Though still increasingly negative, it was negligible for $n \geq 100$ when $K \geq 5$. When $K =  2$, it was still considerable for large $\delta$. As an example, when $\delta = 1, M = 5, K = 2$, and $n = 200$, the bias of the REML estimator was about $-0.03$ at $\tau^2 = 0.5$.

In the additional simulations with $\omega^2 = 4\tau^2$, for $\tau^2=0.02$, the  bias of the  $Q_A$-based estimator was positive and the bias of the REML was negative, both considerably biased for $n=20$.  The bias of the  $Q_A$-based estimator was negligible for $\tau^2 \geq 0.1$ even for $n=20$, whereas REML still had a considerable negative bias.

\subsubsection{ Coverage of interval estimators of $\tau^2$ (Appendices D and I)}

Coverage of interval estimators of $\tau^2$  did not depend much on $\delta$.

In the main series of simulations, when $n = 20$, $M = 5$, and $K = 2$, coverage of the $Q_A$-based intervals was about $97\%$,  and coverage of the PL-based intervals was higher for $\tau^2 \leq 0.2$ and somewhat lower for $\tau^2 = 0.5$.

Coverage of the $Q_A$-based intervals was close to nominal for larger $M$ and $K$ when $n\leq 40$, whereas the PL-based coverage deteriorated to unacceptably low levels. It was only 20\% for $M = 50$ and $K = 20$.

For $n = 100$, coverage of PL improved and typically was above 93\%.  Coverage of $Q_A$ was close to nominal.  For $n \geq 200$, PL coverage was close to nominal, and the $Q_A$-based intervals often had higher coverage, at 96 to 97\%.

In the additional simulations with $\omega^2=4\tau^2$, when $n=20$, coverage of the PL-based intervals was too low for $\tau^2 = 0.02$, $K = 10$, $M = 25$ and for $\tau^2 = 0.1$ and all combinations of  $K = 5,\;10$ and $M = 10,\;25$, staying between 75 and 90\%. Coverage of the $Q_A$-based intervals was close to nominal.

\subsubsection{ Bias of point estimators of $\omega^2$ (Appendices E and I)}

In the main series of simulations, both estimators of $\omega^2$ were somewhat positively biased at zero, the $Q_F$-based estimator somewhat more. For larger $\omega^2$, the REML estimator was typically negatively biased, whereas the $Q_F$-based estimator was somewhat positively biased; both biases increased with $\omega^2$.
When $n = 20$, $M = 5$, and $K = 2$, the maximum bias was 0.02 for REML and up to 0.065 for $Q_F$. For larger $M$ and $K$, $Q_F$ was almost unbiased, and the bias of REML was at worst $-0.06$.

For $n \geq 40$, the pattern was similar, though the bias of REML decreased with $n$. For $n = 100$, the bias of  REML was very small, at worst $-0.015$, and the two estimators did not differ at $n = 1000$.

In the additional simulations with $\omega^2=4\tau^2$, the REML estimator had a considerable negative bias for $n=20$ and $40$. The bias was about 10\% for $n = 20$. ITS bias was negligible for the $Q_F$-based estimator

\subsubsection{ Coverage of interval estimators of $\omega^2$ (Appendices F and I)}

In the main series of simulations, when $n = 20$ and $K = 2$, $Q_F$-based intervals had lower than nominal coverage. Coverage of PL was too high at $\omega^2 = 0$, but close to nominal otherwise.
For $K \geq 5$, coverage of the $Q_F$-based intervals was somewhat high at $\omega^2 = 0$ and close to nominal for $\omega^2 \geq 0.02$. Coverage of PL was higher at $\omega^2 = 0$ and close to nominal for $K \leq 10$, but too low for larger $M$ and $K$; for example, it was below 92\% for $M = 50$ and $K \geq 10$.

For  $n = 40$, the patterns were similar, but coverage of PL was sometimes lower, at 94\%.

For $n \geq 100$, $Q_F$ coverage was too low for $K = 2$ but was similar to and often better than PL for $K = 5$ or $K = 10$. PL coverage was close to nominal for $K = 20$.

The value of $\delta$ did not seem to affect coverage.
In the additional simulations with $\omega^2=4\tau^2$, PL tended to have lower than nominal coverage for $n=20$ and $40$. The coverage of PL and  the $Q_F$-based intervals was rather similar for $n \geq 100$.

\subsubsection{ Bias of point estimators of $\delta$  (Appendices G and I)}

The SSW estimator of $\delta$ was practically unbiased in all scenarios.  The quality of the IV-based estimator was practically the same for $\delta = 0$.

For larger values of $\delta$, the negative bias of the IV estimator increased noticeably, reaching about $-0.07$ for $\delta = 1$ and $n = 20$, regardless of $M$ and $K$.  This was due to increasing correlations between the estimates of $\delta$ and their IV weights.  Typically, for each  $\delta > 0$, a constant difference separated the two estimators for all $M$ and $K$. Differences between the two estimators decreased with $n$, almost disappearing by $n = 200$.

\subsubsection{ Coverage of interval estimators of $\delta$  (Appendices H and I)}

The source of critical values had a large impact. Except when $\tau^2 = 0$, intervals that used normal critical values had coverage that was too low. The situation was especially acute for $M = 5$, where coverage was often below 90\%. In contrast, intervals that used $t$ critical values had coverage close to 95\% (but somewhat higher for $K = 2$). The disparity was smaller when $M = 10$, and it generally shrank as $M$ increased.


The IV-based intervals showed considerable undercoverage for $K \geq 10$, $M \geq 25$, and larger $\delta$. Larger $n$ was required to ameliorate this bias for larger $\delta$ in the vicinity of $\omega^2 = 0$.

For $\delta = 0.2$, $n = 20$ was problematic for $M = 50$, but coverage improved for larger $n$. When $\delta = 0.5$, combinations of $n \leq 40$ with $M = 50$ were of concern. For $\delta = 1$, values of $n$ up to 200 produced undercoverage near $\omega^2 = 0$.

In the additional simulations with $\omega^2 = 4 \tau^2$, the results were similar. The IV-based intervals had too low coverage for $\delta \geq 0.5$ and $n \leq 40$ for $M = 10$ and $25$ when $\tau^2 = 0.02$, and for $M = 25$ when $\tau^2 = 0.1$.

\section{Discussion}\label{sec:Disc}
Multilevel meta-analysis is widely used in various applications. A need for three-level models arises especially often in social sciences and ecology. Examples include labs that each have several studies and studies that reported (mostly) the same set of outcome measures.

We propose a moment estimation approach similar to that in ANOVA, separating the total variation into within and between sums of squares. Importantly, we use fixed, effective-sample-size-based weights. The use of fixed weights allows separate estimation of meta-effects and variance components. This approach avoids convergence issues that require the standard methods to use iterative simultaneous estimation.

The model that we considered readily generalizes to handle cluster-specific between-study variances $\tau_g^2$. For such a model, we would require separate $Q_{A(g)}$ statistics to estimate the $\tau_g^2$.

One can also consider more-complicated variance structures with $\Omega \not = 0$. Then, similar to \cite{Jackson2024}, we would need to estimate the covariance matrix from the residuals $R_F$ (Section~\ref{sec:L3}), instead of using $Q_F$.

For standardized mean difference as the effect measure, our simulation study compared our novel  moment-based method of estimation in a three-level meta-analysis with the standard REML/PL-based methods of
meta-analysis implemented in {\it rma.mv} in {\it metafor}.
Overall, the new methods performed similarly to or better than the standard methods, and we recommend their use in practice.

An important observation is the unacceptably low coverage of the overall effect, regardless of the method used, by the standard confidence intervals based on normal critical values.  A forceful warning about this shortcoming should be prominent in any meta-analysis software in which the $t$-based intervals are not the default.


\clearpage
\bibliography{hierarchical1.bib}

\begin{thebibliography}{14}
\providecommand{\natexlab}[1]{#1}
\providecommand{\url}[1]{\texttt{#1}}
\expandafter\ifx\csname urlstyle\endcsname\relax
  \providecommand{\doi}[1]{doi: #1}\else
  \providecommand{\doi}{doi: \begingroup \urlstyle{rm}\Url}\fi

\bibitem[Cheung(2014)]{Cheung2014}
M.~W.-L. Cheung.
\newblock Modeling dependent effect sizes with three-level meta-analyses: A
  structural equation modeling approach.
\newblock \emph{Psychological Methods}, 19:\penalty0 211--229, 2014.
\newblock \doi{10.1037/a0032968}.

\bibitem[Davies(1980)]{Davies1980}
R.~B. Davies.
\newblock Algorithm {AS} 155: The distribution of a linear combination of
  $\chi^2$ random variables.
\newblock \emph{Journal of the Royal Statistical Society Series C (Applied
  Statistics)}, 29:\penalty0 323--333, 1980.
\newblock \doi{10.2307/2346911}.

\bibitem[DerSimonian and Kacker(2007)]{dersimonian2007random}
Rebecca DerSimonian and Raghu Kacker.
\newblock Random-effects model for meta-analysis of clinical trials: an update.
\newblock \emph{Contemporary Clinical Trials}, 28\penalty0 (2):\penalty0
  105--114, 2007.
\newblock \doi{10.1016/j.cct.2006.04.004}.

\bibitem[Duchesne and {de Micheaux}(2010)]{ComQuadForm}
P.~Duchesne and P.~Lafaye {de Micheaux}.
\newblock Computing the distribution of quadratic forms: Further comparisons
  between the {L}iu-{T}ang-{Z}hang approximation and exact methods.
\newblock \emph{Computational Statistics and Data Analysis}, 54:\penalty0
  858--862, 2010.
\newblock \doi{10.1016/j.csda.2009.11.025}.

\bibitem[Farebrother(1984)]{Farebrother1984}
R.~W. Farebrother.
\newblock Algorithm {AS} 204: The distribution of a positive linear combination
  of $\chi^2$ random variables.
\newblock \emph{Journal of the Royal Statistical Society Series C (Applied
  Statistics)}, 33:\penalty0 332--339, 1984.
\newblock \doi{10.2307/2347721}.

\bibitem[Fernández-Castilla et~al.(2020)Fernández-Castilla, Jamshidi,
  Declercq, Beretvas, Onghena, and den Noortgate]{FernandezCastilla2020}
Bel\'{e}n Fernández-Castilla, Laleh Jamshidi, Lies Declercq, S.~Natasha
  Beretvas, Patrick Onghena, and Wim~Van den Noortgate.
\newblock The application of meta-analytic (multi-level) models with multiple
  random effects: A systematic review.
\newblock \emph{Behavior Research Methods}, 52:\penalty0 2031–2052, 2020.
\newblock \doi{10.3758/s13428-020-01373-9}.

\bibitem[Goldstein et~al.(2000)Goldstein, Yang, Omar, Turner, and
  Thompson]{Goldstein2000}
Harvey Goldstein, Min Yang, Rumana Omar, Rebecca Turner, and Simon Thompson.
\newblock Meta-analysis using multilevel models with an application to the
  study of class size effects.
\newblock \emph{Journal of the Royal Statistical Society Series C (Applied
  Statistics)}, 49:\penalty0 399--412, 2000.
\newblock \doi{10.1111/1467-9876.00200}.

\bibitem[Hardy and Thompson(1996)]{Hardy_1996_StatMed_619}
Rebecca~J. Hardy and Simon~G. Thompson.
\newblock A likelihood approach to meta-analysis with random effects.
\newblock \emph{Statistics in Medicine}, 15\penalty0 (6):\penalty0 619--629,
  1996.
\newblock
  \doi{10.1002/(SICI)1097-0258(19960330)15:6<619::AID-SIM188>3.0.CO;2-A}.

\bibitem[Hedges(1983)]{hedges1983random}
Larry~V. Hedges.
\newblock A random effects model for effect sizes.
\newblock \emph{Psychological Bulletin}, 93\penalty0 (2):\penalty0 388--395,
  1983.
\newblock \doi{10.1037/0033-2909.93.2.388}.

\bibitem[Jackson et~al.(2024)Jackson, Viechtbauer, and van Aert]{Jackson2024}
Dan Jackson, Wolfgang Viechtbauer, and Robbie C.~M. van Aert.
\newblock Multistep estimators of the between-study covariance matrix under the
  multivariate random-effects model for meta-analysis.
\newblock \emph{Statistics in Medicine}, 43\penalty0 (4):\penalty0 756--773,
  2024.
\newblock \doi{10.1002/sim.9985}.

\bibitem[Konstantopoulos(2011)]{Konstantopoulos2011}
Spyros Konstantopoulos.
\newblock Fixed effects and variance components estimation in three-level
  meta-analysis.
\newblock \emph{Research Synthesis Methods}, 2\penalty0 (1):\penalty0 61--76,
  2011.
\newblock \doi{10.1002/jrsm.35}.

\bibitem[{R Core Team}(2016)]{rrr}
{R Core Team}.
\newblock \emph{R: A Language and Environment for Statistical Computing}.
\newblock R Foundation for Statistical Computing, Vienna, Austria, 2016.
\newblock URL \url{https://www.R-project.org/}.

\bibitem[Raudenbush and Bryk(1985)]{Raudenbush1985}
S.~W. Raudenbush and A.~S. Bryk.
\newblock Empirical {B}ayes meta-analysis.
\newblock \emph{Journal of Educational Statistics}, 10\penalty0 (2):\penalty0
  75--98, 1985.
\newblock \doi{10.2307/1164836}.

\bibitem[Viechtbauer(2010)]{metafor}
Wolfgang Viechtbauer.
\newblock Conducting meta-analyses in {R} with the {metafor} package.
\newblock \emph{Journal of Statistical Software}, 36:\penalty0 3, 2010.
\newblock \doi{10.18637/jss.v036.i03}.
\newblock Website: https://www.metafor-project.org.

\end{thebibliography}
\clearpage


\section*{Appendices}
\begin{itemize}
\item Appendix A: Rate of non-convergence for the default {\it rma.mv} REML-based analysis
\item Appendix B: Empirical level of the two $Q$ tests, at $\alpha = .05$, for heterogeneity of SMD
\item Appendix C: Bias in point estimators of the between-study variance $\tau^2$
\item Appendix D: Coverage of 95\% confidence intervals for the between-study variance $\tau^2$
\item Appendix E: Bias in point estimators of the between-cluster variance $\omega^2$
\item Appendix F: Coverage of 95\% confidence intervals for the between-cluster variance $\omega^2$
\item Appendix G: Bias in point estimators of the overall effect $\delta$
\item Appendix H: Coverage of 95\% confidence intervals for the overall effect $\delta$
\item Appendix I: Additional simulations: bias in point estimators of  $\tau^2$, $\omega^2$ and $\delta$ and coverage of their 95\% confidence intervals when $\omega^2=4\tau^2$.
\end{itemize}

\setcounter{figure}{0}
\setcounter{section}{0}
\clearpage

\section*{Appendix A: Rate of non-convergence for the default {\it rma.mv} REML-based analysis}

Each figure corresponds to a value of the number of clusters ($M$ = 5, 10, 25, 50).  \\
For each combination of the number of studies in a cluster ($K$ = 2, 5, 10, 20) and the standardized mean difference ($\delta$ = 0, 0.2, 0.5, 1), a panel plots (versus $\tau^2$ = 0, 0.02, 0.1, 0.2, 0.5) the rate of non-convergence for the {\it rma.mv}-based analysis. \\
The traces in the plot correspond to $n = 20$, $n = 40$, and $n =100$.

\clearpage

\renewcommand{\thefigure}{A.\arabic{figure}}
\begin{figure}[ht]
	\centering
	\includegraphics[scale=0.33]{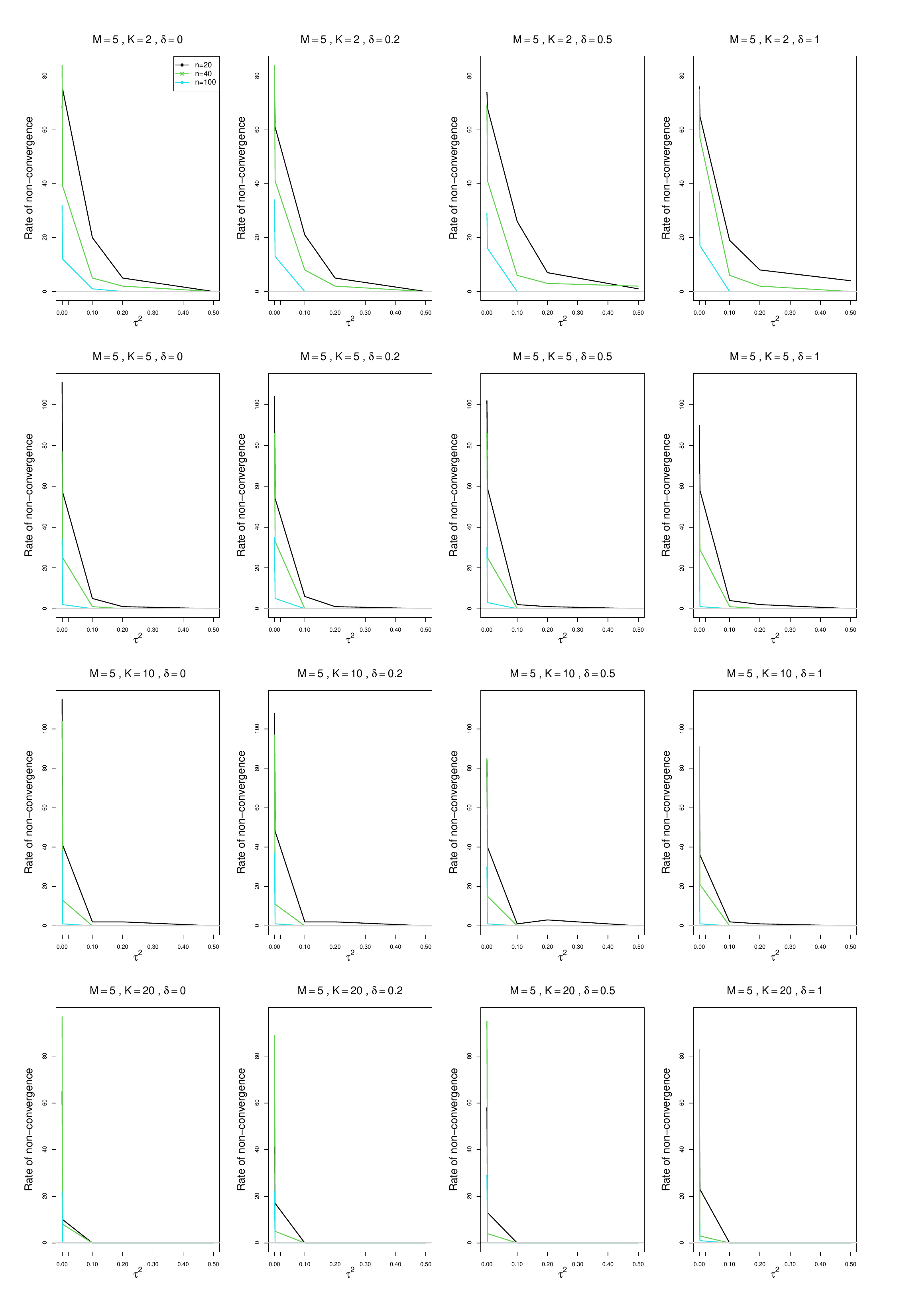}
	\caption{The rate of non-convergence (in units of .001), vs $\tau^2$, for the  {\it rma.mv}-based analysis for $M = 5$ clusters of $K$ studies ($K$ = 2, 5, 10, 20)  and $n$ = 20, 40 and 100, $\delta = 0, 0.2, 0.5$ and $1$.   }
		\label{PlotCountOfMiss_M__5_HIER.pdf}
\end{figure}

\begin{figure}[ht]
	\centering
	\includegraphics[scale=0.33]{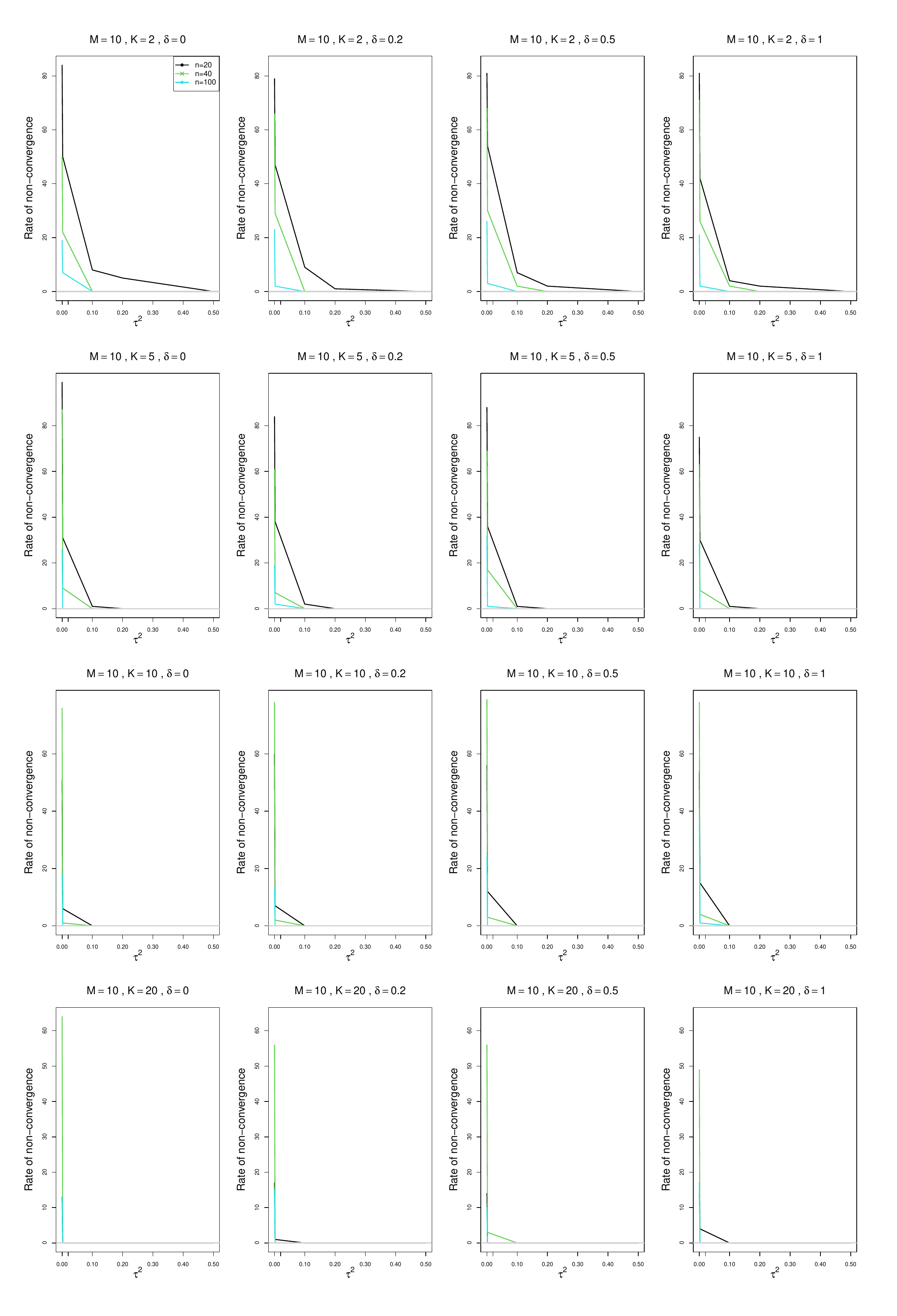}
	\caption{The rate of non-convergence (in units of .001), vs $\tau^2$, for the  {\it rma.mv}-based analysis for $M = 10$ clusters of $K$ studies ($K$ = 2, 5, 10, 20)  and $n$ = 20, 40 and 100, $\delta = 0, 0.2, 0.5$ and $1$.  }
		\label{PlotCountOfMiss_M__10_HIER.pdf}
\end{figure}

\begin{figure}[ht]
	\centering
	\includegraphics[scale=0.33]{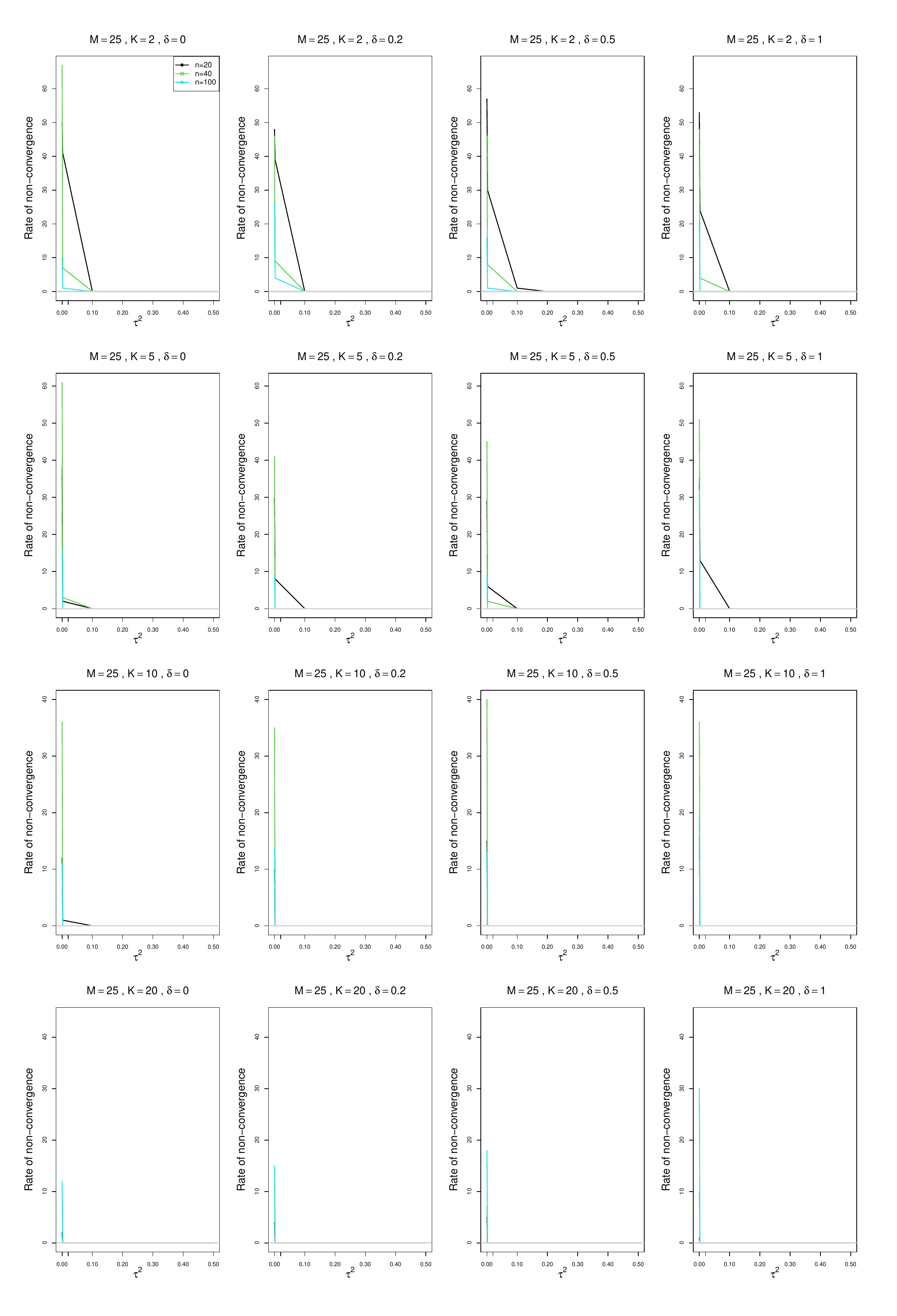}
	\caption{The rate of non-convergence (in units of .001), vs $\tau^2$, for the  {\it rma.mv}-based analysis for $M = 25$ clusters of $K$ studies ($K$ = 2, 5, 10, 20)  and $n$ = 20, 40 and 100, $\delta = 0, 0.2, 0.5$ and $1$.  }
		\label{PlotCountOfMiss_M__25_HIER.pdf}
\end{figure}

\begin{figure}[ht]
	\centering
	\includegraphics[scale=0.33]{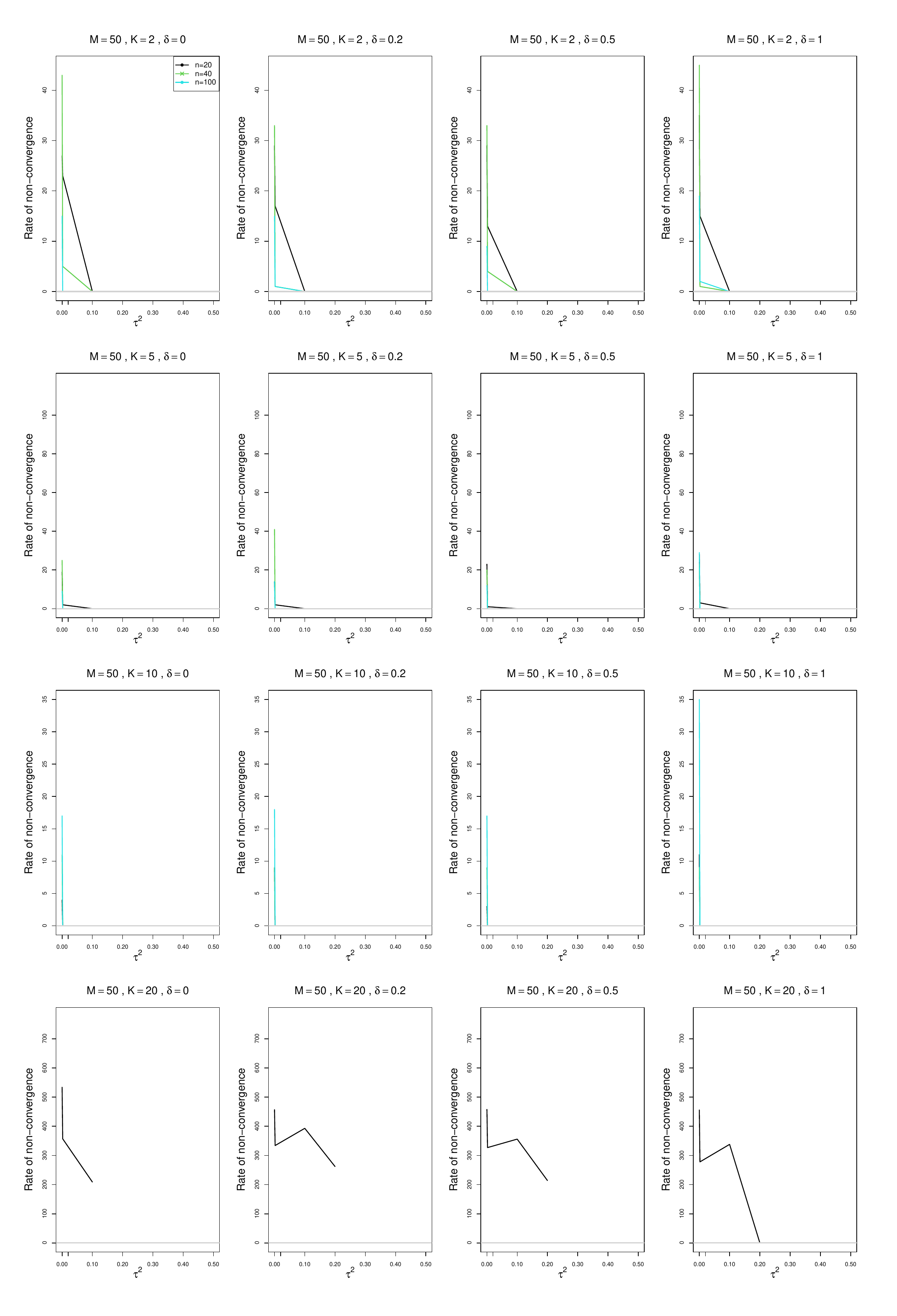}
	\caption{The rate of non-convergence (in units of .001), vs $\tau^2$, for the  {\it rma.mv}-based analysis for $M = 50$ clusters of $K$ studies ($K$ = 2, 5, 10, 20)  and $n$ = 20, 40 and 100, $\delta = 0, 0.2, 0.5$ and $1$. }
		\label{PlotCountOfMiss_M__50_HIER.pdf}
\end{figure}

\clearpage
\section*{Appendix B: Empirical level of the two $Q$ tests, at $\alpha = .05$, for heterogeneity of SMD }

Each figure corresponds to a value of the standardized mean difference ($\delta$ = 0, 0.2, 0.5, 1).

For each combination of the number of studies in a cluster ($K$ = 2, 5, 10, 20) and the number of clusters ($M$ = 5, 10, 25, 50), a panel plots the empirical level of the tests (at the .05 level) versus $n$ (= 20, 40, 100, 200, 1000) for Davies's  approximations to null distributions of $Q_A$ and $Q_F$, which use effective-sample-size weights. \\

\clearpage
\setcounter{figure}{0}
\setcounter{section}{0}
\renewcommand{\thefigure}{B.\arabic{figure}}

\begin{figure}[ht]
	\centering
	\includegraphics[scale=0.33]{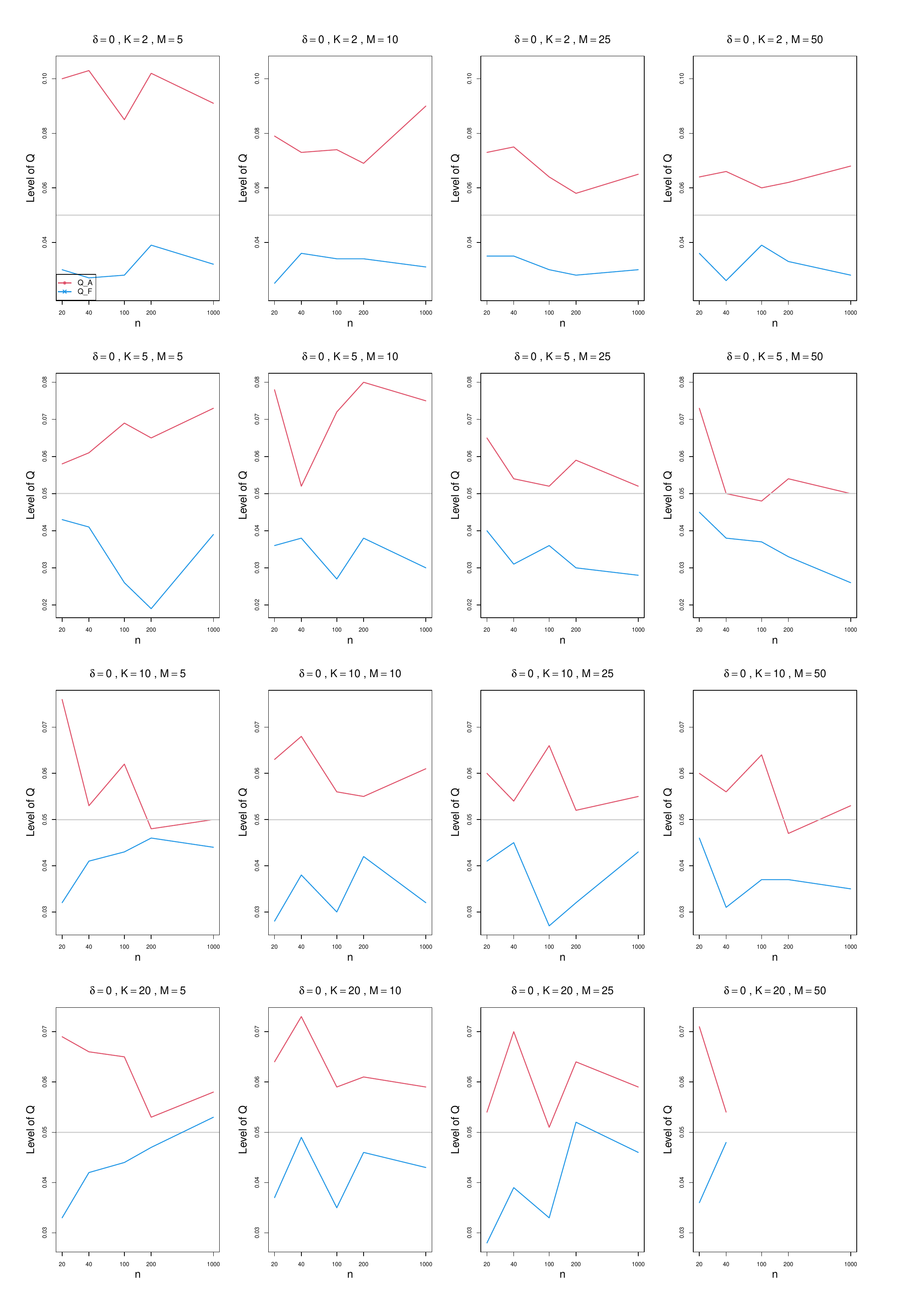}
	\caption{Empirical level (at the .05 level), vs  sample size $n$, for $Q$-statistics-based tests of heterogeneity of SMD: $M$ (= 5, 10, 25, 50) clusters having an equal number of studies $K$ (= 2, 5, 10, 20), $\delta = 0$. }
		\label{PlotLevelsOfQ_0_HIER.pdf}
\end{figure}

\begin{figure}[ht]
	\centering
	\includegraphics[scale=0.33]{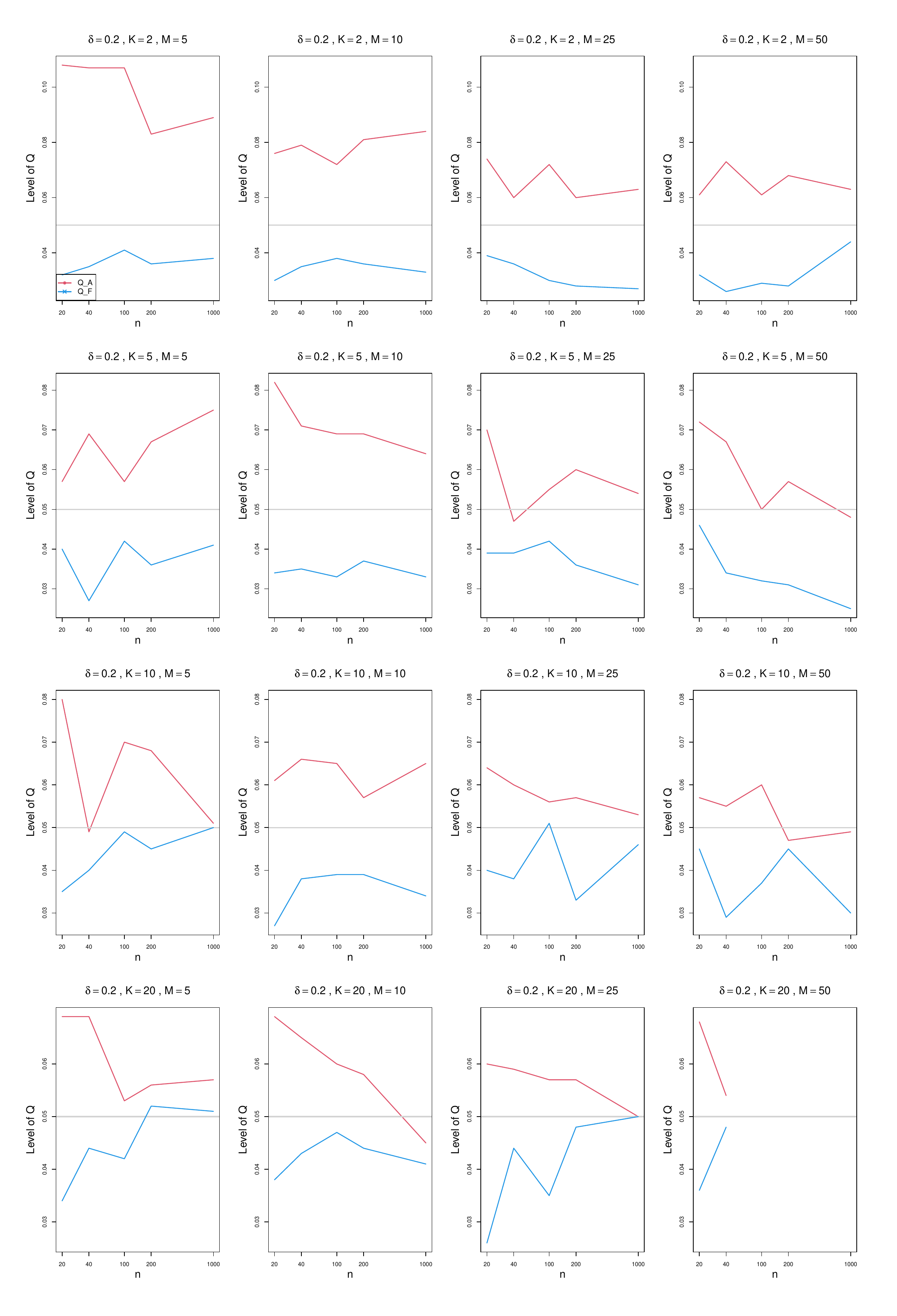}
	\caption{Empirical level (at the .05 level), vs  sample size $n$, for $Q$-statistics-based tests of heterogeneity of SMD: $M$ (= 5, 10, 25, 50) clusters having an equal number of studies $K$ (= 2, 5, 10, 20), $\delta = 0.2$. }
		\label{PlotLevelsOfQ_02_HIER.pdf}
\end{figure}

\begin{figure}[ht]
	\centering
	\includegraphics[scale=0.33]{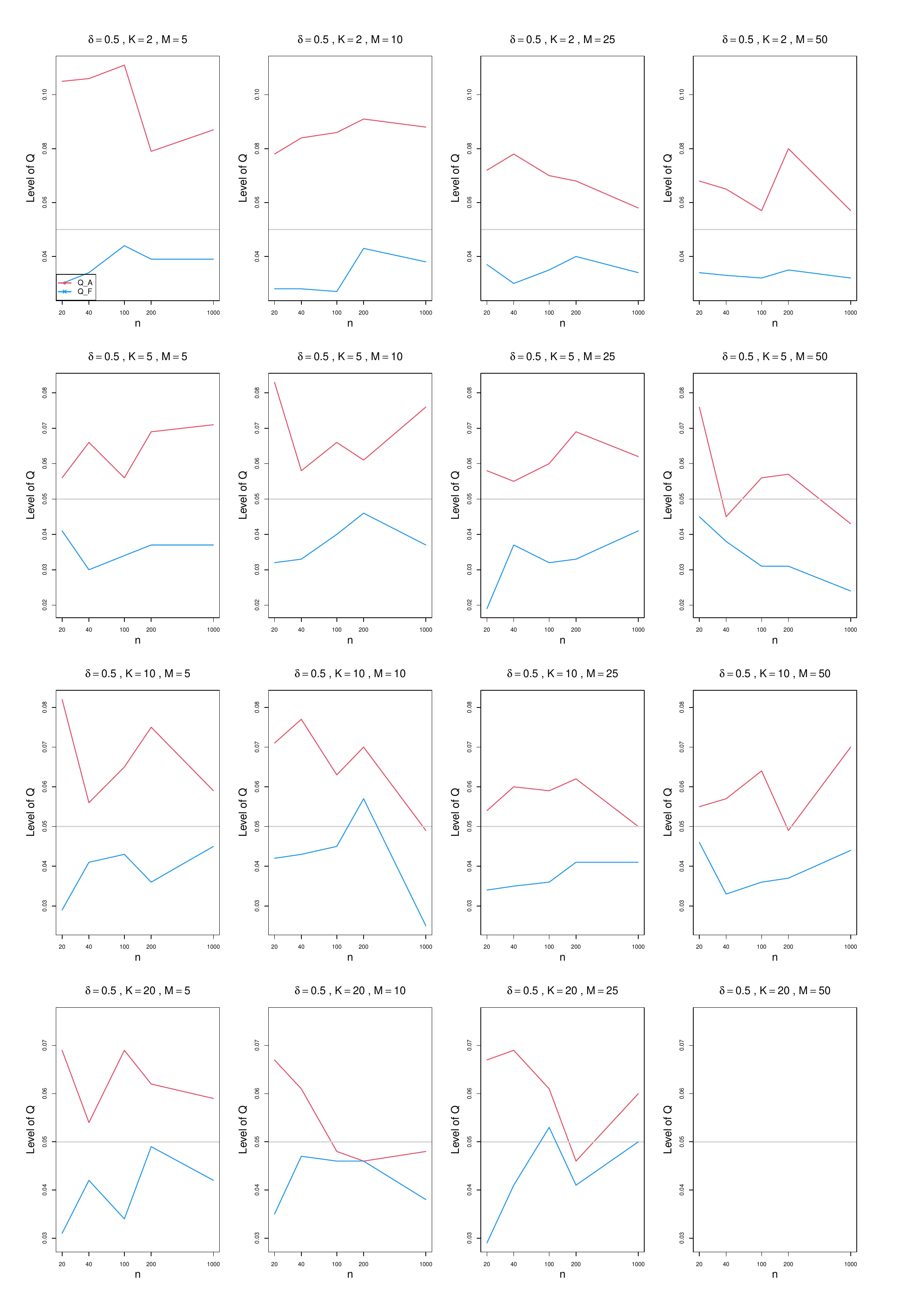}
	\caption{Empirical level (at the .05 level), vs  sample size $n$, for $Q$-statistics-based tests of heterogeneity of SMD: $M$ (= 5, 10, 25, 50) clusters having an equal number of studies $K$ (= 2, 5, 10, 20), $\delta = 0.5$. }
		\label{PlotLevelsOfQ_05_HIER.pdf}
\end{figure}

\begin{figure}[ht]
	\centering
	\includegraphics[scale=0.33]{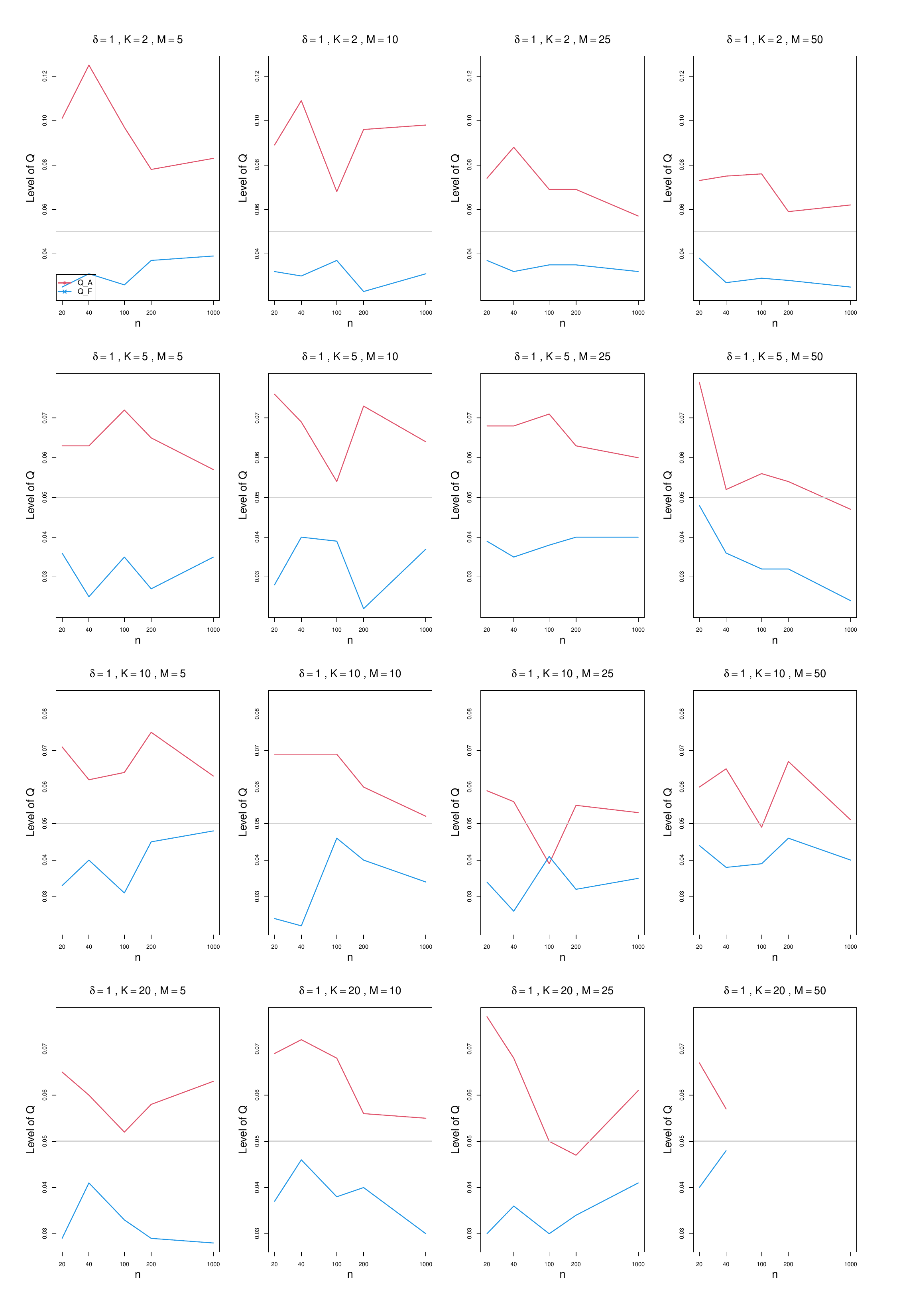}
	\caption{Empirical level (at the .05 level), vs  sample size $n$, for $Q$-statistics-based tests of heterogeneity of SMD: $M$ (= 5, 10, 25, 50) clusters having an equal number of studies $K$ (= 2, 5, 10, 20), $\delta = 1$. }
		\label{PlotLevelsOfQ_1_HIER.pdf}
\end{figure}


\setcounter{figure}{0}
\setcounter{section}{0}
\clearpage

\section*{Appendix C: Bias in point estimators of the between-study variance $\tau^2$}

Each figure corresponds to a value of the standardized mean difference ($\delta$  = 0, 0.2, 0.5, 1) and a value of the study sample size ($n$ = 20, 40, 100, 200, 1000).\\
For each combination of the number of studies in a cluster ($K$ = 2, 5, 10, 20) and the number of clusters ($M$ = 5, 10, 25, 50), a panel plots bias versus $\tau^2$ (= 0, 0.02, 0.1, 0.2, 0.5).\\
The two variance components are held equal ($\tau^2 = \omega^2$).\\
The point estimators of $\tau^2$ are
\begin{itemize}
\item REML method, inverse-variance weights, {\it  rma.mv} in {\it metafor}
\item $Q_A$ (conditional moment-based method, effective-sample-size weights)
\end{itemize}

\clearpage
\renewcommand{\thefigure}{C.\arabic{figure}}


\begin{figure}[ht]
	\centering
	\includegraphics[scale=0.33]{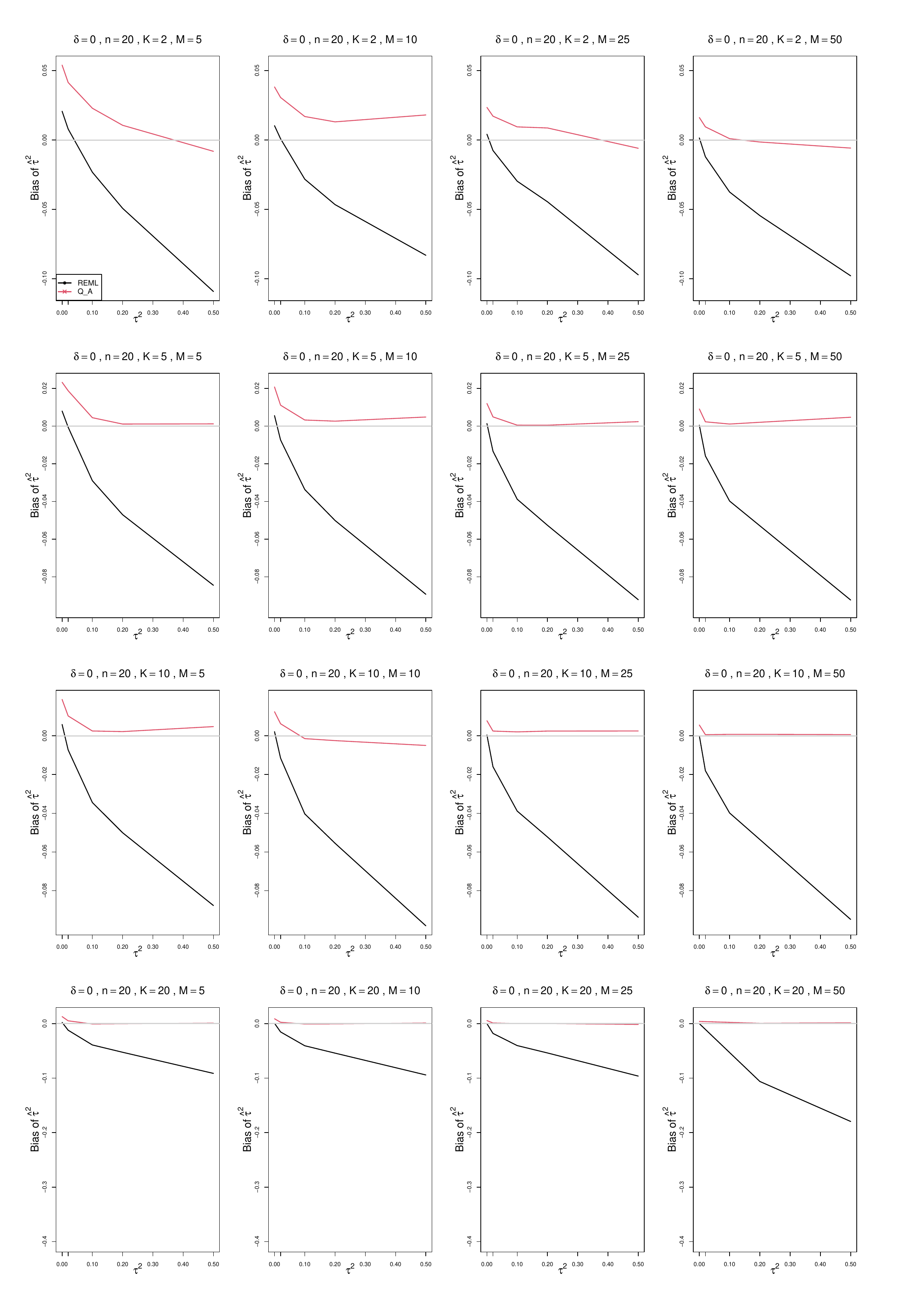}
	\caption{Bias of estimators of between-study variance of SMD (REML and $Q_A$) vs $\tau^2$, for $K$ = 2, 5, 10, and 20 studies per cluster and $M$ = 5, 10, 25, and 50 clusters; $\delta = 0$, and the sample size $n$ = 20 in each study.   }
	\label{PlotBiasOfTau2_20_0_HIER.pdf}
\end{figure}

\begin{figure}[ht]
	\centering
	\includegraphics[scale=0.33]{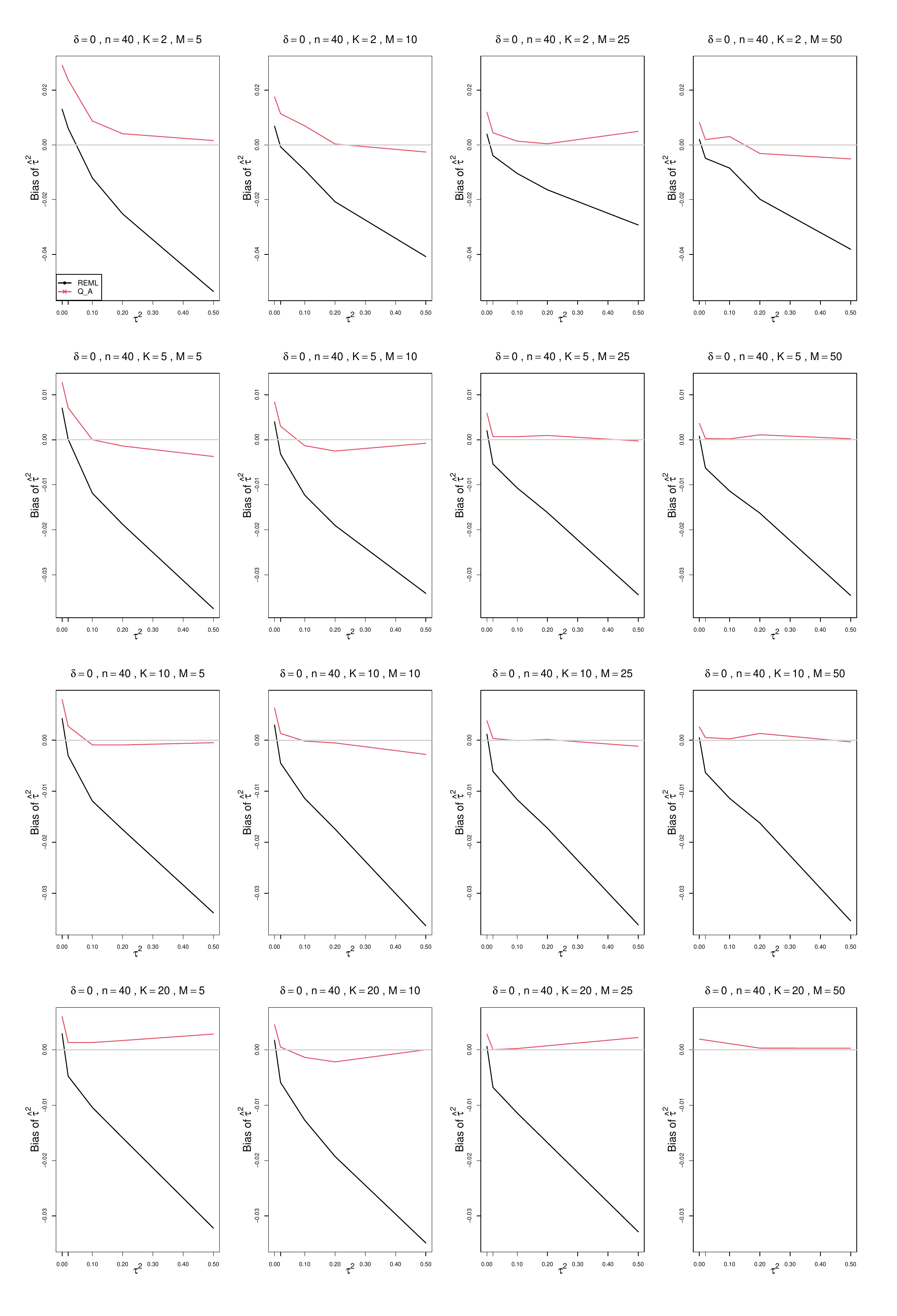}
	\caption{ Bias of estimators of between-study variance of SMD (REML and $Q_A$) vs $\tau^2$, for $K$ = 2, 5, 10, and 20 studies per cluster and $M$ = 5, 10, 25, and 50 clusters; $\delta = 0$, and the sample size $n$ = 40 in each study.  }
	\label{PlotBiasOfTau2_40_0_HIER.pdf}
\end{figure}

\begin{figure}[ht]
	\centering
	\includegraphics[scale=0.33]{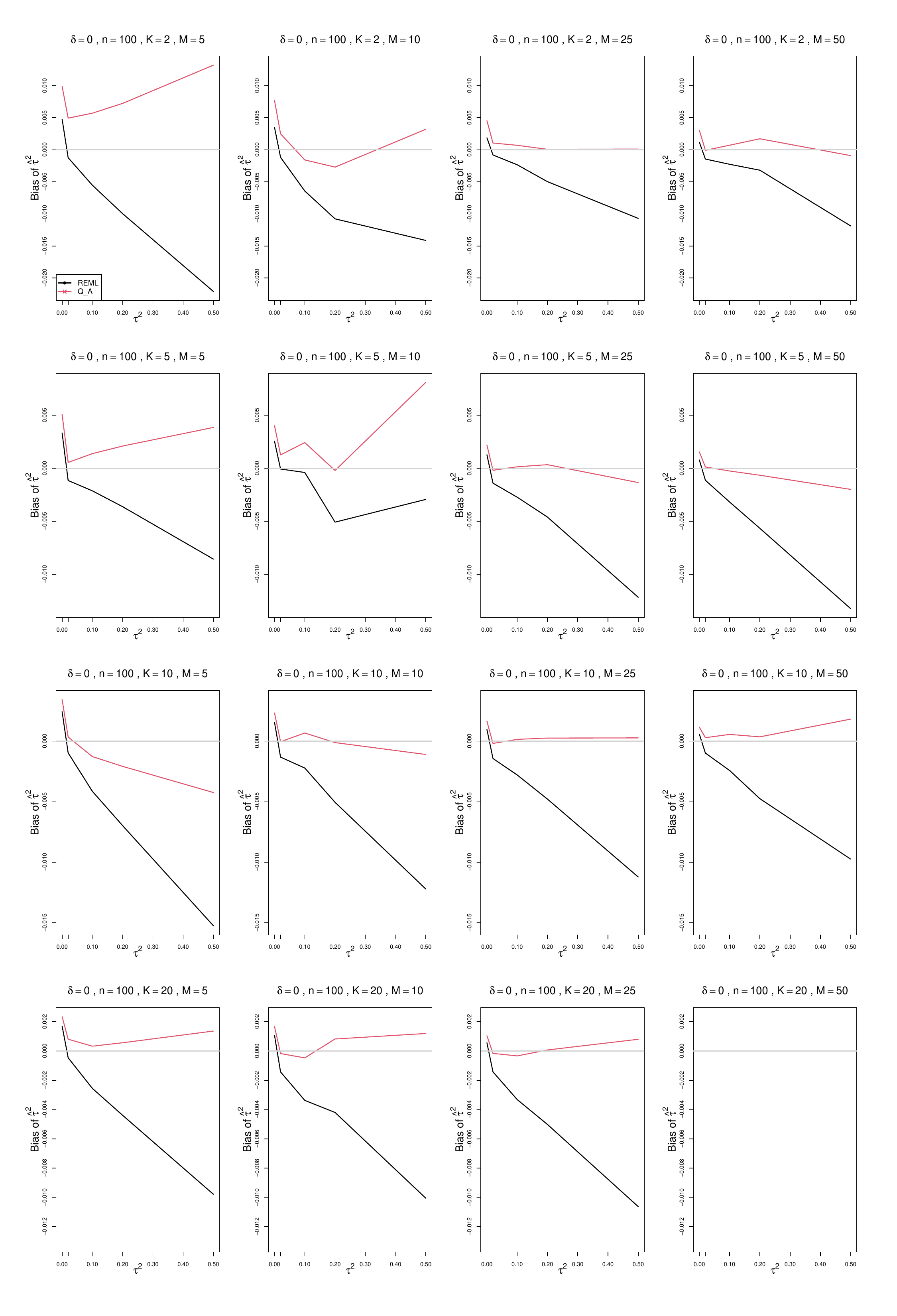}
	\caption{ Bias of estimators of between-study variance of SMD (REML and $Q_A$) vs $\tau^2$, for $K$ = 2, 5, 10, and 20 studies per cluster and $M$ = 5, 10, 25, and 50 clusters; $\delta = 0$, and the sample size $n$ = 100 in each study.   }
	\label{PlotBiasOfTau2_100_0_HIER.pdf}
\end{figure}

\begin{figure}[ht]
	\centering
	\includegraphics[scale=0.33]{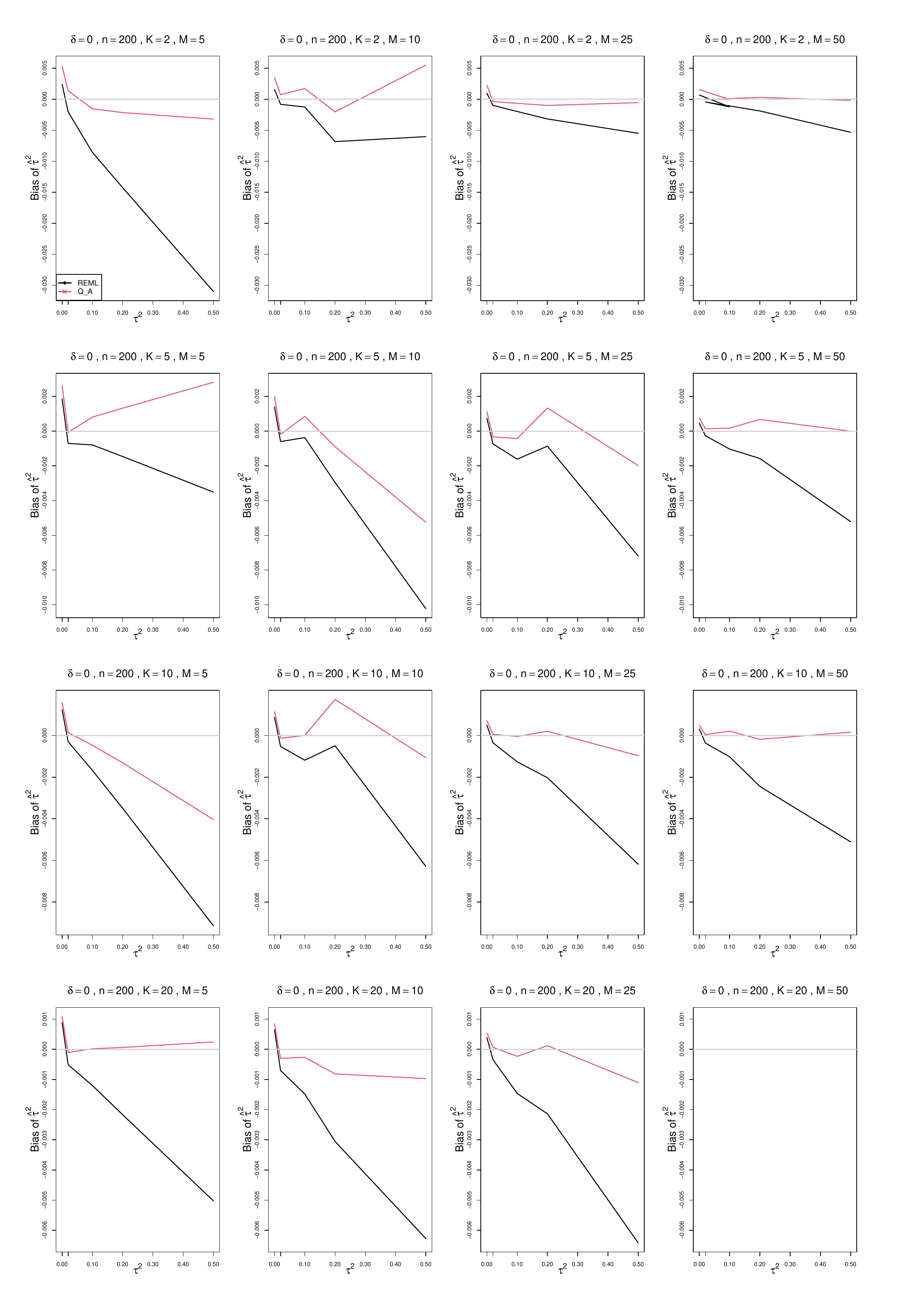}
	\caption{ Bias of estimators of between-study variance of SMD (REML and $Q_A$) vs $\tau^2$, for $K$ = 2, 5, 10, and 20 studies per cluster and $M$ = 5, 10, 25, and 50 clusters; $\delta = 0$, and the sample size $n$ = 200 in each study.  }
	\label{PlotBiasOfTau2_200_0_HIER.pdf}
\end{figure}

\begin{figure}[ht]
	\centering
	\includegraphics[scale=0.33]{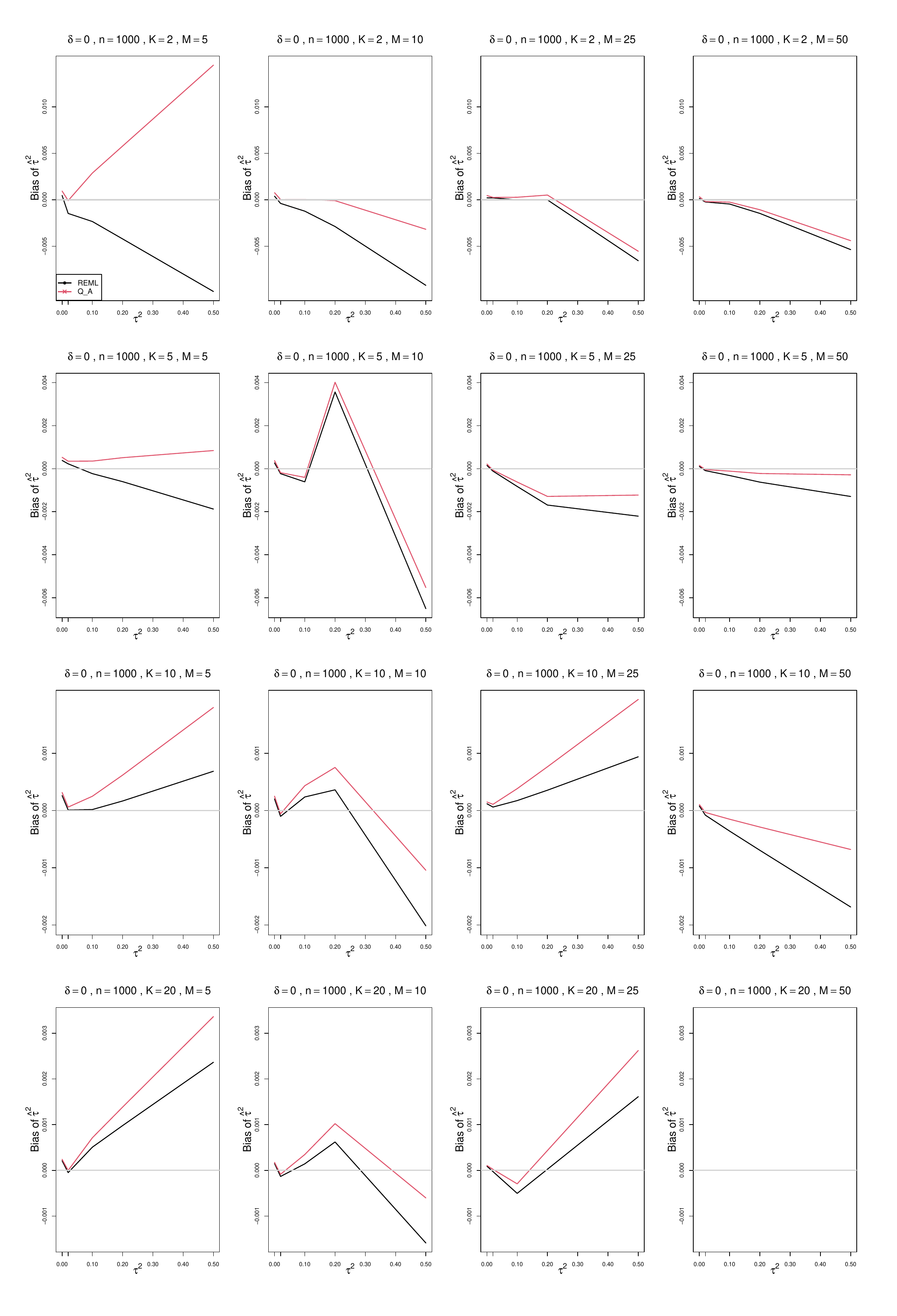}
	\caption{ Bias of estimators of between-study variance of SMD (REML and $Q_A$) vs $\tau^2$, for $K$ = 2, 5, 10, and 20 studies per cluster and $M$ = 5, 10, 25, and 50 clusters; $\delta = 0$, and the sample size $n$ = 1000 in each study.   }
	\label{PlotBiasOfTau2_1000_0_HIER.pdf}
\end{figure}

\begin{figure}[ht]
	\centering
	\includegraphics[scale=0.33]{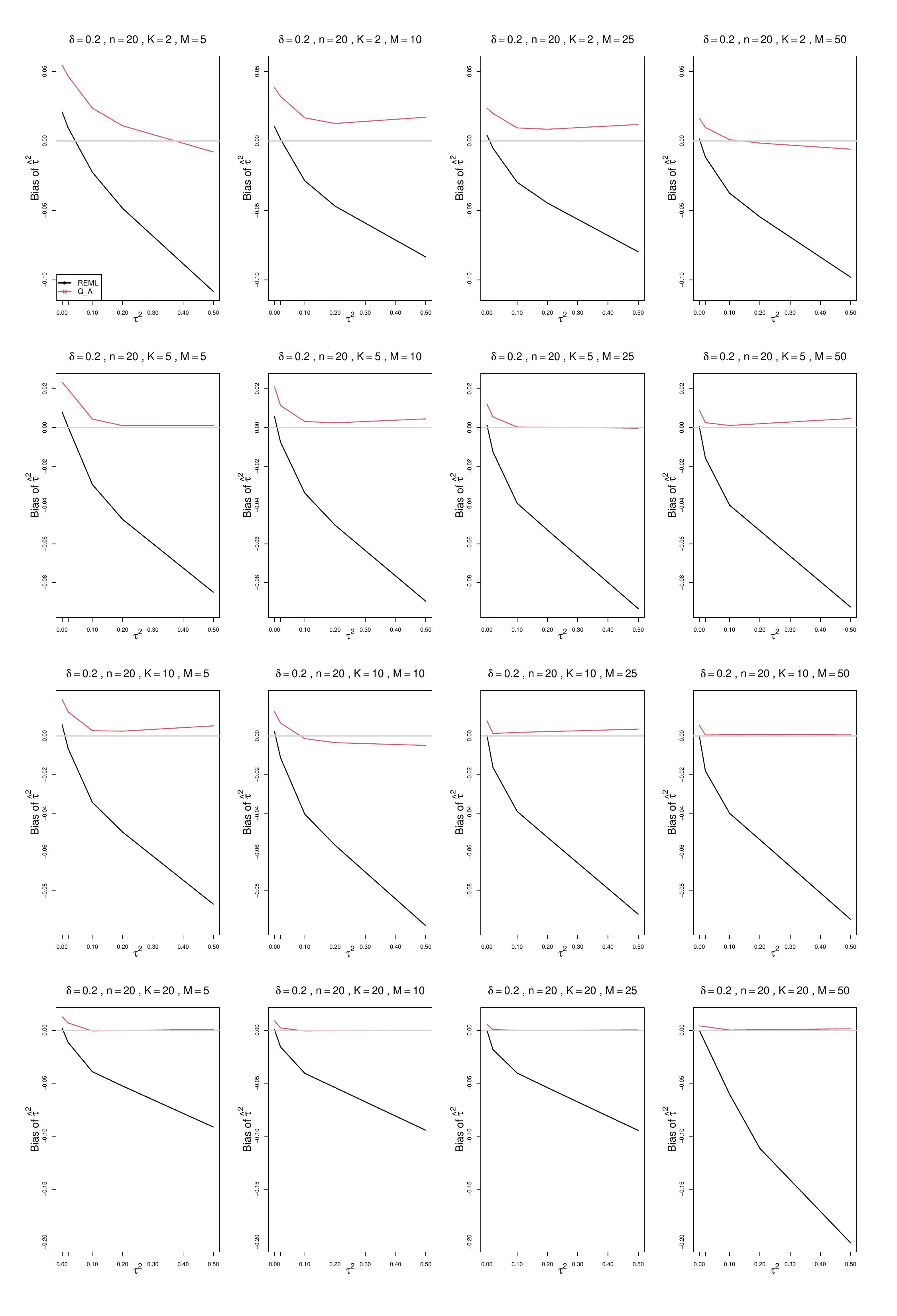}
	\caption{ Bias of estimators of between-study variance of SMD (REML and $Q_A$) vs $\tau^2$, for $K$ = 2, 5, 10, and 20 studies per cluster and $M$ = 5, 10, 25, and 50 clusters; $\delta = 0.2$, and the sample size $n$ = 20 in each study.   }
	\label{PlotBiasOfTau2_20_02_HIER.pdf}
\end{figure}

\begin{figure}[ht]
	\centering
	\includegraphics[scale=0.33]{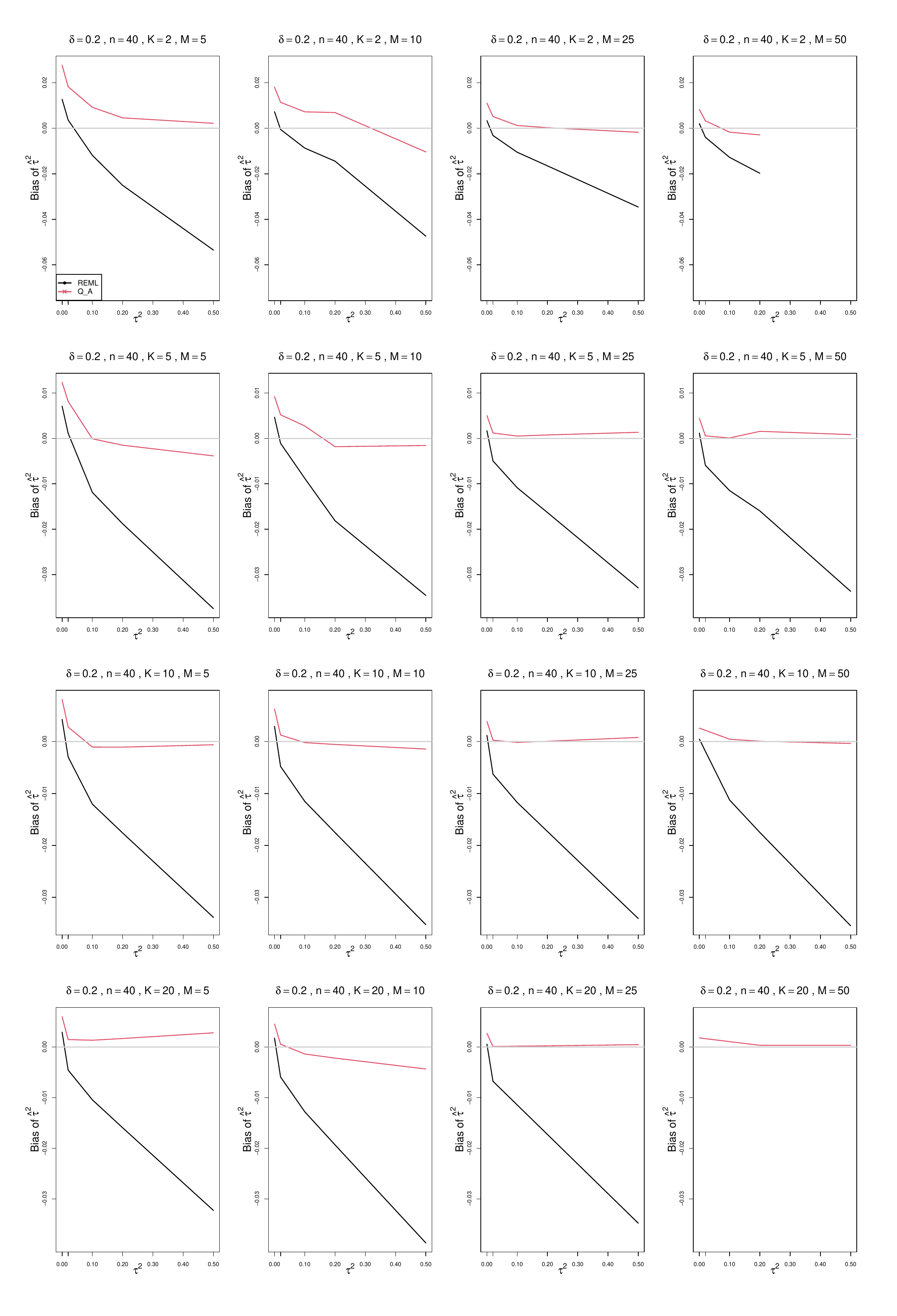}
	\caption{ Bias of estimators of between-study variance of SMD (REML and $Q_A$) vs $\tau^2$, for $K$ = 2, 5, 10, and 20 studies per cluster and $M$ = 5, 10, 25, and 50 clusters; $\delta = 0.2$, and the sample size $n$ = 40 in each study.  }
	\label{PlotBiasOfTau2_40_02_HIER.pdf}
\end{figure}

\begin{figure}[ht]
	\centering
	\includegraphics[scale=0.33]{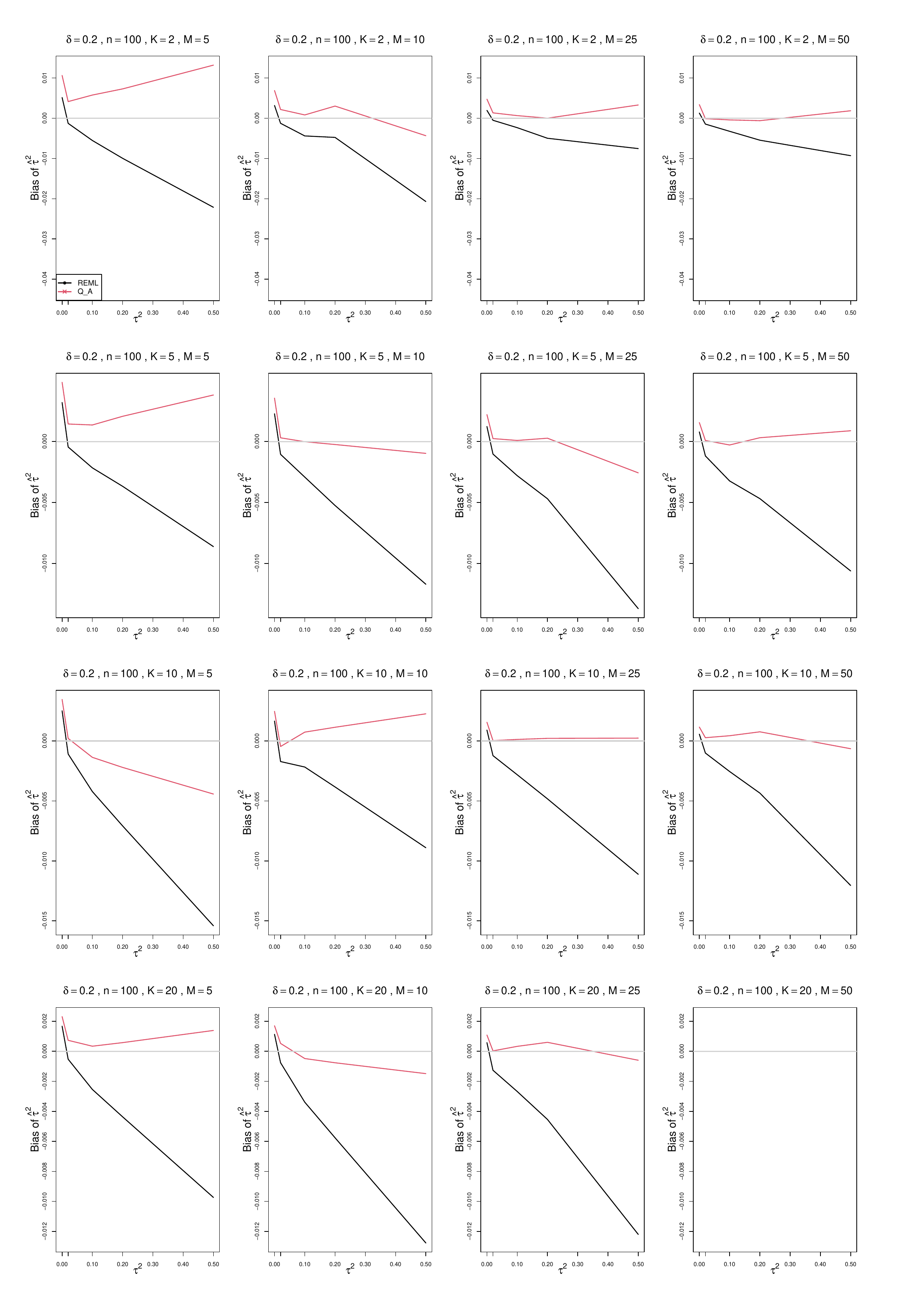}
	\caption{ Bias of estimators of between-study variance of SMD (REML and $Q_A$) vs $\tau^2$, for $K$ = 2, 5, 10, and 20 studies per cluster and $M$ = 5, 10, 25, and 50 clusters; $\delta = 0.2$, and the sample size $n$ = 100 in each study.  }
	\label{PlotBiasOfTau2_100_02_HIER.pdf}
\end{figure}

\begin{figure}[ht]
	\centering
	\includegraphics[scale=0.33]{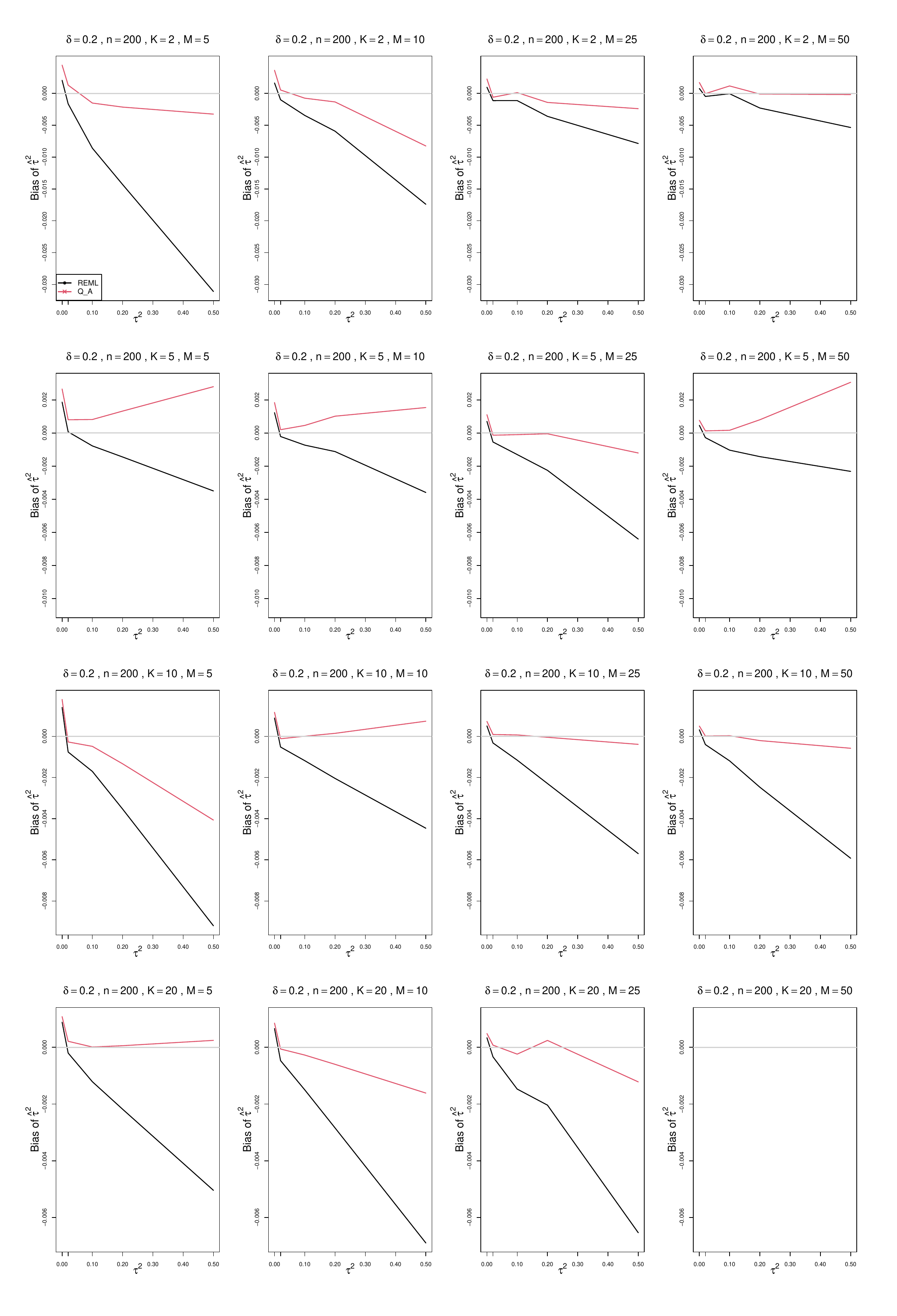}
	\caption{ Bias of estimators of between-study variance of SMD (REML and $Q_A$) vs $\tau^2$, for $K$ = 2, 5, 10, and 20 studies per cluster and $M$ = 5, 10, 25, and 50 clusters; $\delta = 0.2$, and the sample size $n$ = 200 in each study.  }
	\label{PlotBiasOfTau2_200_02_HIER.pdf}
\end{figure}

\begin{figure}[ht]
	\centering
	\includegraphics[scale=0.33]{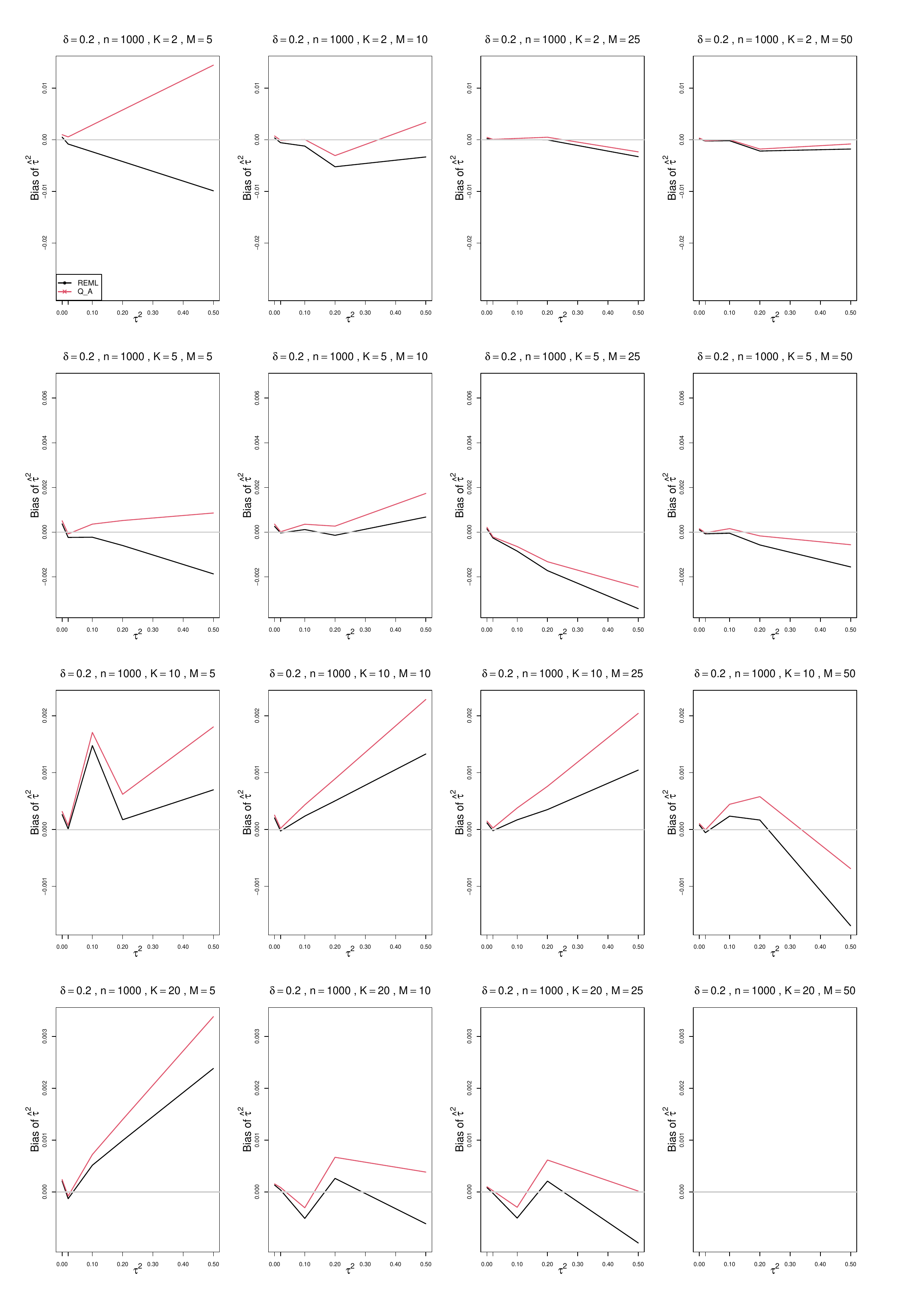}
	\caption{ Bias of estimators of between-study variance of SMD (REML and $Q_A$) vs $\tau^2$, for $K$ = 2, 5, 10, and 20 studies per cluster and $M$ = 5, 10, 25, and 50 clusters; $\delta = 0.2$, and the sample size $n$ = 1000 in each study.  }
	\label{PlotBiasOfTau2_1000_02_HIER.pdf}
\end{figure}

\begin{figure}[ht]
	\centering
	\includegraphics[scale=0.33]{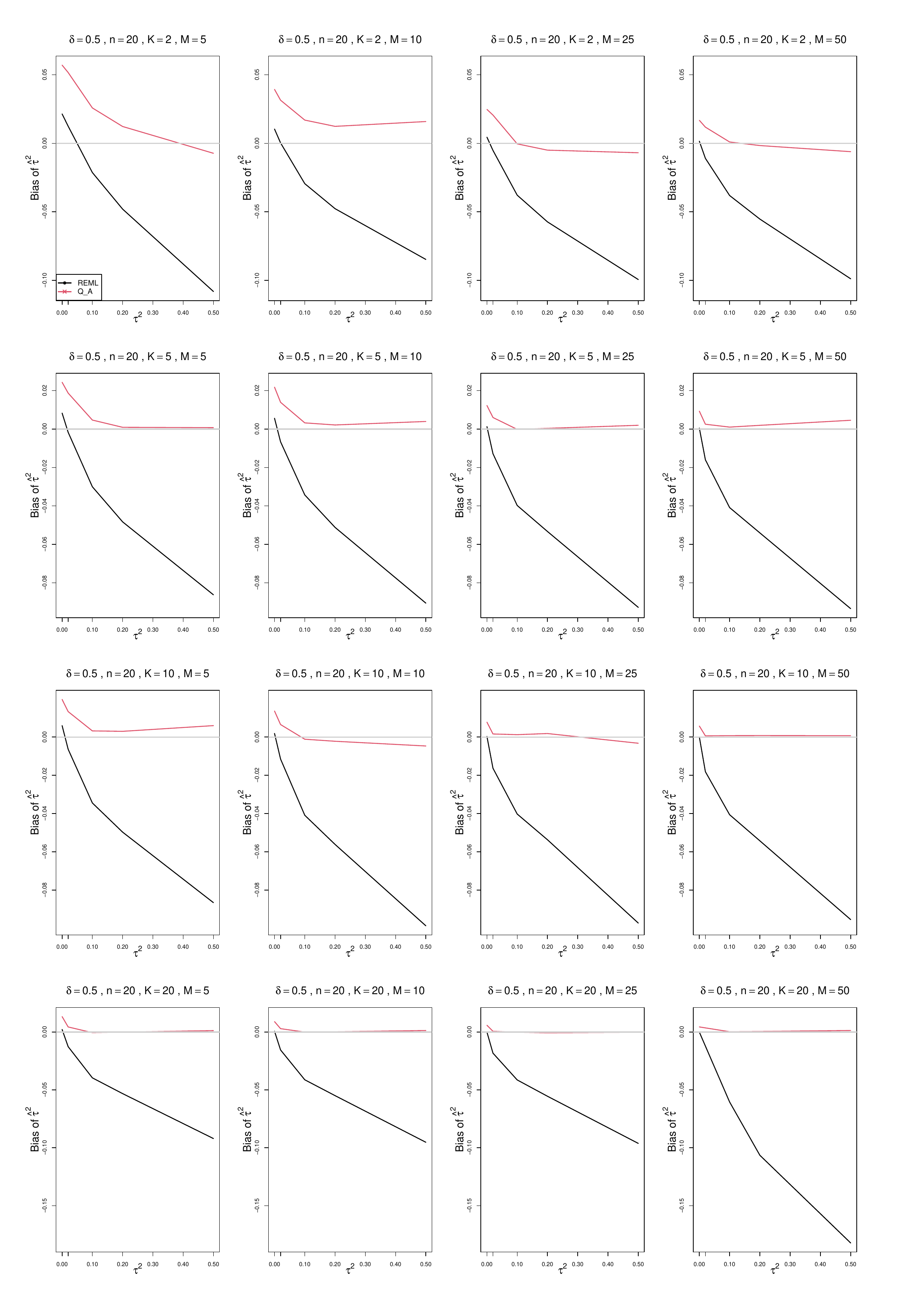}
	\caption{ Bias of estimators of between-study variance of SMD (REML and $Q_A$) vs $\tau^2$, for $K$ = 2, 5, 10, and 20 studies per cluster and $M$ = 5, 10, 25, and 50 clusters; $\delta = 0.5$, and the sample size $n$ = 20 in each study.  }
	\label{PlotBiasOfTau2_20_05_HIER.pdf}
\end{figure}

\begin{figure}[ht]
	\centering
	\includegraphics[scale=0.33]{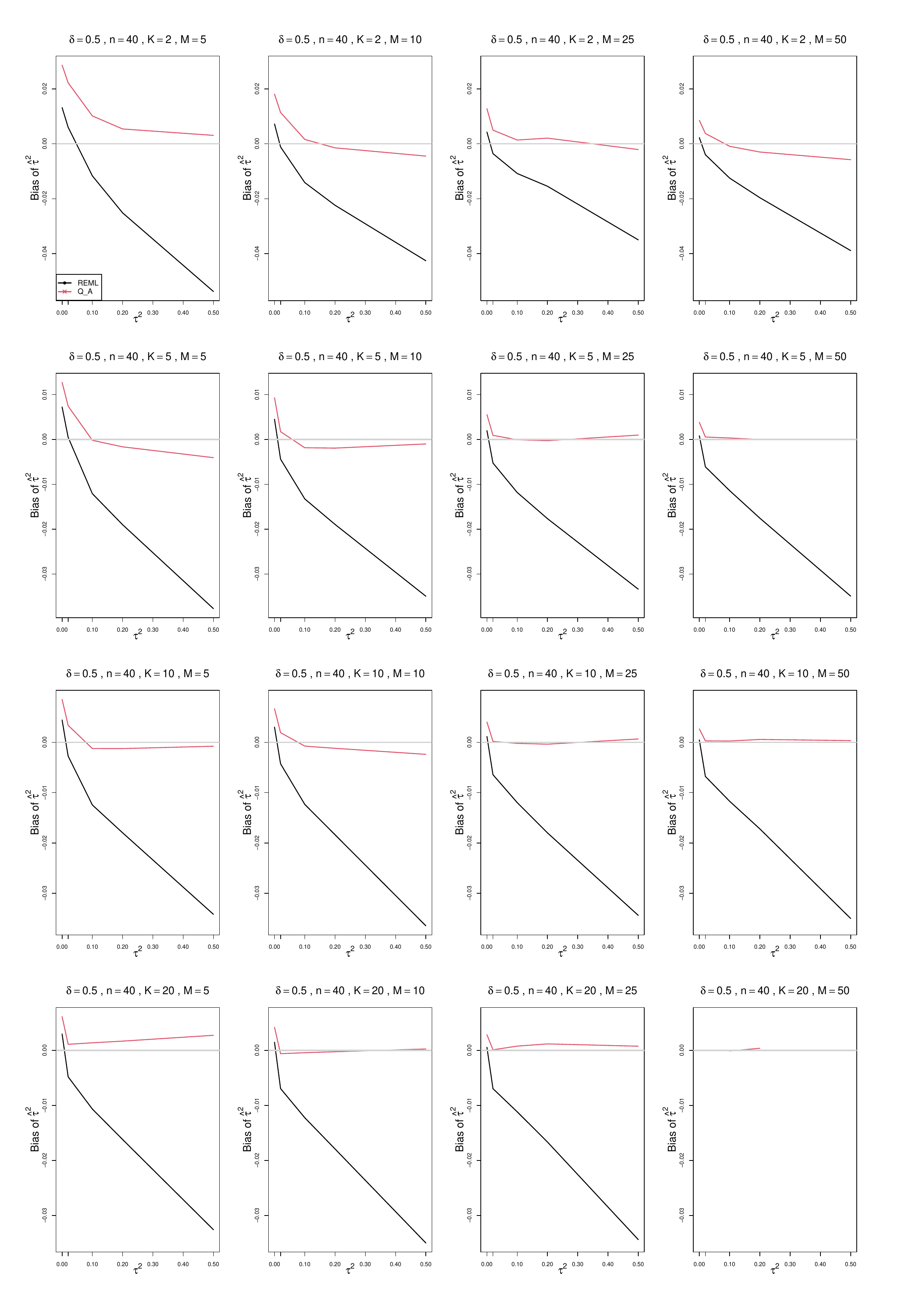}
	\caption{ Bias of estimators of between-study variance of SMD (REML and $Q_A$) vs $\tau^2$, for $K$ = 2, 5, 10, and 20 studies per cluster and $M$ = 5, 10, 25, and 50 clusters; $\delta = 0.5$, and the sample size $n$ = 40 in each study.  }
	\label{PlotBiasOfTau2_40_05_HIER.pdf}
\end{figure}

\begin{figure}[ht]
	\centering
	\includegraphics[scale=0.33]{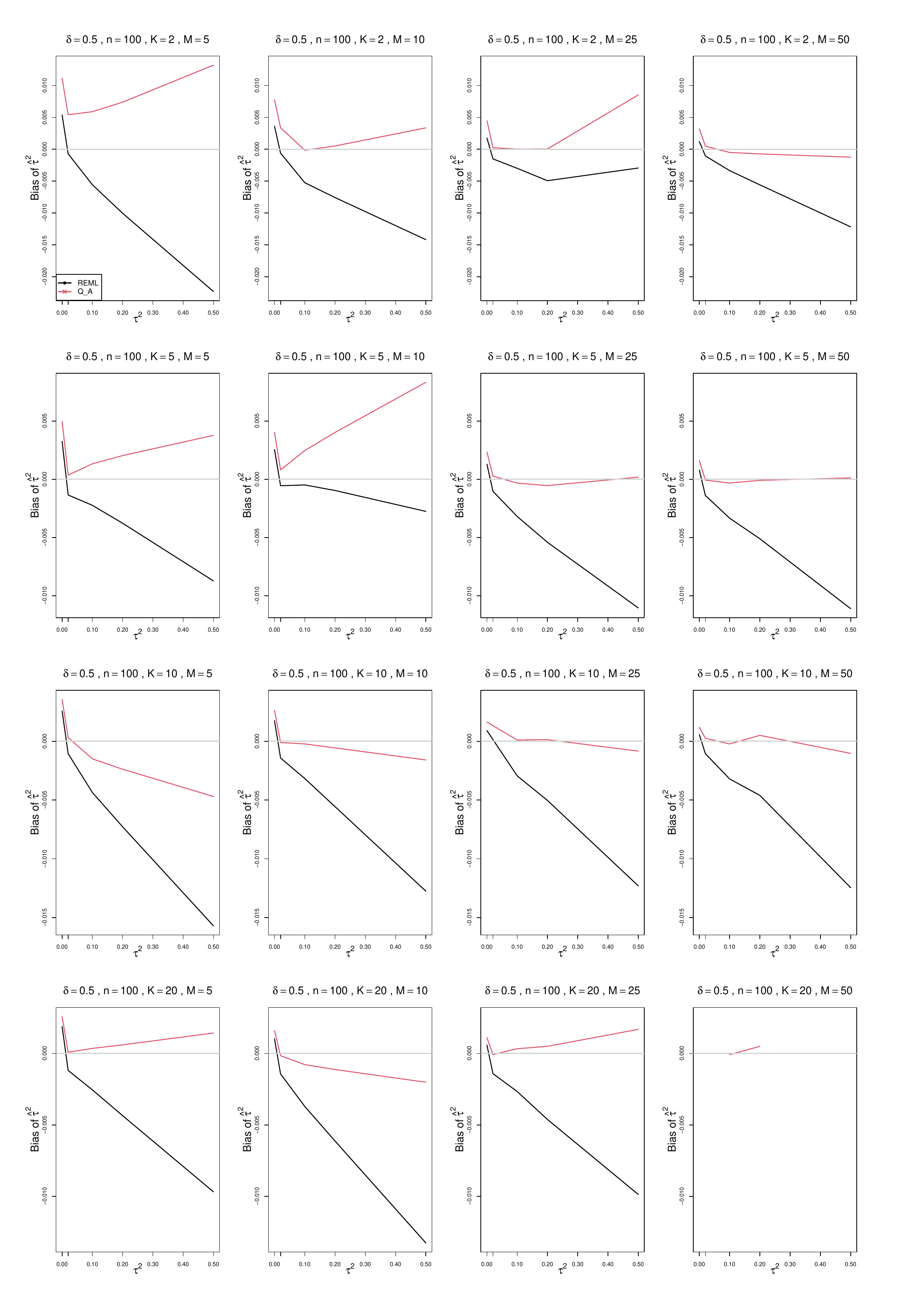}
	\caption{ Bias of estimators of between-study variance of SMD (REML and $Q_A$) vs $\tau^2$, for $K$ = 2, 5, 10, and 20 studies per cluster and $M$ = 5, 10, 25, and 50 clusters; $\delta = 0.5$, and the sample size $n$ = 100 in each study.  }
	\label{PlotBiasOfTau2_100_05_HIER.pdf}
\end{figure}

\begin{figure}[ht]
	\centering
	\includegraphics[scale=0.33]{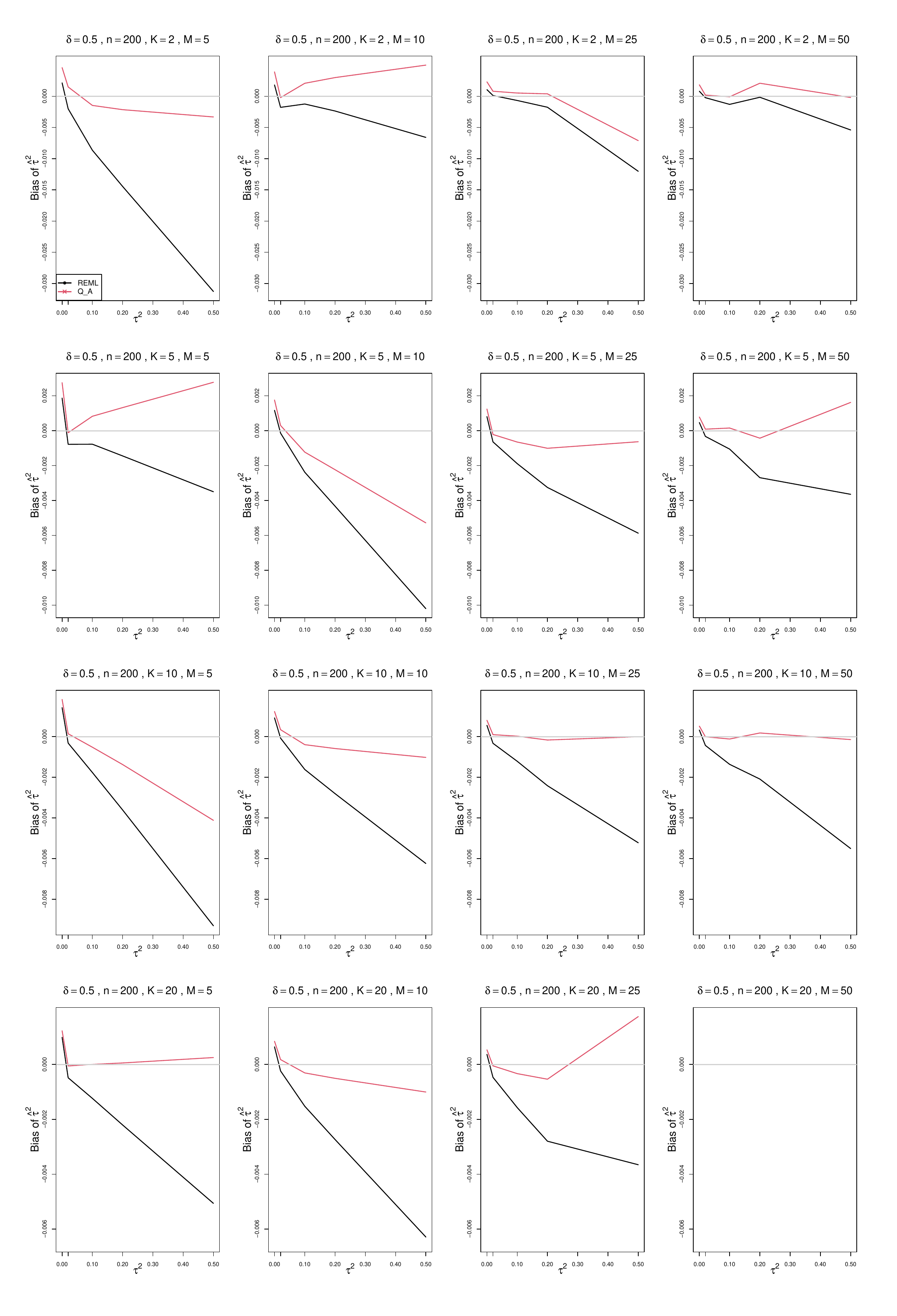}
	\caption{ Bias of estimators of between-study variance of SMD (REML and $Q_A$) vs $\tau^2$, for $K$ = 2, 5, 10, and 20 studies per cluster and $M$ = 5, 10, 25, and 50 clusters; $\delta = 0.5$, and the sample size $n$ = 200 in each study.  }
	\label{PlotBiasOfTau2_200_05_HIER.pdf}
\end{figure}

\begin{figure}[ht]
	\centering
	\includegraphics[scale=0.33]{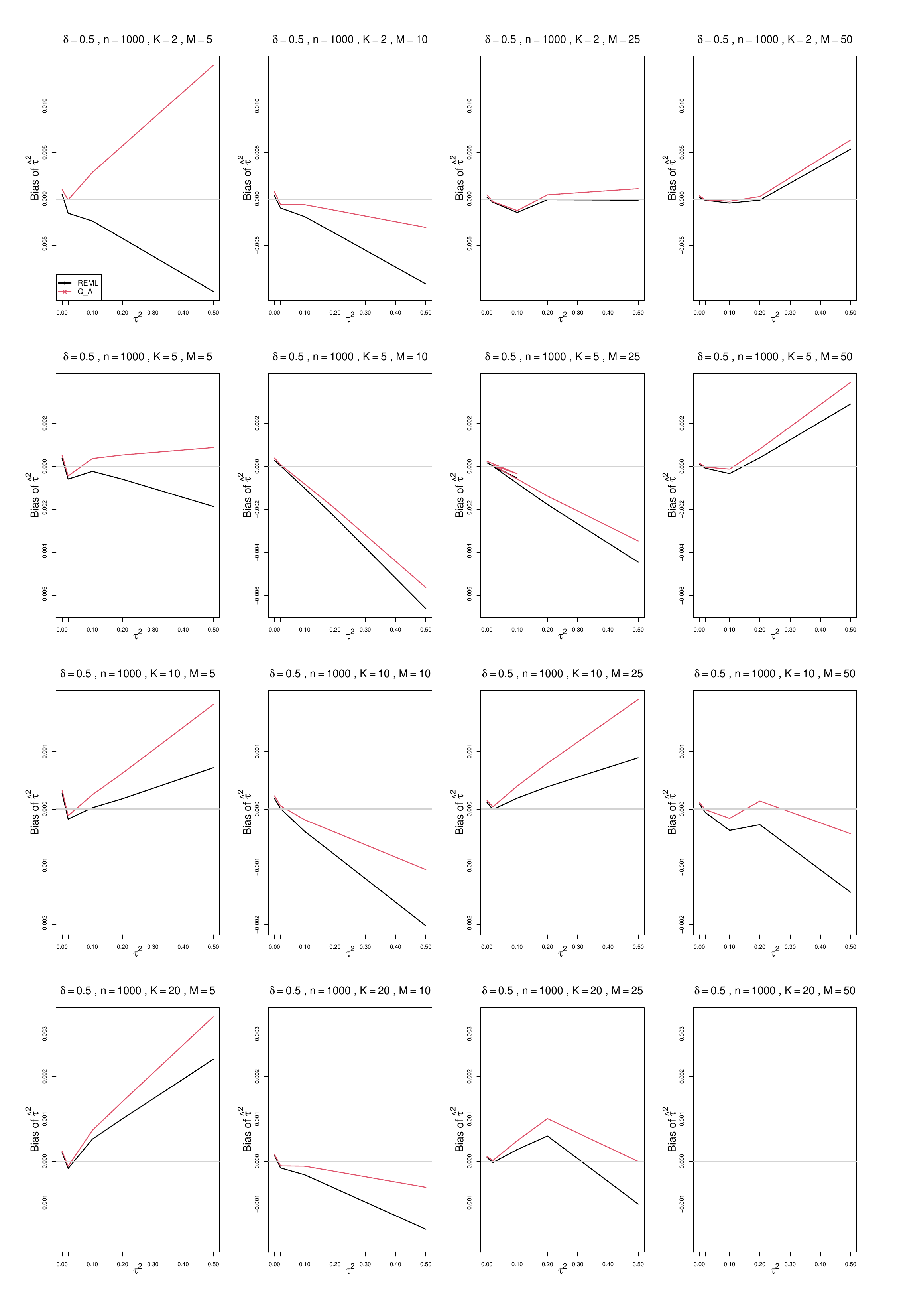}
	\caption{ Bias of estimators of between-study variance of SMD (REML and $Q_A$) vs $\tau^2$, for $K$ = 2, 5, 10, and 20 studies per cluster and $M$ = 5, 10, 25, and 50 clusters; $\delta = 0.5$, and the sample size $n$ = 1000 in each study. }
	\label{PlotBiasOfTau2_1000_05_HIER.pdf}
\end{figure}

\begin{figure}[ht]
	\centering
	\includegraphics[scale=0.33]{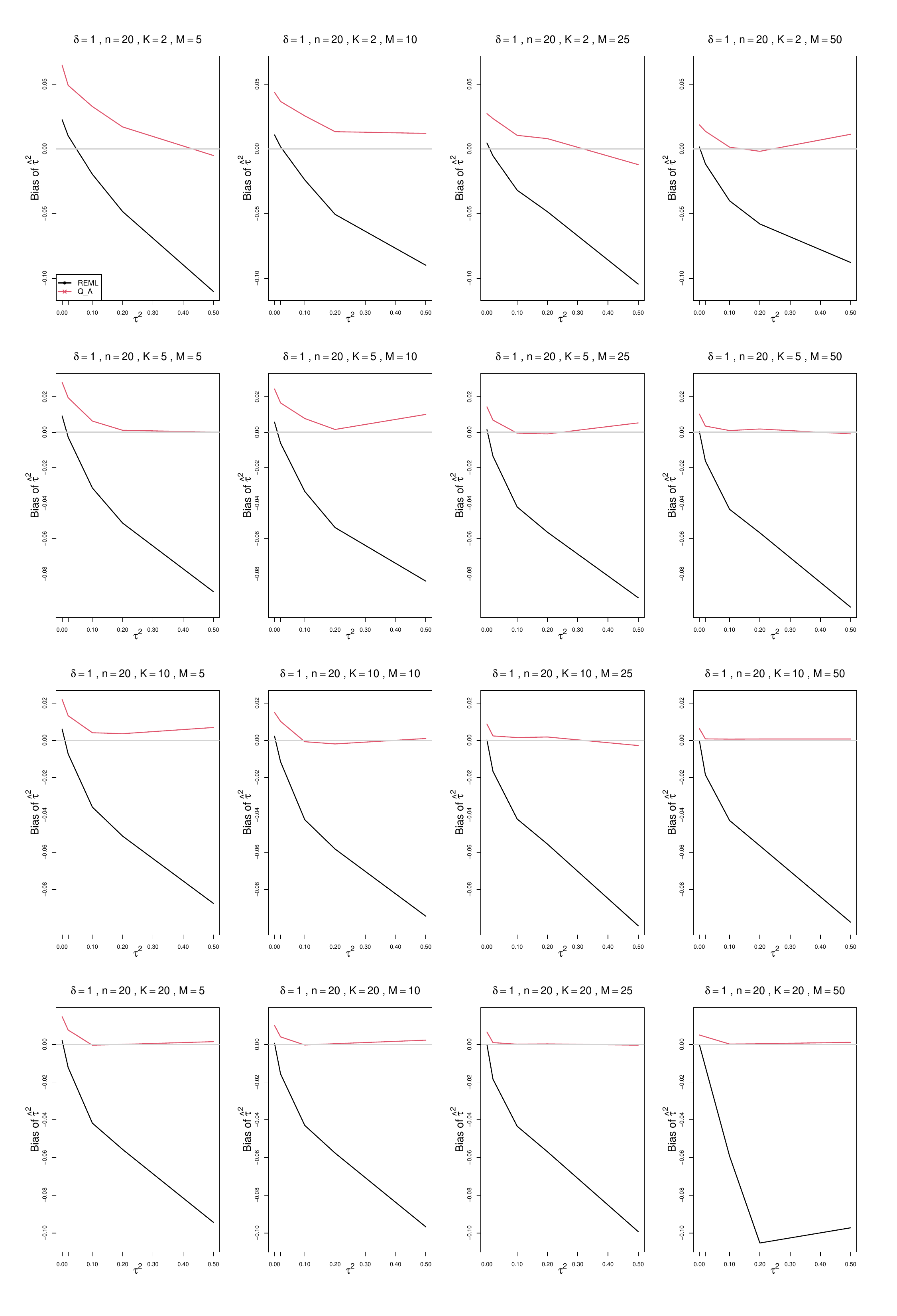}
	\caption{ Bias of estimators of between-study variance of SMD (REML and $Q_A$) vs $\tau^2$, for $K$ = 2, 5, 10, and 20 studies per cluster and $M$ = 5, 10, 25, and 50 clusters; $\delta = 1$, and the sample size $n$ = 20 in each study.  }
	\label{PlotBiasOfTau2_20_1_HIER.pdf}
\end{figure}

\begin{figure}[ht]
	\centering
	\includegraphics[scale=0.33]{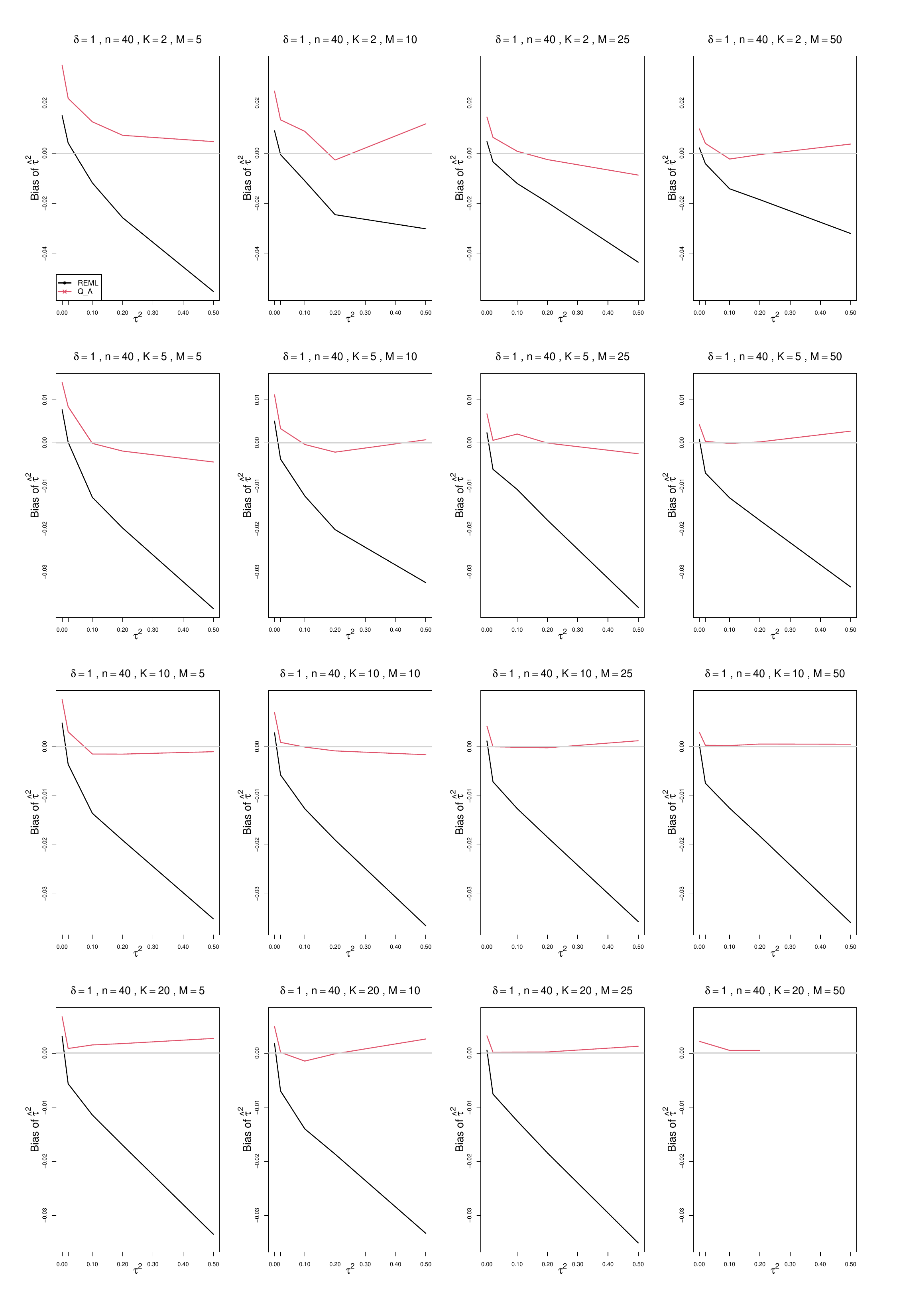}
	\caption{ Bias of estimators of between-study variance of SMD (REML and $Q_A$) vs $\tau^2$, for $K$ = 2, 5, 10, and 20 studies per cluster and $M$ = 5, 10, 25, and 50 clusters; $\delta = 1$, and the sample size $n$ = 40 in each study.  }
	\label{PlotBiasOfTau2_40_1_HIER.pdf}
\end{figure}

\begin{figure}[ht]
	\centering
	\includegraphics[scale=0.33]{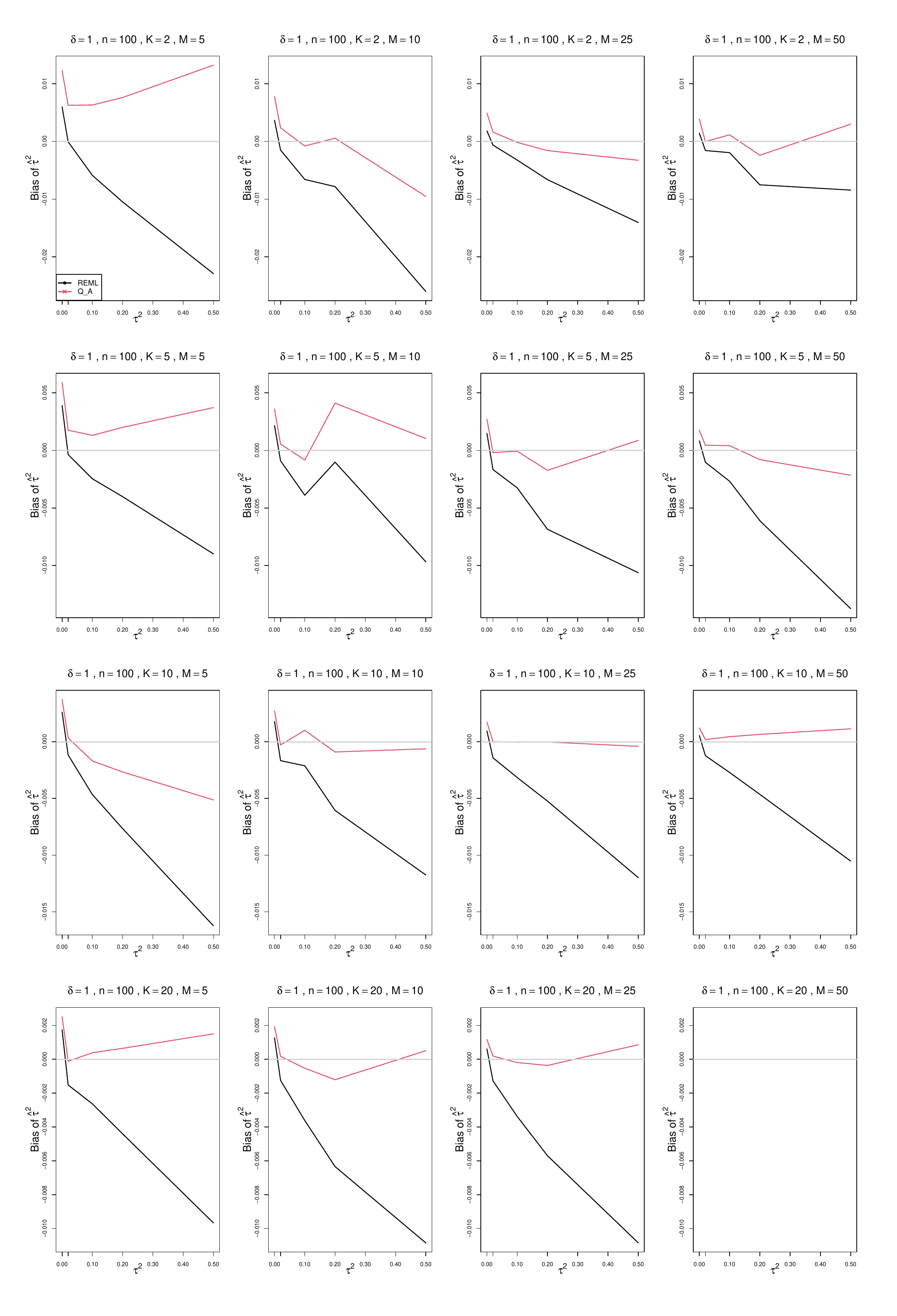}
	\caption{ Bias of estimators of between-study variance of SMD (REML and $Q_A$) vs $\tau^2$, for $K$ = 2, 5, 10, and 20 studies per cluster and $M$ = 5, 10, 25, and 50 clusters; $\delta = 1$, and the sample size $n$ = 100 in each study.  }
	\label{PlotBiasOfTau2_100_1_HIER.pdf}
\end{figure}

\begin{figure}[ht]
	\centering
	\includegraphics[scale=0.33]{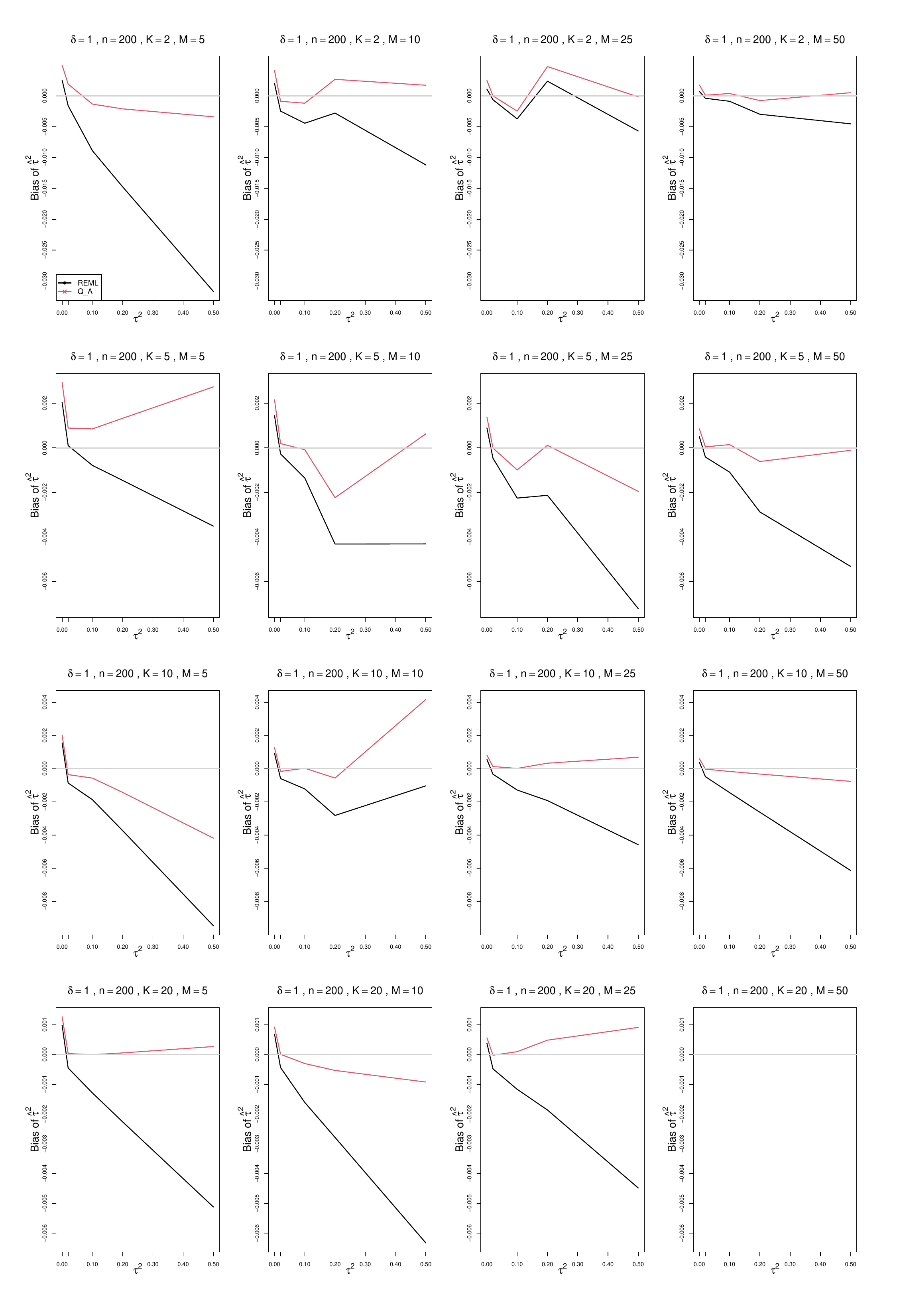}
	\caption{ Bias of estimators of between-study variance of SMD (REML and $Q_A$) vs $\tau^2$, for $K$ = 2, 5, 10, and 20 studies per cluster and $M$ = 5, 10, 25, and 50 clusters; $\delta = 1$, and the sample size $n$ = 200 in each study. }
	\label{PlotBiasOfTau2_200_1_HIER.pdf}
\end{figure}

\begin{figure}[ht]
	\centering
	\includegraphics[scale=0.33]{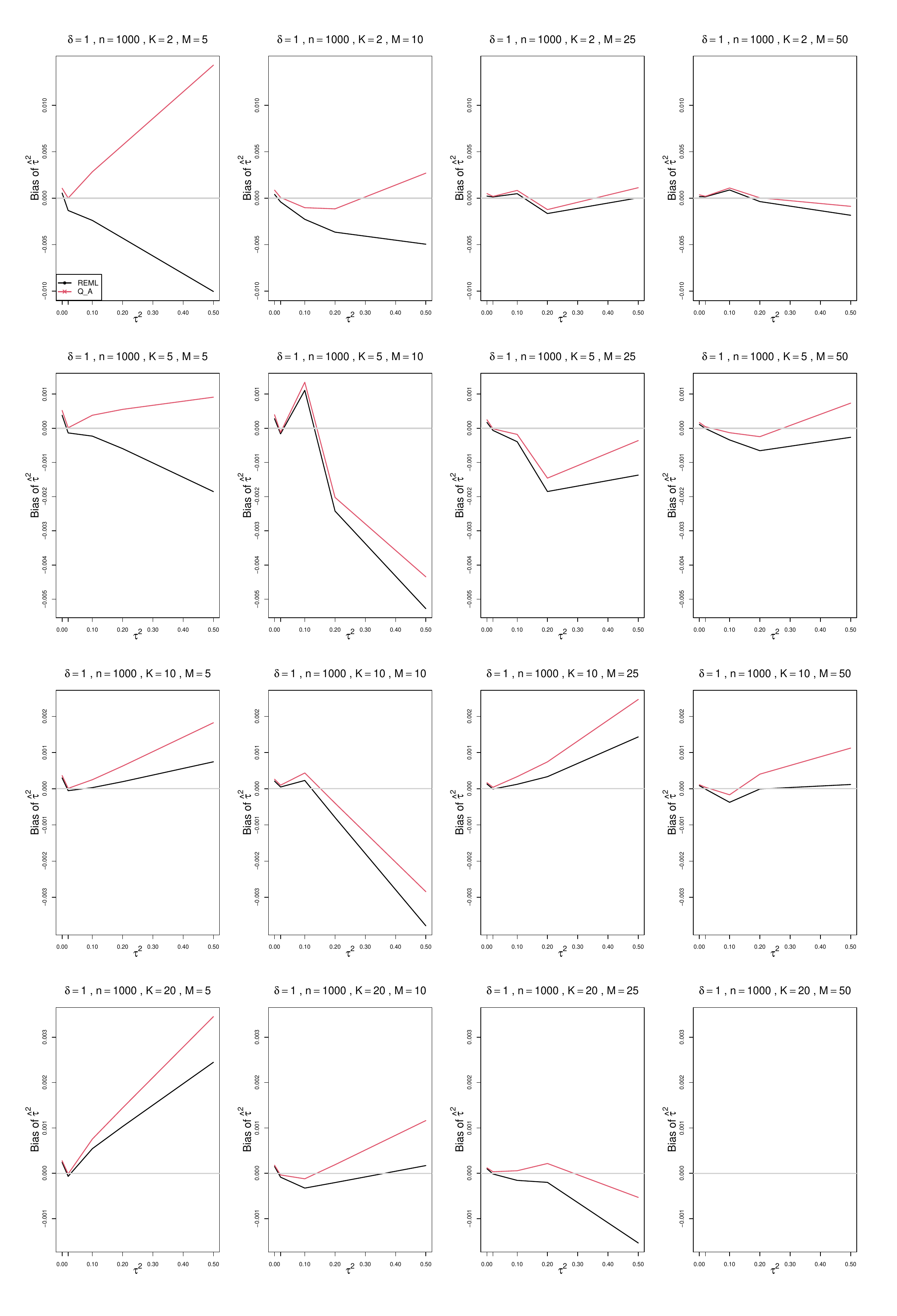}
	\caption{ Bias of estimators of between-study variance of SMD (REML and $Q_A$) vs $\tau^2$, for $K$ = 2, 5, 10, and 20 studies per cluster and $M$ = 5, 10, 25, and 50 clusters; $\delta = 1$, and the sample size $n$ = 1000 in each study.  }
	\label{PlotBiasOfTau2_1000_1_HIER.pdf}
\end{figure}


\clearpage
\renewcommand{\thefigure}{D.\arabic{figure}}

\section*{Appendix D: Coverage of 95\% confidence intervals for the between-study variance $\tau^2$}

Each figure corresponds to a value of the standardized mean difference ($\delta$ = 0, 0.2, 0.5, 1) and a value of the study sample size ($n$ = 20, 40, 100, 200, 1000).\\
For each combination of the number of studies in a cluster ($K$ = 2, 5, 10, 20) and the number of clusters ($M$ = 5, 10, 25, 50),  a panel plots coverage of $\tau^2$ versus $\tau^2$ (= 0, 0.02, 0.1, 0.2, 0.5).\\
The two variance components are held equal ($\tau^2 = \omega^2$).\\
The interval estimators of $\tau^2$ are
\begin{itemize}
\item PL (Profile-Likelihood method, inverse-variance weights,  {\it  rma.mv} in {\it metafor})
\item $Q_A$ (conditional moment-based method, effective-sample-size weights, Davies's approximation to the distribution)
\end{itemize}

\clearpage
\setcounter{figure}{0}
\renewcommand{\thefigure}{D.\arabic{figure}}

\begin{figure}[ht]
	\centering
	\includegraphics[scale=0.33]{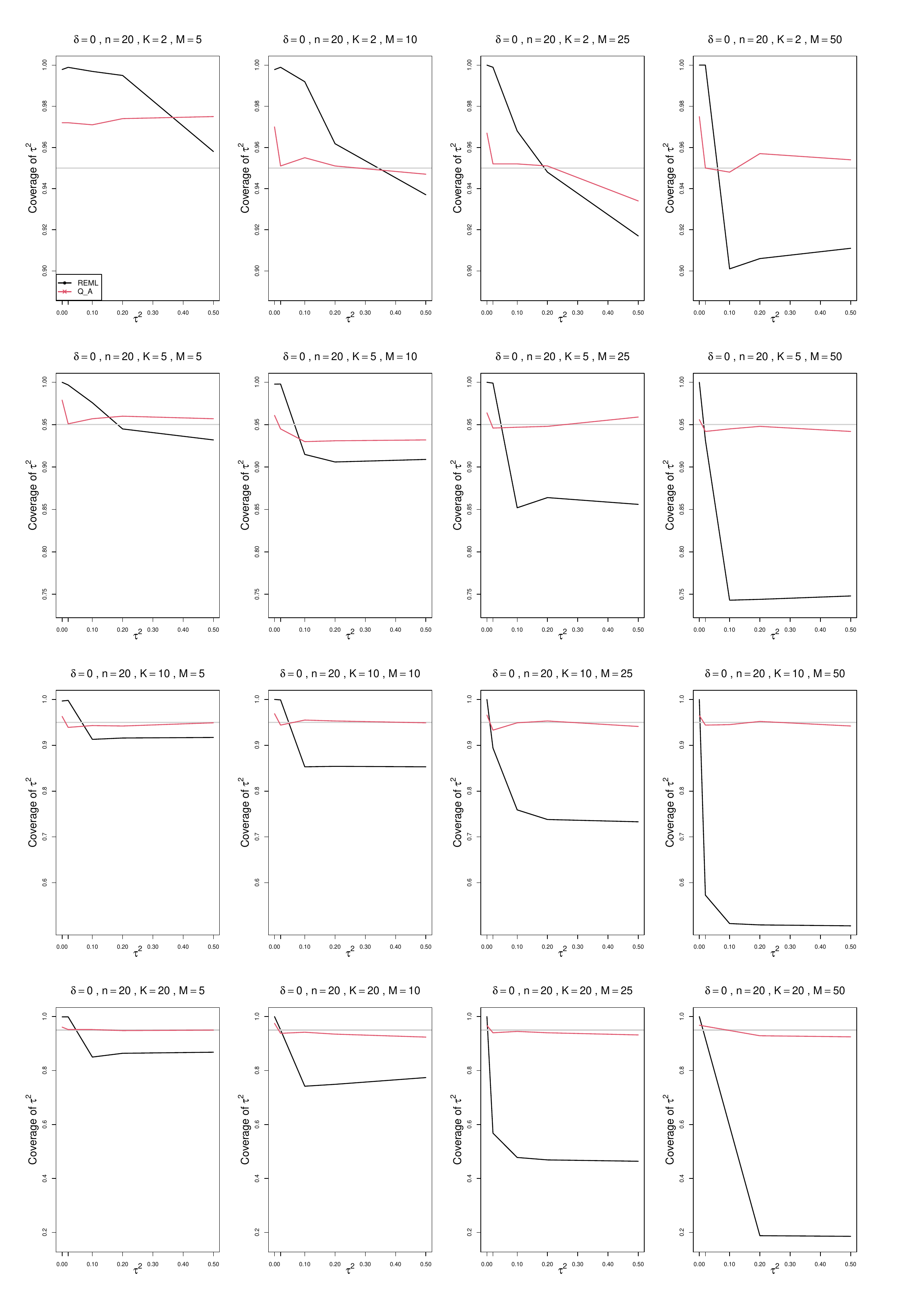}
	\caption{ Coverage of 95\% confidence intervals for between-study variance of SMD (REML and $Q_A$) vs $\tau^2$, for $K$ = 2, 5, 10, and 20 studies per cluster and $M$ = 5, 10, 25, and 50 clusters; $\delta = 0$, and the sample size $n$ = 20 in each study.  }
	\label{PlotCoverageOfTau2_20_0_HIER.pdf}
\end{figure}

\begin{figure}[ht]
	\centering
	\includegraphics[scale=0.33]{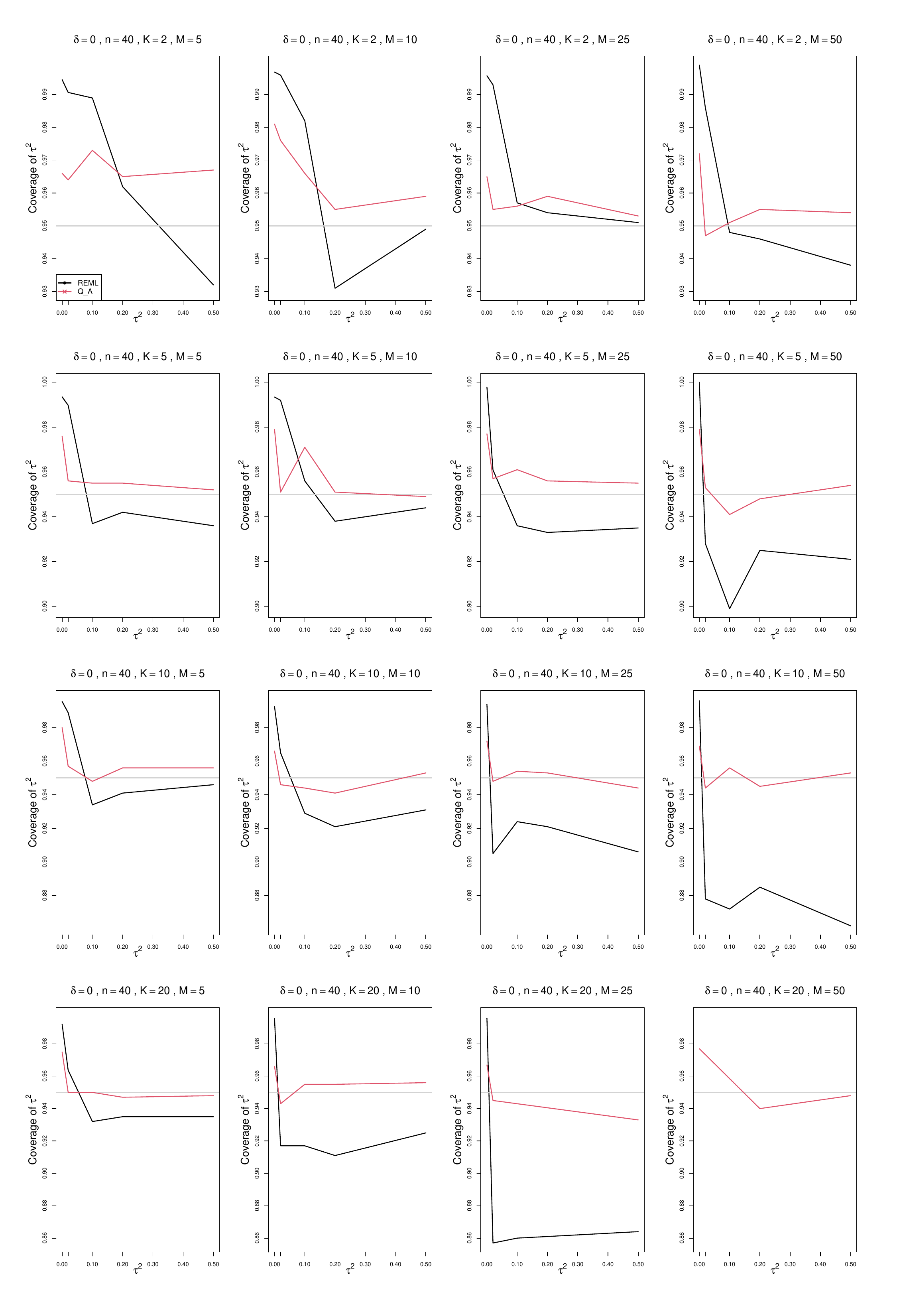}
	\caption{ Coverage of 95\% confidence intervals for between-study variance of SMD (REML and $Q_A$) vs $\tau^2$, for $K$ = 2, 5, 10, and 20 studies per cluster and $M$ = 5, 10, 25, and 50 clusters; $\delta = 0$, and the sample size $n$ = 40 in each study.  }
	\label{PlotCoverageOfTau2_40_0_HIER.pdf}
\end{figure}

\begin{figure}[ht]
	\centering
	\includegraphics[scale=0.33]{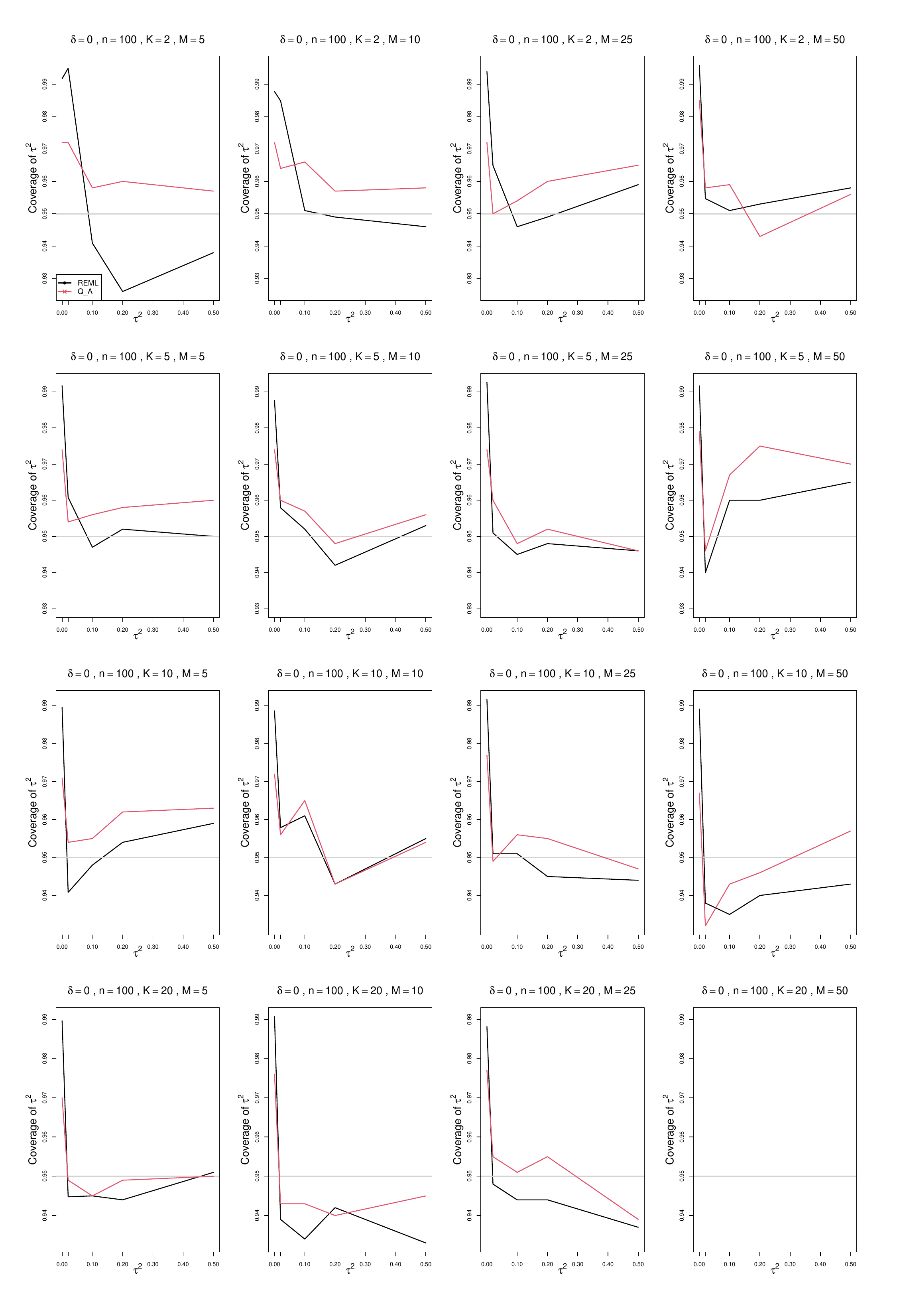}
	\caption{ Coverage of 95\% confidence intervals for between-study variance of SMD (REML and $Q_A$) vs $\tau^2$, for $K$ = 2, 5, 10, and 20 studies per cluster and $M$ = 5, 10, 25, and 50 clusters; $\delta = 0$, and the sample size $n$ = 100 in each study.  }
	\label{PlotCoverageOfTau2_100_0_HIER.pdf}
\end{figure}

\begin{figure}[ht]
	\centering
	\includegraphics[scale=0.33]{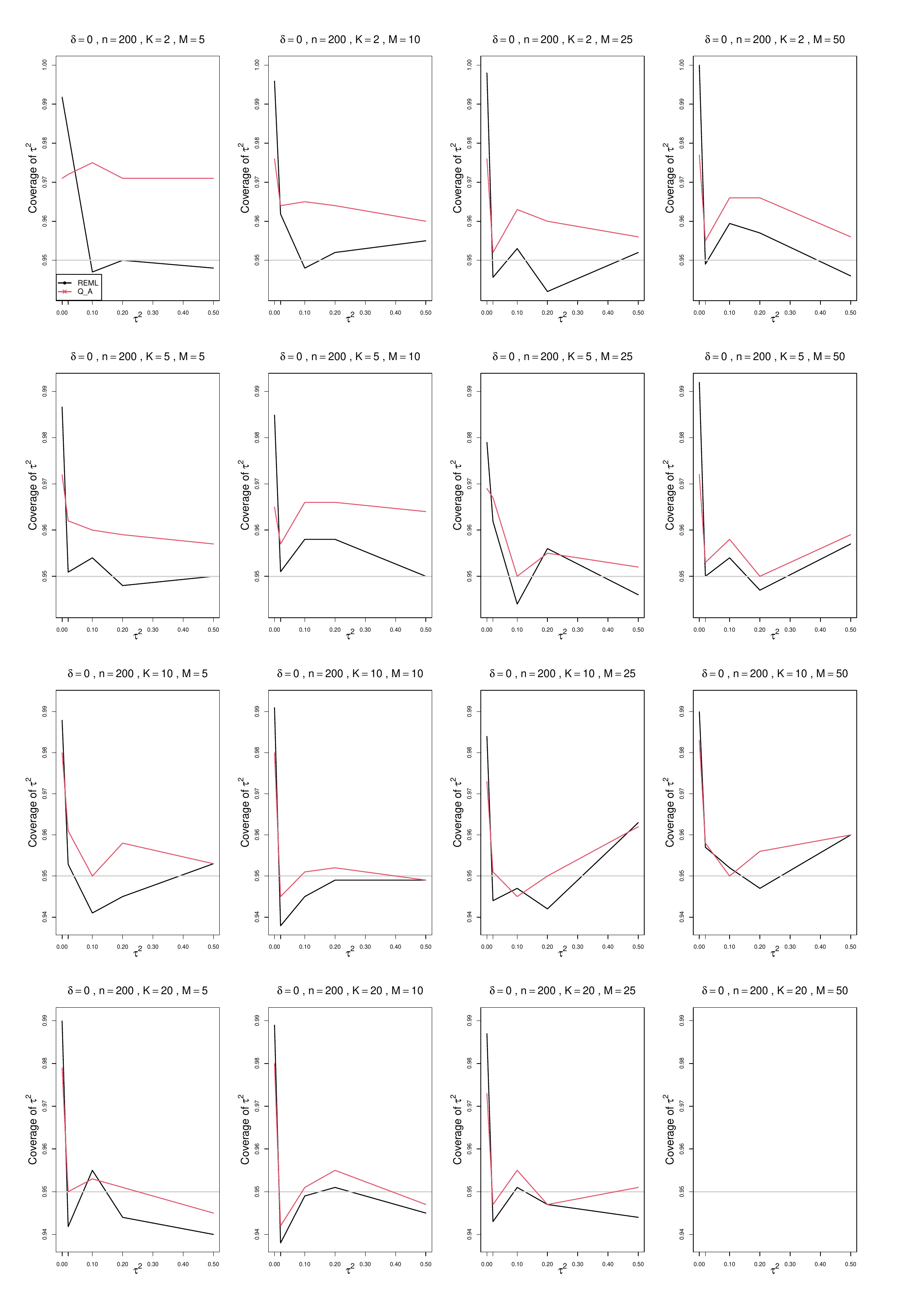}
	\caption{ Coverage of 95\% confidence intervals for between-study variance of SMD (REML and $Q_A$) vs $\tau^2$, for $K$ = 2, 5, 10, and 20 studies per cluster and $M$ = 5, 10, 25, and 50 clusters; $\delta = 0$, and the sample size $n$ = 200 in each study.  }
	\label{PlotCoverageOfTau2_200_0_HIER.pdf}
\end{figure}

\begin{figure}[ht]
	\centering
	\includegraphics[scale=0.33]{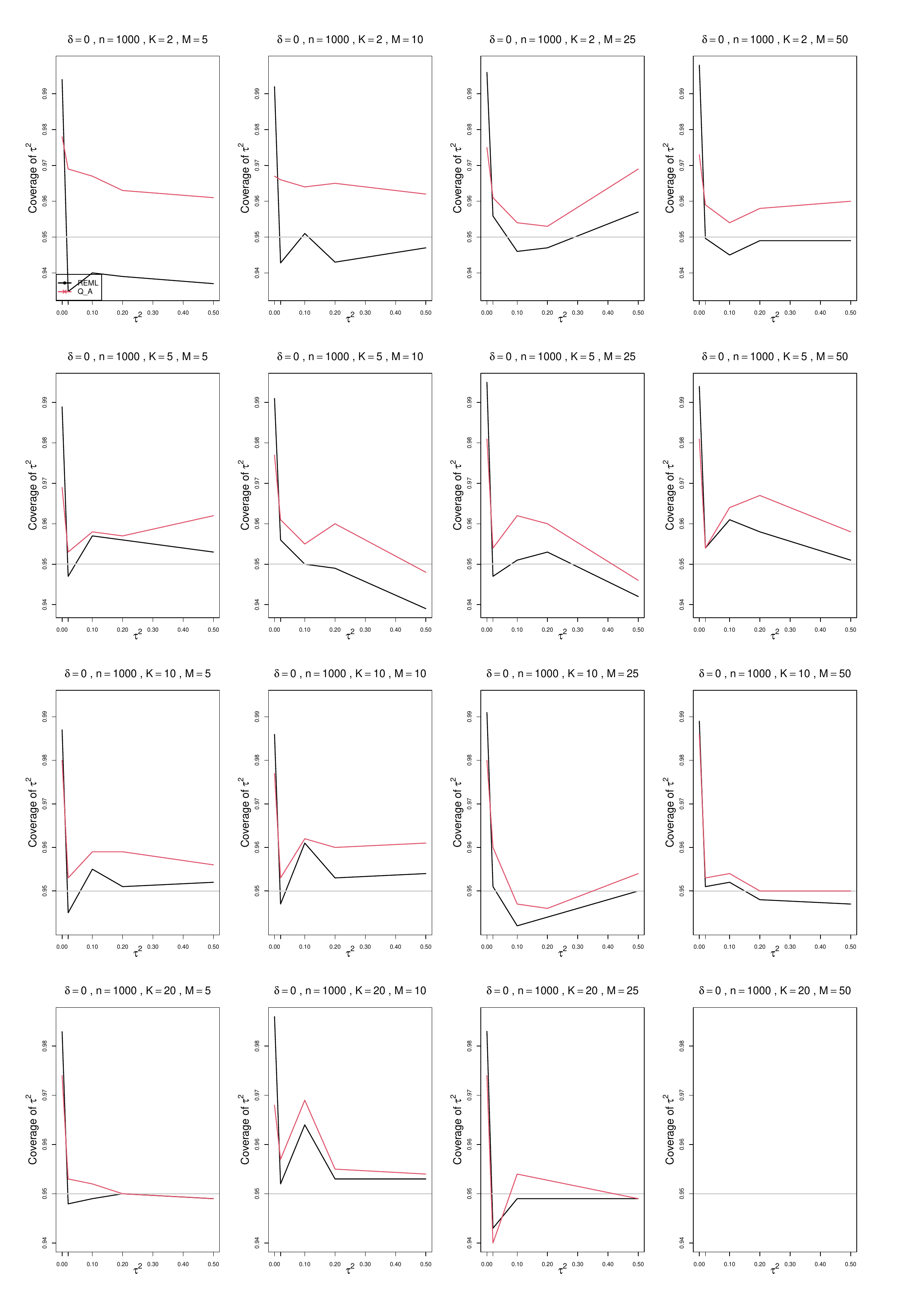}
	\caption{ Coverage of 95\% confidence intervals for between-study variance of SMD (REML and $Q_A$) vs $\tau^2$, for $K$ = 2, 5, 10, and 20 studies per cluster and $M$ = 5, 10, 25, and 50 clusters; $\delta = 0$, and the sample size $n$ = 1000 in each study. }
	\label{PlotCoverageOfTau2_1000_0_HIER.pdf}
\end{figure}

\begin{figure}[ht]
	\centering
	\includegraphics[scale=0.33]{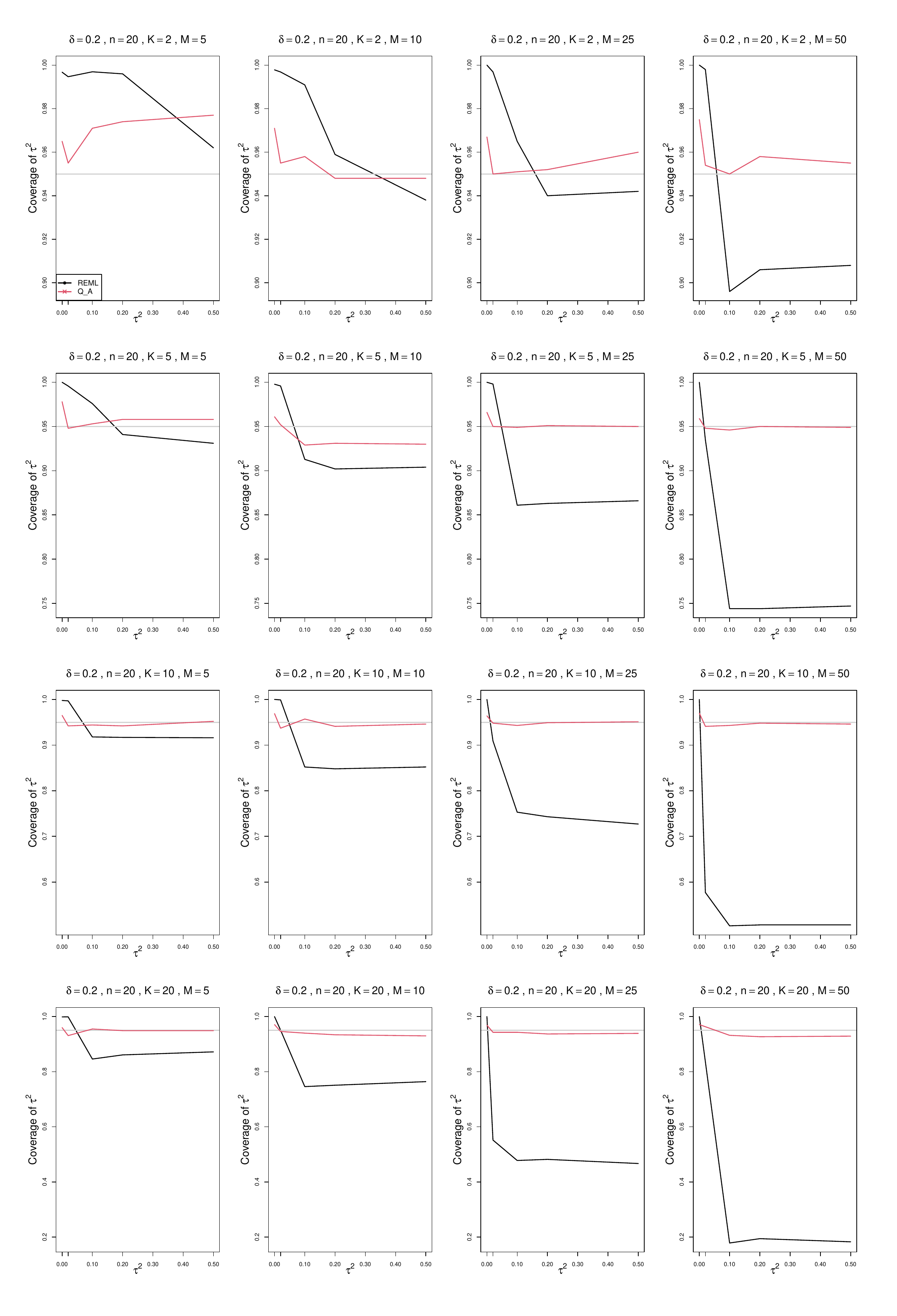}
	\caption{ Coverage of 95\% confidence intervals for between-study variance of SMD (REML and $Q_A$) vs $\tau^2$, for $K$ = 2, 5, 10, and 20 studies per cluster and $M$ = 5, 10, 25, and 50 clusters; $\delta = 0.2$, and the sample size $n$ = 20 in each study. }
	\label{PlotCoverageOfTau2_20_02_HIER.pdf}
\end{figure}

\begin{figure}[ht]
	\centering
	\includegraphics[scale=0.33]{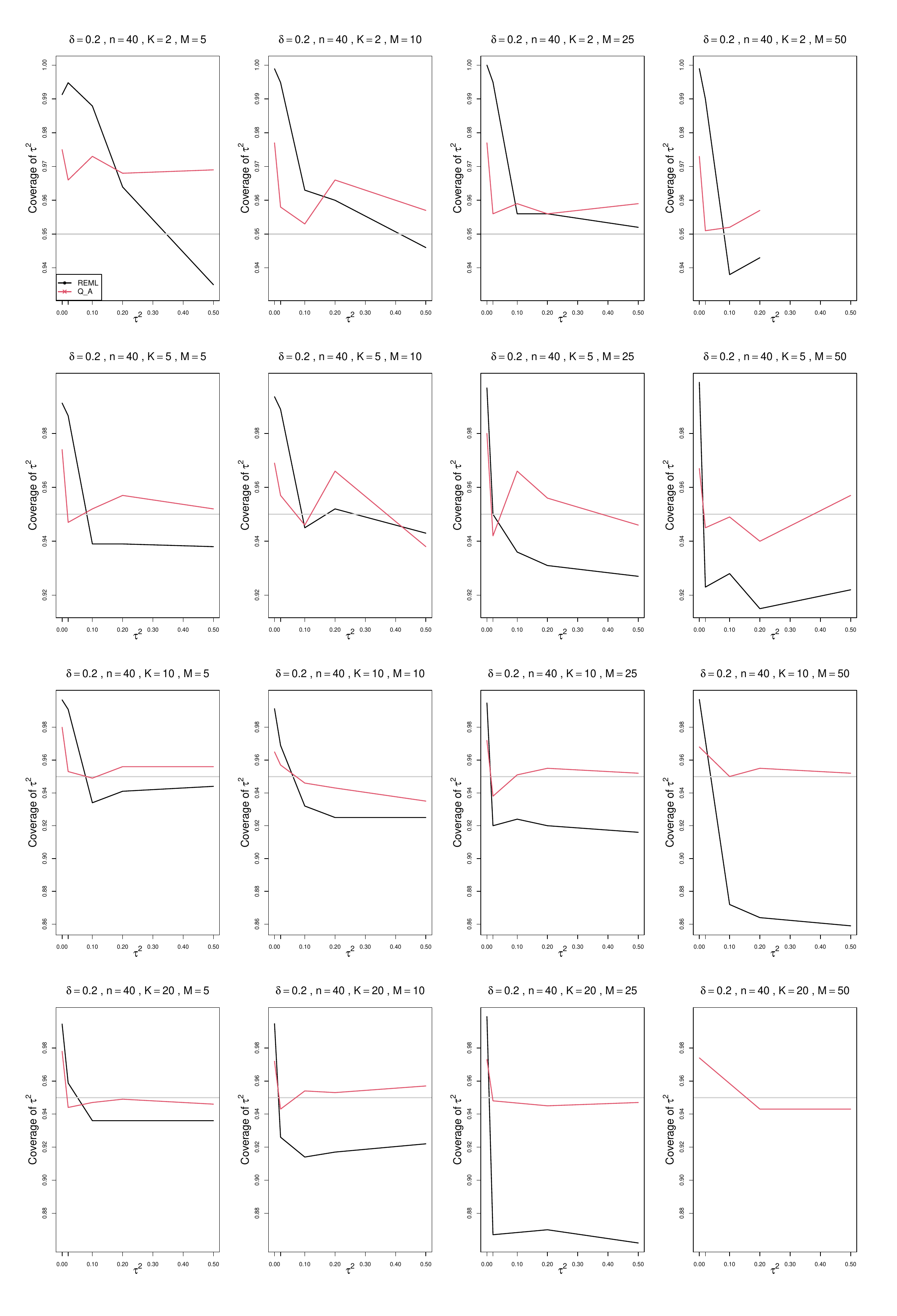}
	\caption{ Coverage of 95\% confidence intervals for between-study variance of SMD (REML and $Q_A$) vs $\tau^2$, for $K$ = 2, 5, 10, and 20 studies per cluster and $M$ = 5, 10, 25, and 50 clusters; $\delta = 0.2$, and the sample size $n$ = 40 in each study.  }
	\label{PlotCoverageOfTau2_40_02_HIER.pdf}
\end{figure}

\begin{figure}[ht]
	\centering
	\includegraphics[scale=0.33]{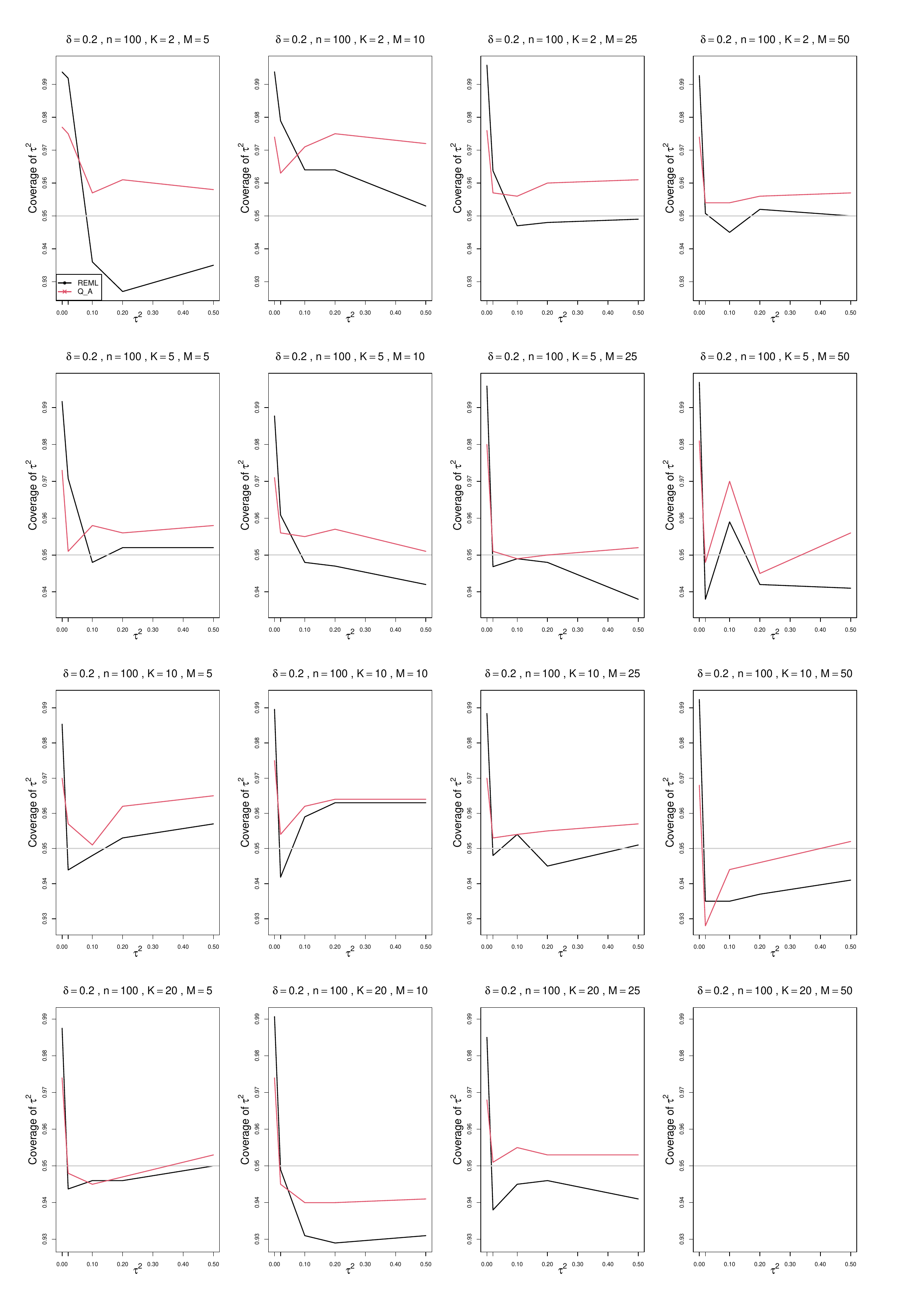}
	\caption{ Coverage of 95\% confidence intervals for between-study variance of SMD (REML and $Q_A$) vs $\tau^2$, for $K$ = 2, 5, 10, and 20 studies per cluster and $M$ = 5, 10, 25, and 50 clusters; $\delta = 0.2$, and the sample size $n$ = 100 in each study.  }
	\label{PlotCoverageOfTau2_100_02_HIER.pdf}
\end{figure}

\begin{figure}[ht]
	\centering
	\includegraphics[scale=0.33]{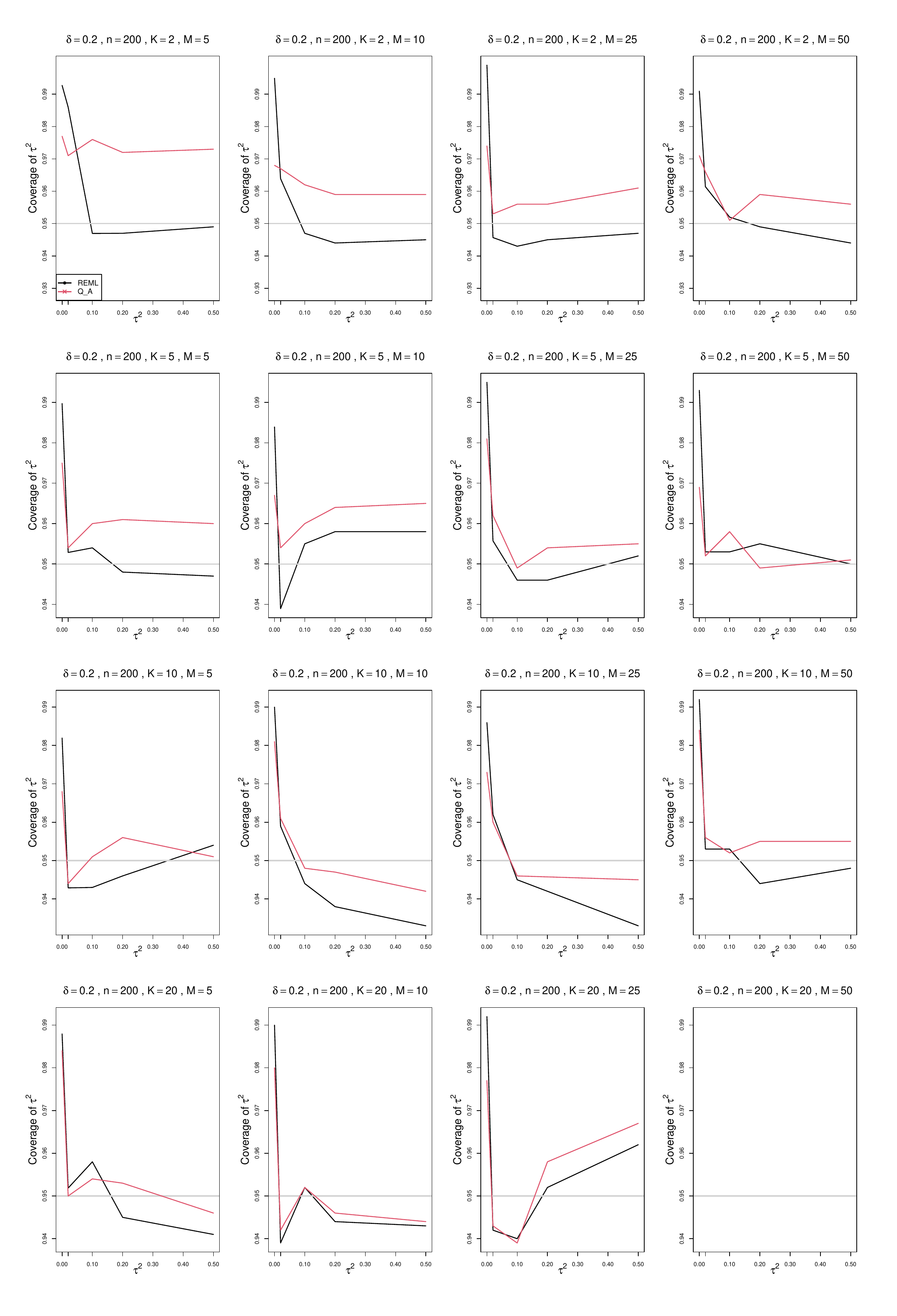}
	\caption{ Coverage of 95\% confidence intervals for between-study variance of SMD (REML and $Q_A$) vs $\tau^2$, for $K$ = 2, 5, 10, and 20 studies per cluster and $M$ = 5, 10, 25, and 50 clusters; $\delta = 0.2$, and the sample size $n$ = 200 in each study.  }
	\label{PlotCoverageOfTau2_200_02_HIER.pdf}
\end{figure}

\begin{figure}[ht]
	\centering
	\includegraphics[scale=0.33]{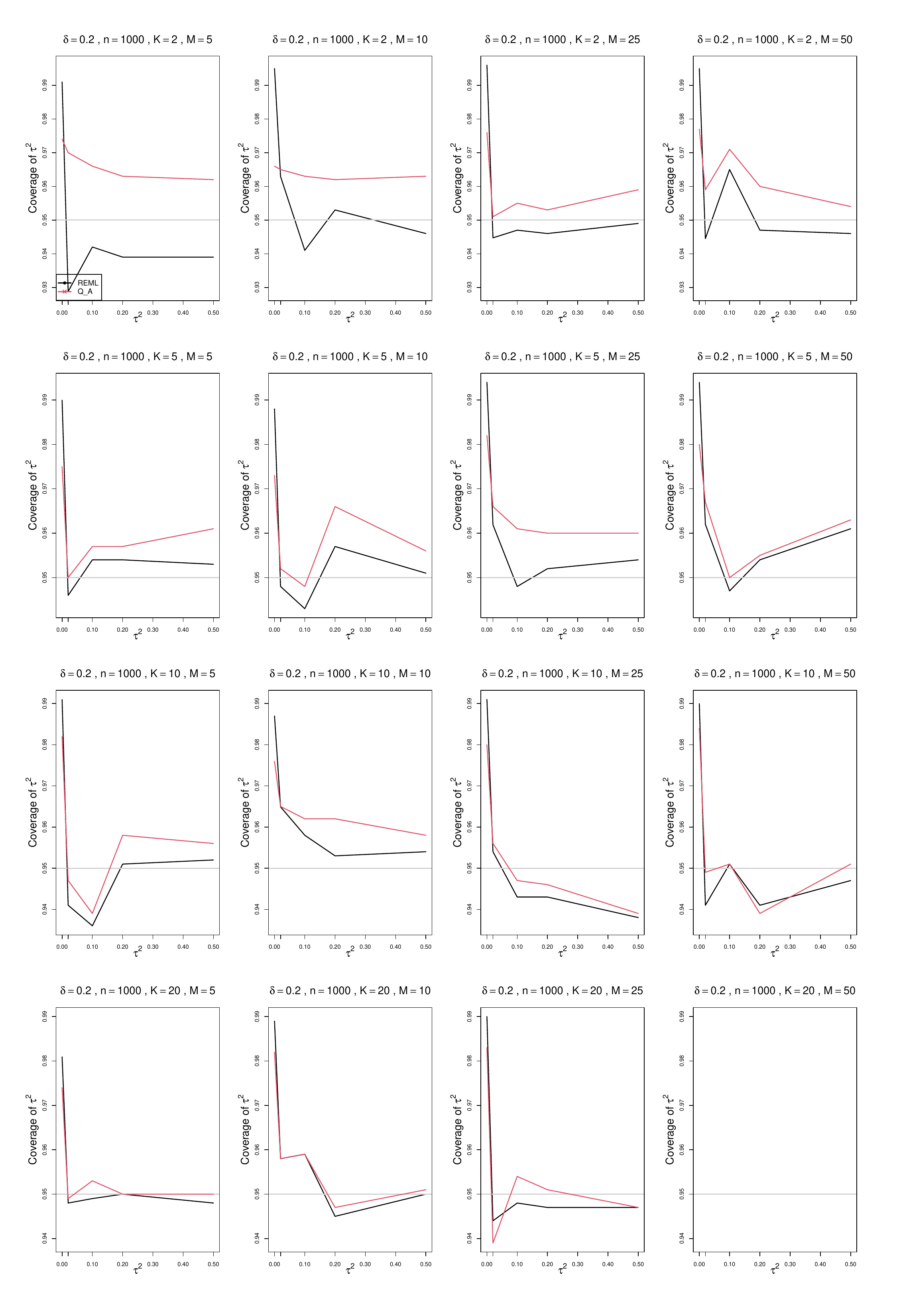}
	\caption{ Coverage of 95\% confidence intervals for between-study variance of SMD (REML and $Q_A$) vs $\tau^2$, for $K$ = 2, 5, 10, and 20 studies per cluster and $M$ = 5, 10, 25, and 50 clusters; $\delta = 0.2$, and the sample size $n$ = 1000 in each study.  }
	\label{PlotCoverageOfTau2_1000_02_HIER.pdf}
\end{figure}

\begin{figure}[ht]
	\centering
	\includegraphics[scale=0.33]{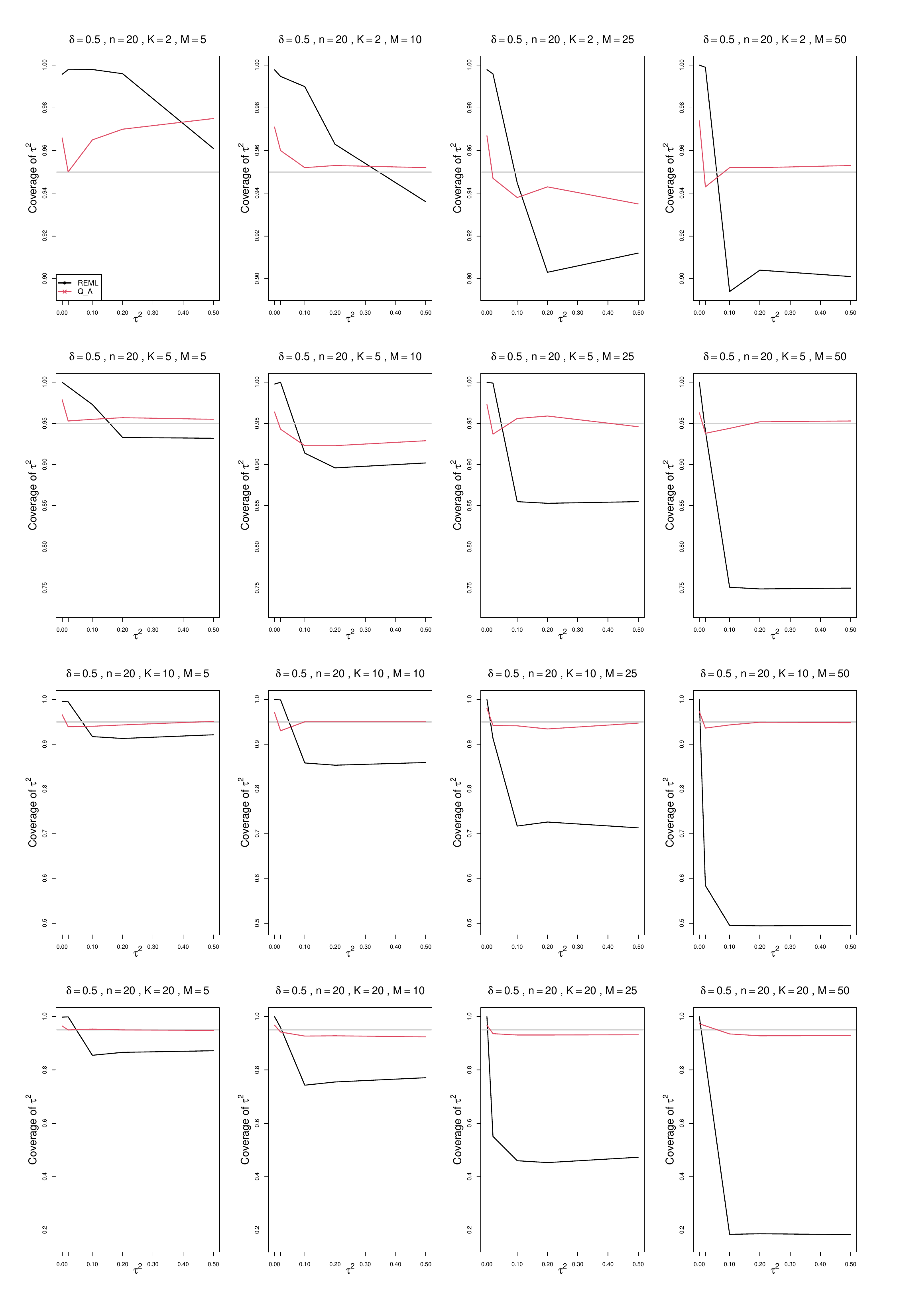}
	\caption{ Coverage of 95\% confidence intervals for between-study variance of SMD (REML and $Q_A$) vs $\tau^2$, for $K$ = 2, 5, 10, and 20 studies per cluster and $M$ = 5, 10, 25, and 50 clusters; $\delta = 0.5$, and the sample size $n$ = 20 in each study.  }
	\label{PlotCoverageOfTau2_20_05_HIER.pdf}
\end{figure}

\begin{figure}[ht]
	\centering
	\includegraphics[scale=0.33]{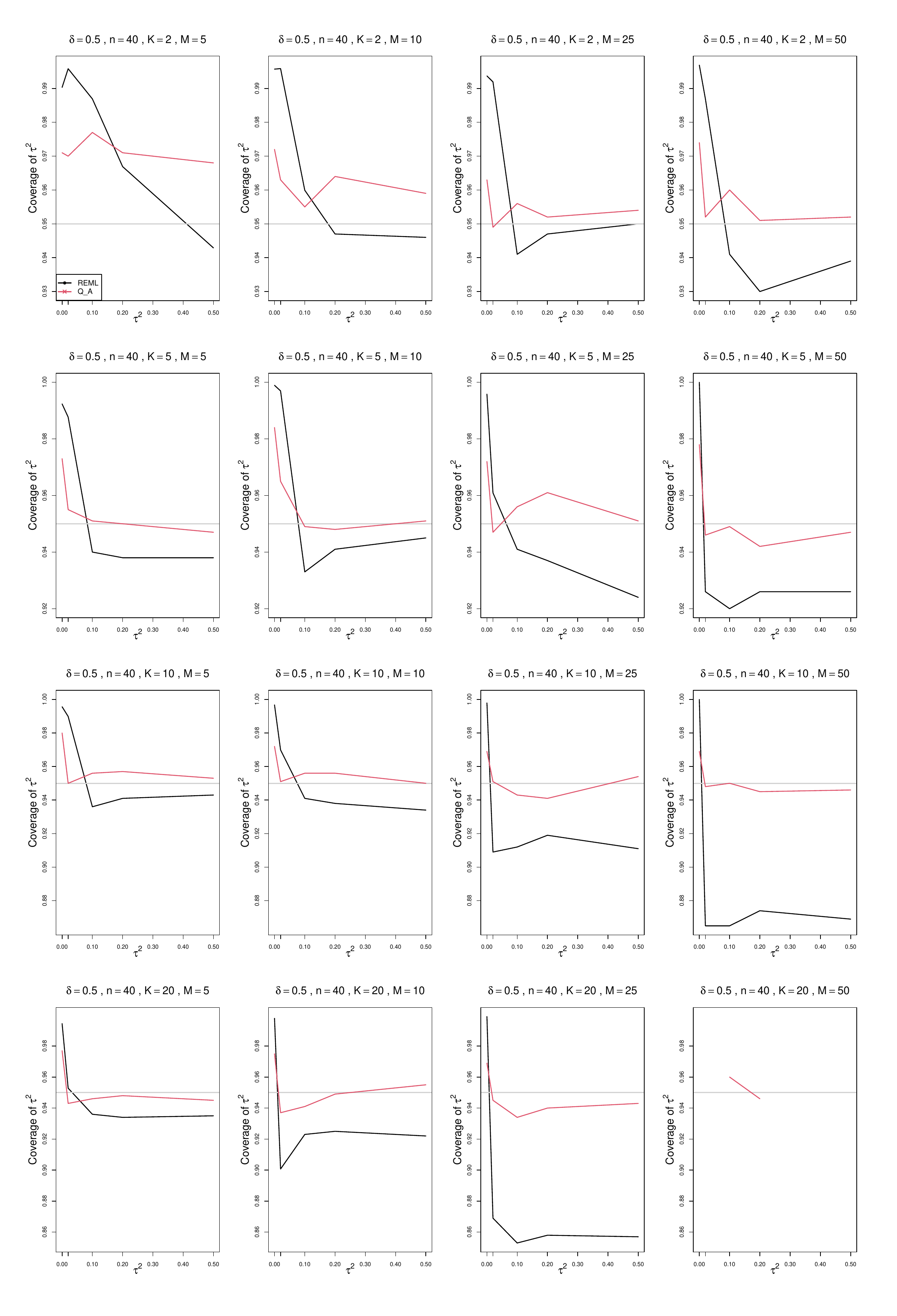}
	\caption{ Coverage of 95\% confidence intervals for between-study variance of SMD (REML and $Q_A$) vs $\tau^2$, for $K$ = 2, 5, 10, and 20 studies per cluster and $M$ = 5, 10, 25, and 50 clusters; $\delta = 0.5$, and the sample size $n$ = 40 in each study.  }
	\label{PlotCoverageOfTau2_40_05_HIER.pdf}
\end{figure}

\begin{figure}[ht]
	\centering
	\includegraphics[scale=0.33]{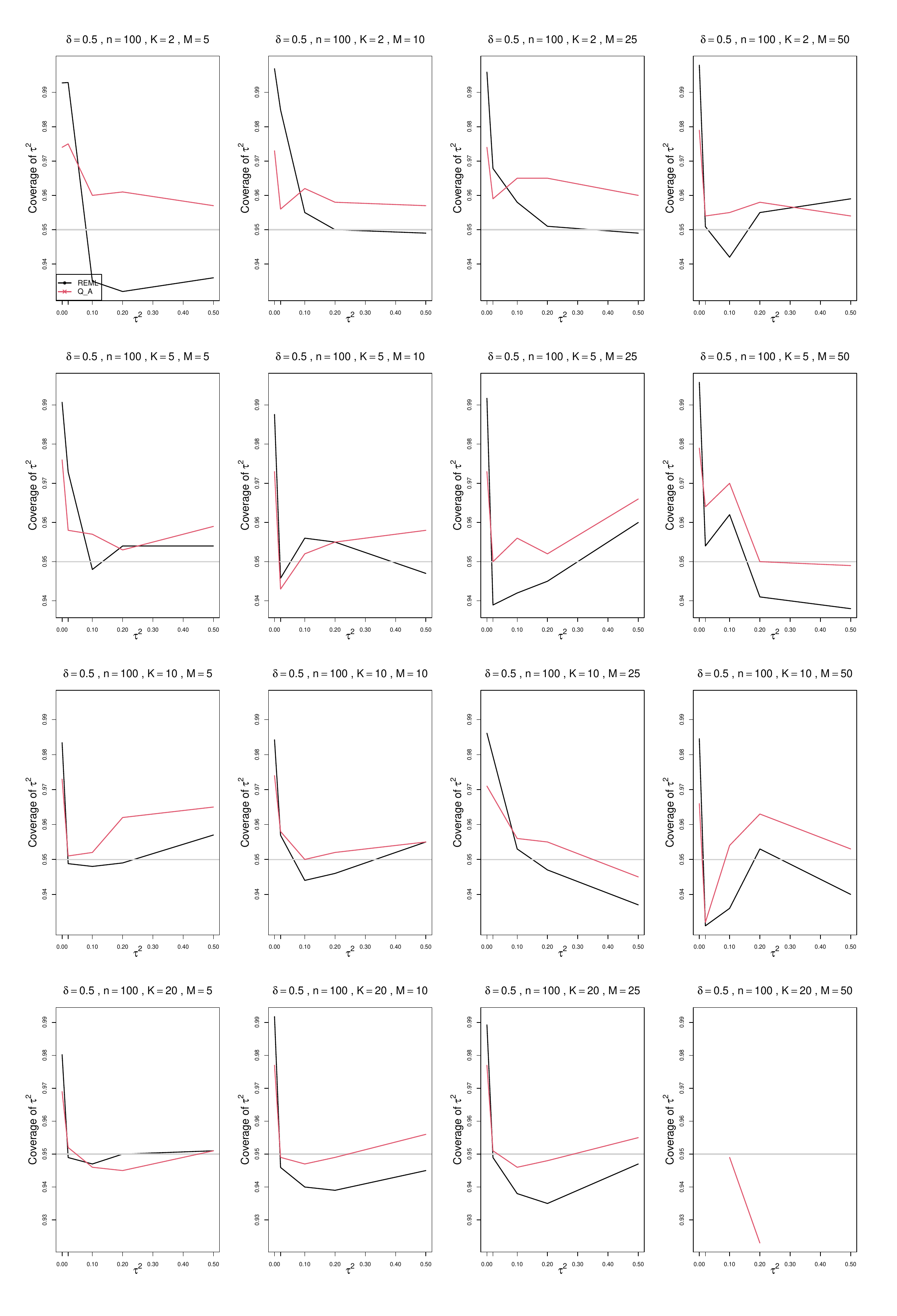}
	\caption{ Coverage of 95\% confidence intervals for between-study variance of SMD (REML and $Q_A$) vs $\tau^2$, for $K$ = 2, 5, 10, and 20 studies per cluster and $M$ = 5, 10, 25, and 50 clusters; $\delta = 0.5$, and the sample size $n$ = 100 in each study.  }
	\label{PlotCoverageOfTau2_100_05_HIER.pdf}
\end{figure}

\begin{figure}[ht]
	\centering
	\includegraphics[scale=0.33]{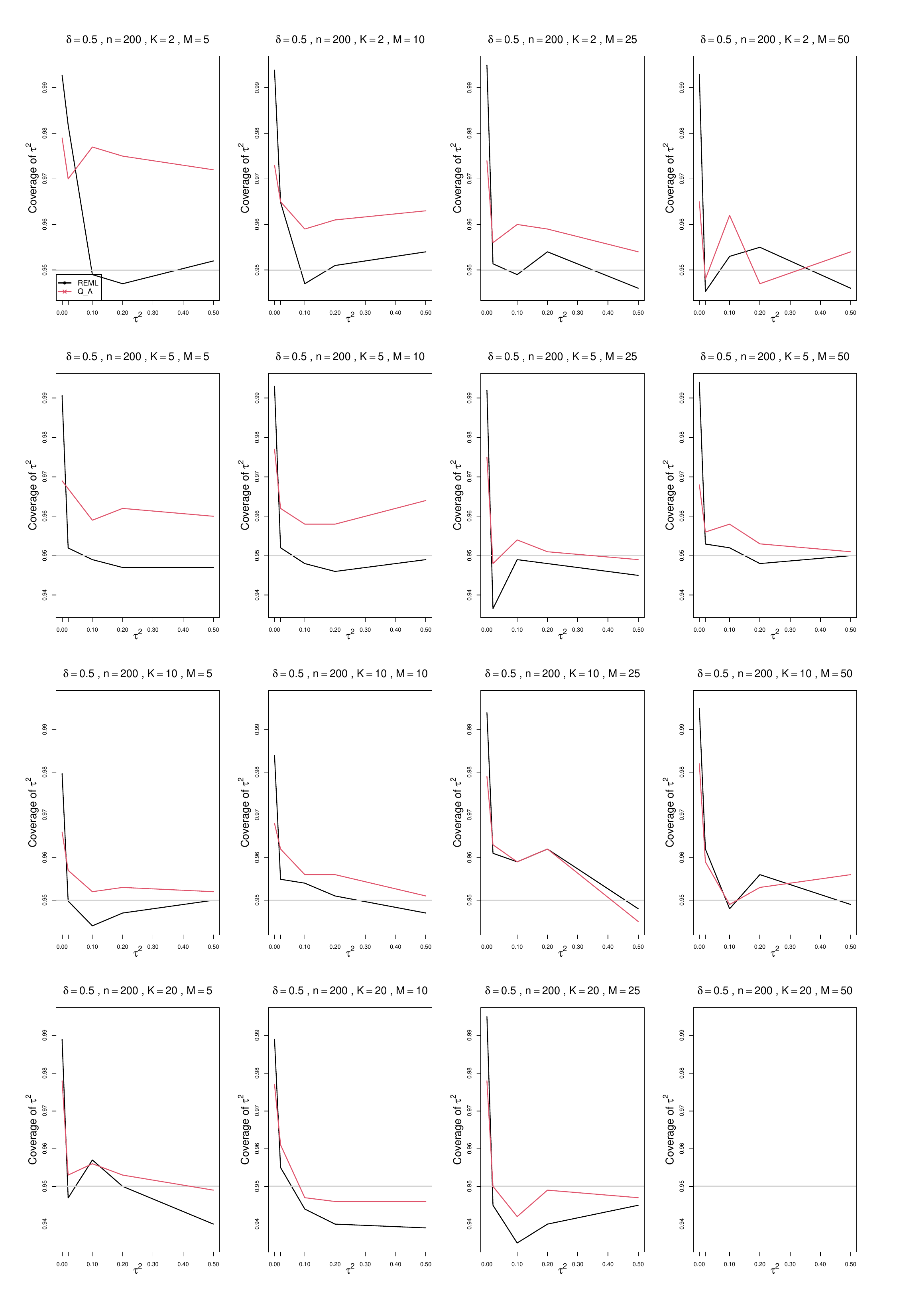}
	\caption{ Coverage of 95\% confidence intervals for between-study variance of SMD (REML and $Q_A$) vs $\tau^2$, for $K$ = 2, 5, 10, and 20 studies per cluster and $M$ = 5, 10, 25, and 50 clusters; $\delta = 0.5$, and the sample size $n$ = 200 in each study. }
	\label{PlotCoverageOfTau2_200_05_HIER.pdf}
\end{figure}

\begin{figure}[ht]
	\centering
	\includegraphics[scale=0.33]{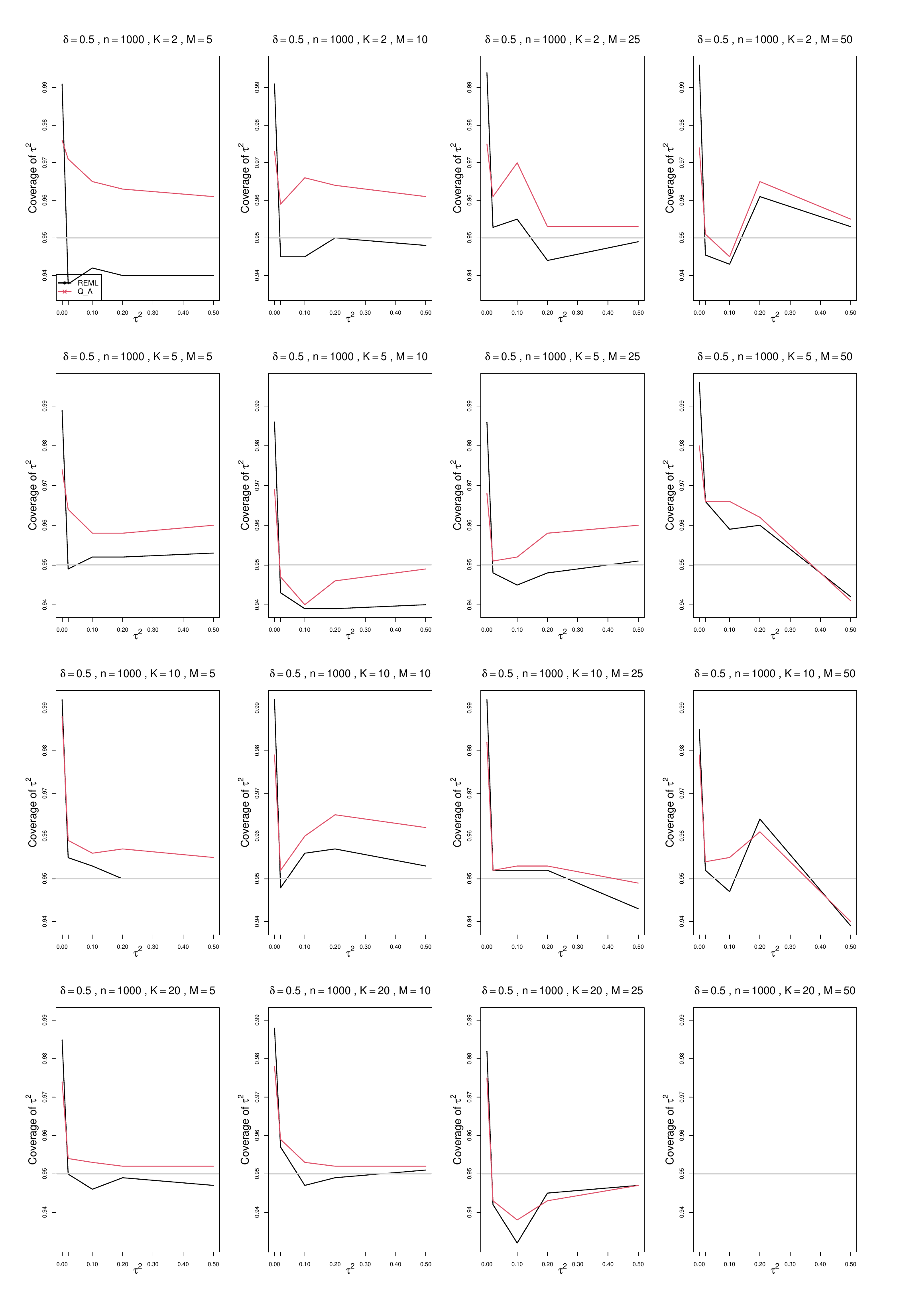}
	\caption{ Coverage of 95\% confidence intervals for between-study variance of SMD (REML and $Q_A$) vs $\tau^2$, for $K$ = 2, 5, 10, and 20 studies per cluster and $M$ = 5, 10, 25, and 50 clusters; $\delta = 0.5$, and the sample size $n$ = 1000 in each study.  }
	\label{PlotCoverageOfTau2_1000_05_HIER.pdf}
\end{figure}

\begin{figure}[ht]
	\centering
	\includegraphics[scale=0.33]{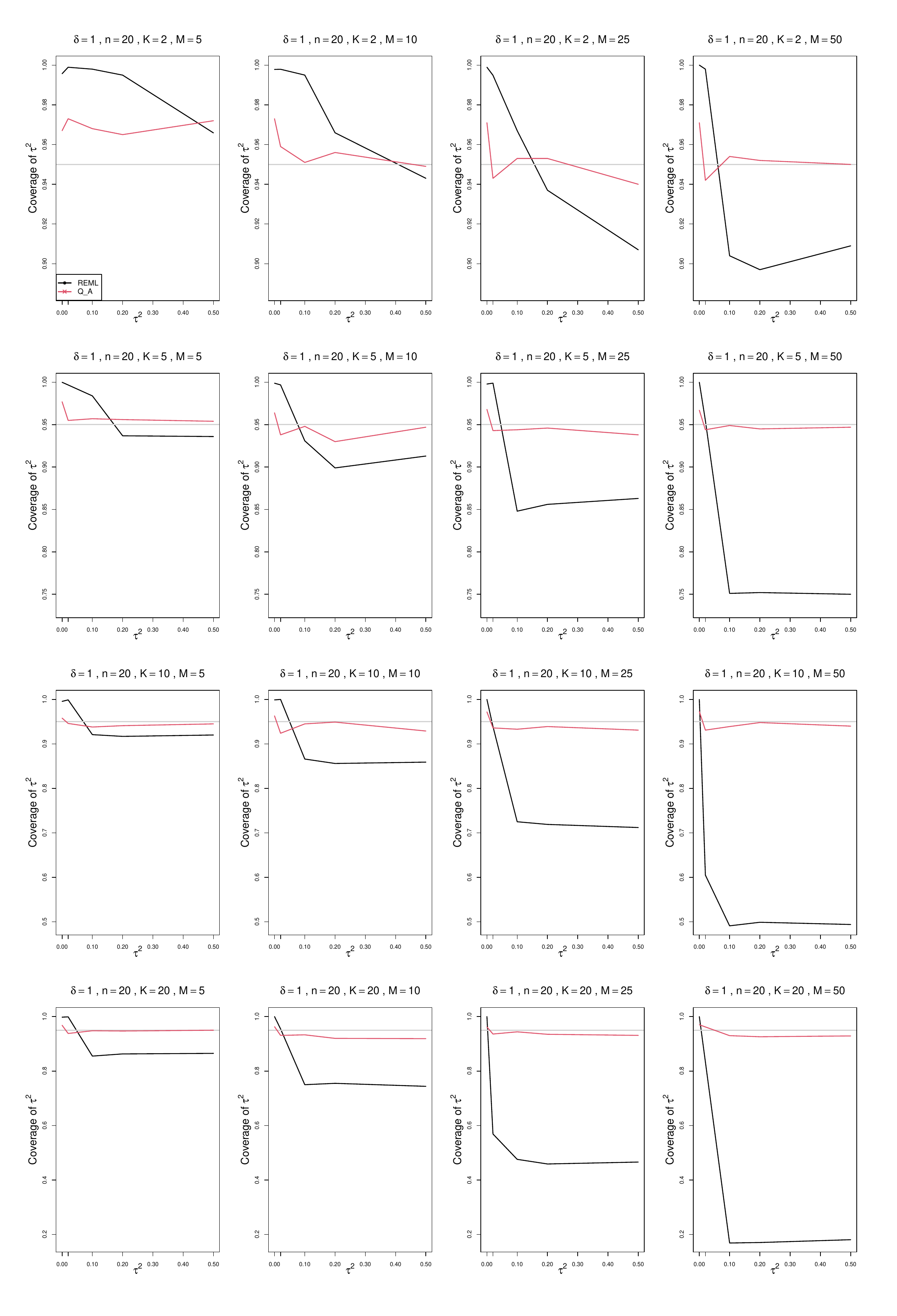}
	\caption{ Coverage of 95\% confidence intervals for between-study variance of SMD (REML and $Q_A$) vs $\tau^2$, for $K$ = 2, 5, 10, and 20 studies per cluster and $M$ = 5, 10, 25, and 50 clusters; $\delta = 1$, and the sample size $n$ = 20 in each study.  }
	\label{PlotCoverageOfTau2_20_1_HIER.pdf}
\end{figure}

\begin{figure}[ht]
	\centering
	\includegraphics[scale=0.33]{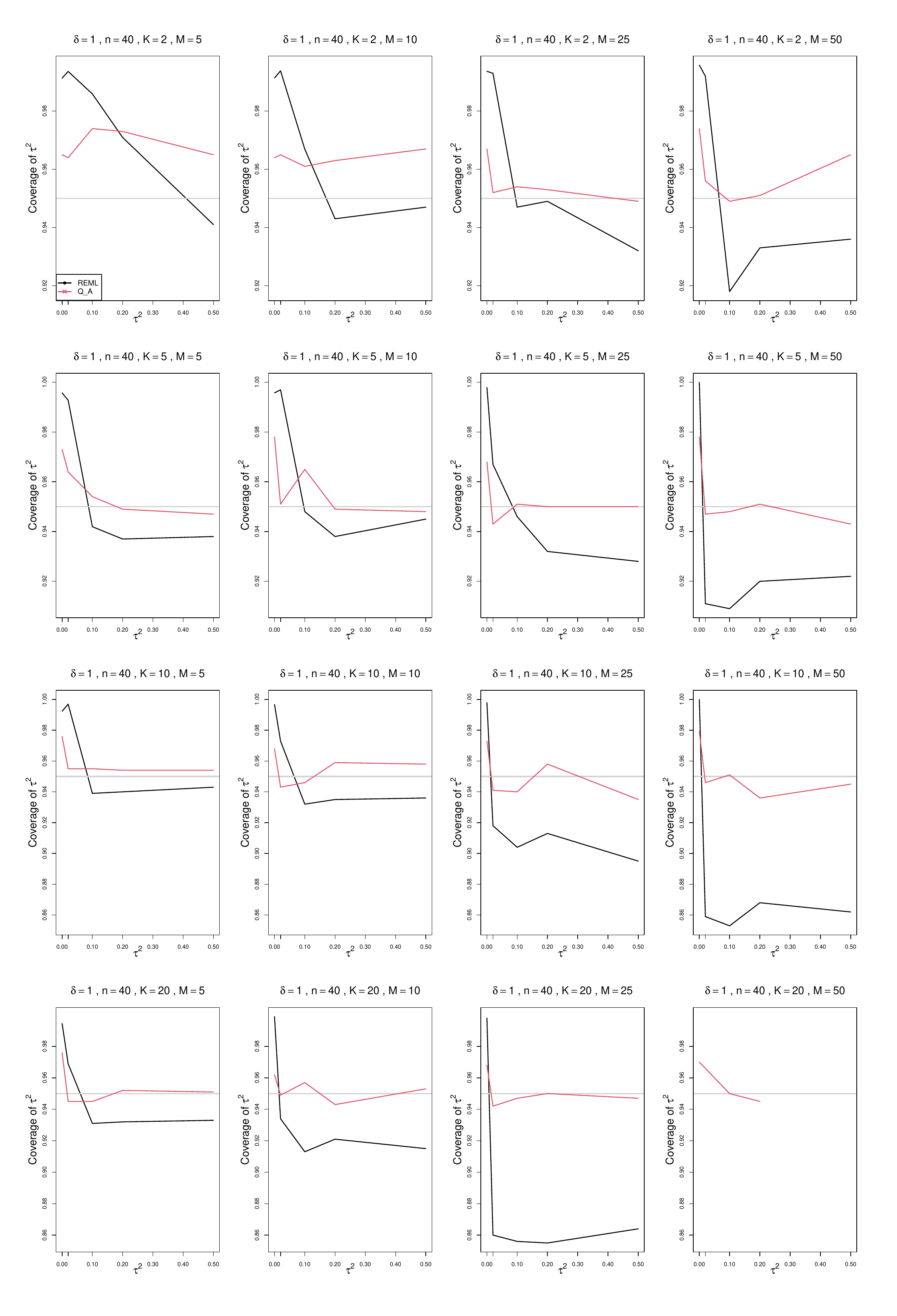}
	\caption{ Coverage of 95\% confidence intervals for between-study variance of SMD (REML and $Q_A$) vs $\tau^2$, for $K$ = 2, 5, 10, and 20 studies per cluster and $M$ = 5, 10, 25, and 50 clusters; $\delta = 1$, and the sample size $n$ = 40 in each study.  }
	\label{PlotCoverageOfTau2_40_1_HIER.pdf}
\end{figure}
\begin{figure}[ht]
	\centering
	\includegraphics[scale=0.33]{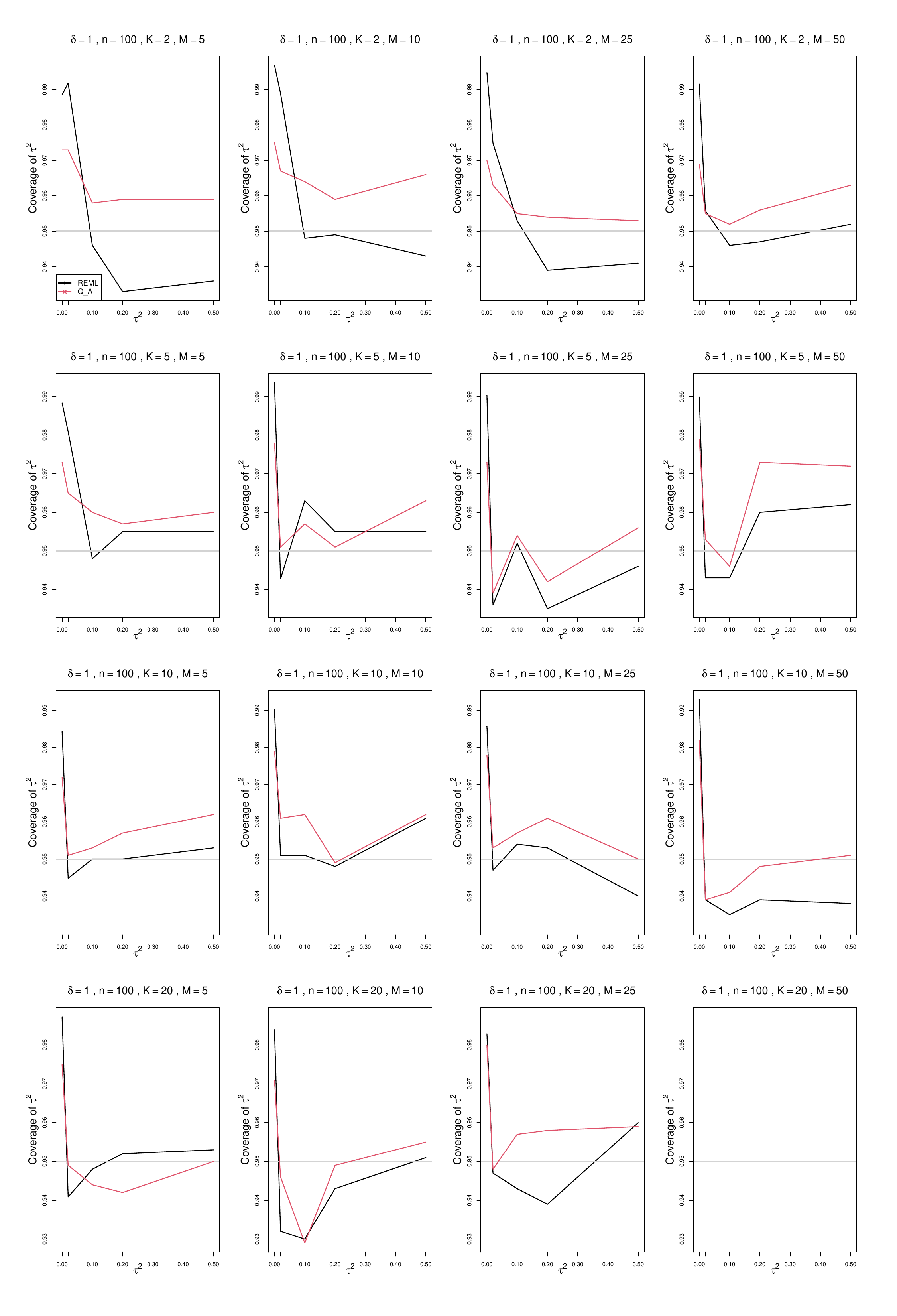}
	\caption{ Coverage of 95\% confidence intervals for between-study variance of SMD (REML and $Q_A$) vs $\tau^2$, for $K$ = 2, 5, 10, and 20 studies per cluster and $M$ = 5, 10, 25, and 50 clusters; $\delta = 1$, and the sample size $n$ = 100 in each study.   }
	\label{PlotCoverageOfTau2_100_1_HIER.pdf}
\end{figure}

\begin{figure}[ht]
	\centering
	\includegraphics[scale=0.33]{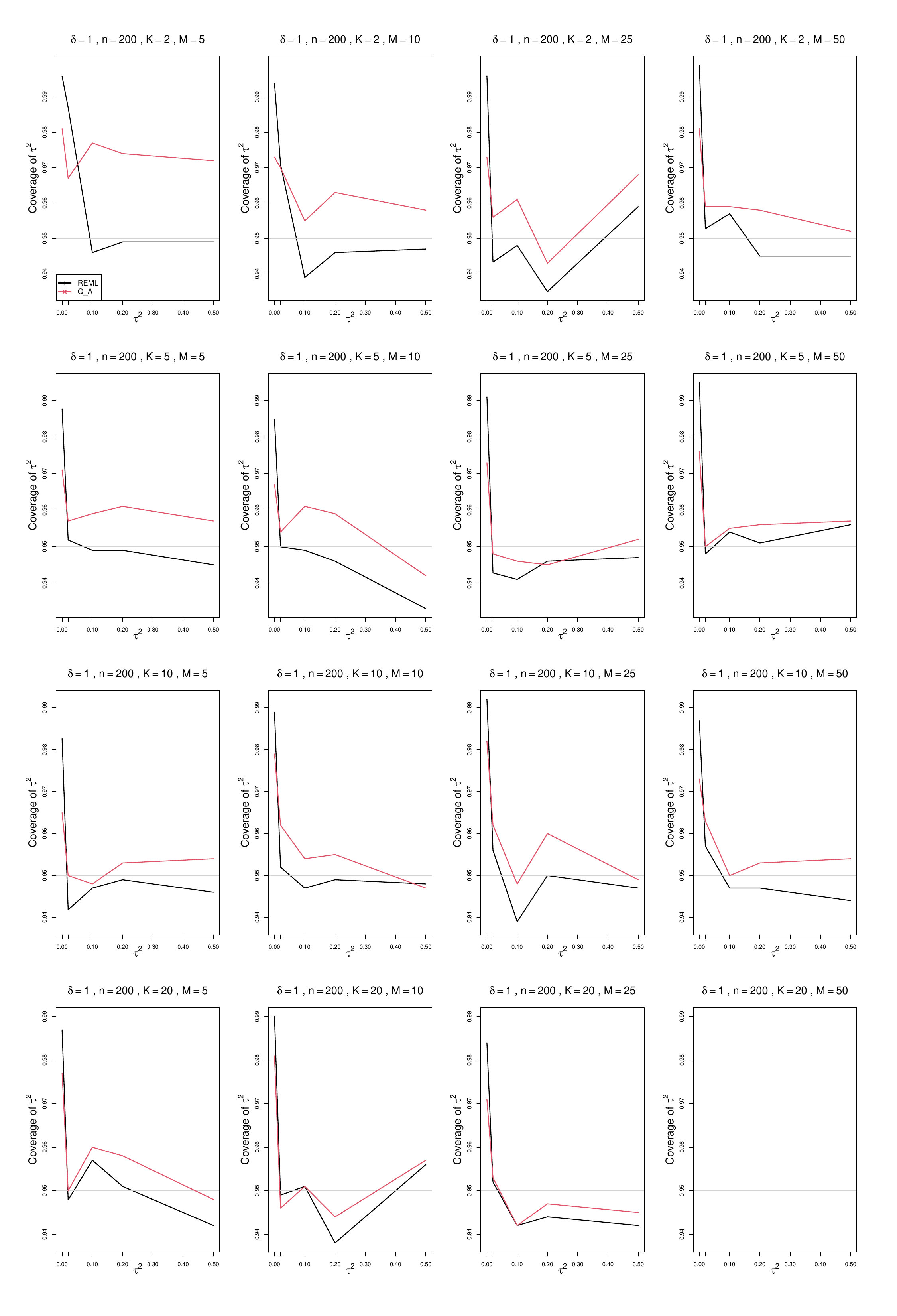}
	\caption{ Coverage of 95\% confidence intervals for between-study variance of SMD (REML and $Q_A$) vs $\tau^2$, for $K$ = 2, 5, 10, and 20 studies per cluster and $M$ = 5, 10, 25, and 50 clusters; $\delta = 1$, and the sample size $n$ = 200 in each study.  }
	\label{PlotCoverageOfTau2_200_1_HIER.pdf}
\end{figure}

\begin{figure}[ht]
	\centering
	\includegraphics[scale=0.33]{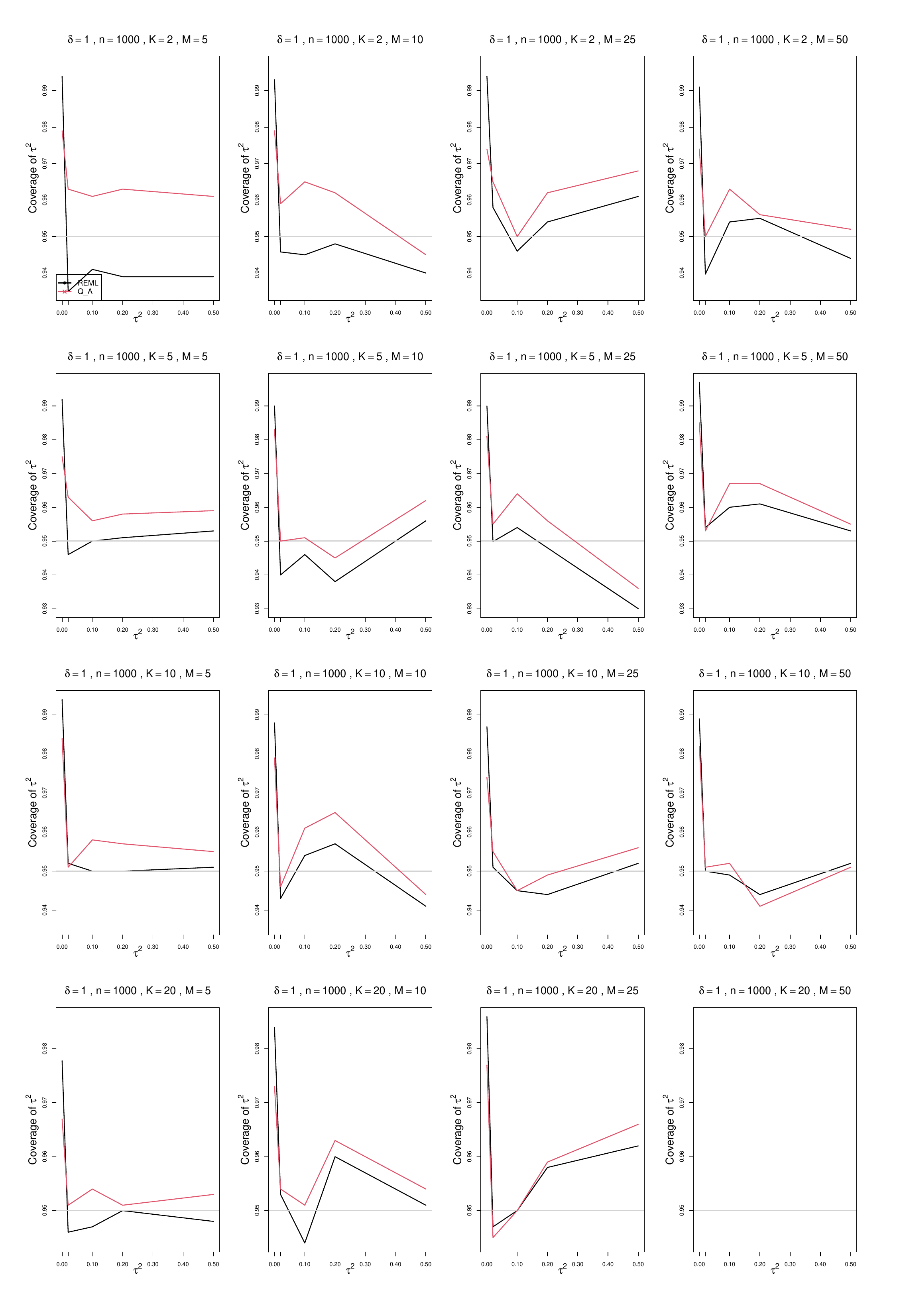}
	\caption{ Coverage of 95\% confidence intervals for between-study variance of SMD (REML and $Q_A$) vs $\tau^2$, for $K$ = 2, 5, 10, and 20 studies per cluster and $M$ = 5, 10, 25, and 50 clusters; $\delta = 1$, and the sample size $n$ = 1000 in each study.  }
	\label{PlotCoverageOfTau2_1000_1_HIER.pdf}
\end{figure}

\clearpage

\section*{Appendix E: Bias in point estimators of the between-cluster variance $\omega^2$}

Each figure corresponds to a value of the standardized mean difference ($\delta$  = 0, 0.2, 0.5, 1) and a value of the study sample size ($n$ = 20, 40, 100, 200, 1000).\\
For each combination of the number of studies in a cluster ($K$ = 2, 5, 10, 20) and the number of clusters ($M$ = 5, 10, 25, 50), a panel plots bias versus $\omega^2$ (= 0, 0.02, 0.1, 0.2, 0.5).\\
The two variance components are held equal ($\tau^2 = \omega^2$). \\
The point estimators of $\omega^2$ are
\begin{itemize}
\item REML method, inverse-variance weights,  {\it  rma.mv} in {\it metafor})
\item $Q_F$ (conditional moment-based method, effective-sample-size weights)
\end{itemize}

\clearpage
\setcounter{figure}{0}
\renewcommand{\thefigure}{E.\arabic{figure}}


\begin{figure}[ht]
	\centering
	\includegraphics[scale=0.33]{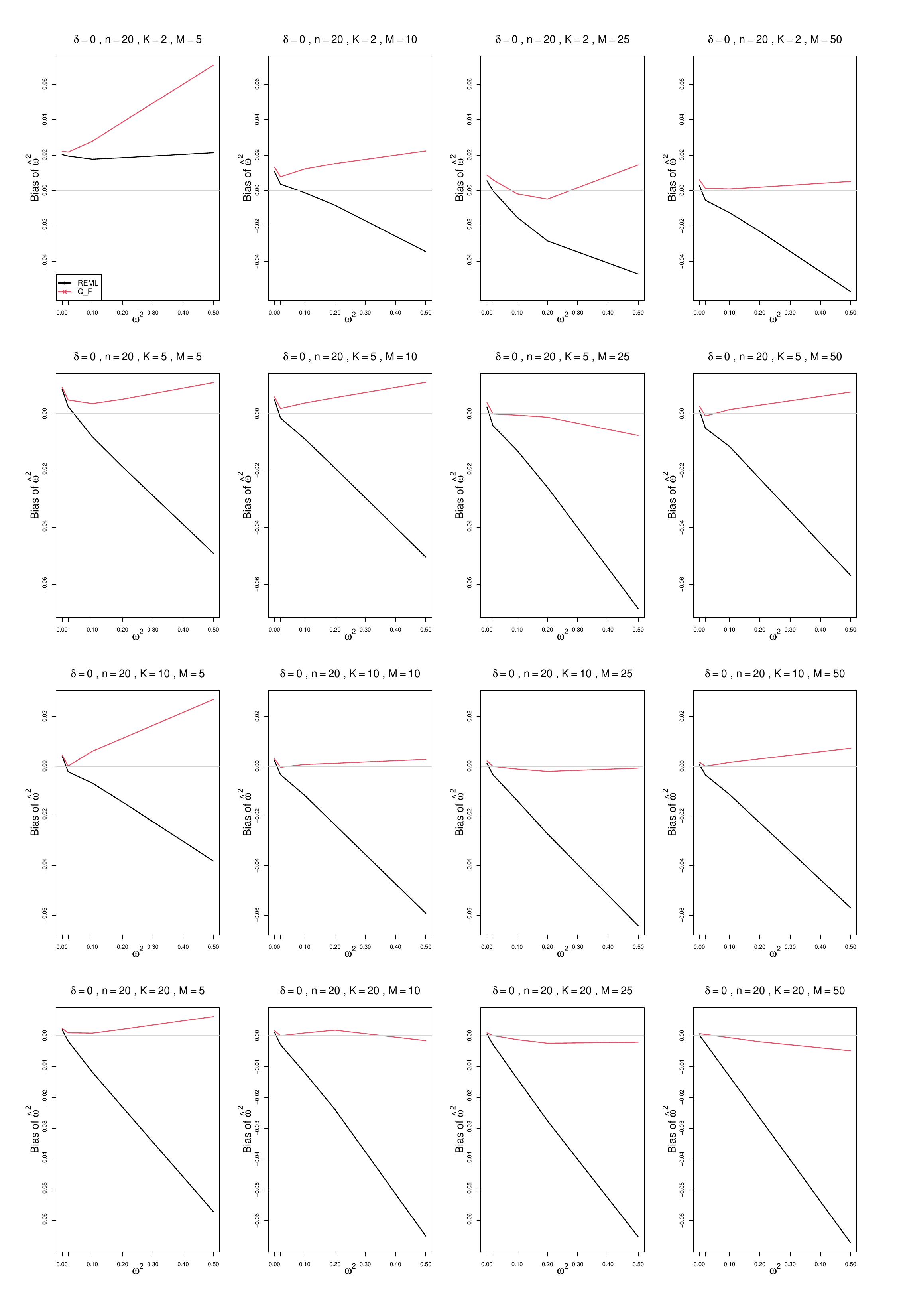}
	\caption{Bias of estimators of between-cluster  variance of SMD (REML and $Q_F$ ) vs $\omega^2$, for $K$ = 2, 5, 10, and 20 studies per cluster and $M$ = 5, 10, 25, and 50 clusters; $\delta = 0$, and the sample size $n$ = 20 in each study.  }
	\label{PlotBiasOfOmega2_20_0_HIER.pdf}
\end{figure}

\begin{figure}[ht]
	\centering
	\includegraphics[scale=0.33]{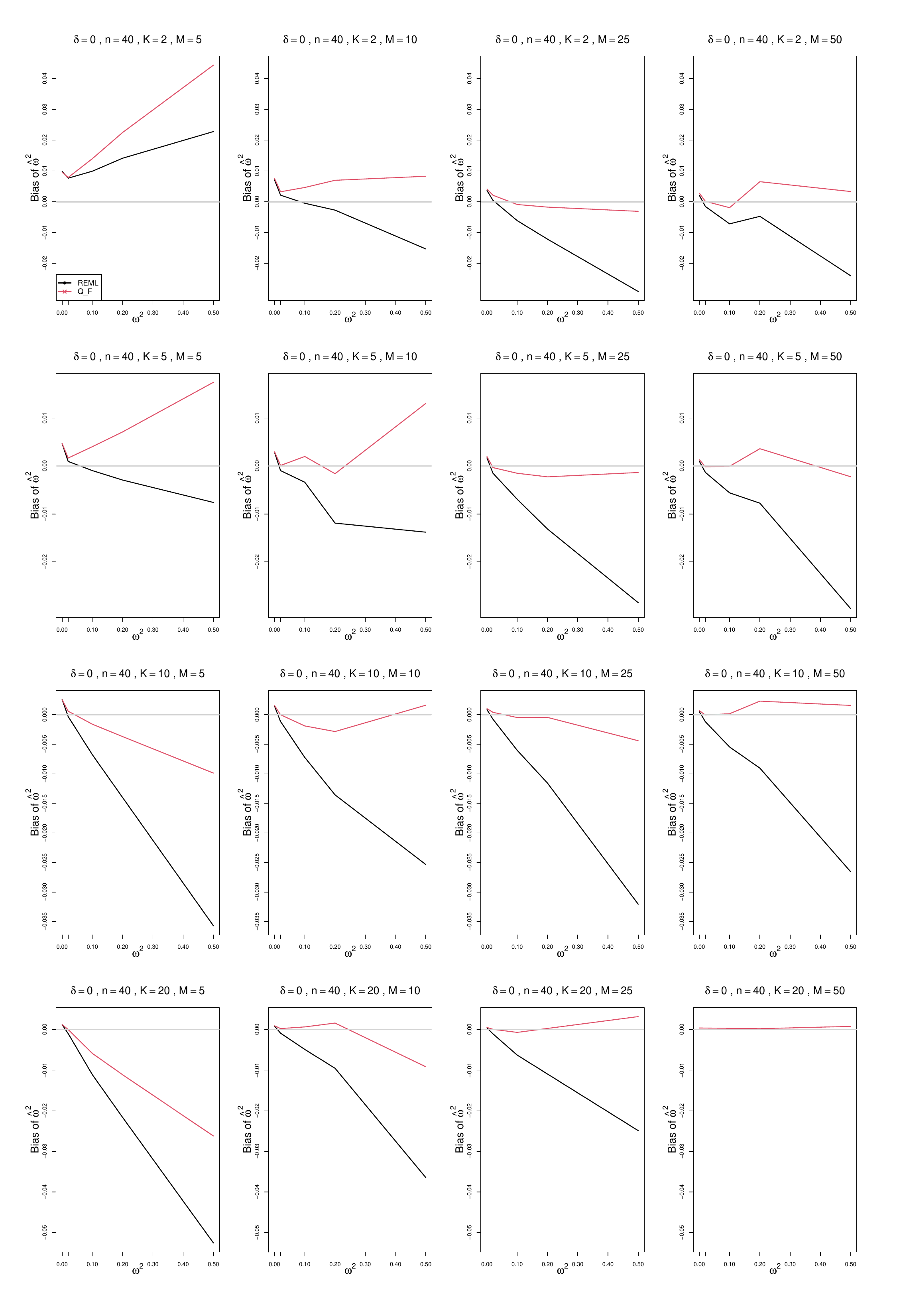}
	\caption{Bias  of estimators of between-cluster  variance of SMD (REML and $Q_F$ ) vs $\omega^2$, for $K$ = 2, 5, 10, and 20 studies per cluster and $M$ = 5, 10, 25, and 50 clusters; $\delta = 0$, and the sample size $n$ = 40 in each study.  }
	\label{PlotBiasOfOmega2_40_0_HIER.pdf}
\end{figure}

\begin{figure}[ht]
	\centering
	\includegraphics[scale=0.33]{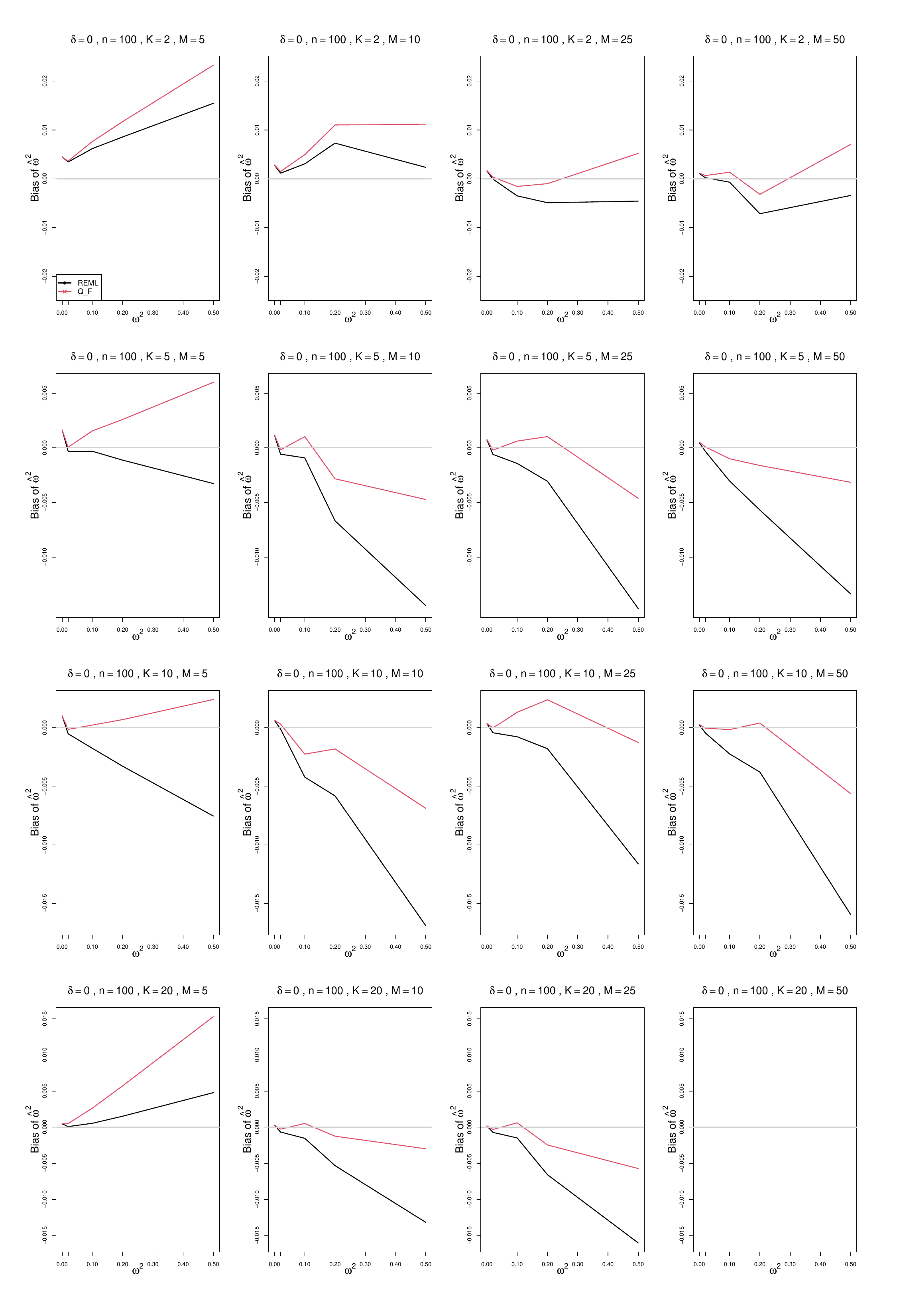}
	\caption{Bias  of estimators of between-cluster  variance of SMD (REML and $Q_F$ ) vs $\omega^2$, for $K$ = 2, 5, 10, and 20 studies per cluster and $M$ = 5, 10, 25, and 50 clusters; $\delta = 0$, and the sample size $n$ = 100 in each study. }
	\label{PlotBiasOfOmega2_100_0_HIER.pdf}
\end{figure}

\begin{figure}[ht]
	\centering
	\includegraphics[scale=0.33]{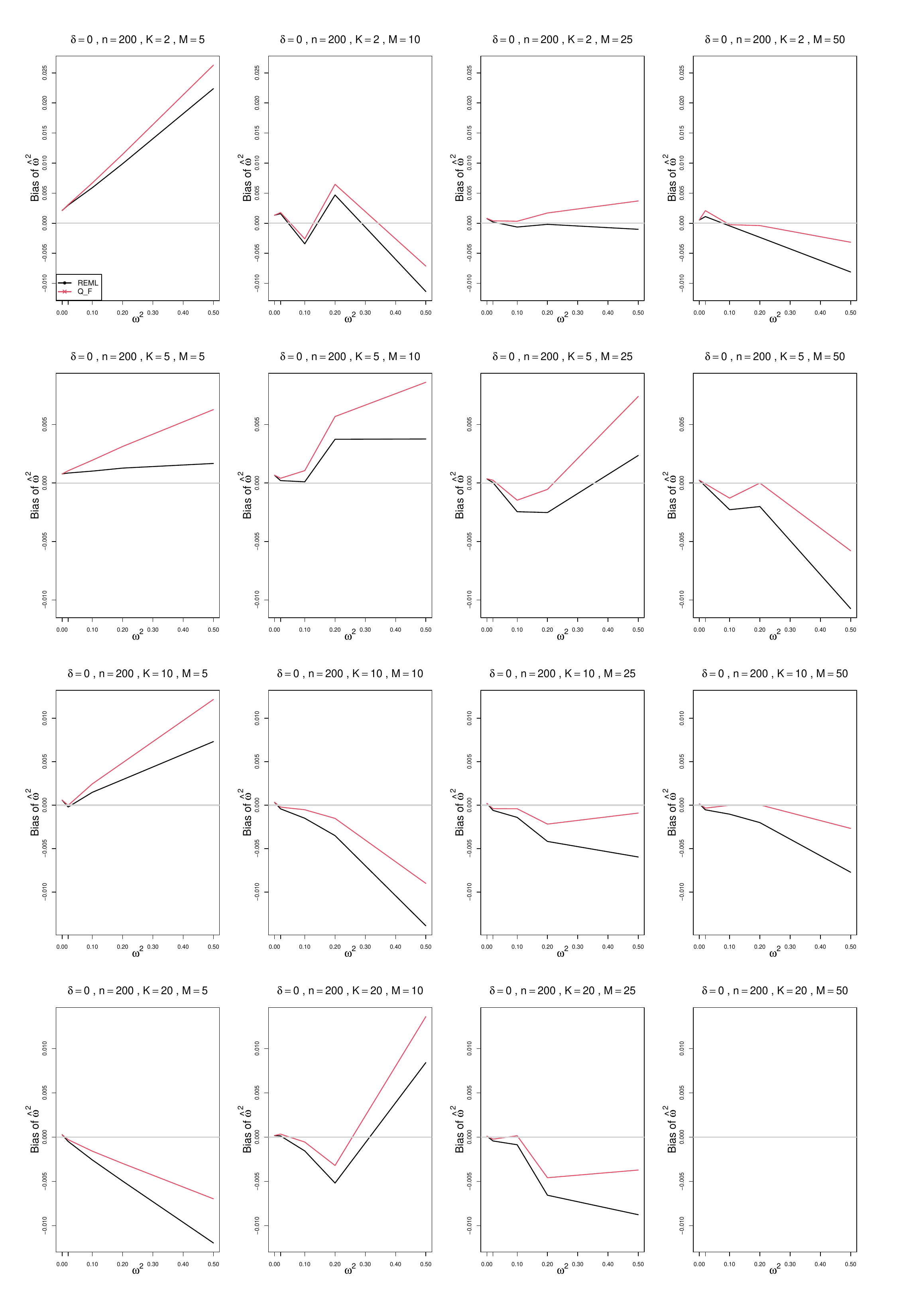}
	\caption{Bias  of estimators of between-cluster  variance of SMD (REML and $Q_F$ ) vs $\omega^2$, for $K$ = 2, 5, 10, and 20 studies per cluster and $M$ = 5, 10, 25, and 50 clusters; $\delta = 0$, and the sample size $n$ = 200 in each study.  }
	\label{PlotBiasOfOmega2_200_0_HIER.pdf}
\end{figure}

\begin{figure}[ht]
	\centering
	\includegraphics[scale=0.33]{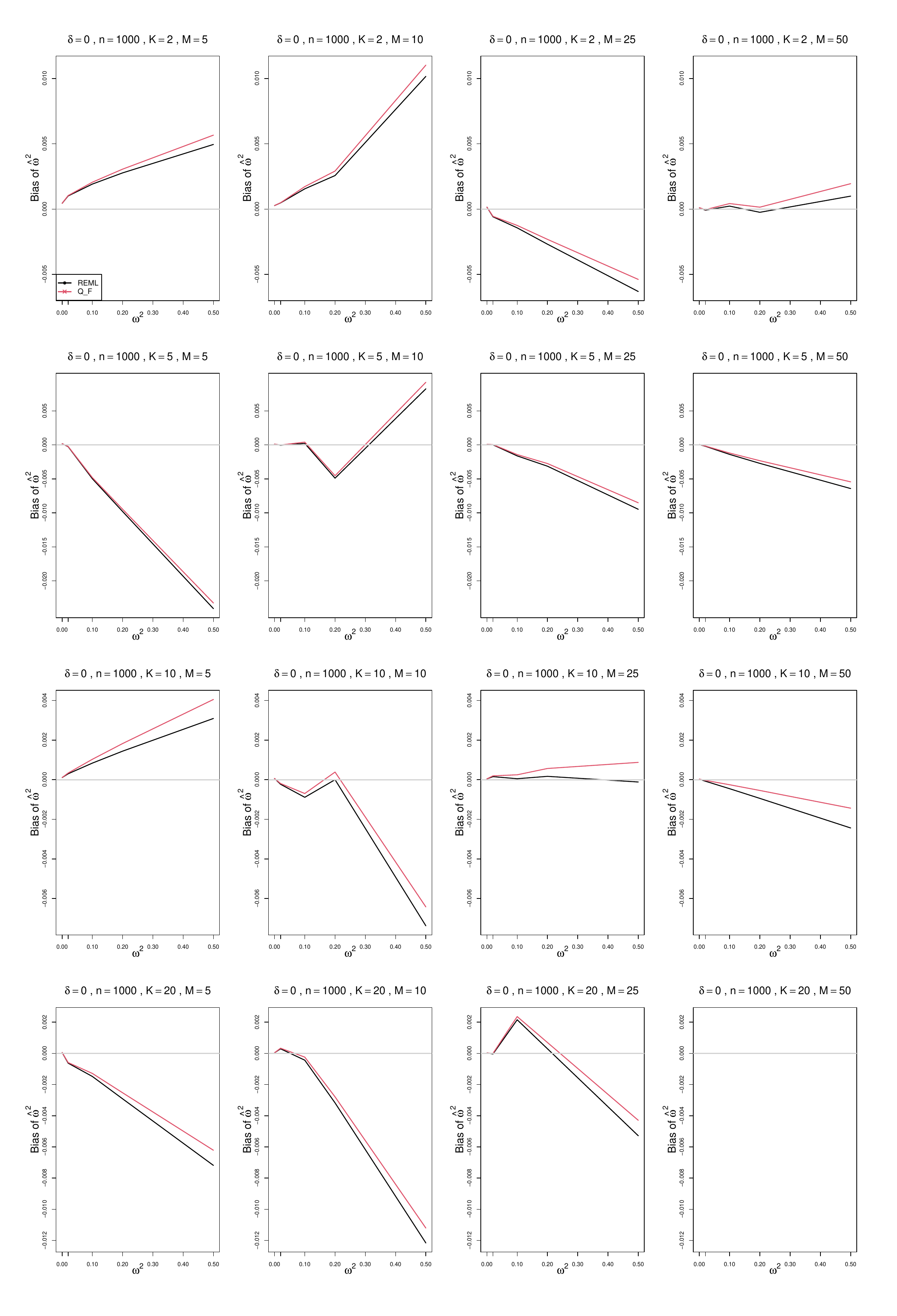}
	\caption{Bias  of estimators of between-cluster  variance of SMD (REML and $Q_F$ ) vs $\omega^2$, for $K$ = 2, 5, 10, and 20 studies per cluster and $M$ = 5, 10, 25, and 50 clusters; $\delta = 0$, and the sample size $n$ = 1000 in each study.  }
	\label{PlotBiasOfOmega2_1000_0_HIER.pdf}
\end{figure}

\begin{figure}[ht]
	\centering
	\includegraphics[scale=0.33]{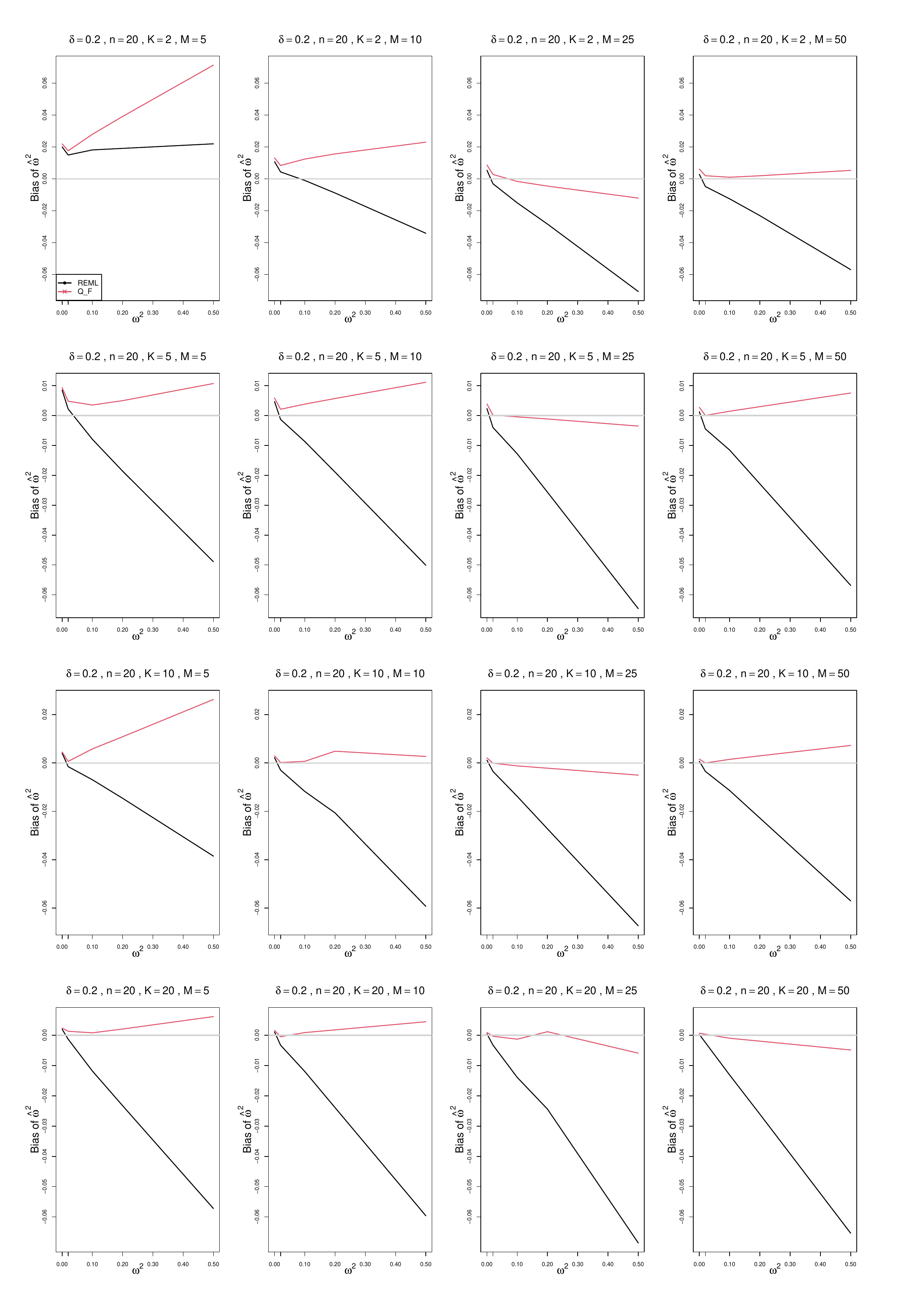}
	\caption{Bias  of estimators of between-cluster  variance of SMD (REML and $Q_F$ ) vs $\omega^2$, for $K$ = 2, 5, 10, and 20 studies per cluster and $M$ = 5, 10, 25, and 50 clusters; $\delta = 0.2$, and the sample size $n$ = 20 in each study.  }
	\label{PlotBiasOfOmega2_20_02_HIER.pdf}
\end{figure}

\begin{figure}[ht]
	\centering
	\includegraphics[scale=0.33]{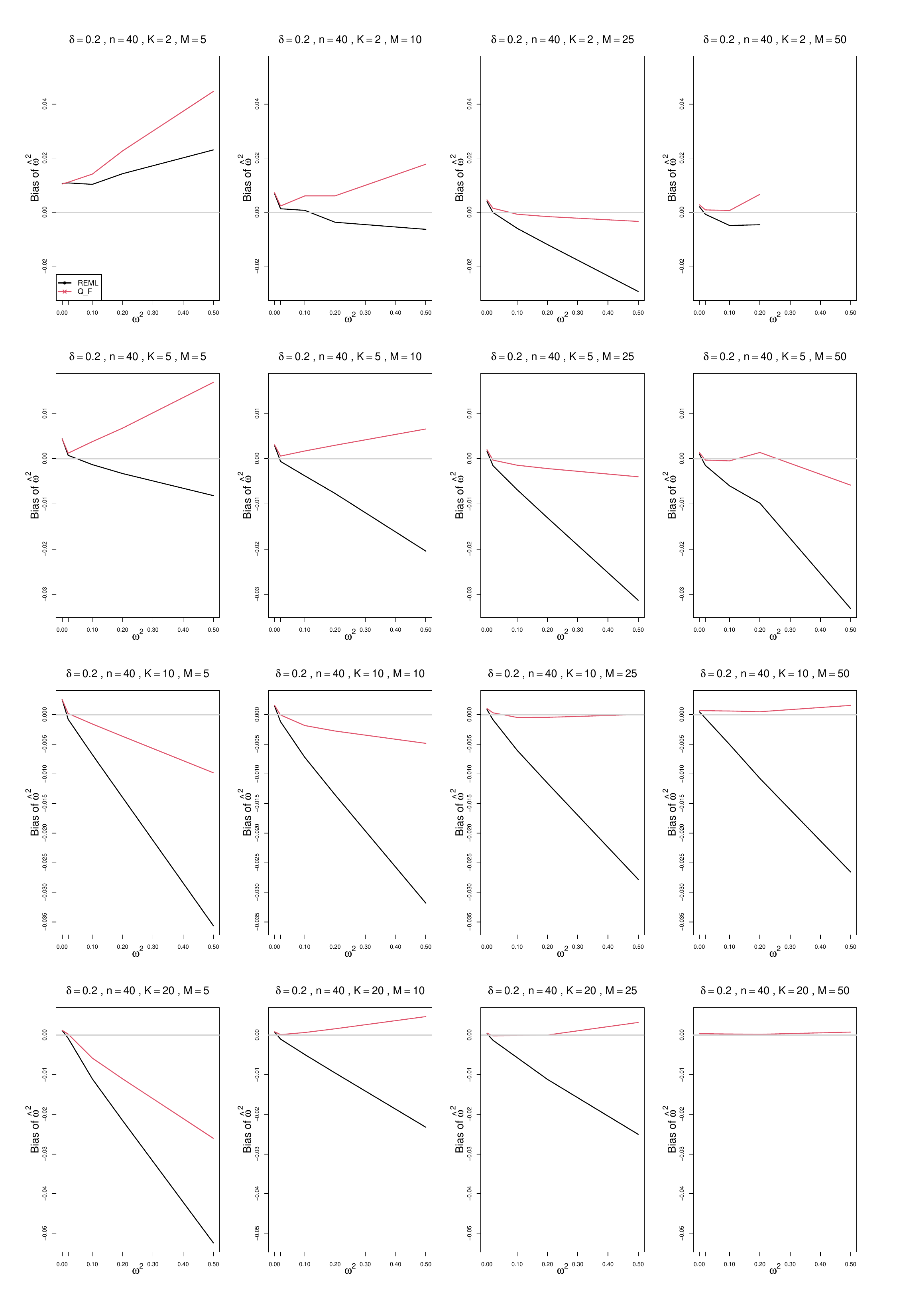}
	\caption{Bias  of estimators of between-cluster  variance of SMD (REML and $Q_F$ ) vs $\omega^2$, for $K$ = 2, 5, 10, and 20 studies per cluster and $M$ = 5, 10, 25, and 50 clusters; $\delta = 0.2$, and the sample size $n$ = 40 in each study.  }
	\label{PlotBiasOfOmega2_40_02_HIER.pdf}
\end{figure}

\begin{figure}[ht]
	\centering
	\includegraphics[scale=0.33]{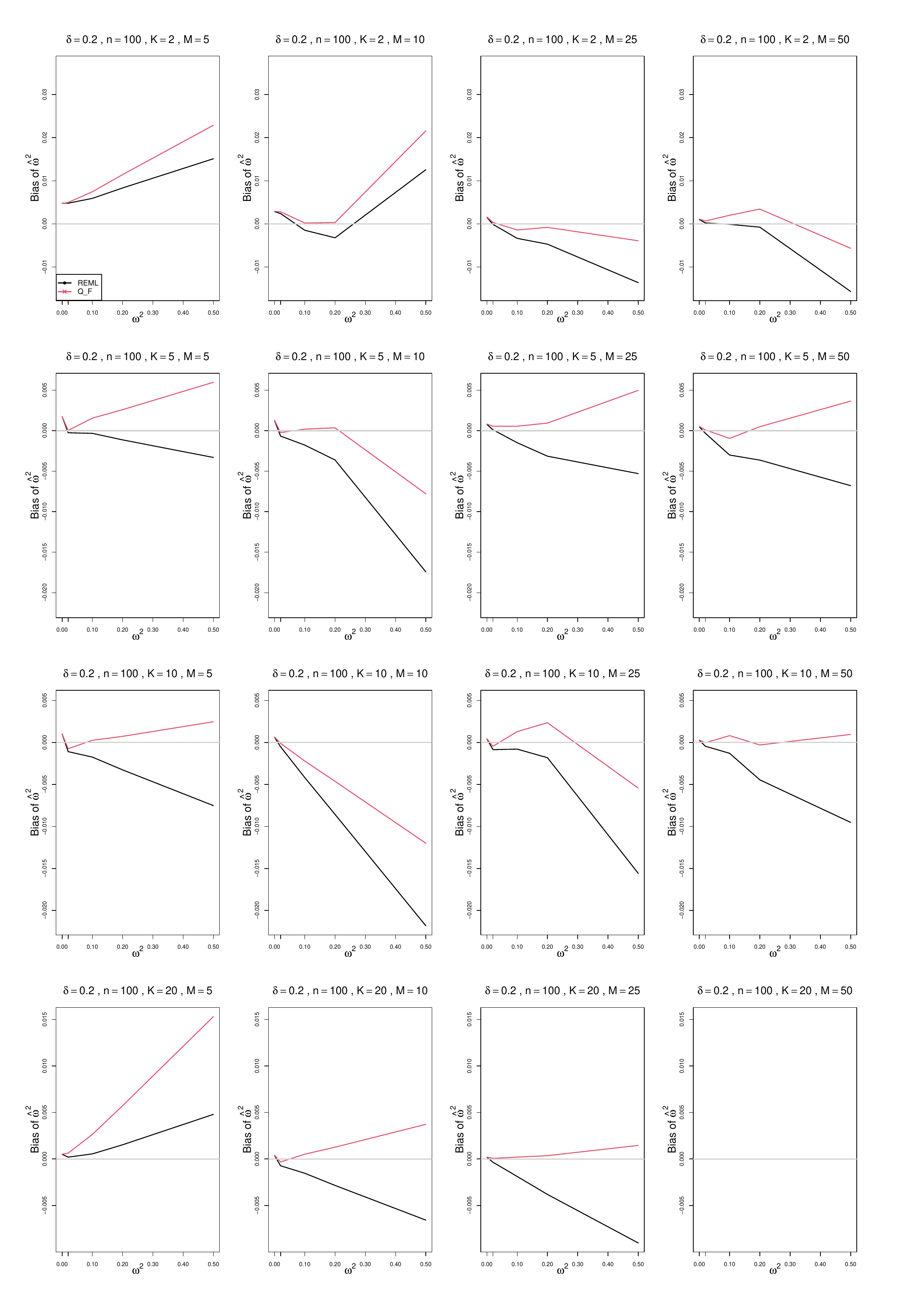}
	\caption{Bias  of estimators of between-cluster  variance of SMD (REML and $Q_F$ ) vs $\omega^2$, for $K$ = 2, 5, 10, and 20 studies per cluster and $M$ = 5, 10, 25, and 50 clusters; $\delta = 0.2$, and the sample size $n$ = 100 in each study.  }
	\label{PlotBiasOfOmega2_100_02_HIER.pdf}
\end{figure}

\begin{figure}[ht]
	\centering
	\includegraphics[scale=0.33]{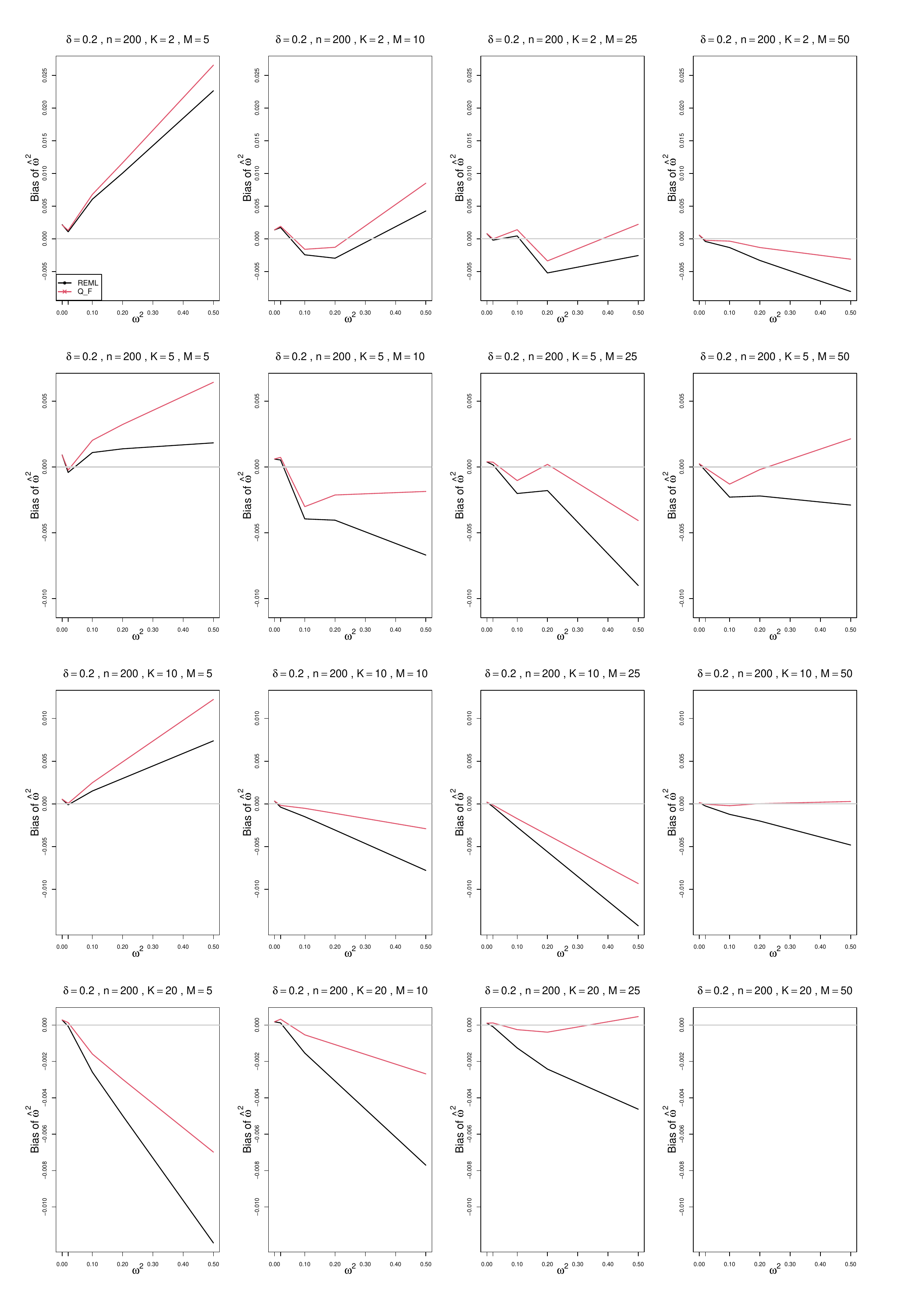}
	\caption{Bias  of estimators of between-cluster  variance of SMD (REML and $Q_F$ ) vs $\omega^2$, for $K$ = 2, 5, 10, and 20 studies per cluster and $M$ = 5, 10, 25, and 50 clusters; $\delta = 0.2$, and the sample size $n$ = 200 in each study.   }
	\label{PlotBiasOfOmega2_200_02_HIER.pdf}
\end{figure}

\begin{figure}[ht]
	\centering
	\includegraphics[scale=0.33]{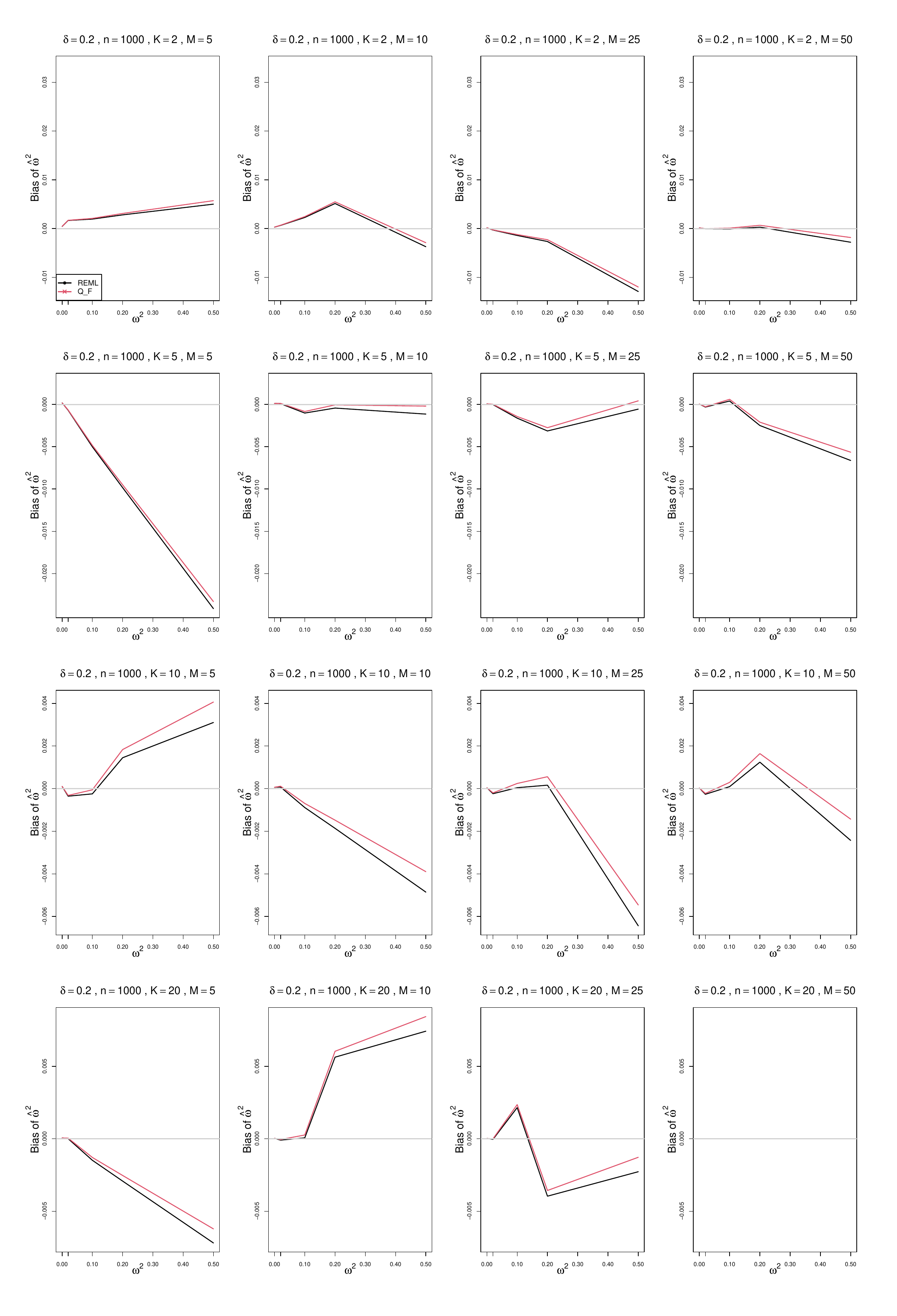}
	\caption{Bias  of estimators of between-cluster  variance of SMD (REML and $Q_F$ ) vs $\omega^2$, for $K$ = 2, 5, 10, and 20 studies per cluster and $M$ = 5, 10, 25, and 50 clusters; $\delta = 0.2$, and the sample size $n$ = 1000 in each study.  }
	\label{PlotBiasOfOmega2_1000_02_HIER.pdf}
\end{figure}

\begin{figure}[ht]
	\centering
	\includegraphics[scale=0.33]{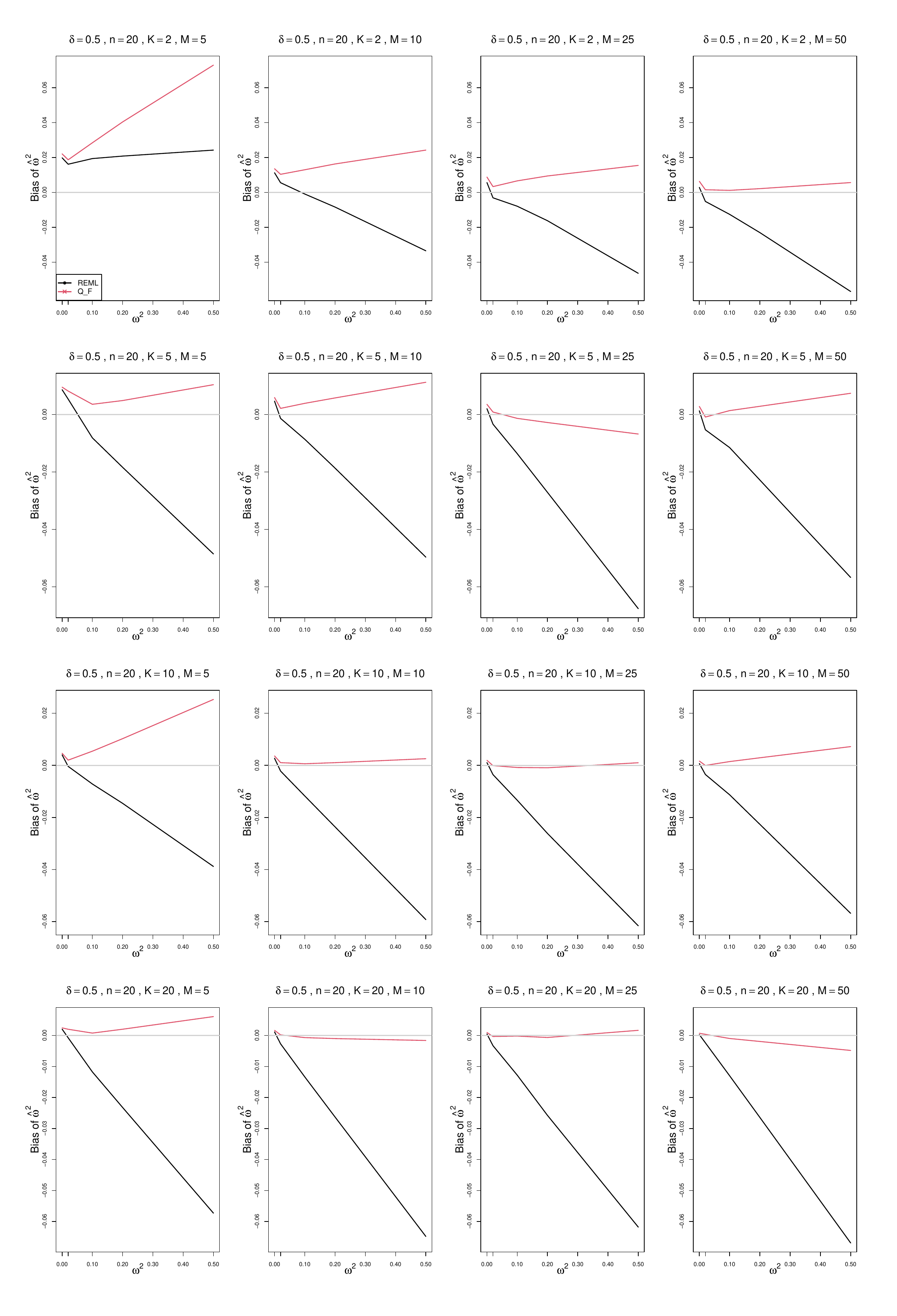}
	\caption{Bias  of estimators of between-cluster  variance of SMD (REML and $Q_F$ ) vs $\omega^2$, for $K$ = 2, 5, 10, and 20 studies per cluster and $M$ = 5, 10, 25, and 50 clusters; $\delta = 0.5$, and the sample size $n$ = 20 in each study.  }
	\label{PlotBiasOfOmega2_20_05_HIER.pdf}
\end{figure}

\begin{figure}[ht]
	\centering
	\includegraphics[scale=0.33]{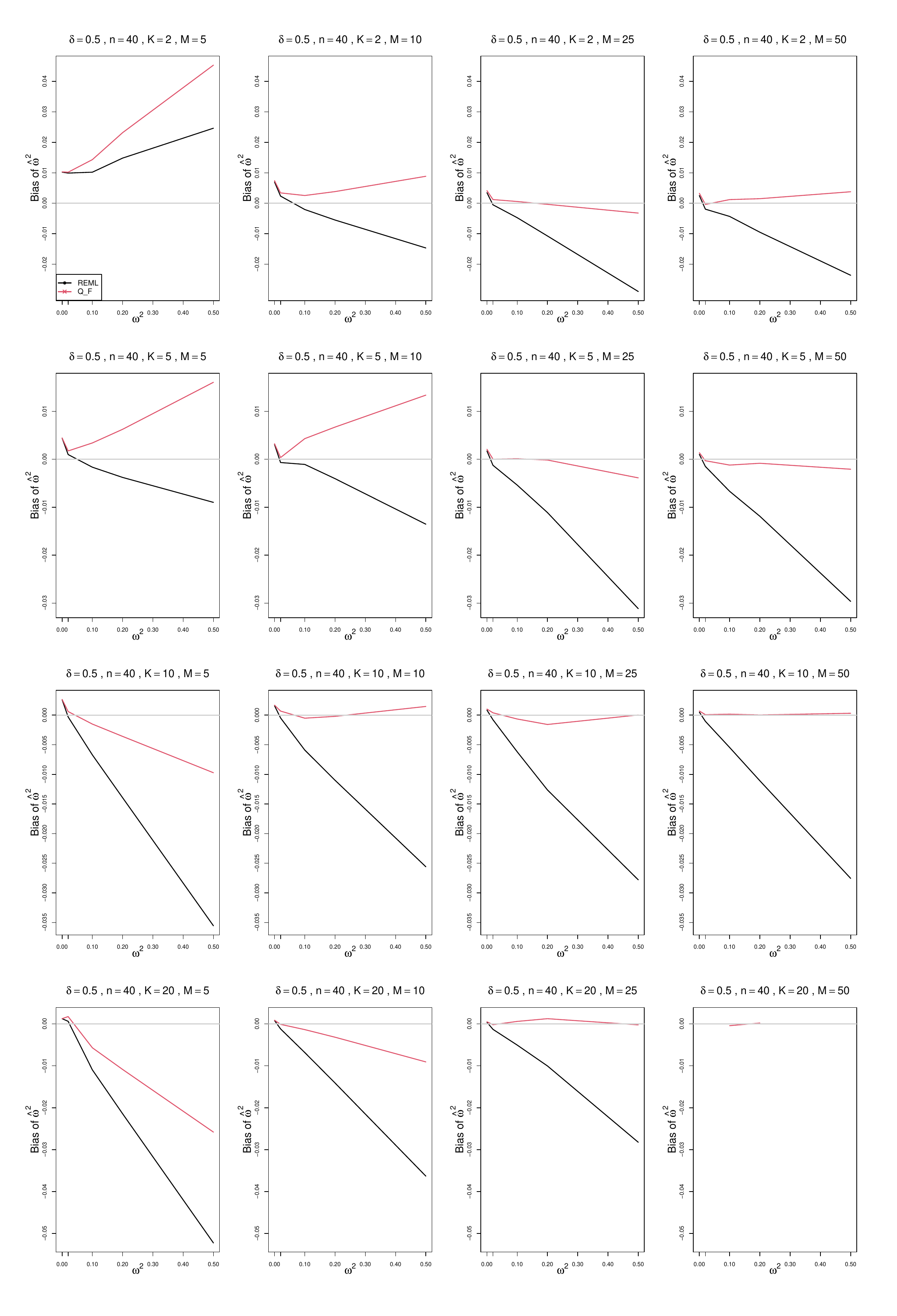}
	\caption{Bias  of estimators of between-cluster  variance of SMD (REML and $Q_F$ ) vs $\omega^2$, for $K$ = 2, 5, 10, and 20 studies per cluster and $M$ = 5, 10, 25, and 50 clusters; $\delta = 0.5$, and the sample size $n$ = 40 in each study.  }
	\label{PlotBiasOfOmega2_40_05_HIER.pdf}
\end{figure}

\begin{figure}[ht]
	\centering
	\includegraphics[scale=0.33]{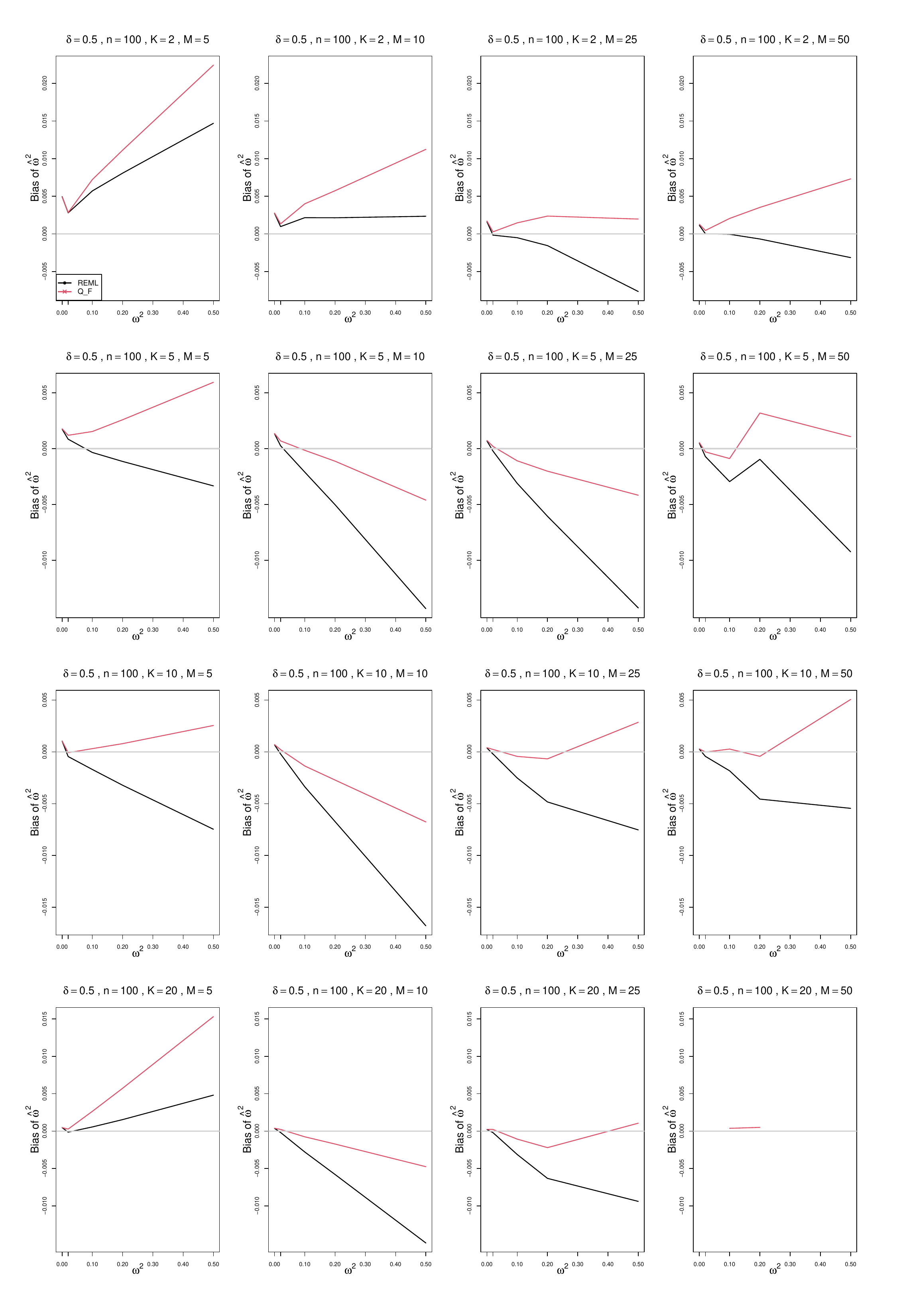}
	\caption{Bias  of estimators of between-cluster  variance of SMD (REML and $Q_F$ ) vs $\omega^2$, for $K$ = 2, 5, 10, and 20 studies per cluster and $M$ = 5, 10, 25, and 50 clusters; $\delta = 0.5$, and the sample size $n$ = 100 in each study.  }
	\label{PlotBiasOfOmega2_100_05_HIER.pdf}
\end{figure}

\begin{figure}[ht]
	\centering
	\includegraphics[scale=0.33]{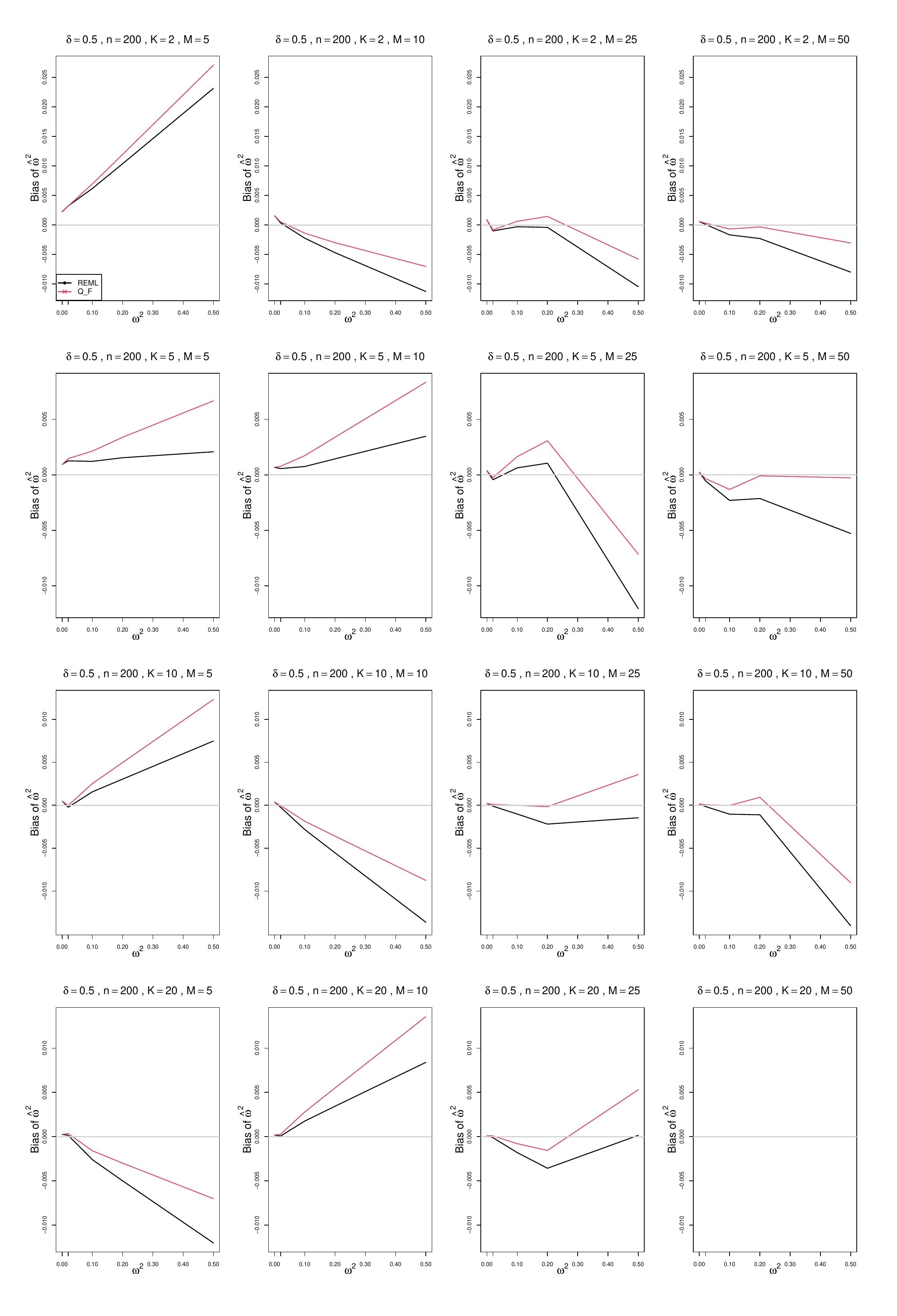}
	\caption{Bias  of estimators of between-cluster  variance of SMD (REML and $Q_F$) vs $\omega^2$, for $K$ = 2, 5, 10, and 20 studies per cluster and $M$ = 5, 10, 25, and 50 clusters; $\delta = 0.5$, and the sample size $n$ = 200 in each study.  }
	\label{PlotBiasOfOmega2_200_05_HIER.pdf}
\end{figure}

\begin{figure}[ht]
	\centering
	\includegraphics[scale=0.33]{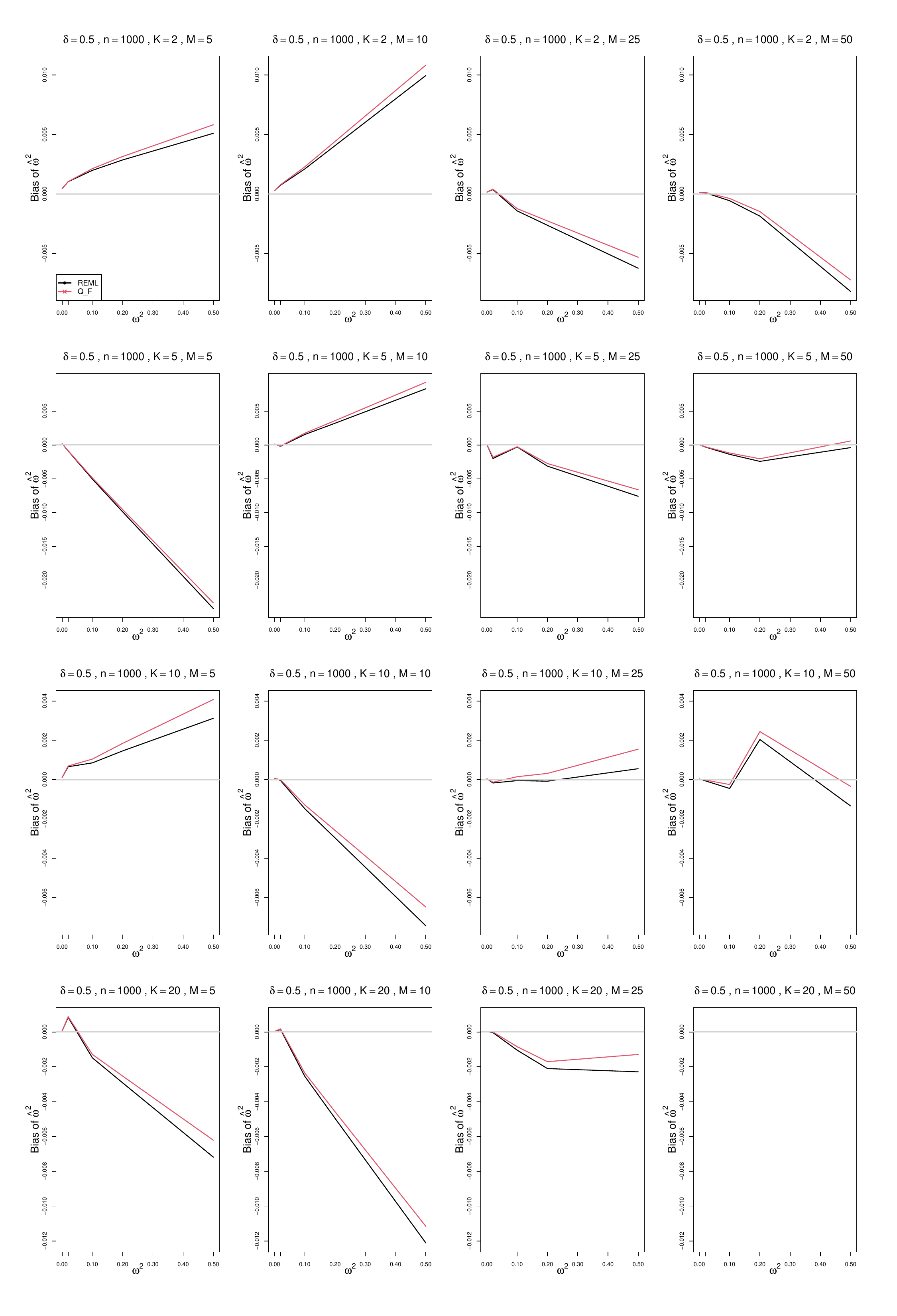}
	\caption{Bias  of estimators of between-cluster  variance of SMD (REML and $Q_F$ ) vs $\omega^2$, for $K$ = 2, 5, 10, and 20 studies per cluster and $M$ = 5, 10, 25, and 50 clusters; $\delta = 0.5$, and the sample size $n$ = 1000 in each study.  }
	\label{PlotBiasOfOmega2_1000_05_HIER.pdf}
\end{figure}

\begin{figure}[ht]
	\centering
	\includegraphics[scale=0.33]{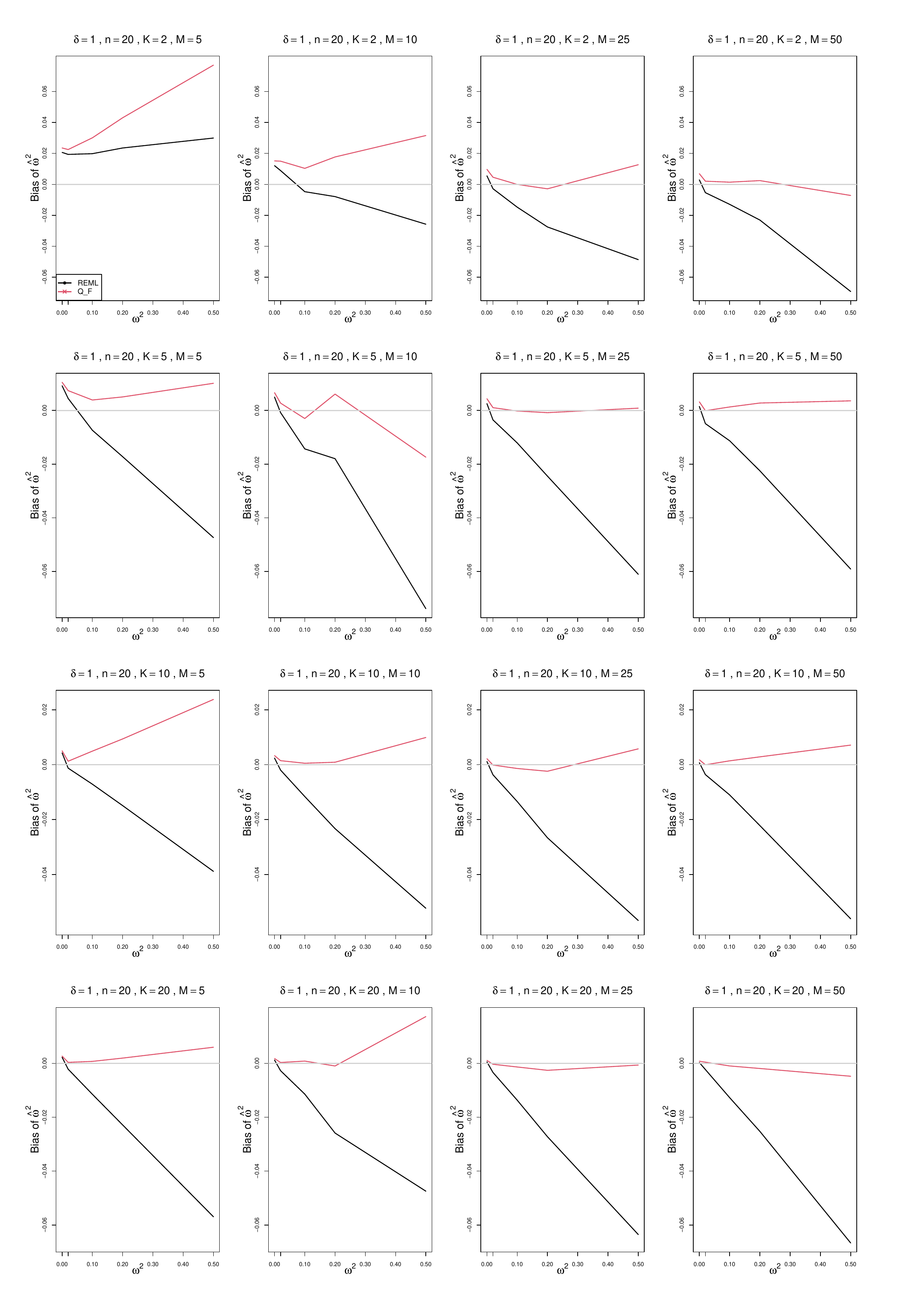}
	\caption{Bias of estimators of between-cluster  variance of SMD (REML and $Q_F$ ) vs $\omega^2$, for $K$ = 2, 5, 10, and 20 studies per cluster and $M$ = 5, 10, 25, and 50 clusters; $\delta = 1$, and the sample size $n$ = 20 in each study.  }
	\label{PlotBiasOfOmega2_20_1_HIER.pdf}
\end{figure}

\begin{figure}[ht]
	\centering
	\includegraphics[scale=0.33]{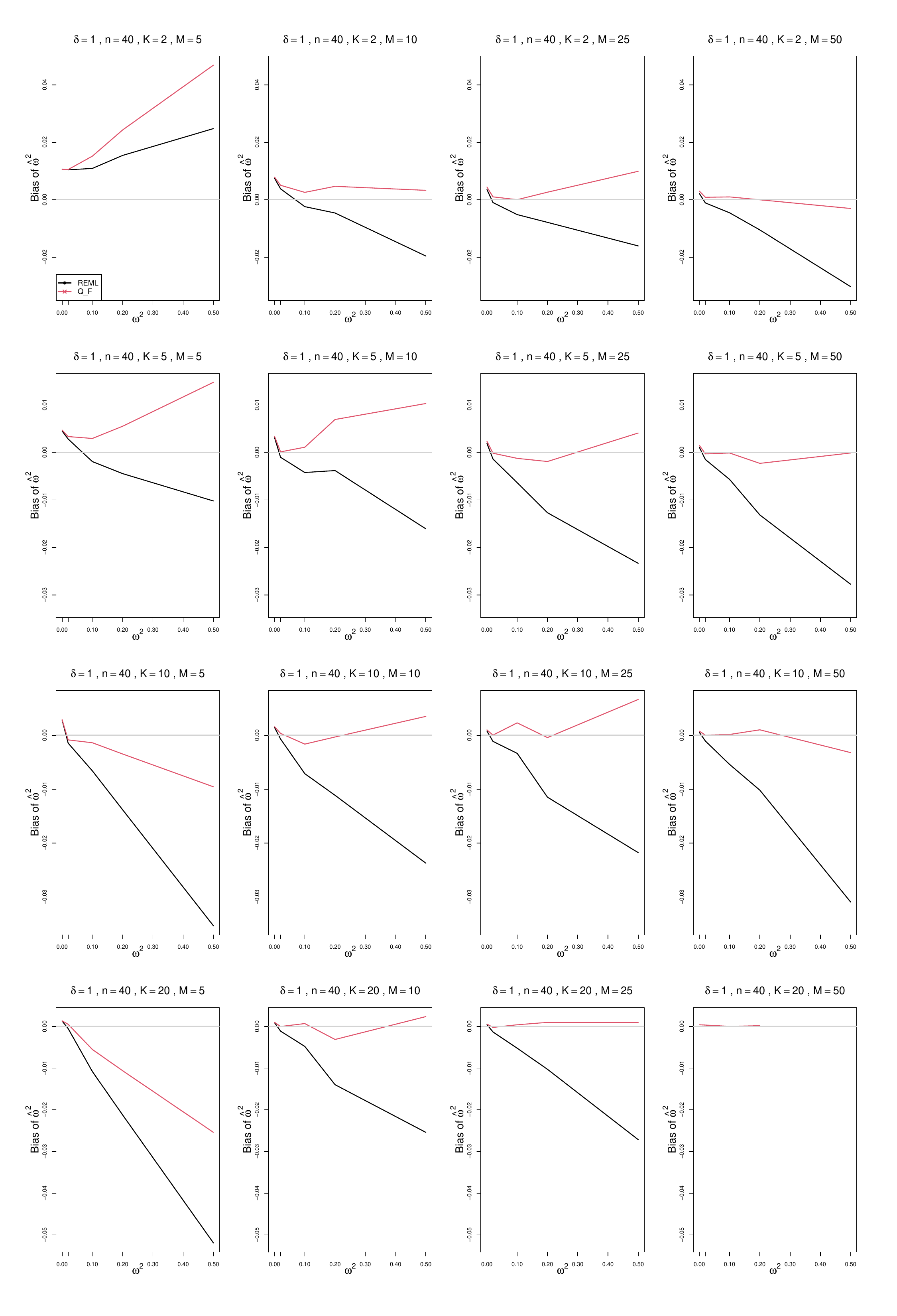}
	\caption{Bias  of estimators of between-cluster  variance of SMD (REML and $Q_F$ ) vs $\omega^2$, for $K$ = 2, 5, 10, and 20 studies per cluster and $M$ = 5, 10, 25, and 50 clusters; $\delta = 1$, and the sample size $n$ = 40 in each study. }
	\label{PlotBiasOfOmega2_40_1_HIER.pdf}
\end{figure}

\begin{figure}[ht]
	\centering
	\includegraphics[scale=0.33]{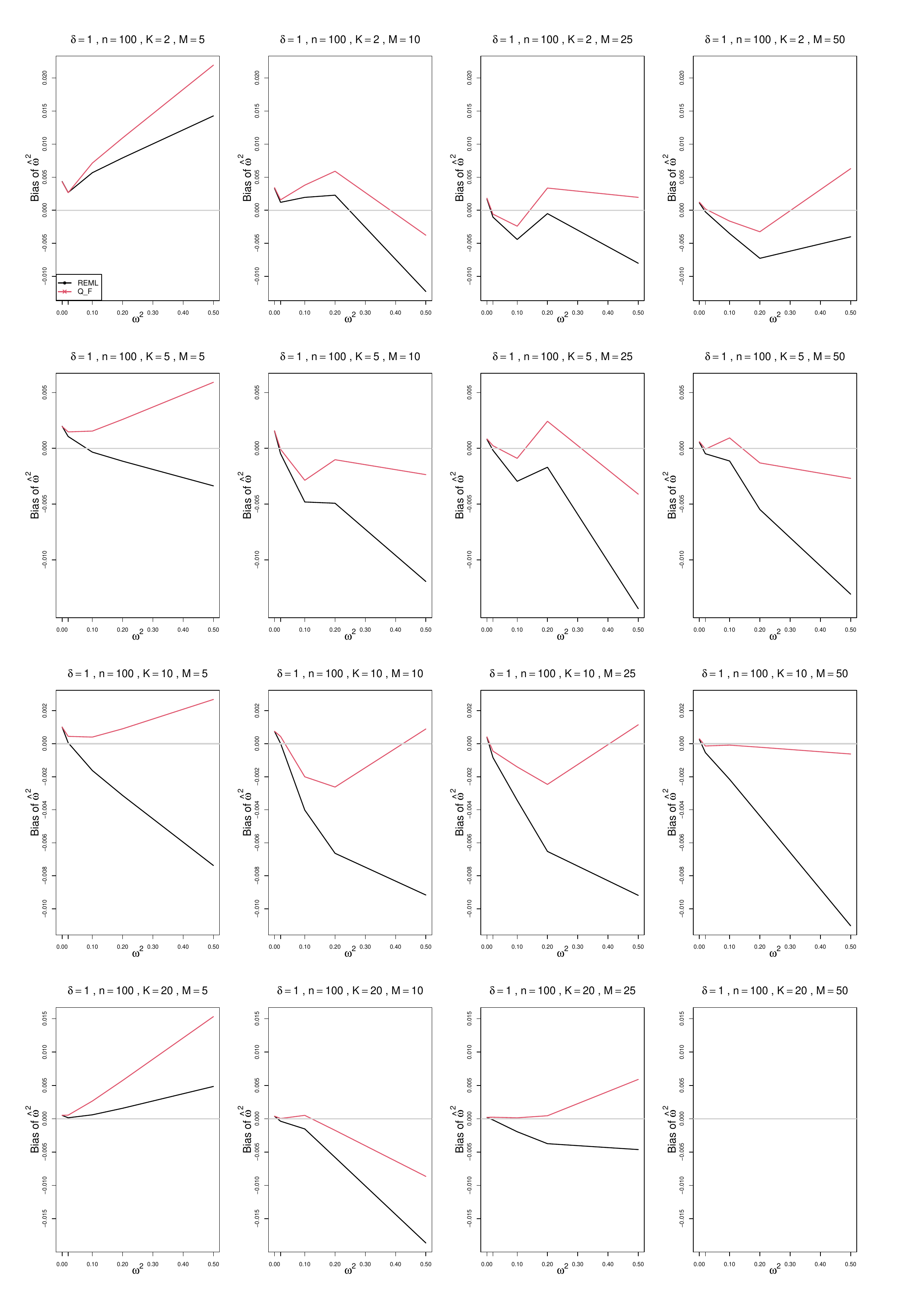}
	\caption{Bias  of estimators of between-cluster  variance of SMD (REML and $Q_F$ ) vs $\omega^2$, for $K$ = 2, 5, 10, and 20 studies per cluster and $M$ = 5, 10, 25, and 50 clusters; $\delta = 1$, and the sample size $n$ = 100 in each study.  }
	\label{PlotBiasOfOmega2_100_1_HIER.pdf}
\end{figure}

\begin{figure}[ht]
	\centering
	\includegraphics[scale=0.33]{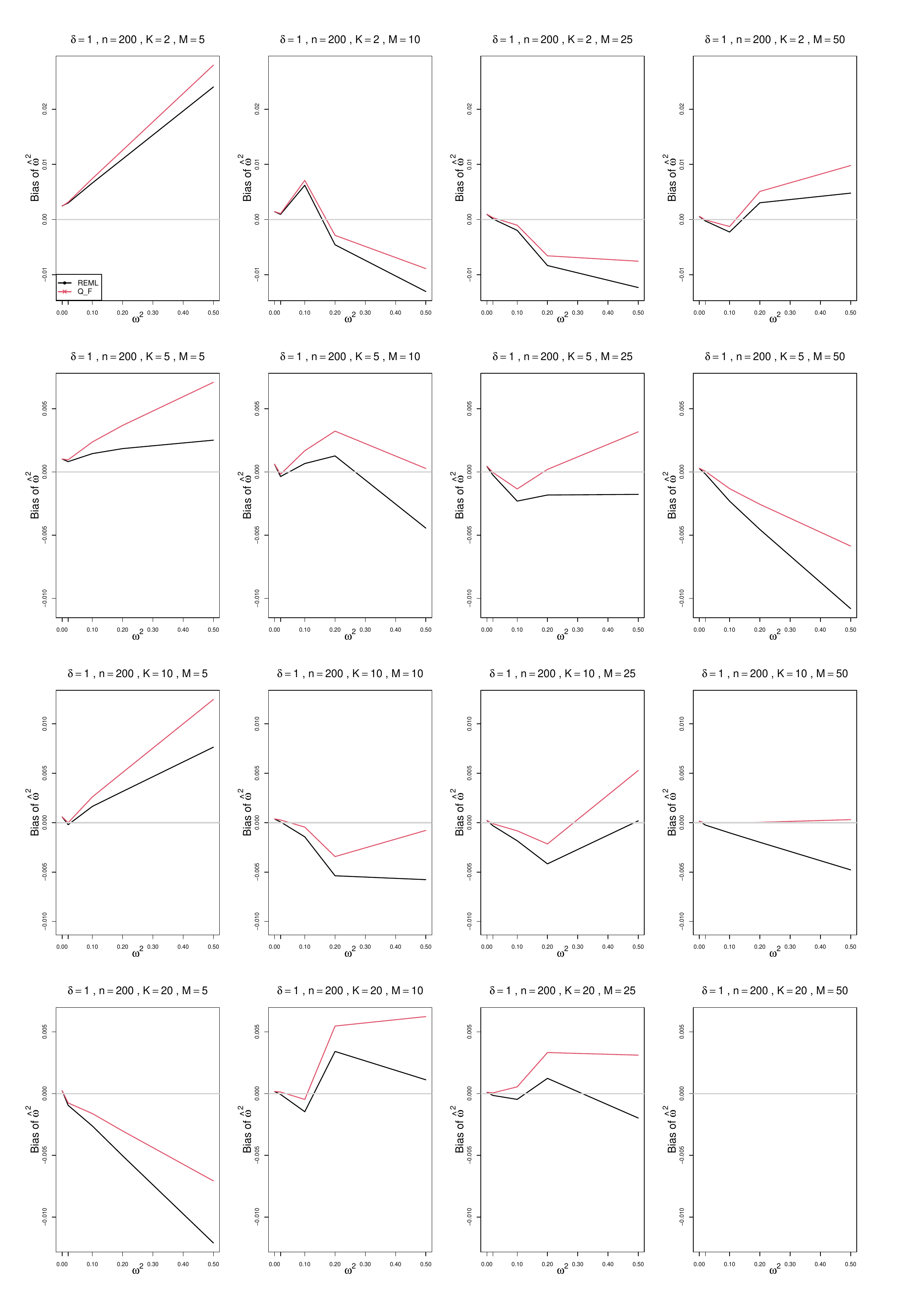}
	\caption{Bias  of estimators of between-cluster  variance of SMD (REML and $Q_F$ ) vs $\omega^2$, for $K$ = 2, 5, 10, and 20 studies per cluster and $M$ = 5, 10, 25, and 50 clusters; $\delta = 1$, and the sample size $n$ = 200 in each study. }
	\label{PlotBiasOfOmega2_200_1_HIER.pdf}
\end{figure}

\begin{figure}[ht]
	\centering
	\includegraphics[scale=0.33]{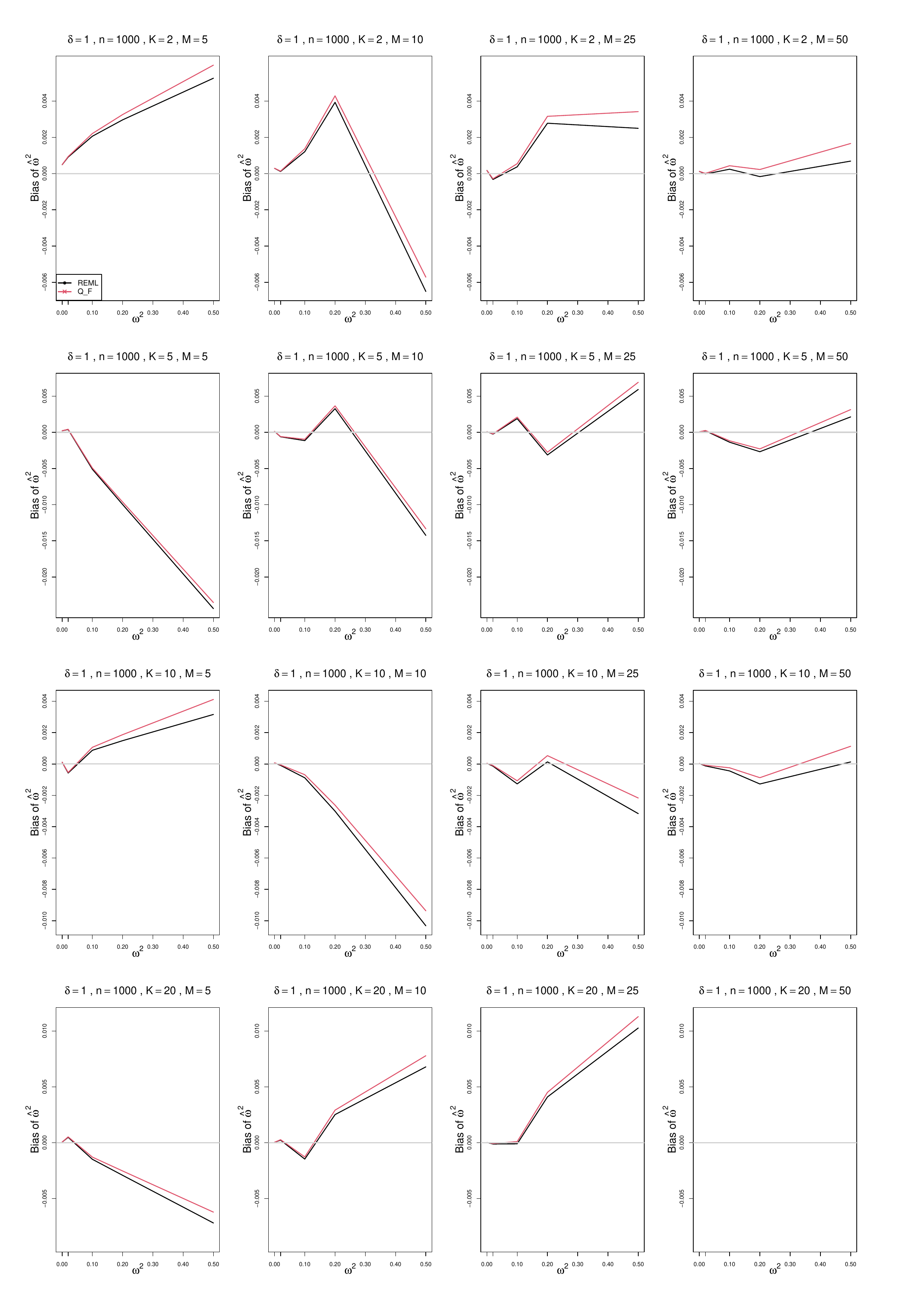}
	\caption{Bias  of estimators of between-cluster  variance of SMD (REML and $Q_F$ ) vs $\omega^2$, for $K$ = 2, 5, 10, and 20 studies per cluster and $M$ = 5, 10, 25, and 50 clusters; $\delta = 1$, and the sample size $n$ = 1000 in each study.  }
	\label{PlotBiasOfOmega2_1000_1_HIER.pdf}
\end{figure}


\clearpage

\section*{Appendix F: Coverage of 95\% confidence intervals for the between-cluster variance}

Each figure corresponds to a value of the standardized mean difference ($\delta$  = 0, 0.2, 0.5, 1) and a value of the study sample size ($n$ = 20, 40, 100, 200, 1000).\\
For each combination of the number of studies in a cluster ($K$ = 2, 5, 10, 20) and the number of clusters ($M$ = 5, 10, 25, 50), a panel plots coverage of $\omega^2$ at the 95\% nominal level versus $\omega^2$ (= 0, 0.02, 0.1, 0.2, 0.5).\\
The two variance components are held equal ($\tau^2 = \omega^2$).\\
The interval estimators of $\omega^2$ are
\begin{itemize}
\item PL (Profile-Likelihood method, inverse-variance weights,  {\it  rma.mv} in {\it metafor})
\item $Q_F$ (conditional moment-based method, effective-sample-size weights, Davies's approximation to the distribution)
\end{itemize}

\clearpage
\setcounter{figure}{0}
\renewcommand{\thefigure}{F.\arabic{figure}}

\begin{figure}[ht]
	\centering
	\includegraphics[scale=0.33]{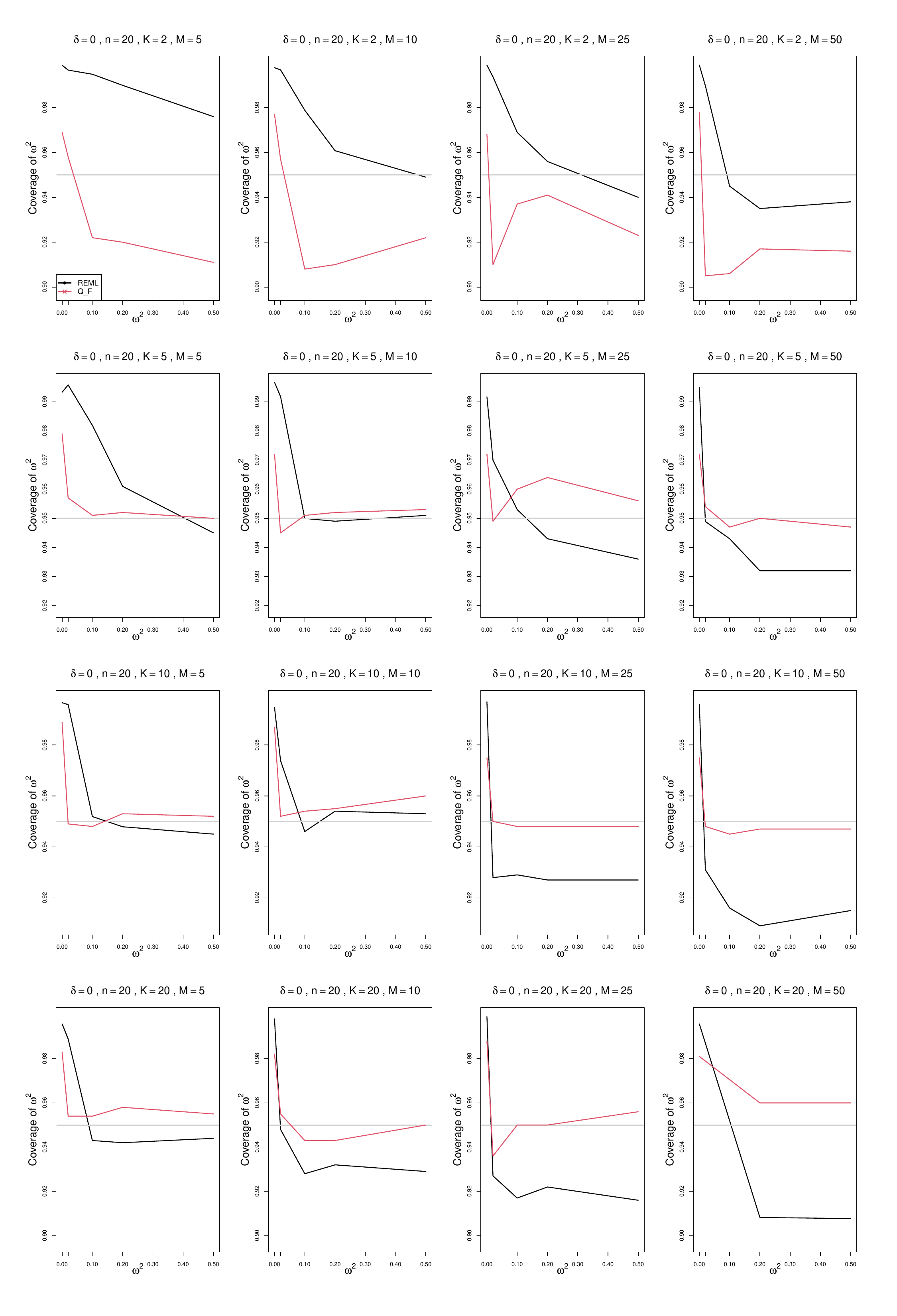}
	\caption{Coverage of 95\% confidence intervals for between-cluster variance of SMD (PL and $Q_F$ ) vs $\omega^2$, for $K$ = 2, 5, 10, and 20 studies per cluster and $M$ = 5, 10, 25, and 50 clusters; $\delta = 0$, and the sample size $n$ = 20 in each study.  }
	\label{PlotCoverageOfOmega2_20_0_HIER.pdf}
\end{figure}

\begin{figure}[ht]
	\centering
	\includegraphics[scale=0.33]{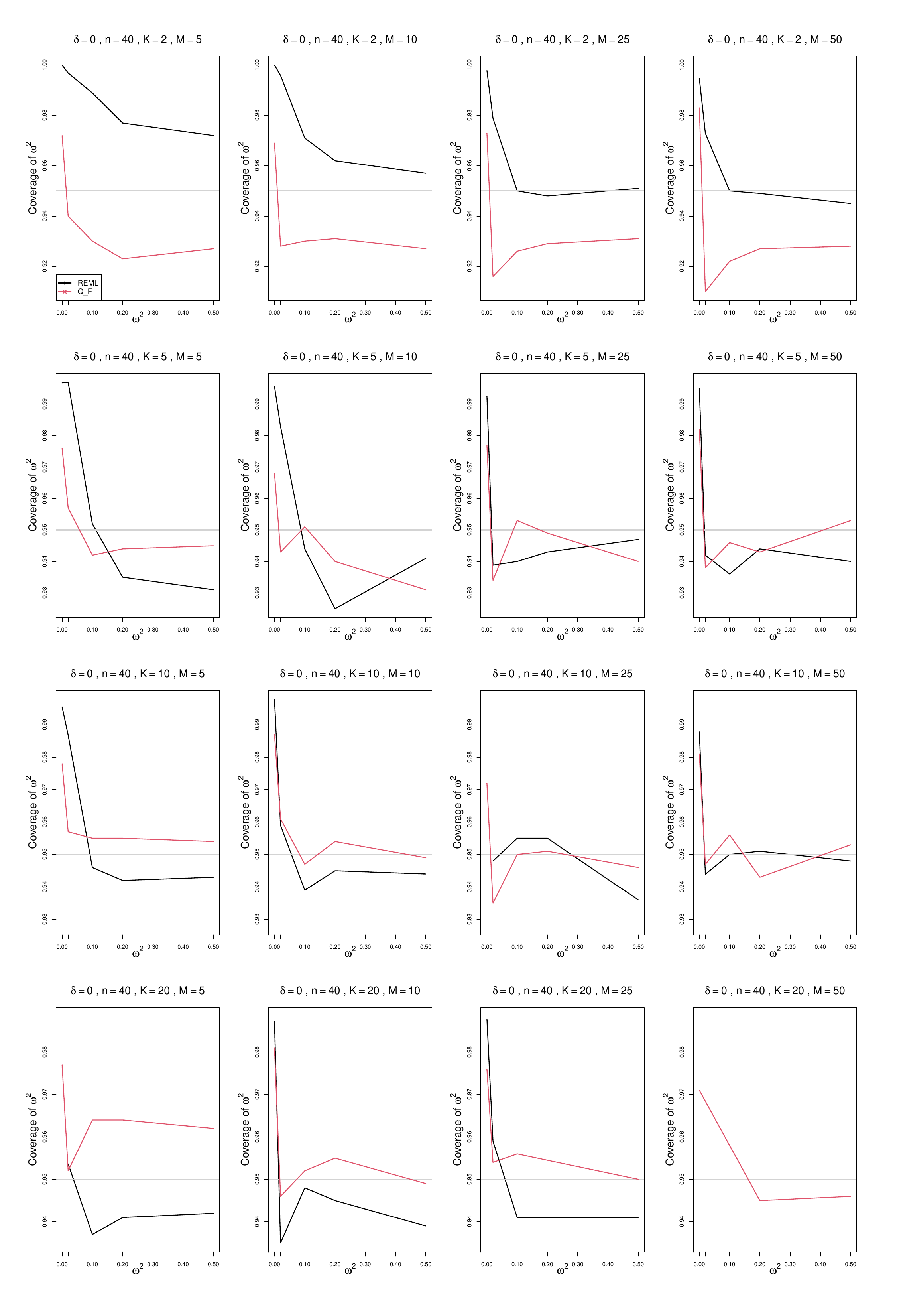}
	\caption{Coverage of 95\% confidence intervals for between-study variance of SMD (PL and $Q_F$ ) vs $\omega^2$, for $K$ = 2, 5, 10, and 20 studies per cluster and $M$ = 5, 10, 25, and 50 clusters; $\delta = 0$, and the sample size $n$ = 40 in each study.  }
	\label{PlotCoverageOfOmega2_40_0_HIER.pdf}
\end{figure}

\begin{figure}[ht]
	\centering
	\includegraphics[scale=0.33]{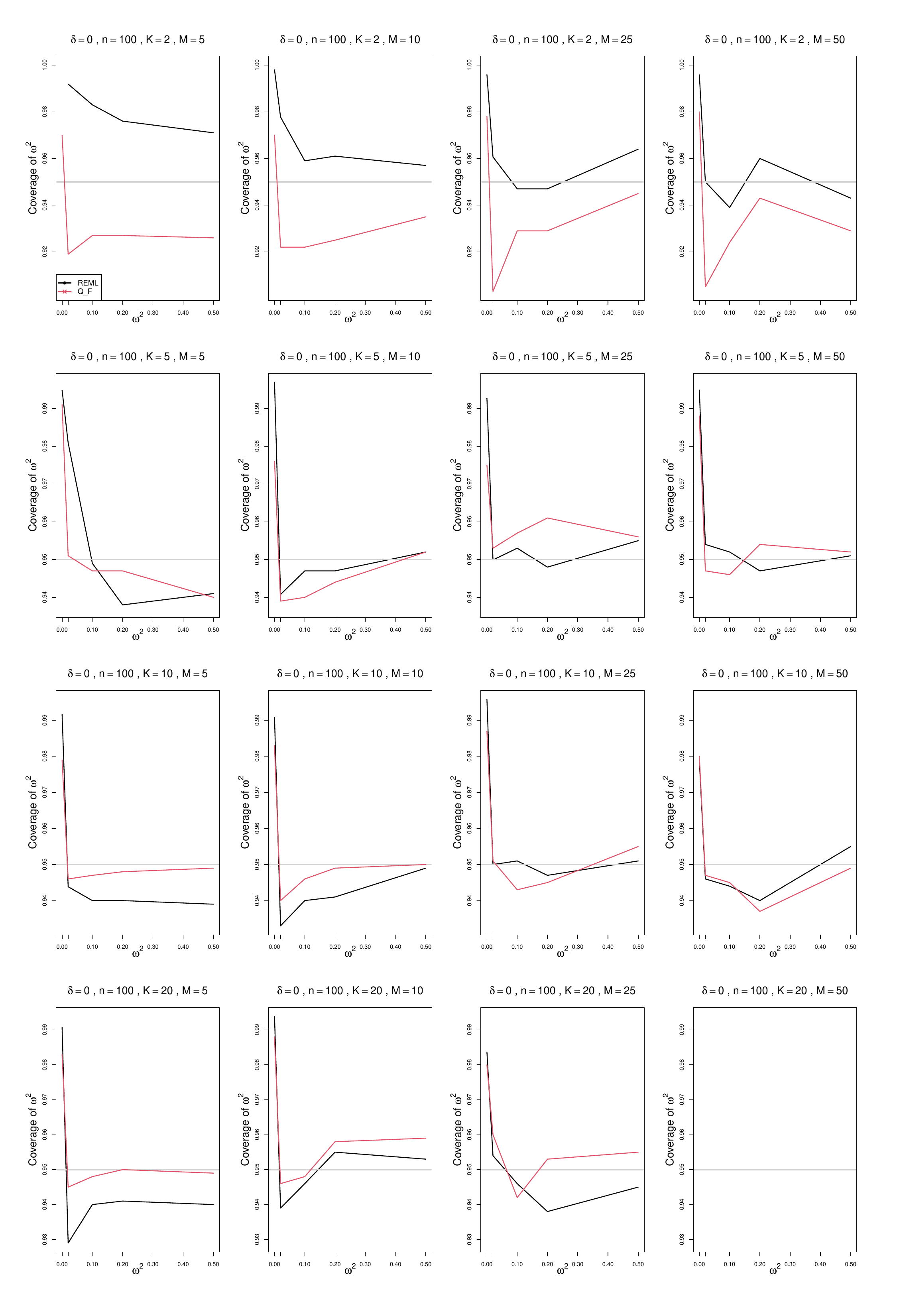}
	\caption{Coverage of 95\% confidence intervals for between-study variance of SMD (PL and $Q_F$ ) vs $\omega^2$, for $K$ = 2, 5, 10, and 20 studies per cluster and $M$ = 5, 10, 25, and 50 clusters; $\delta = 0$, and the sample size $n$ = 100 in each study.  }
	\label{PlotCoverageOfOmega2_100_0_HIER.pdf}
\end{figure}

\begin{figure}[ht]
	\centering
	\includegraphics[scale=0.33]{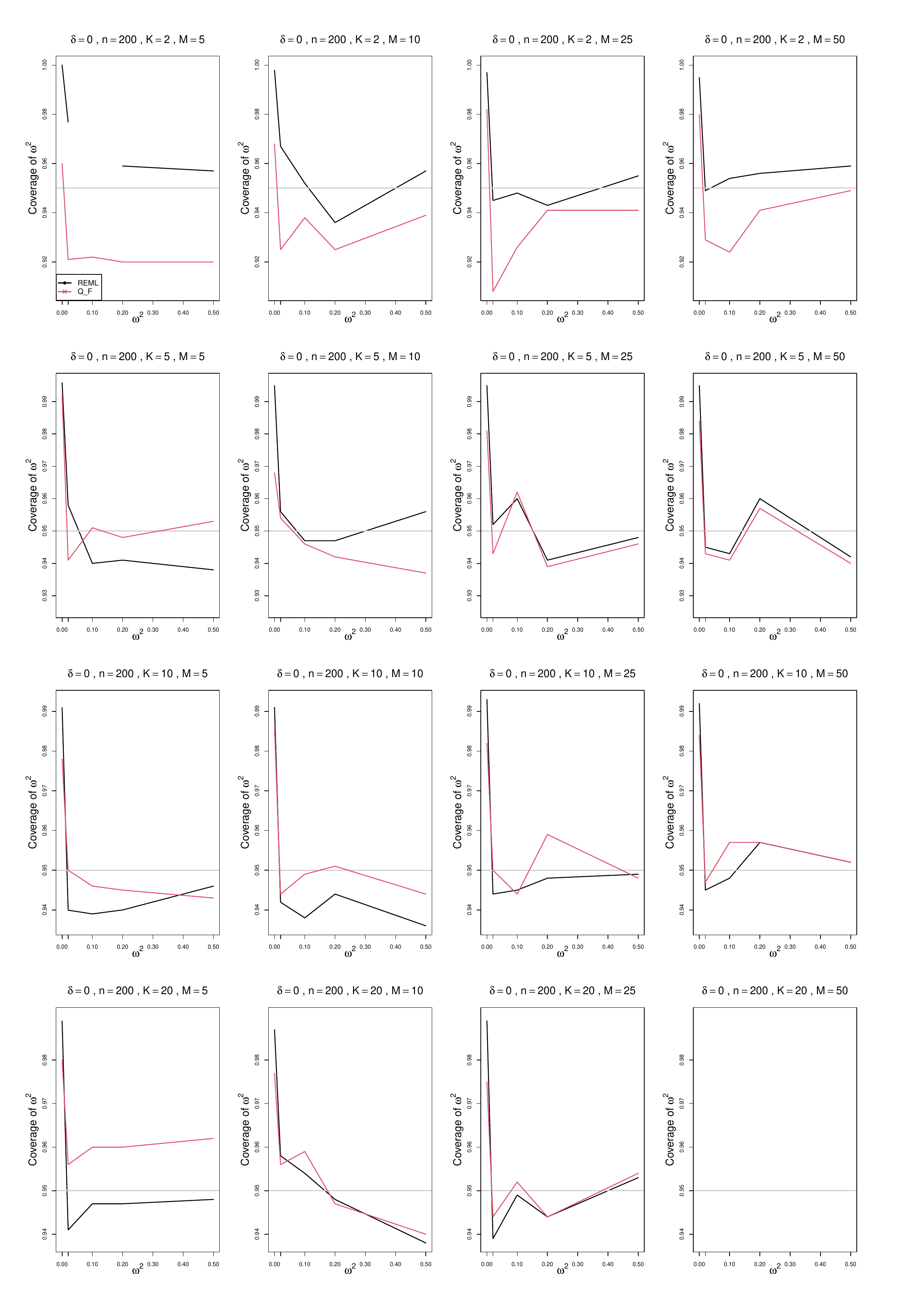}
	\caption{Coverage of 95\% confidence intervals for between-study variance of SMD (PL and $Q_A$ ) vs $\omega^2$, for $K$ = 2, 5, 10, and 20 studies per cluster and $M$ = 5, 10, 25, and 50 clusters; $\delta = 0$, and the sample size $n$ = 200 in each study.  }
	\label{PlotCoverageOfOmega2_200_0_HIER.pdf}
\end{figure}

\begin{figure}[ht]
	\centering
	\includegraphics[scale=0.33]{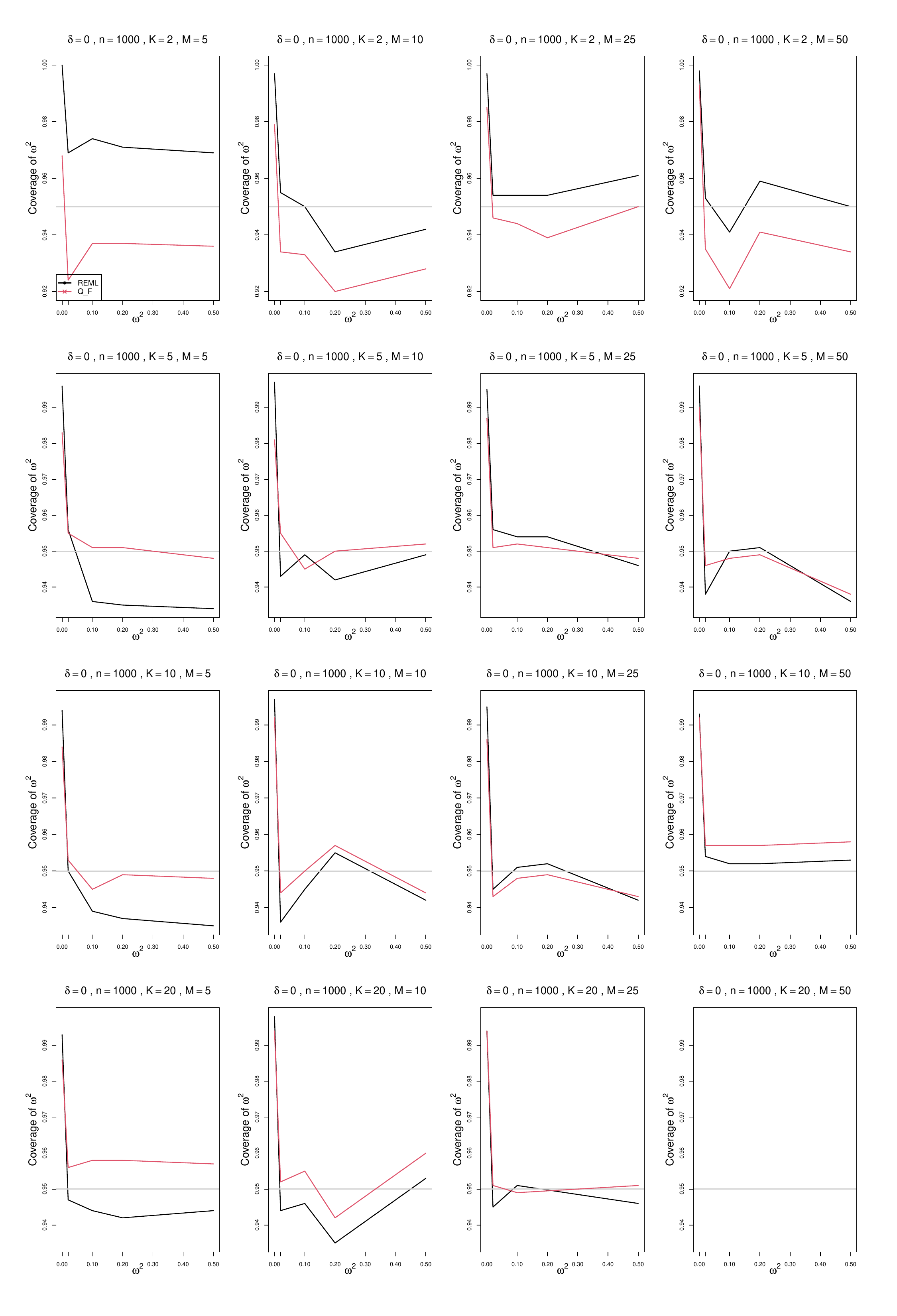}
	\caption{Coverage of 95\% confidence intervals for between-study variance of SMD (PL and $Q_F$ ) vs $\omega^2$, for $K$ = 2, 5, 10, and 20 studies per cluster and $M$ = 5, 10, 25, and 50 clusters; $\delta = 0$, and the sample size $n$ = 1000 in each study.  }
	\label{PlotCoverageOfOmega2_1000_0_HIER.pdf}
\end{figure}

\begin{figure}[ht]
	\centering
	\includegraphics[scale=0.33]{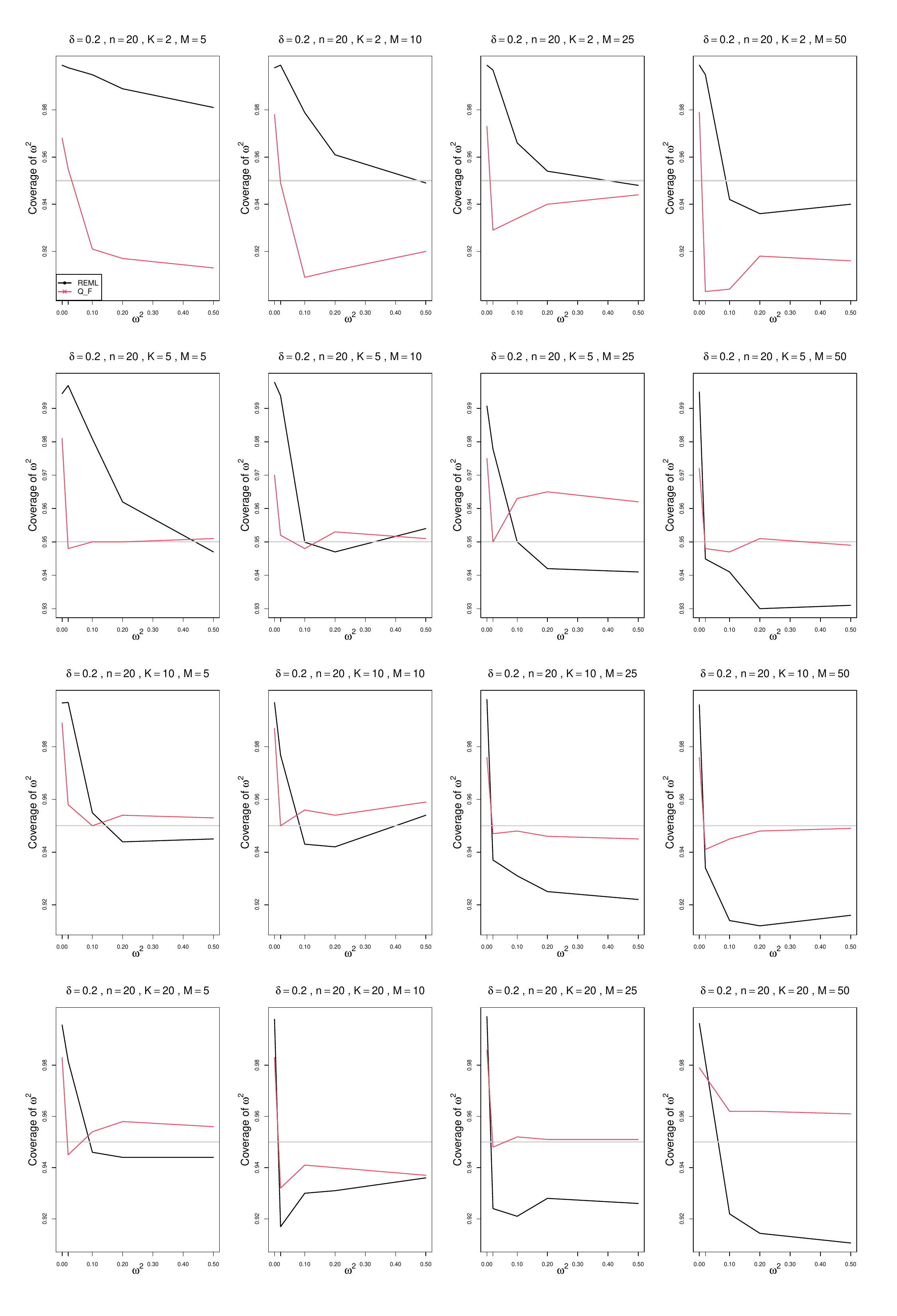}
	\caption{Coverage of 95\% confidence intervals for between-study variance of SMD (PL and $Q_F$ ) vs $\omega^2$, for $K$ = 2, 5, 10, and 20 studies per cluster and $M$ = 5, 10, 25, and 50 clusters; $\delta = 0.2$, and the sample size $n$ = 20 in each study.  }
	\label{PlotCoverageOfOmega2_20_02_HIER.pdf}
\end{figure}

\begin{figure}[ht]
	\centering
	\includegraphics[scale=0.33]{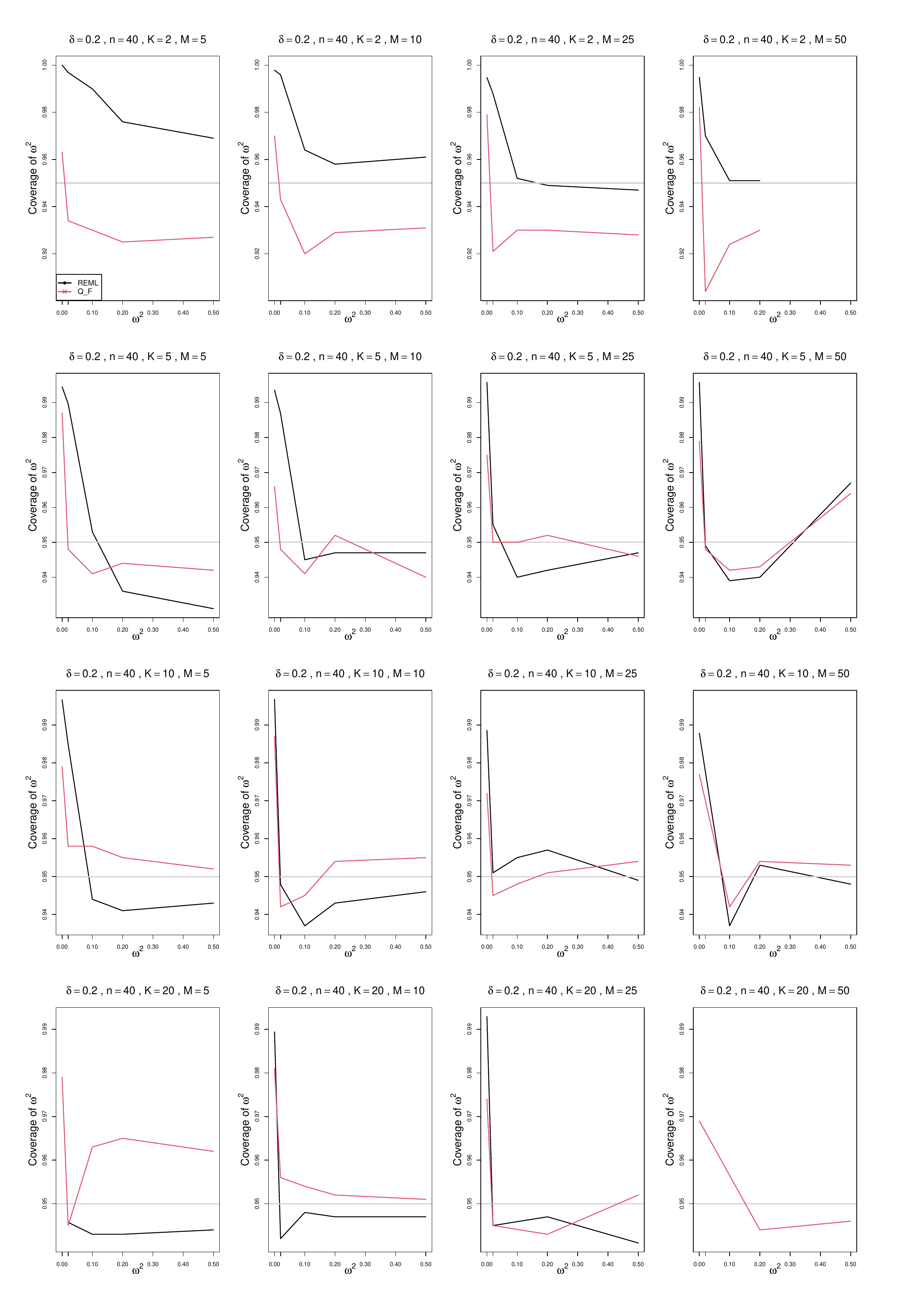}
	\caption{Coverage of 95\% confidence intervals for between-study variance of SMD (PL and $Q_F$ ) vs $\omega^2$, for $K$ = 2, 5, 10, and 20 studies per cluster and $M$ = 5, 10, 25, and 50 clusters; $\delta = 0.2$, and the sample size $n$ = 40 in each study. }
	\label{PlotCoverageOfOmega2_40_02_HIER.pdf}
\end{figure}

\begin{figure}[ht]
	\centering
	\includegraphics[scale=0.33]{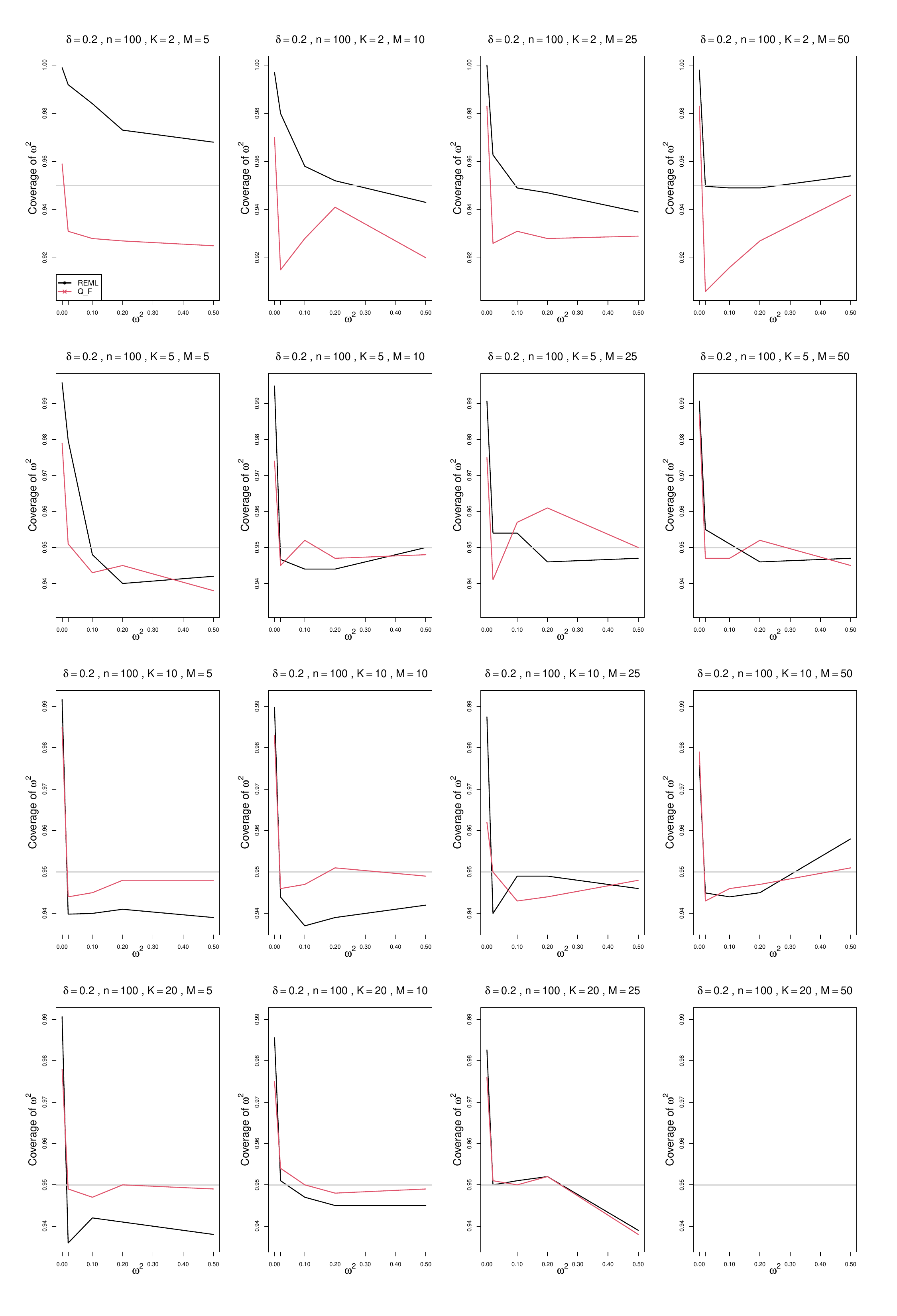}
	\caption{Coverage of 95\% confidence intervals for between-study variance of SMD (PL and $Q_F$ ) vs $\omega^2$, for $K$ = 2, 5, 10, and 20 studies per cluster and $M$ = 5, 10, 25, and 50 clusters; $\delta = 0.2$, and the sample size $n$ = 100 in each study. }
	\label{PlotCoverageOfOmega2_100_02_HIER.pdf}
\end{figure}

\begin{figure}[ht]
	\centering
	\includegraphics[scale=0.33]{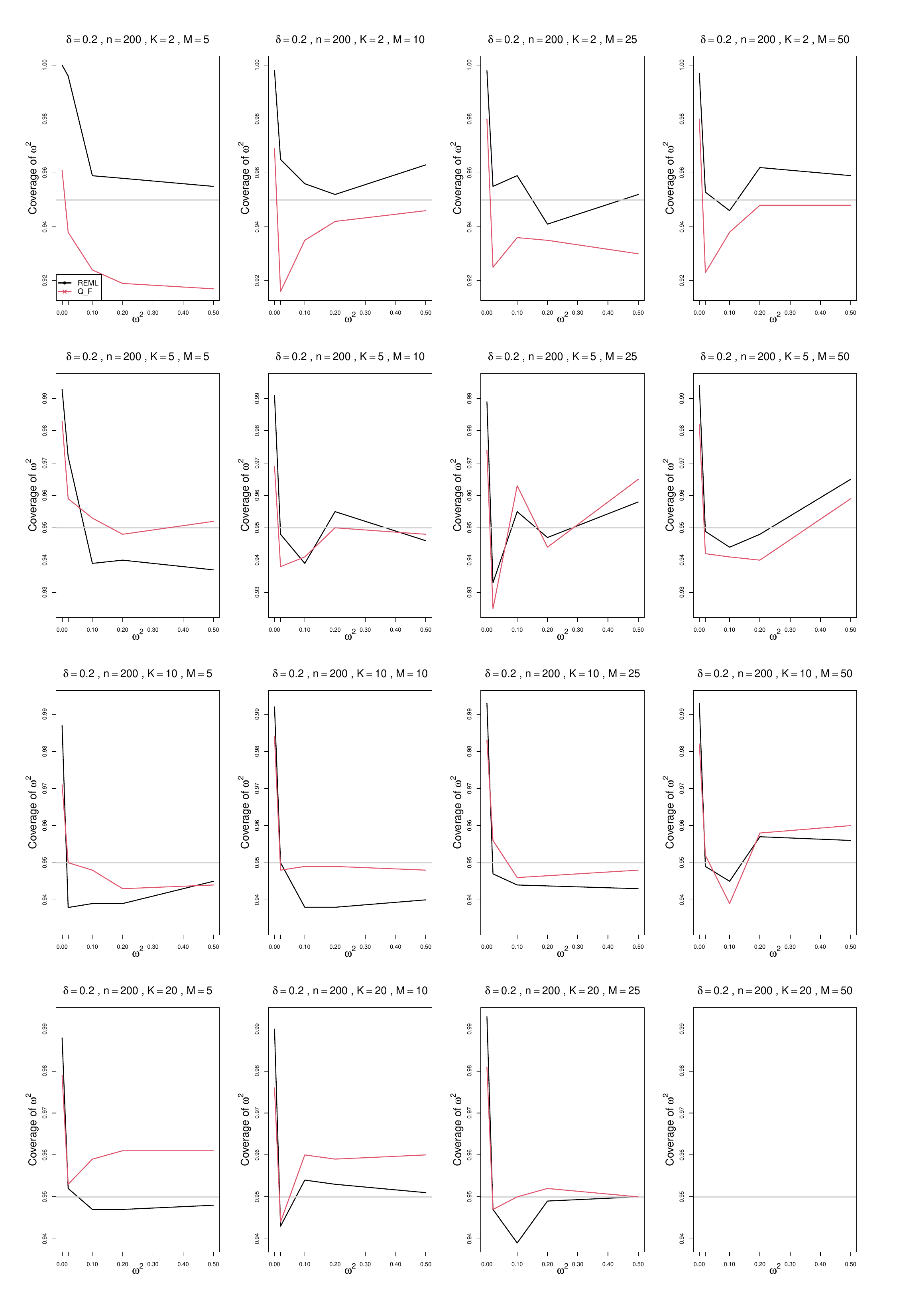}
	\caption{Coverage of 95\% confidence intervals for between-study variance of SMD (PL and $Q_F$ )  vs $\omega^2$, for $K$ = 2, 5, 10, and 20 studies per cluster and $M$ = 5, 10, 25, and 50 clusters; $\delta = 0.2$, and the sample size $n$ = 200 in each study.  }
	\label{PlotCoverageOfOmega2_200_02_HIER.pdf}
\end{figure}

\begin{figure}[ht]
	\centering
	\includegraphics[scale=0.33]{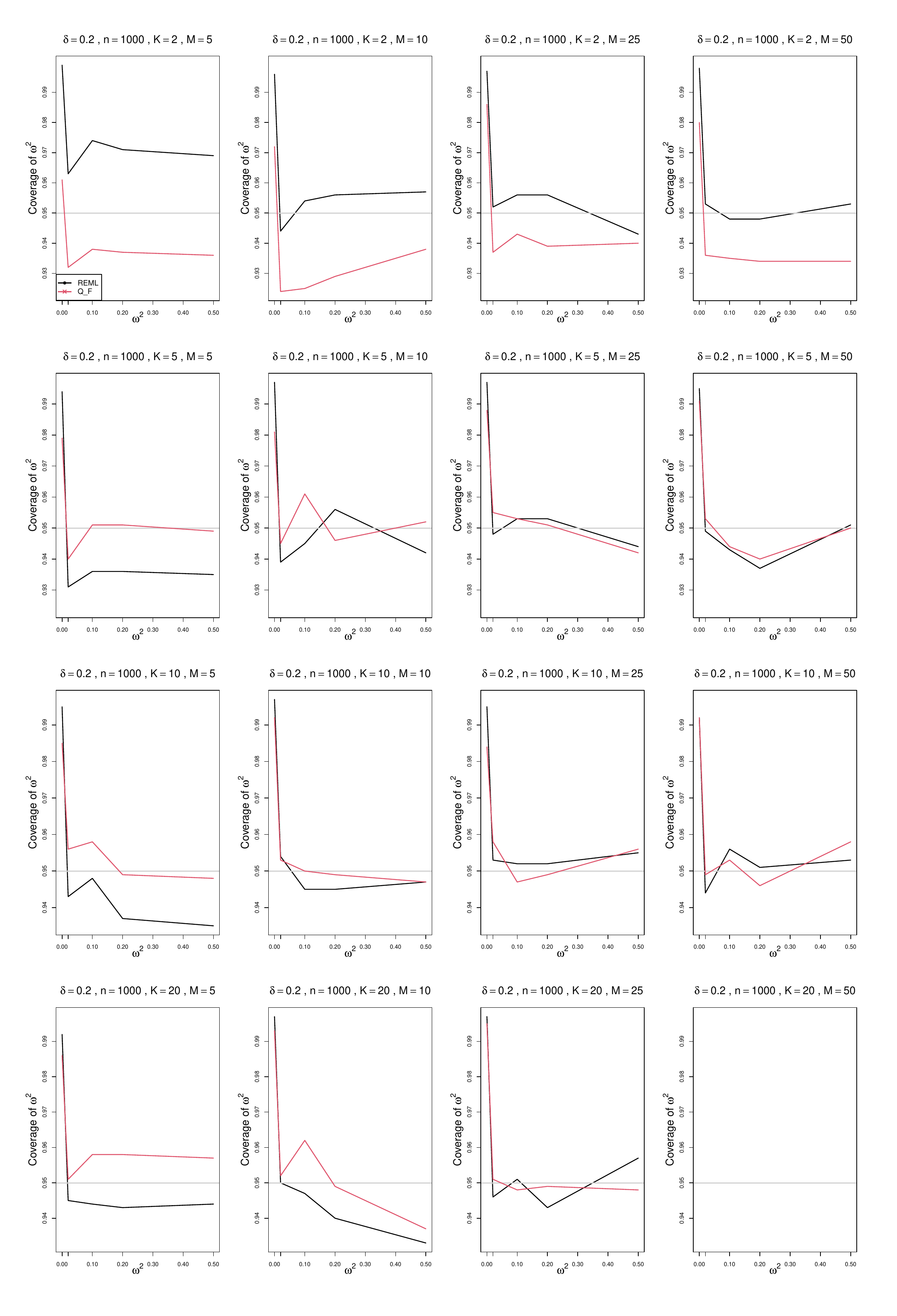}
	\caption{Coverage of 95\% confidence intervals for between-study variance of SMD (PL and $Q_F$ ) vs $\omega^2$, for $K$ = 2, 5, 10, and 20 studies per cluster and $M$ = 5, 10, 25, and 50 clusters; $\delta = 0.2$, and the sample size $n$ = 1000 in each study. }
	\label{PlotCoverageOfOmega2_1000_02_HIER.pdf}
\end{figure}

\begin{figure}[ht]
	\centering
	\includegraphics[scale=0.33]{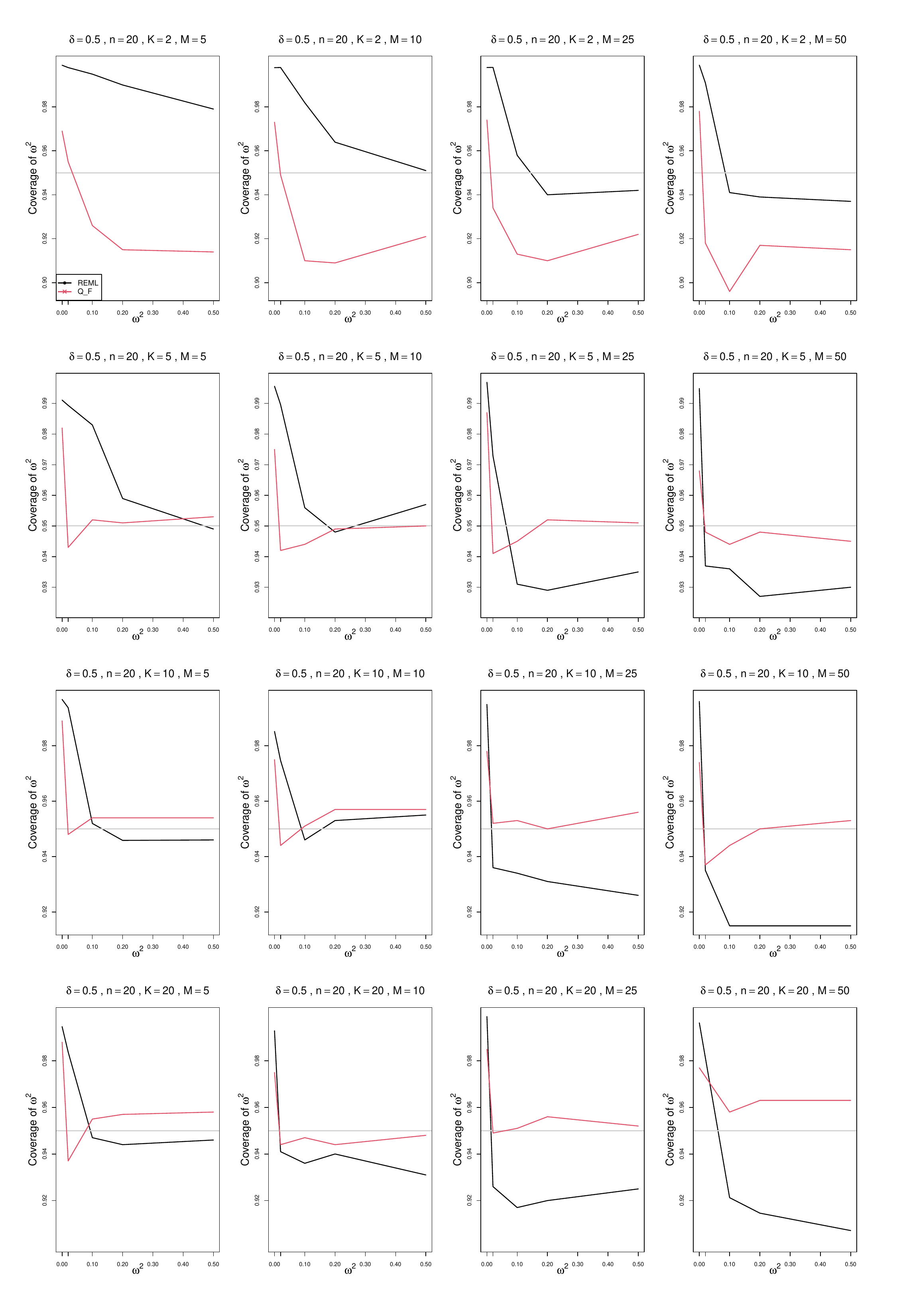}
	\caption{Coverage of 95\% confidence intervals for between-study variance of SMD (PL and $Q_F$ ) vs $\omega^2$, for $K$ = 2, 5, 10, and 20 studies per cluster and $M$ = 5, 10, 25, and 50 clusters; $\delta = 0.5$, and the sample size $n$ = 20 in each study. }
	\label{PlotCoverageOfOmega2_20_05_HIER.pdf}
\end{figure}

\begin{figure}[ht]
	\centering
	\includegraphics[scale=0.33]{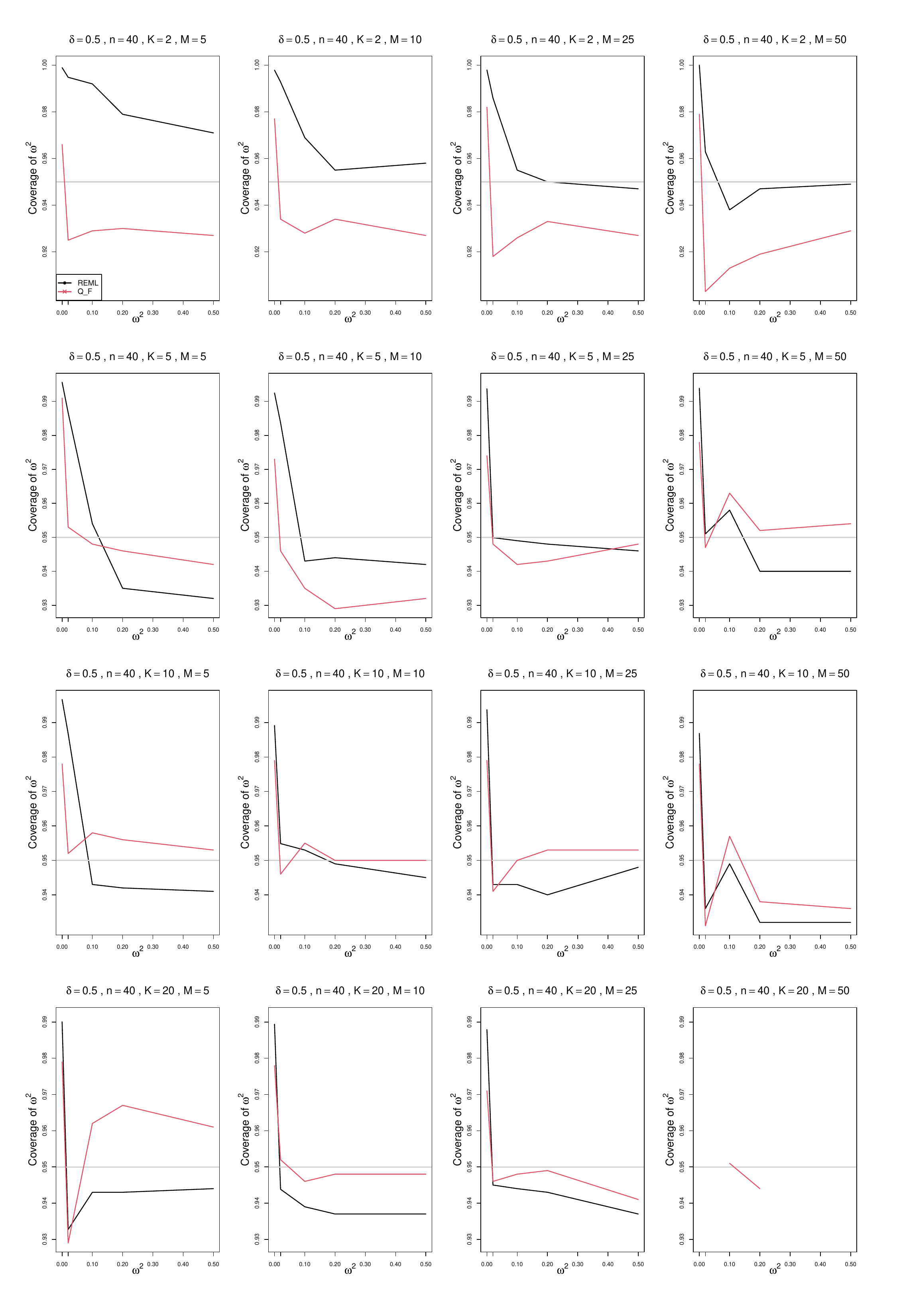}
	\caption{Coverage of 95\% confidence intervals for between-study variance of SMD (PL and $Q_F$ ) vs $\omega^2$, for $K$ = 2, 5, 10, and 20 studies per cluster and $M$ = 5, 10, 25, and 50 clusters; $\delta = 0.5$, and the sample size $n$ = 40 in each study.  }
	\label{PlotCoverageOfOmega2_40_05_HIER.pdf}
\end{figure}

\begin{figure}[ht]
	\centering
	\includegraphics[scale=0.33]{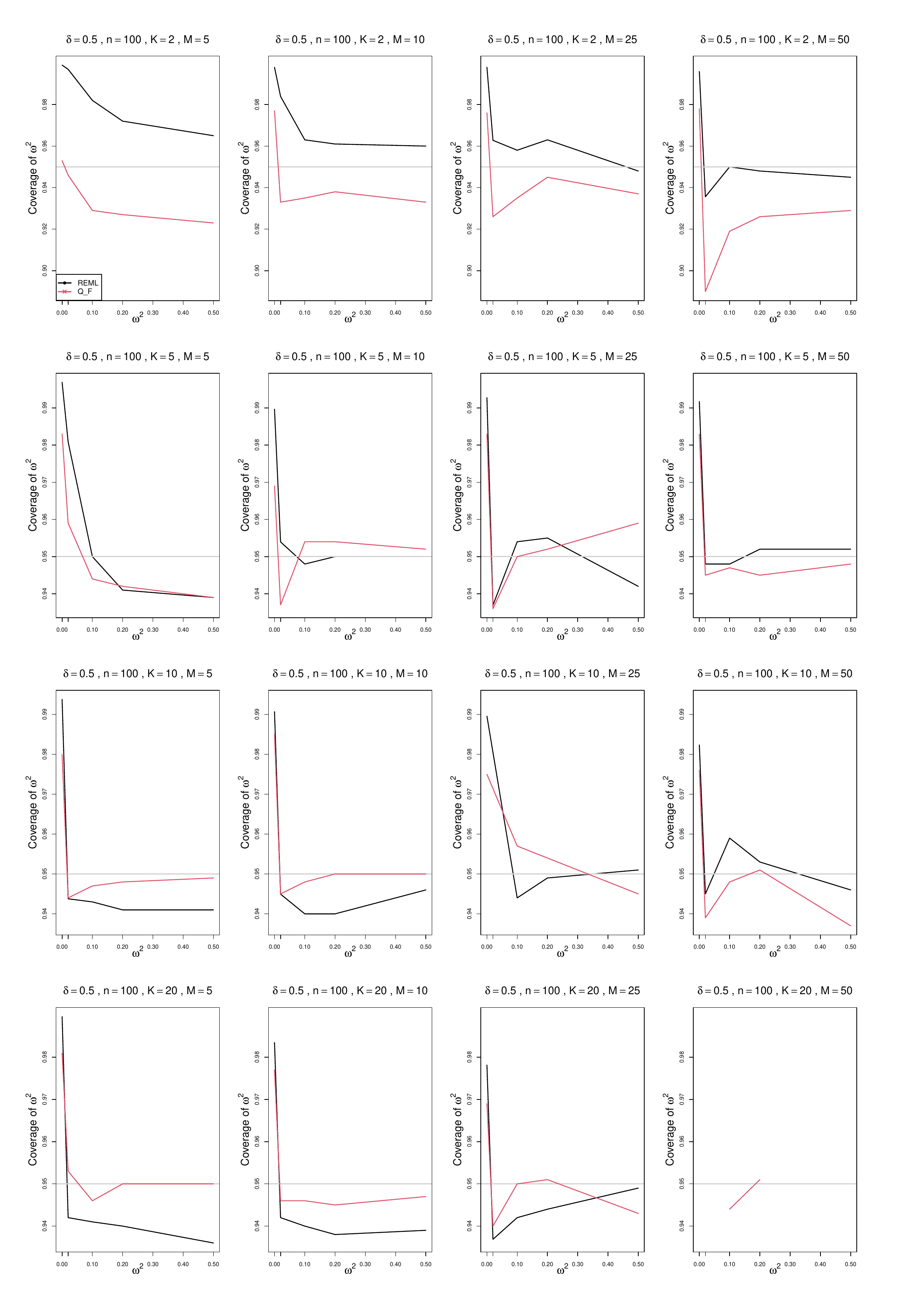}
	\caption{Coverage of 95\% confidence intervals for between-study variance of SMD (PL and $Q_F$ ) vs $\omega^2$, for $K$ = 2, 5, 10, and 20 studies per cluster and $M$ = 5, 10, 25, and 50 clusters; $\delta = 0.5$, and the sample size $n$ = 100 in each study.  }
	\label{PlotCoverageOfOmega2_100_05_HIER.pdf}
\end{figure}

\begin{figure}[ht]
	\centering
	\includegraphics[scale=0.33]{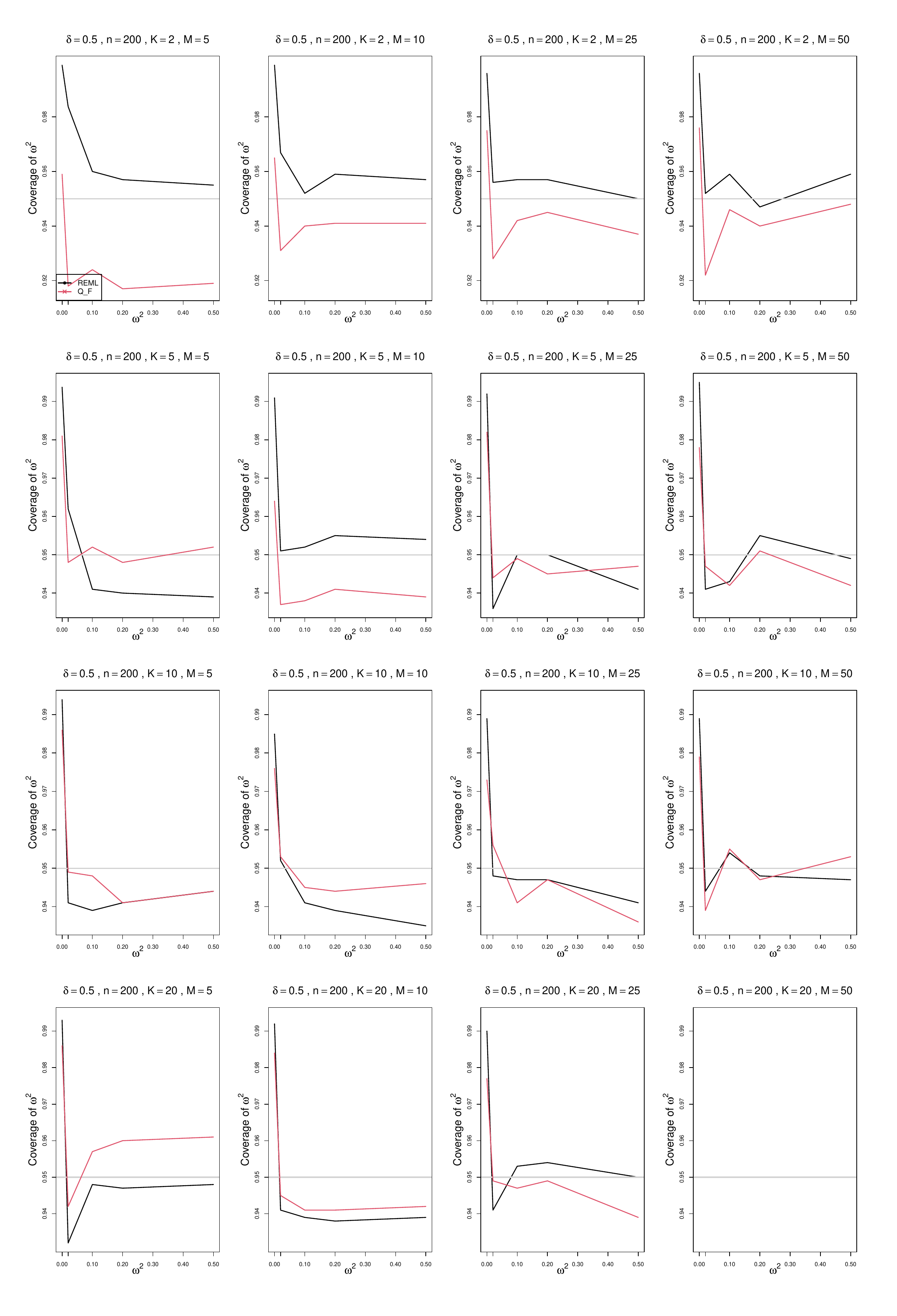}
	\caption{Coverage of 95\% confidence intervals for between-study variance of SMD (PL and $Q_F$ ) vs $\omega^2$, for $K$ = 2, 5, 10, and 20 studies per cluster and $M$ = 5, 10, 25, and 50 clusters; $\delta = 0.5$, and the sample size $n$ = 200 in each study.  }
	\label{PlotCoverageOfOmega2_200_05_HIER.pdf}
\end{figure}

\begin{figure}[ht]
	\centering
	\includegraphics[scale=0.33]{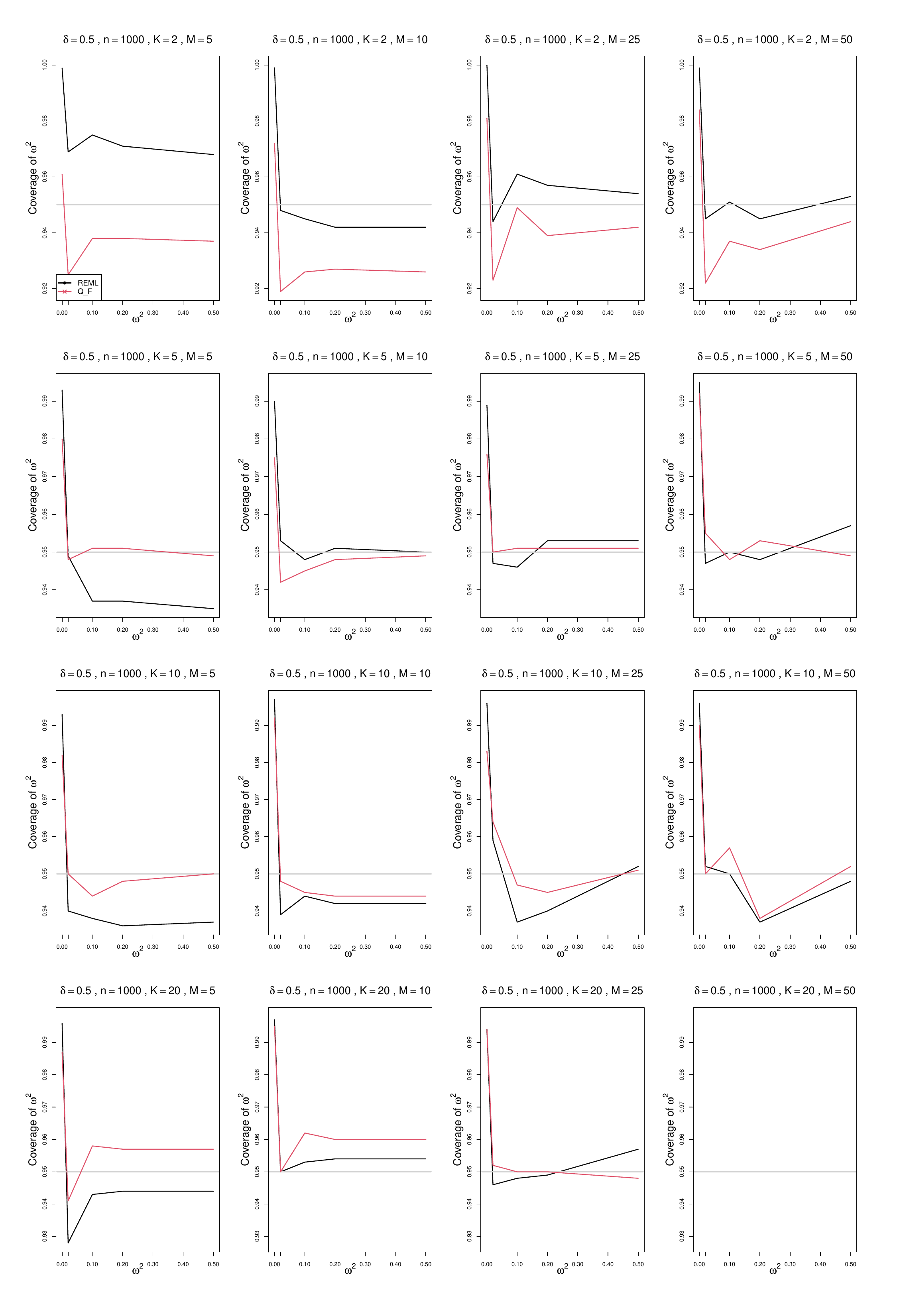}
	\caption{Coverage of 95\% confidence intervals for between-study variance of SMD (PL and $Q_F$ ) vs $\omega^2$, for $K$ = 2, 5, 10, and 20 studies per cluster and $M$ = 5, 10, 25, and 50 clusters; $\delta = 0.5$, and the sample size $n$ = 1000 in each study. }
	\label{PlotCoverageOfOmega2_1000_05_HIER.pdf}
\end{figure}

\begin{figure}[ht]
	\centering
	\includegraphics[scale=0.33]{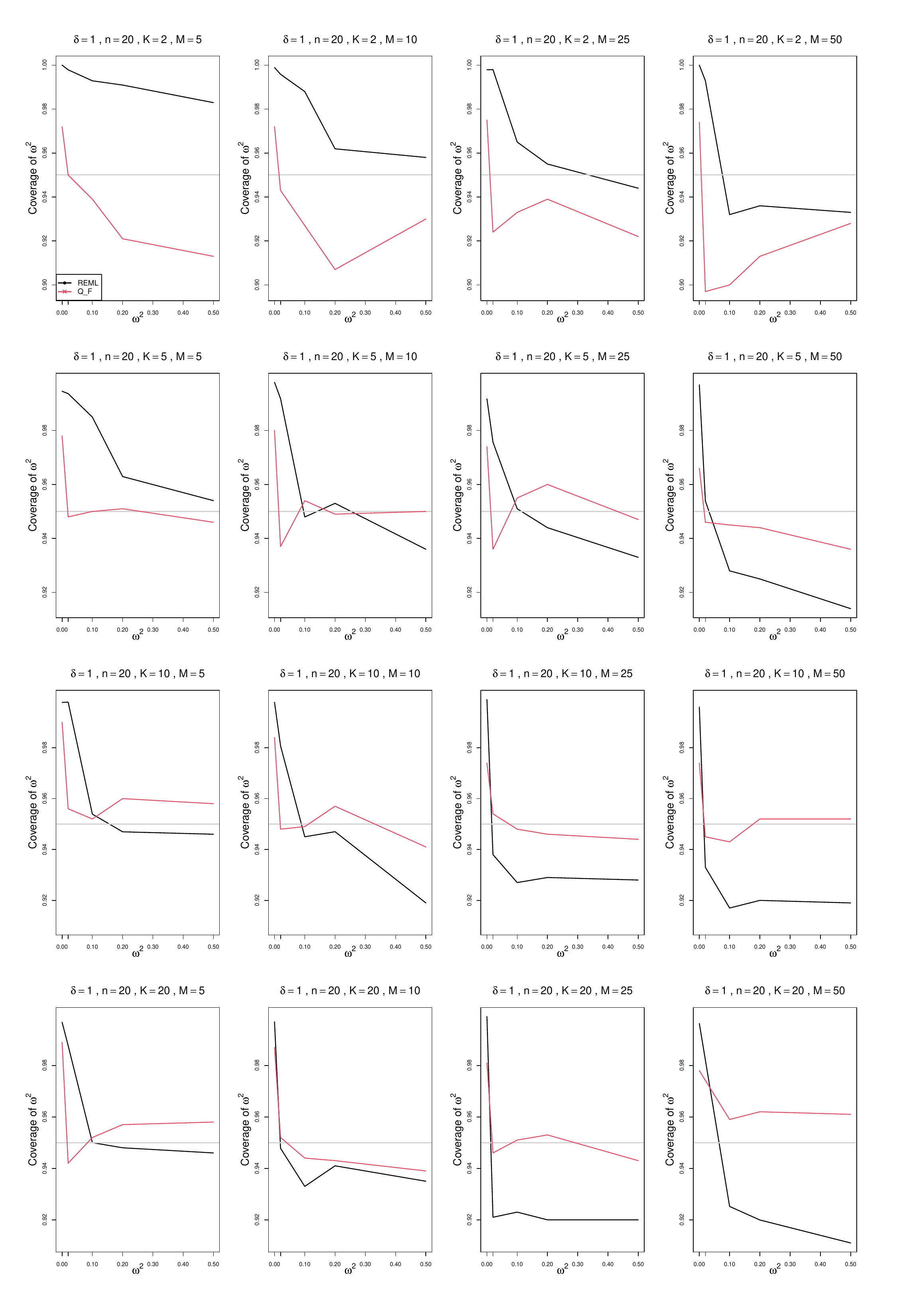}
	\caption{Coverage of 95\% confidence intervals for between-study variance of SMD (PL and $Q_F$ ) vs $\omega^2$, for $K$ = 2, 5, 10, and 20 studies per cluster and $M$ = 5, 10, 25, and 50 clusters; $\delta = 1$, and the sample size $n$ = 20 in each study.  }
	\label{PlotCoverageOfOmega2_20_1_HIER.pdf}
\end{figure}

\begin{figure}[ht]
	\centering
	\includegraphics[scale=0.33]{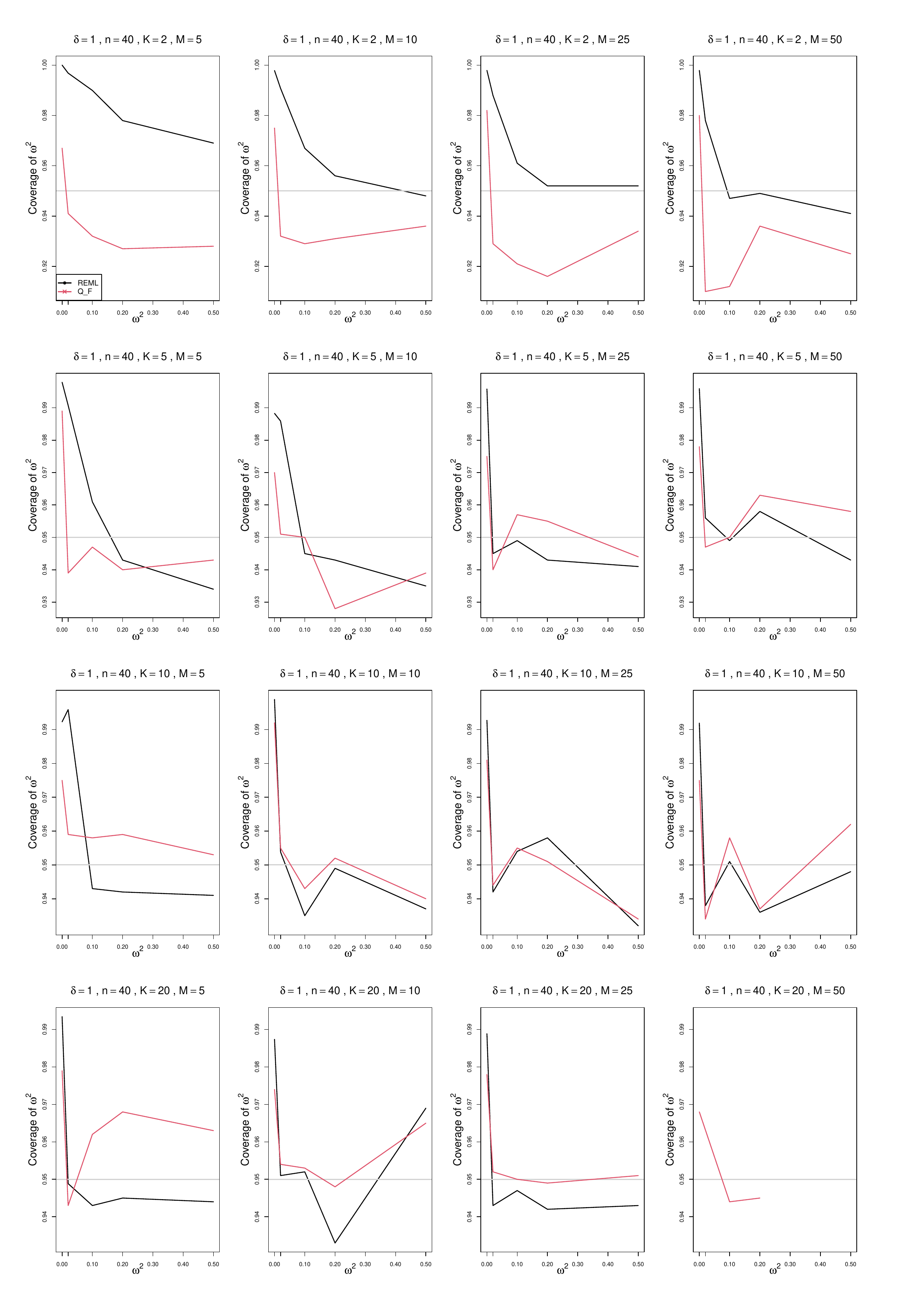}
	\caption{Coverage of 95\% confidence intervals for between-study variance of SMD (PL and $Q_F$ ) vs $\omega^2$, for $K$ = 2, 5, 10, and 20 studies per cluster and $M$ = 5, 10, 25, and 50 clusters; $\delta = 1$, and the sample size $n$ = 40 in each study. }
	\label{PlotCoverageOfOmega2_40_1_HIER.pdf}
\end{figure}

\begin{figure}[ht]
	\centering
	\includegraphics[scale=0.33]{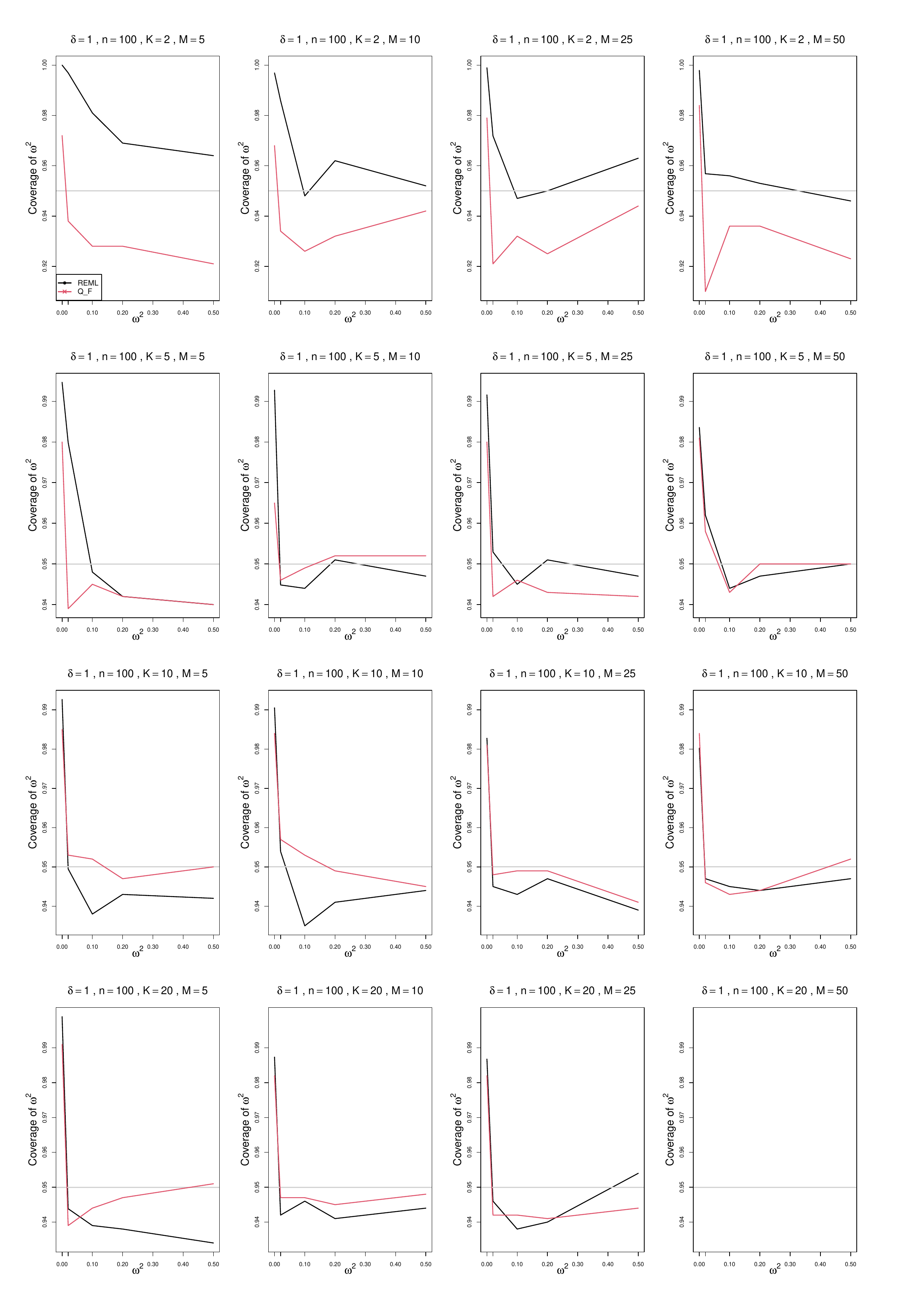}
	\caption{Coverage of 95\% confidence intervals for between-study variance of SMD (PL and $Q_F$ ) vs $\omega^2$, for $K$ = 2, 5, 10, and 20 studies per cluster and $M$ = 5, 10, 25, and 50 clusters; $\delta = 1$, and the sample size $n$ = 100 in each study.  }
	\label{PlotCoverageOfOmega2_100_1_HIER.pdf}
\end{figure}

\begin{figure}[ht]
	\centering
	\includegraphics[scale=0.33]{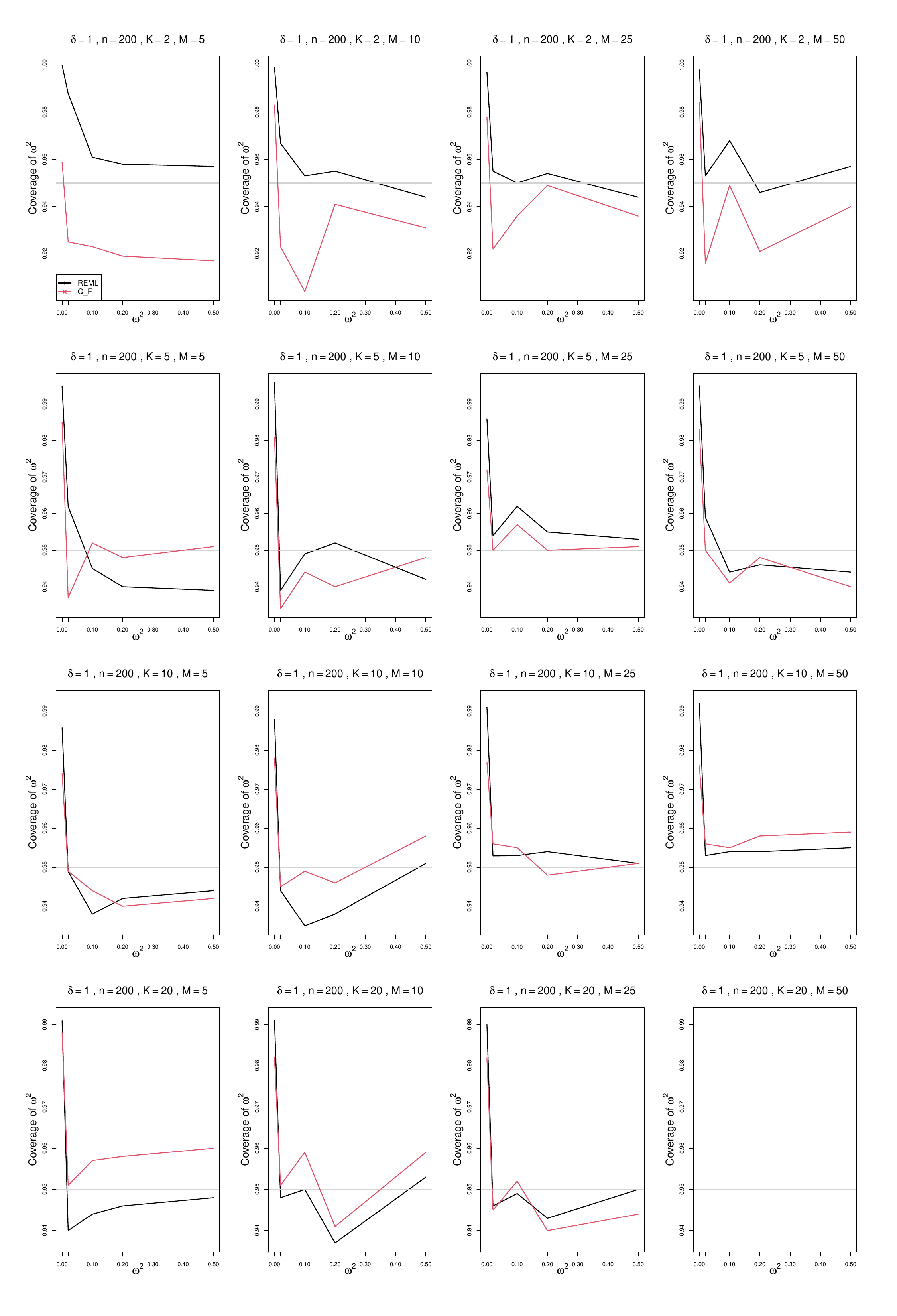}
	\caption{Coverage of 95\% confidence intervals for between-study variance of SMD (PL and $Q_F$ ) vs $\omega^2$, for $K$ = 2, 5, 10, and 20 studies per cluster and $M$ = 5, 10, 25, and 50 clusters; $\delta = 1$, and the sample size $n$ = 200 in each study.  }
	\label{PlotCoverageOfOmega2_200_1_HIER.pdf}
\end{figure}

\begin{figure}[ht]
	\centering
	\includegraphics[scale=0.33]{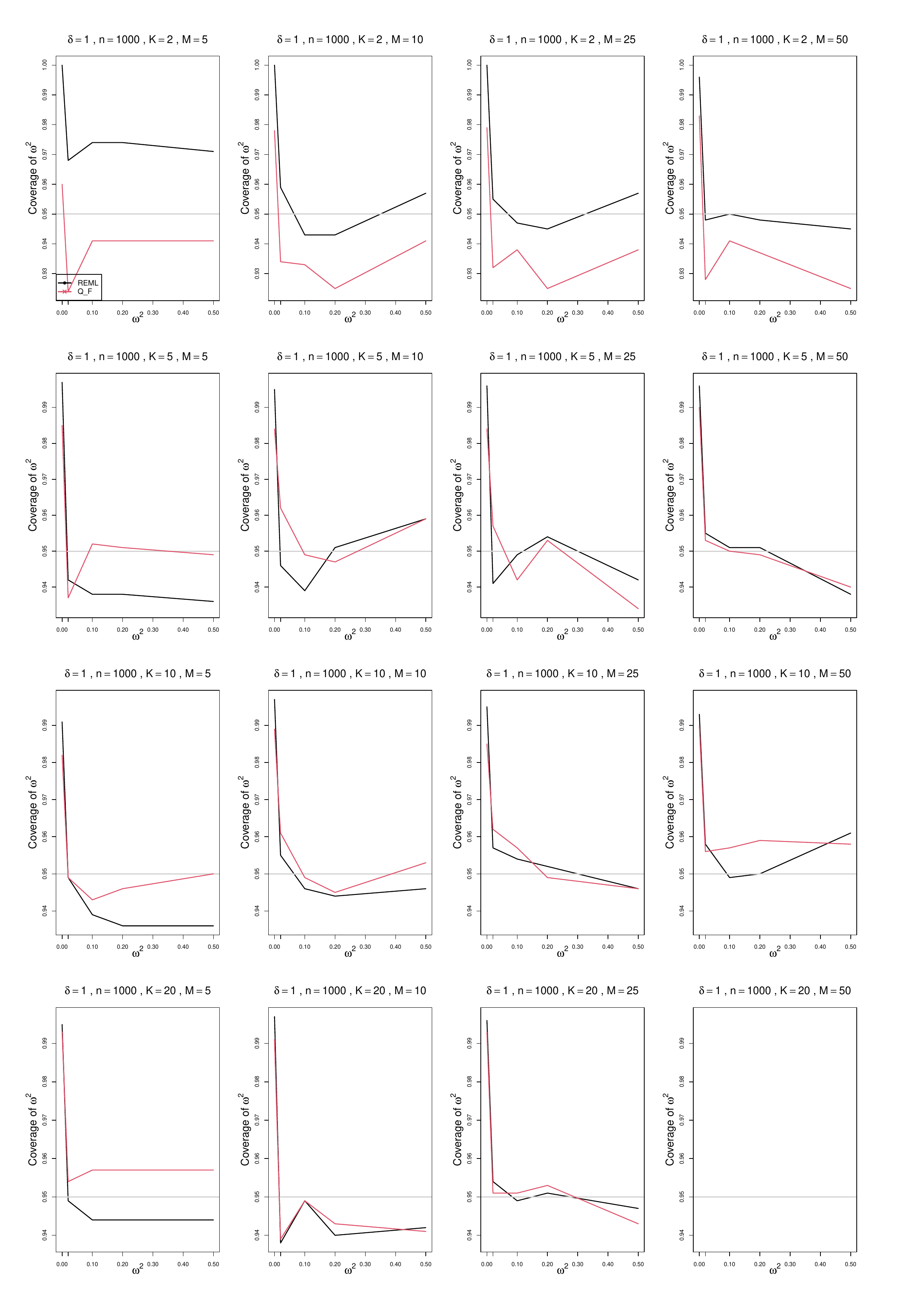}
	\caption{Coverage of 95\% confidence intervals for between-study variance of SMD (PL and $Q_F$ ) vs $\omega^2$, for $K$ = 2, 5, 10, and 20 studies per cluster and $M$ = 5, 10, 25, and 50 clusters; $\delta = 1$, and the sample size $n$ = 1000 in each study.  }
	\label{PlotCoverageOfOmega2_1000_1_HIER.pdf}
\end{figure}


\clearpage

\section*{Appendix G: Bias in point estimation of the overall effect $\delta$}

Each figure corresponds to a value of the standardized mean difference ($\delta$ = 0, 0.2, 0.5, 1) and a value of the study sample size ($n$ = 20, 40, 100, 200, 1000).\\
For each combination of the number of studies in a cluster ($K$ = 2, 5, 10, 20) and the number of clusters ($M$ = 5, 10, 25, 50), a panel plots bias of $\hat\delta$ versus $\tau^2$ (= 0, 0.02, 0.1, 0.2, 0.5).\\
The two variance components are held equal ($\tau^2 = \omega^2$).\\
The point estimators of $\delta$ are
\begin{itemize}
\item REML method, inverse-variance weights,  {\it  rma.mv} in {\it metafor})
\item SSW (conditional moment-based method, effective-sample-size weights)
\end{itemize}

\clearpage
\setcounter{figure}{0}
\renewcommand{\thefigure}{G.\arabic{figure}}


\begin{figure}[ht]
	\centering
	\includegraphics[scale=0.33]{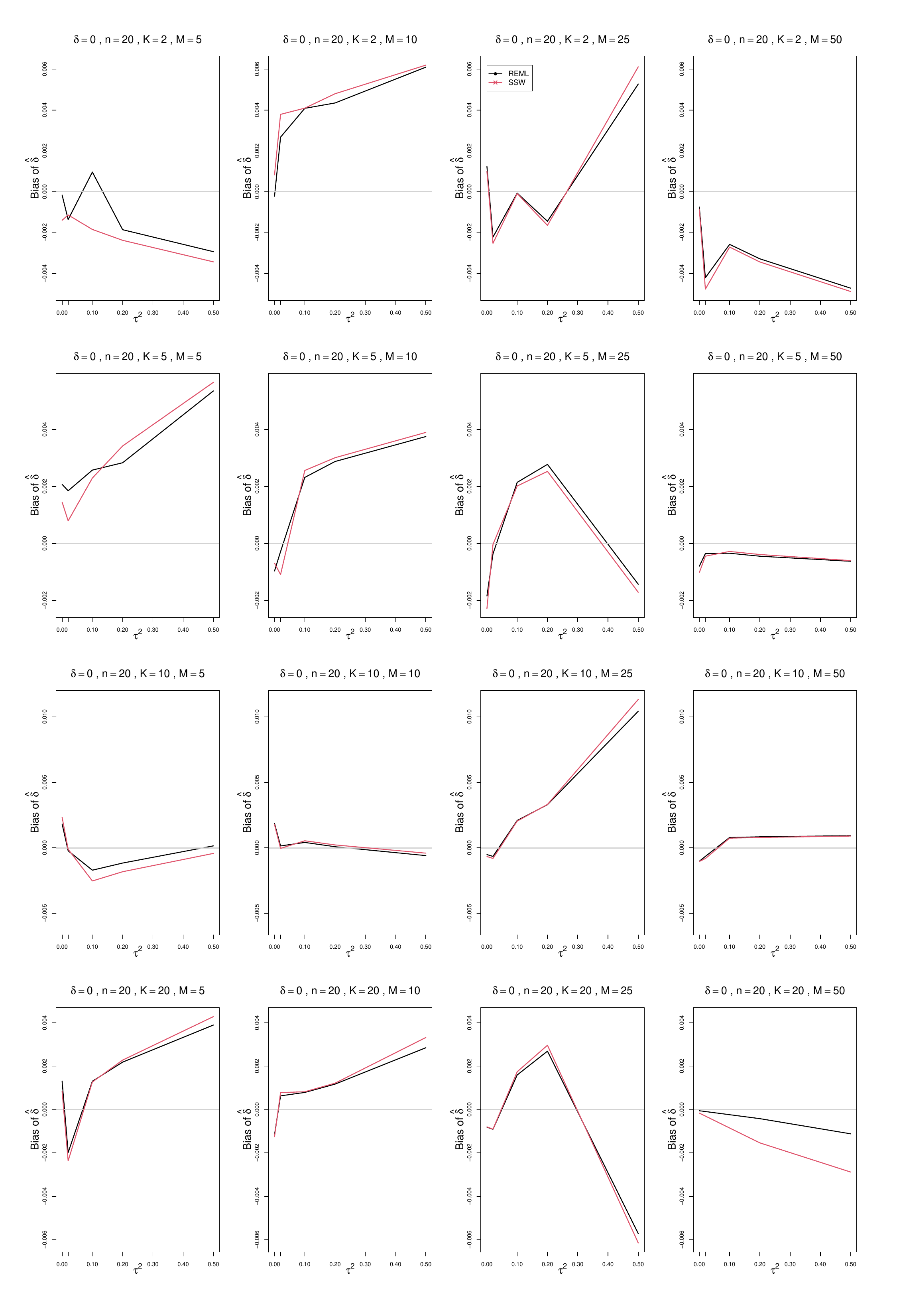}
	\caption{Bias of estimators of overall effect $\delta$  (REML and SSW ) vs $\tau^2$, for $K$ = 2, 5, 10, and 20 studies per cluster and $M$ = 5, 10, 25, and 50 clusters; $\delta = 0$, and the sample size $n$ = 20 in each study.  }
	\label{PlotBiasOfDelta_20_0_HIER.pdf}
\end{figure}

\begin{figure}[ht]
	\centering
	\includegraphics[scale=0.33]{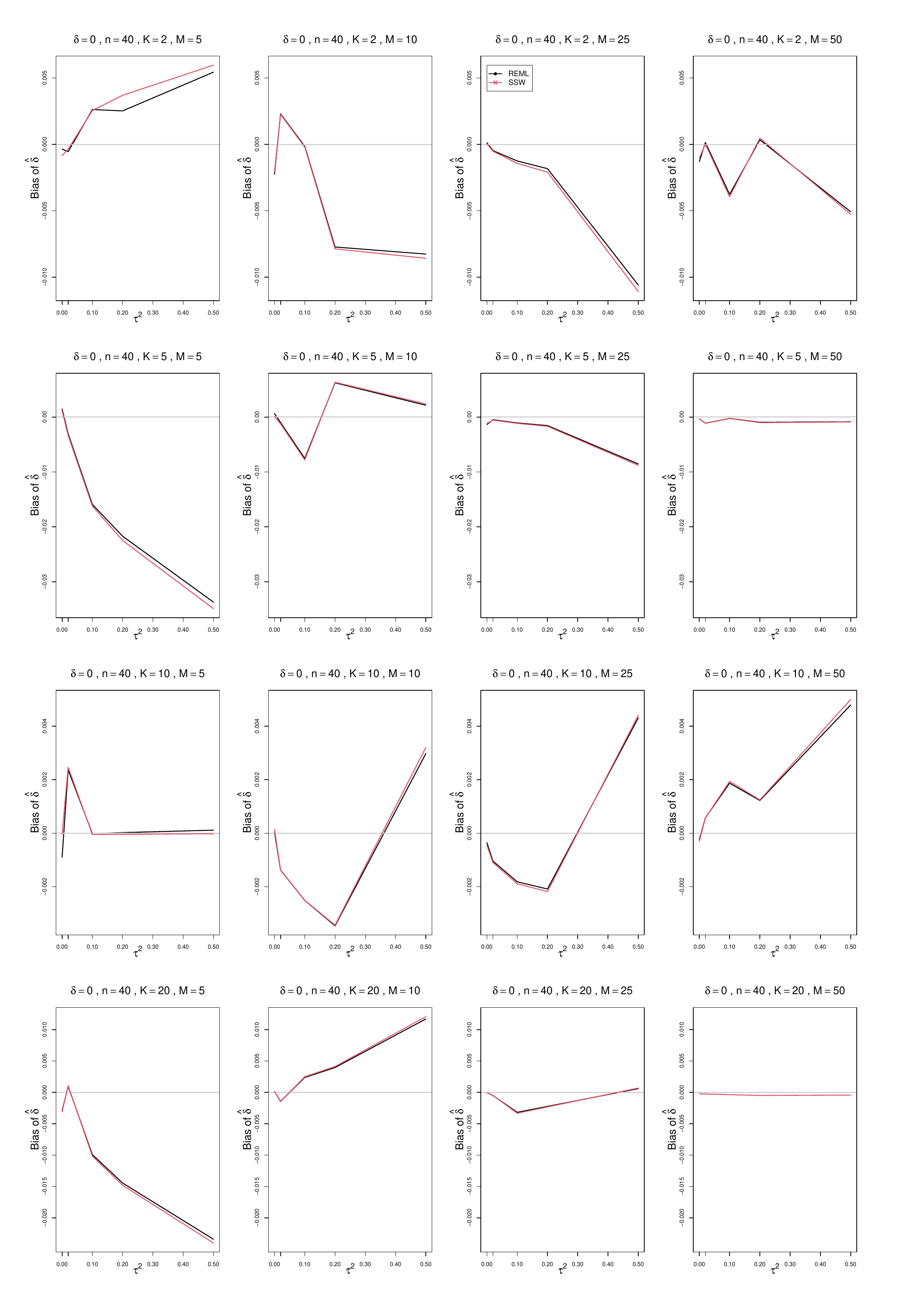}
	\caption{Bias  of estimators of  overall effect $\delta$ (REML and SSW ) vs $\tau^2$, for $K$ = 2, 5, 10, and 20 studies per cluster and $M$ = 5, 10, 25, and 50 clusters; $\delta = 0$, and the sample size $n$ = 40 in each study.  }
	\label{PlotBiasOfDelta_40_0_HIER.pdf}
\end{figure}

\begin{figure}[ht]
	\centering
	\includegraphics[scale=0.33]{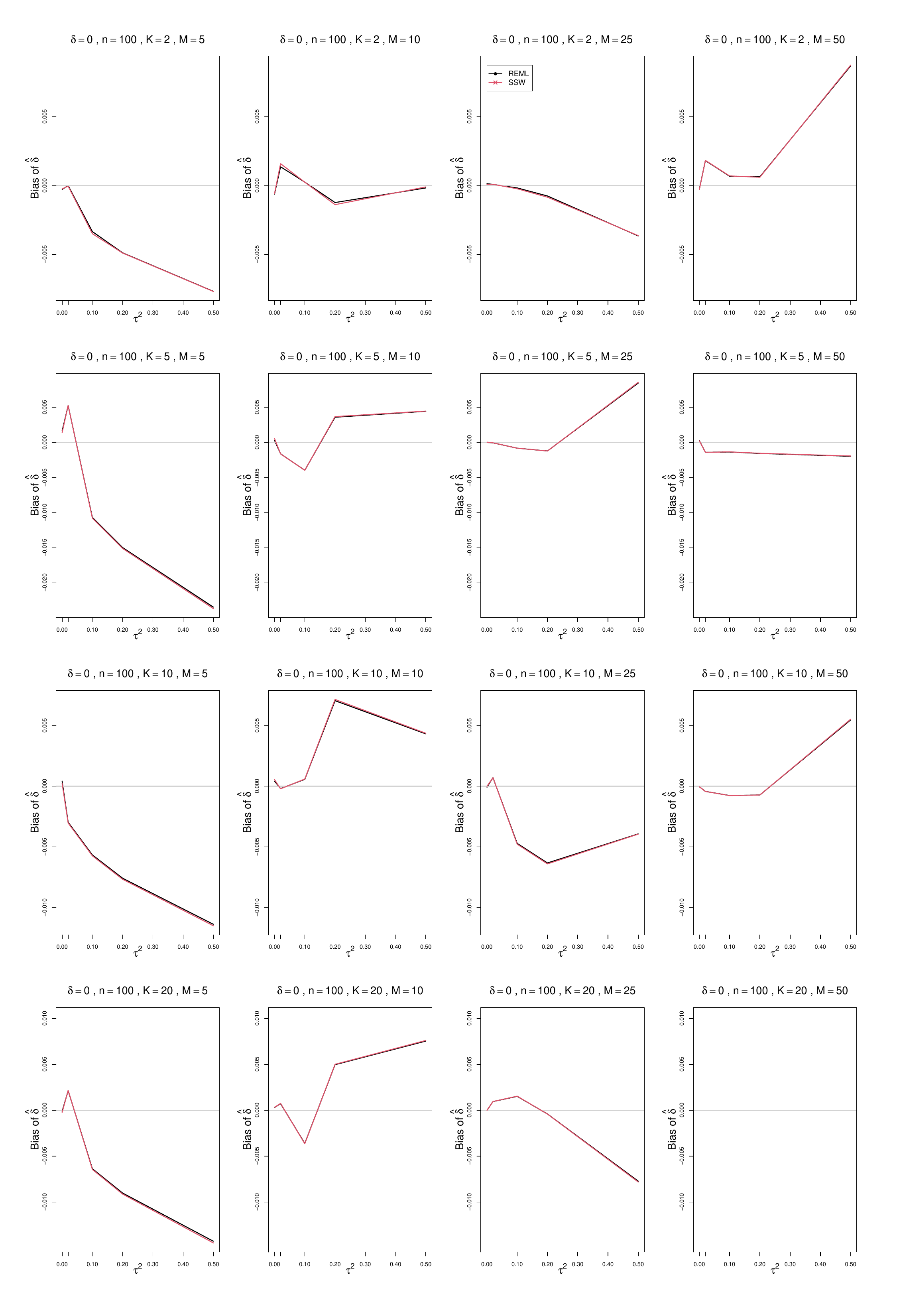}
	\caption{Bias  of estimators of  overall effect $\delta$ (REML and SSW ) vs $\tau^2$, for $K$ = 2, 5, 10, and 20 studies per cluster and $M$ = 5, 10, 25, and 50 clusters; $\delta = 0$, and the sample size $n$ = 100 in each study.   }
	\label{PlotBiasOfDelta_100_0_HIER.pdf}
\end{figure}

\begin{figure}[ht]
	\centering
	\includegraphics[scale=0.33]{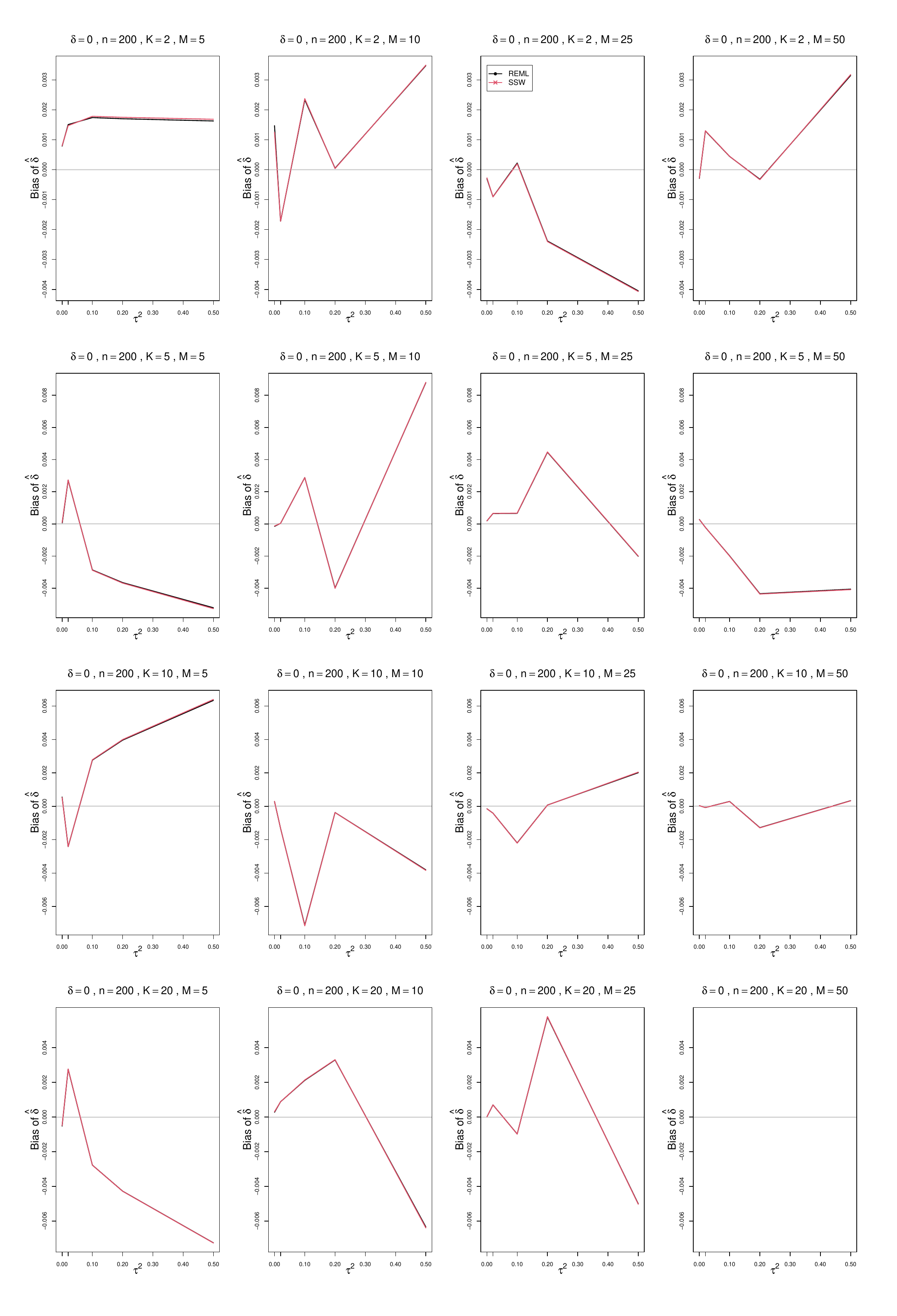}
	\caption{Bias  of estimators of overall effect $\delta$ (REML and SSW ) vs $\tau^2$, for $K$ = 2, 5, 10, and 20 studies per cluster and $M$ = 5, 10, 25, and 50 clusters; $\delta = 0$, and the sample size $n$ = 200 in each study.  }
	\label{PlotBiasOfDelta_200_0_HIER.pdf}
\end{figure}

\begin{figure}[ht]
	\centering
	\includegraphics[scale=0.33]{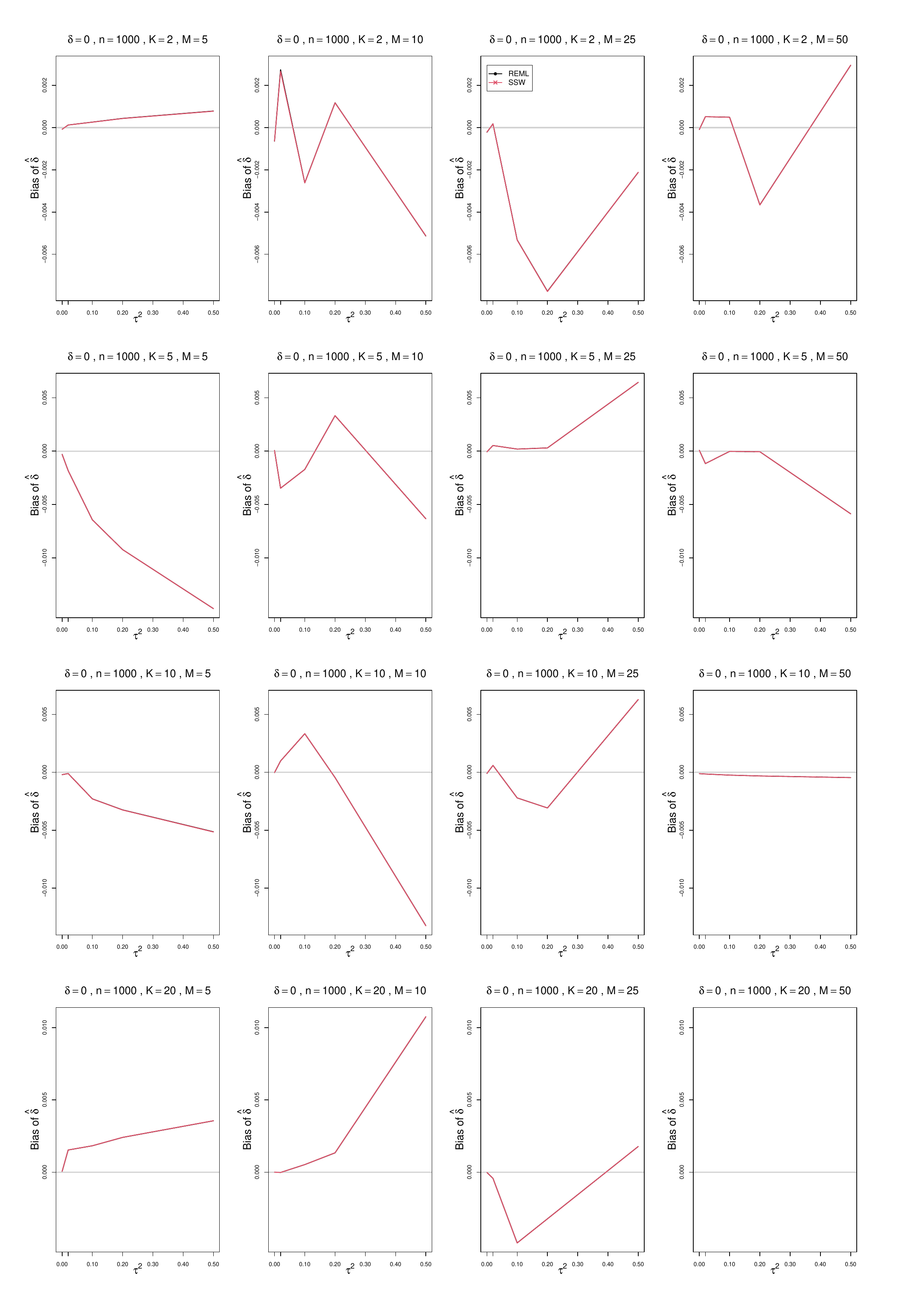}
	\caption{Bias  of estimators of overall effect $\delta$ (REML and SSW ) vs $\tau^2$, for $K$ = 2, 5, 10, and 20 studies per cluster and $M$ = 5, 10, 25, and 50 clusters; $\delta = 0$, and the sample size $n$ = 1000 in each study.  }
	\label{PlotBiasOfDelta_1000_0_HIER.pdf}
\end{figure}

\begin{figure}[ht]
	\centering
	\includegraphics[scale=0.33]{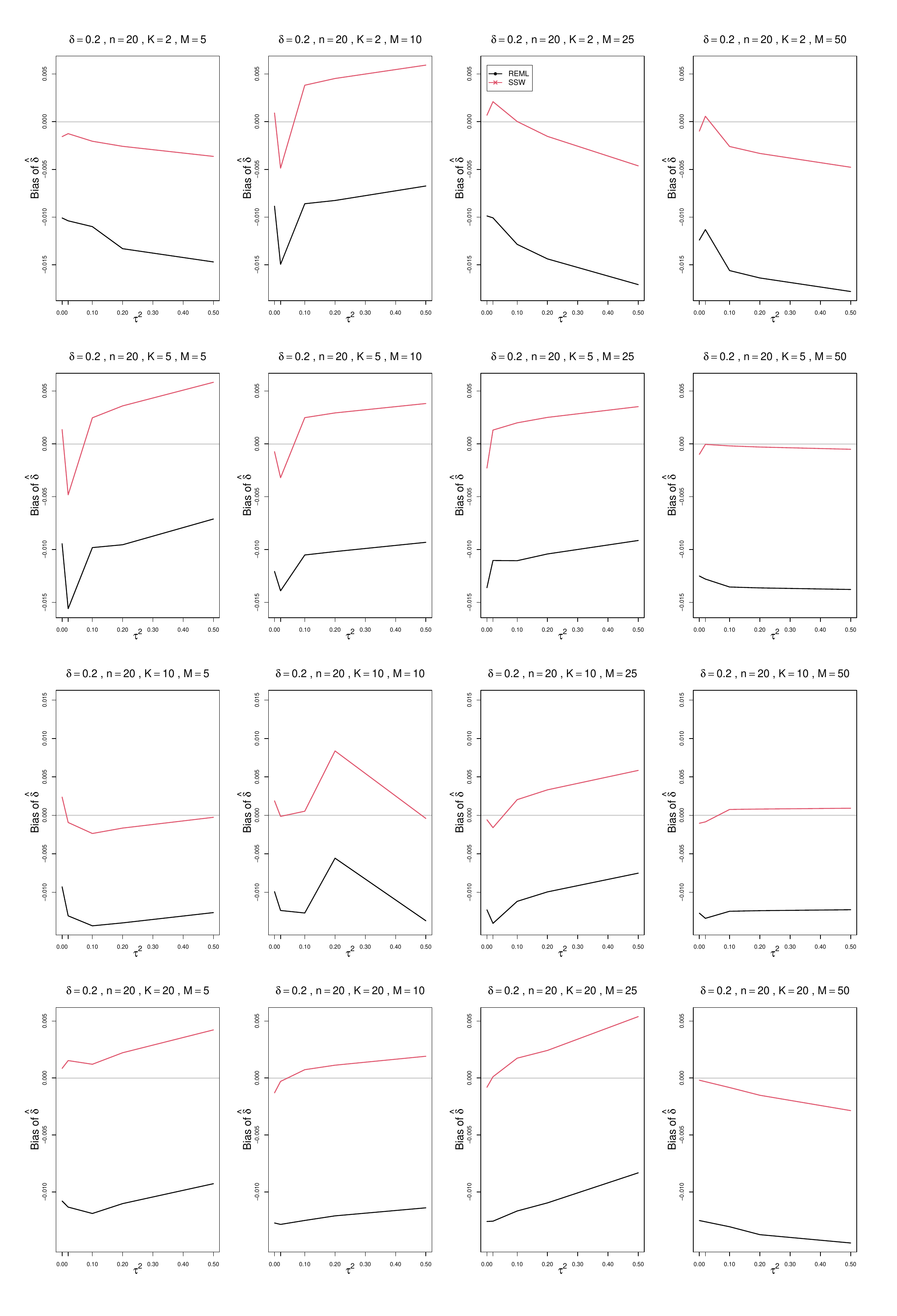}
	\caption{Bias  of estimators of overall effect $\delta$ (REML and SSW ) vs $\tau^2$, for $K$ = 2, 5, 10, and 20 studies per cluster and $M$ = 5, 10, 25, and 50 clusters; $\delta = 0.2$, and the sample size $n$ = 20 in each study.   }
	\label{PlotBiasOfDelta_20_02_HIER.pdf}
\end{figure}

\begin{figure}[ht]
	\centering
	\includegraphics[scale=0.33]{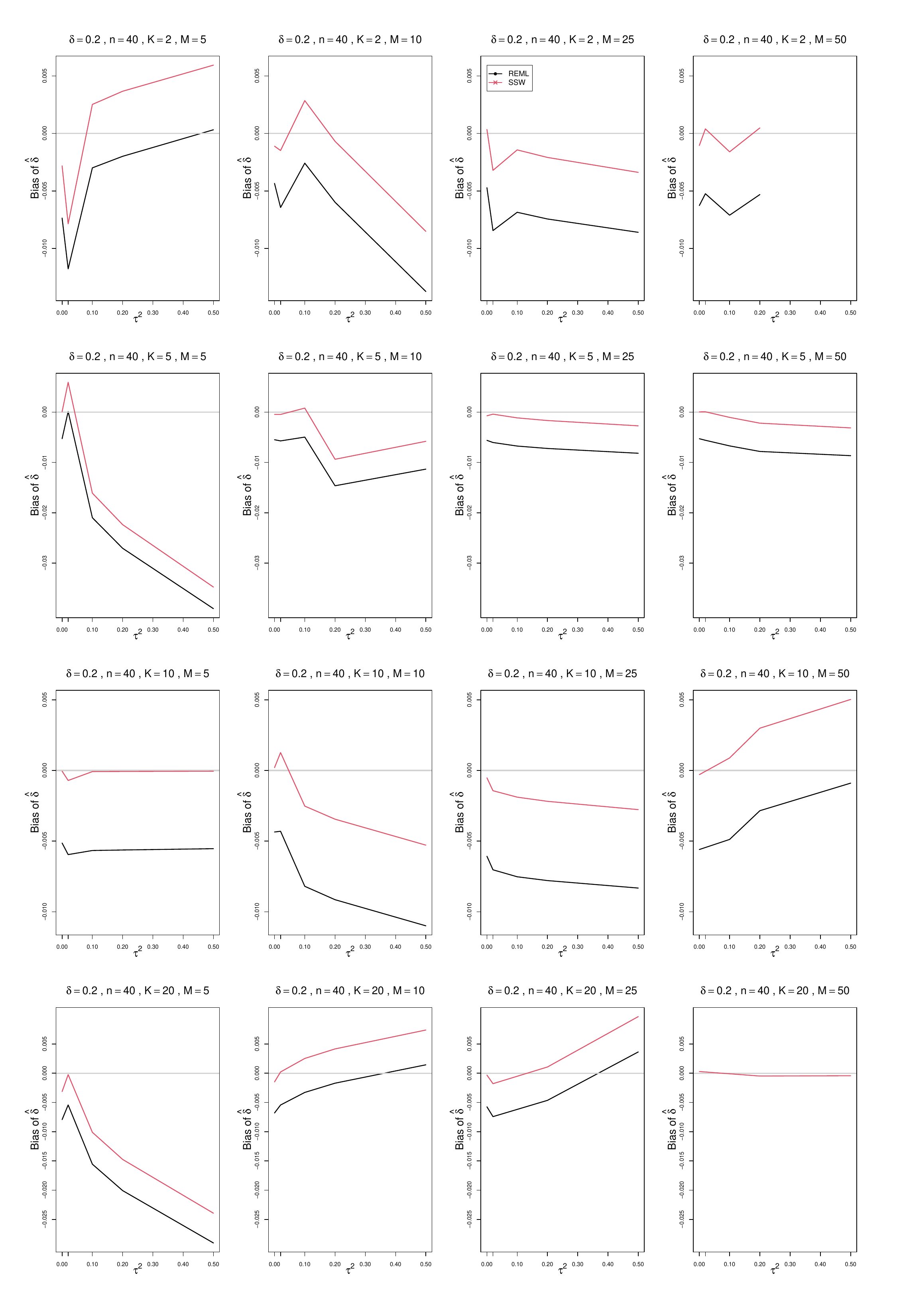}
	\caption{Bias  of estimators of overall effect $\delta$ (REML and SSW ) vs $\tau^2$, for $K$ = 2, 5, 10, and 20 studies per cluster and $M$ = 5, 10, 25, and 50 clusters; $\delta = 0.2$, and the sample size $n$ = 40 in each study.  }
	\label{PlotBiasOfDelta_40_02_HIER.pdf}
\end{figure}

\begin{figure}[ht]
	\centering
	\includegraphics[scale=0.33]{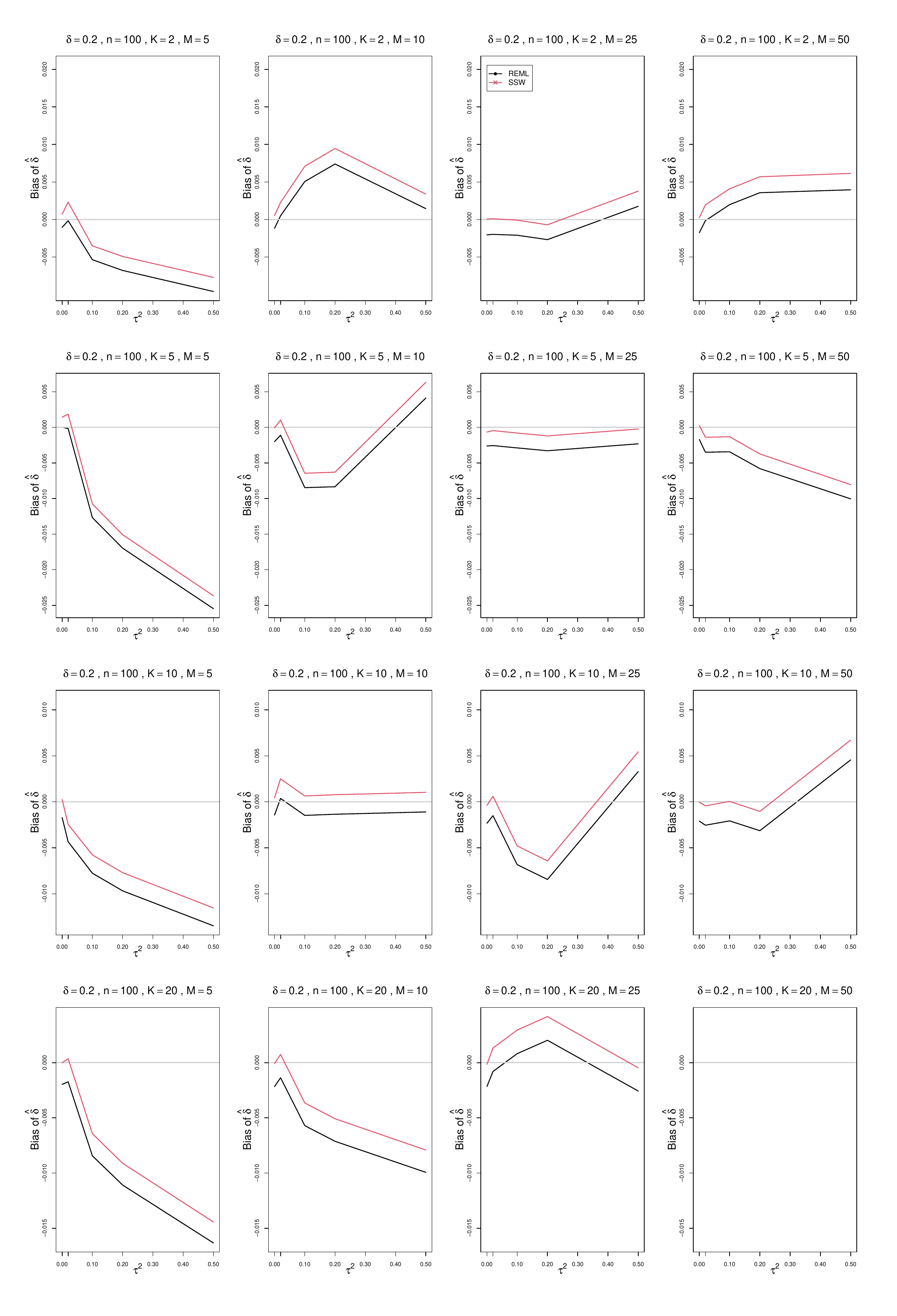}
	\caption{Bias  of estimators of overall effect $\delta$ (REML and SSW ) vs $\tau^2$, for $K$ = 2, 5, 10, and 20 studies per cluster and $M$ = 5, 10, 25, and 50 clusters; $\delta = 0.2$, and the sample size $n$ = 100 in each study.  }
	\label{PlotBiasOfDelta_100_02_HIER.pdf}
\end{figure}

\begin{figure}[ht]
	\centering
	\includegraphics[scale=0.33]{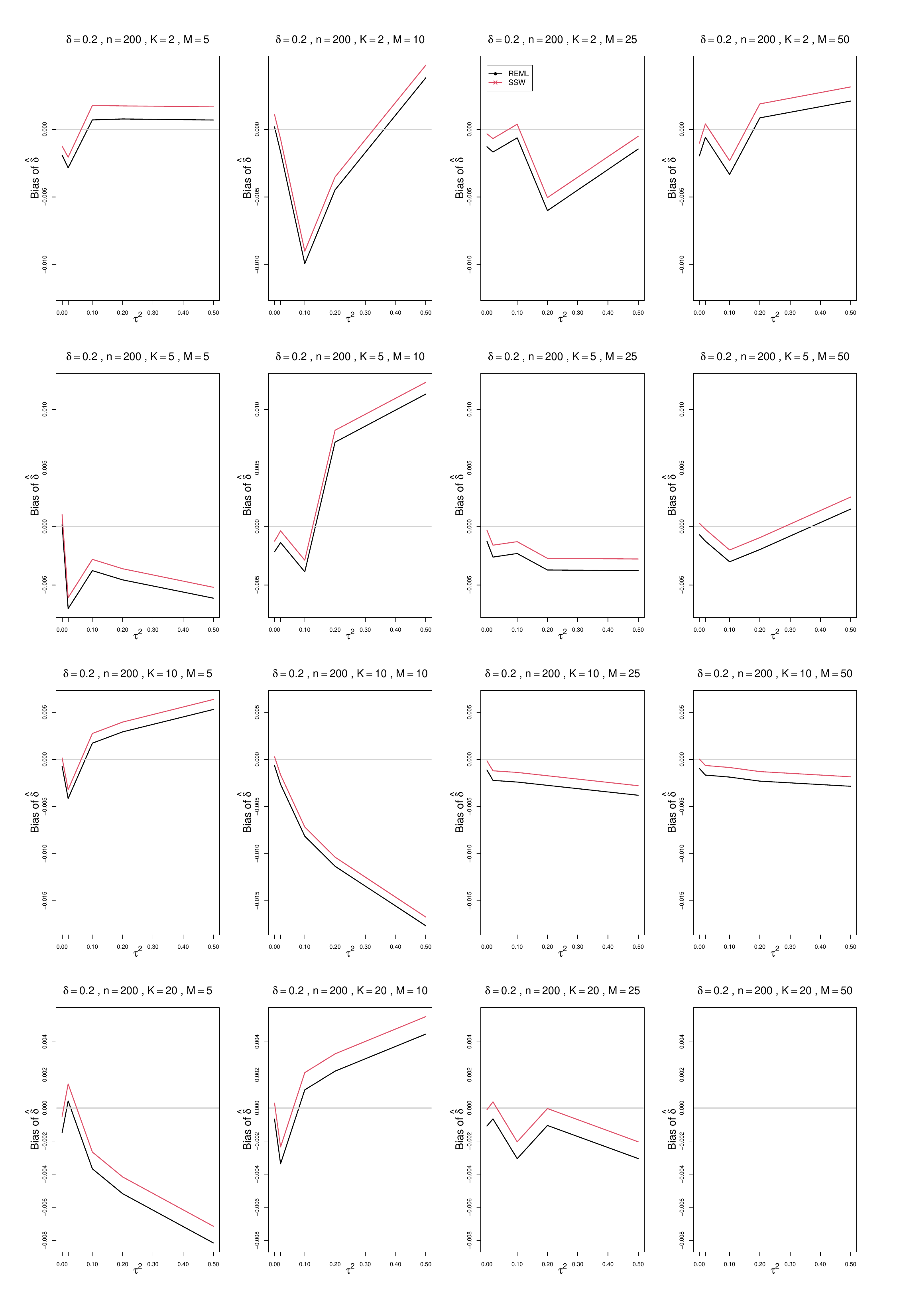}
	\caption{Bias  of estimators of overall effect $\delta$ (REML and SSW ) vs $\tau^2$, for $K$ = 2, 5, 10, and 20 studies per cluster and $M$ = 5, 10, 25, and 50 clusters; $\delta = 0.2$, and the sample size $n$ = 200 in each study.   }
	\label{PlotBiasOfDelta_200_02_HIER.pdf}
\end{figure}

\begin{figure}[ht]
	\centering
	\includegraphics[scale=0.33]{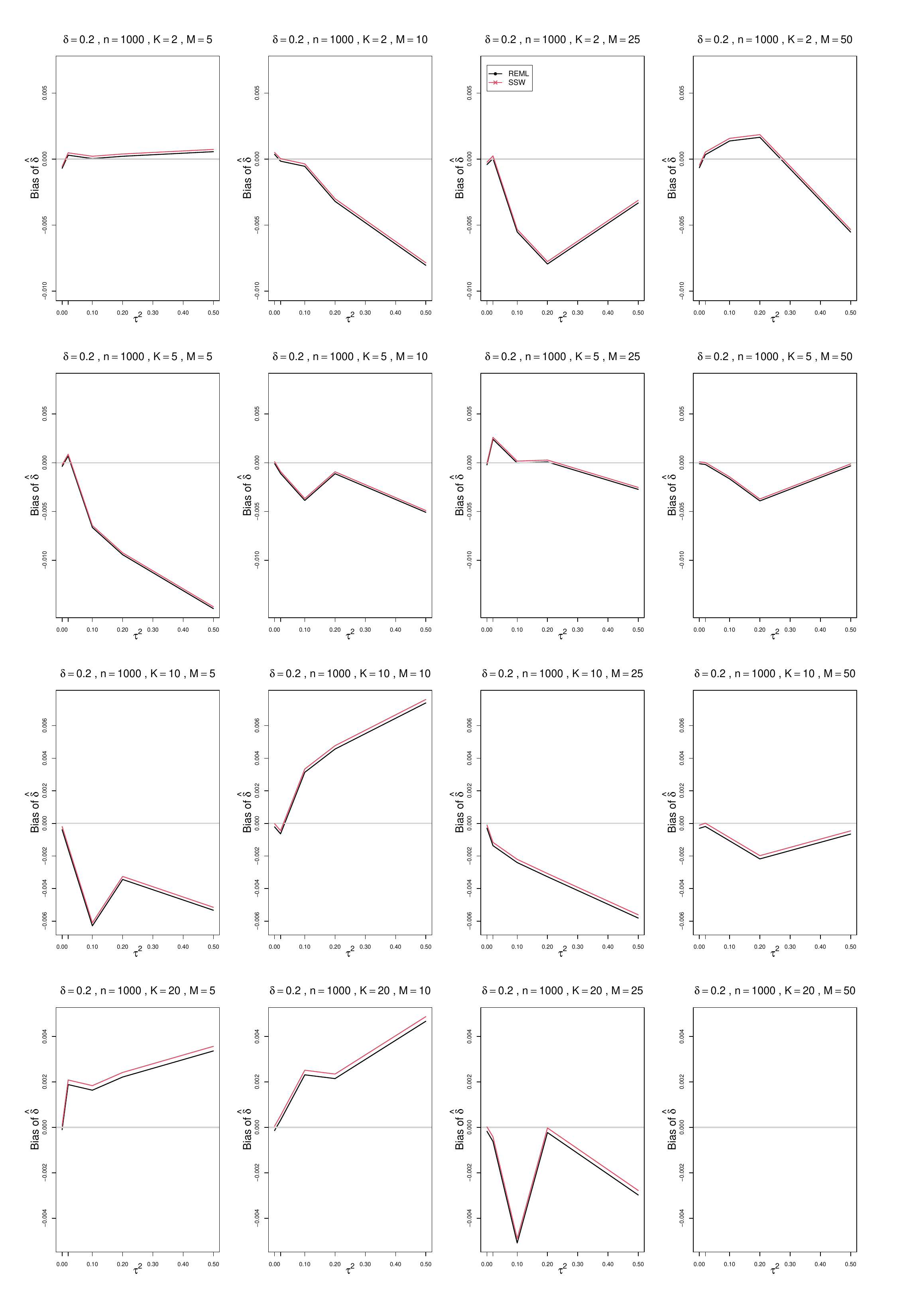}
	\caption{Bias  of estimators of overall effect $\delta$ (REML and SSW ) vs $\tau^2$, for $K$ = 2, 5, 10, and 20 studies per cluster and $M$ = 5, 10, 25, and 50 clusters; $\delta = 0.2$, and the sample size $n$ = 1000 in each study.  }
	\label{PlotBiasOfDelta_1000_02_HIER.pdf}
\end{figure}

\begin{figure}[ht]
	\centering
	\includegraphics[scale=0.33]{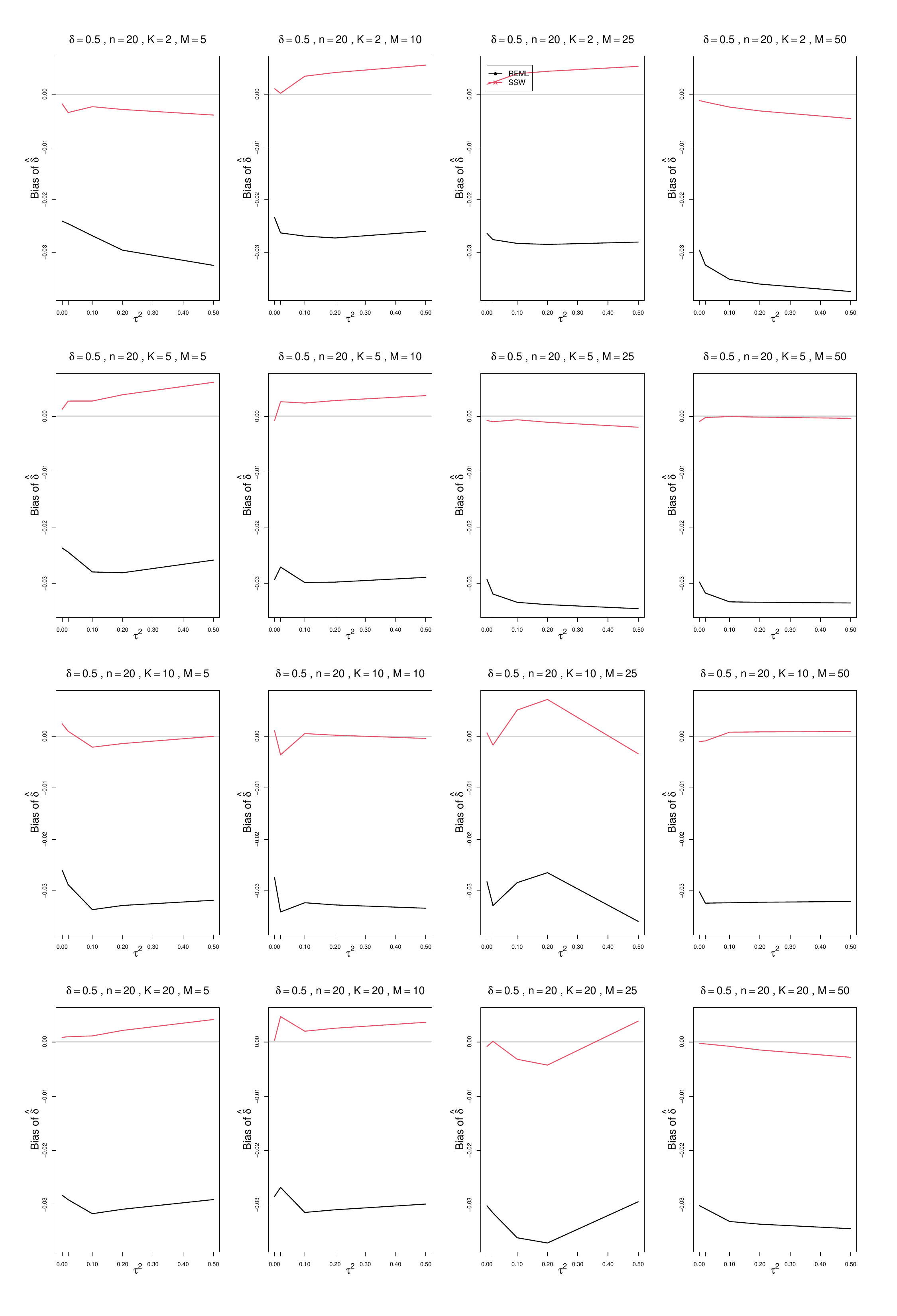}
	\caption{Bias  of estimators of overall effect $\delta$ (REML and SSW ) vs $\tau^2$, for $K$ = 2, 5, 10, and 20 studies per cluster and $M$ = 5, 10, 25, and 50 clusters; $\delta = 0.5$, and the sample size $n$ = 20 in each study.  }
	\label{PlotBiasOfDelta_20_05_HIER.pdf}
\end{figure}

\begin{figure}[ht]
	\centering
	\includegraphics[scale=0.33]{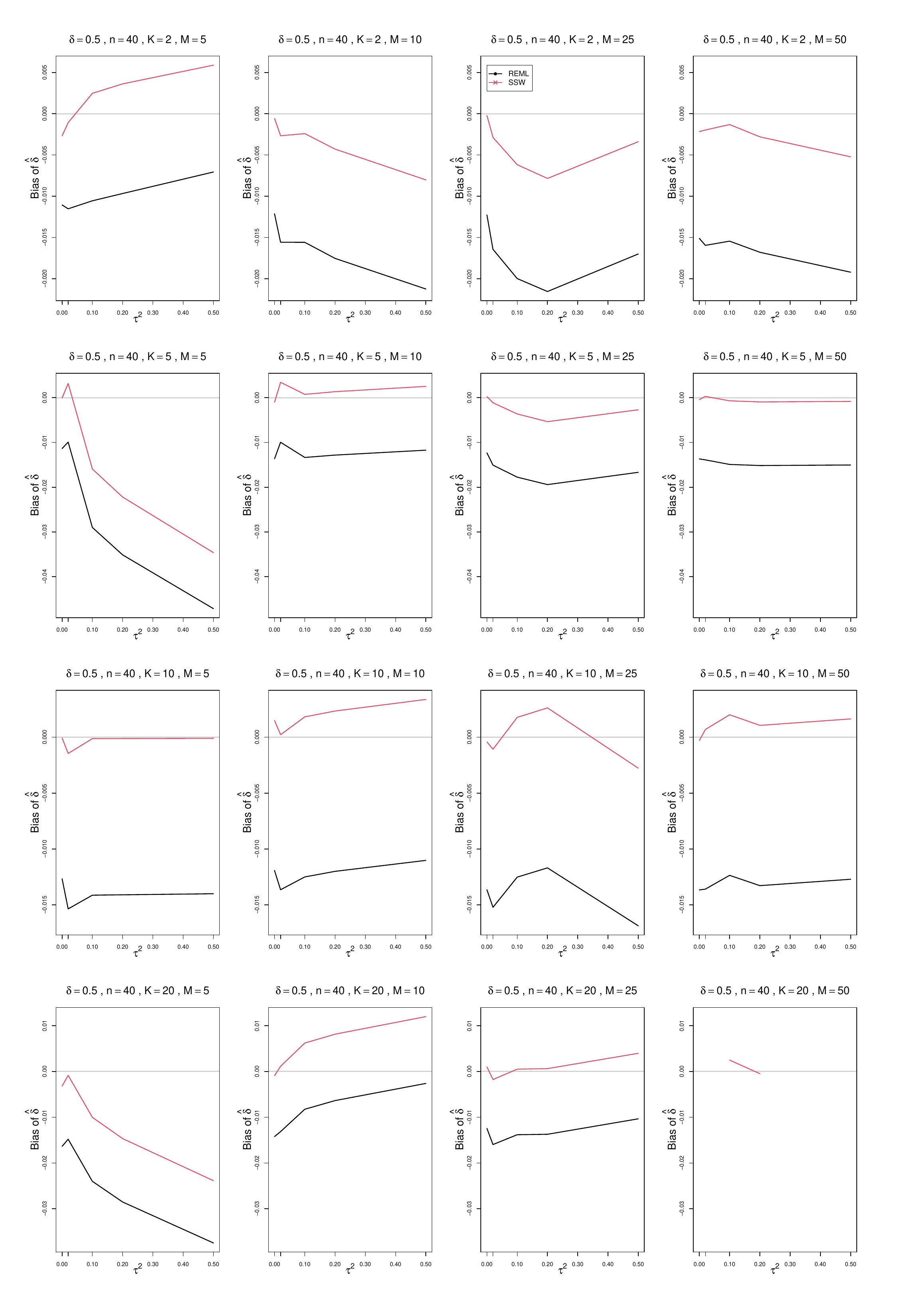}
	\caption{Bias  of estimators of overall effect $\delta$ (REML and SSW ) vs $\tau^2$, for $K$ = 2, 5, 10, and 20 studies per cluster and $M$ = 5, 10, 25, and 50 clusters; $\delta = 0.5$, and the sample size $n$ = 40 in each study. }
	\label{PlotBiasOfDelta_40_05_HIER.pdf}
\end{figure}

\begin{figure}[ht]
	\centering
	\includegraphics[scale=0.33]{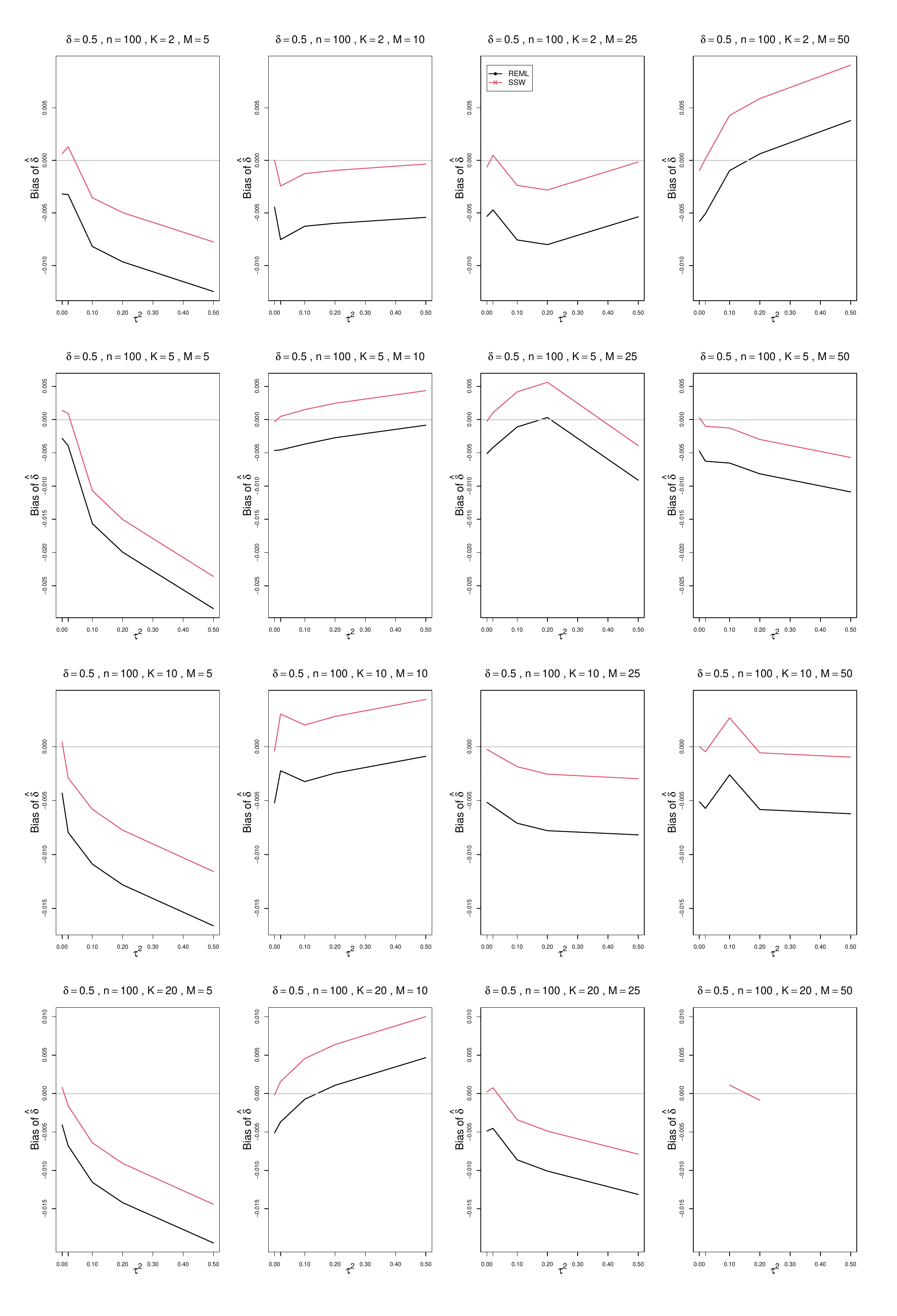}
	\caption{Bias  of estimators of overall effect $\delta$ (REML and SSW ) vs $\tau^2$, for $K$ = 2, 5, 10, and 20 studies per cluster and $M$ = 5, 10, 25, and 50 clusters; $\delta = 0.5$, and the sample size $n$ = 100 in each study.  }
	\label{PlotBiasOfDelta_100_05_HIER.pdf}
\end{figure}

\begin{figure}[ht]
	\centering
	\includegraphics[scale=0.33]{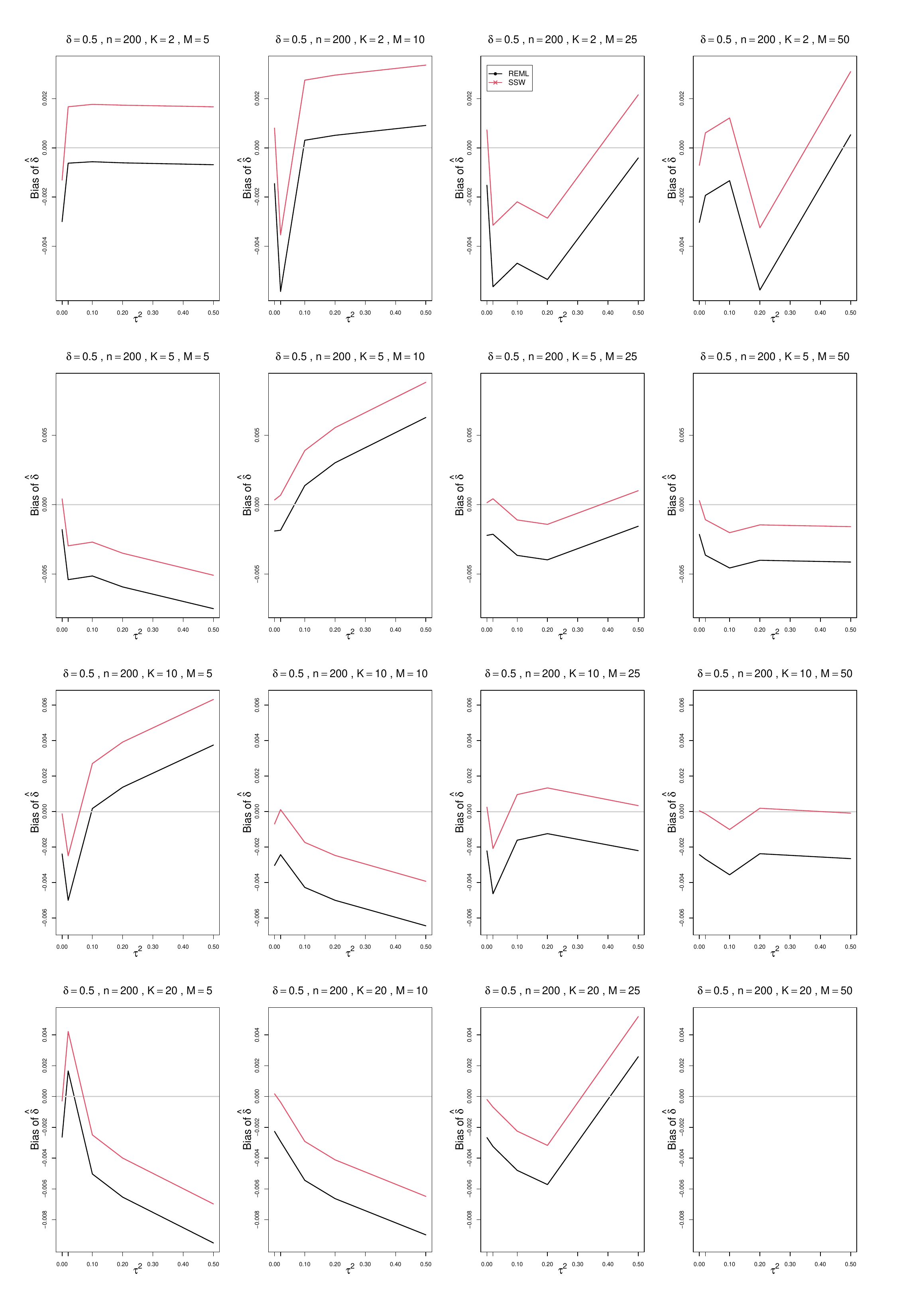}
	\caption{Bias  of estimators of overall effect $\delta$ (REML and $Q_A$ ) vs $\tau^2$, for $K$ = 2, 5, 10, and 20 studies per cluster and $M$ = 5, 10, 25, and 50 clusters; $\delta = 0.5$, and the sample size $n$ = 200 in each study.   }
	\label{PlotBiasOfDelta_200_05_HIER.pdf}
\end{figure}

\begin{figure}[ht]
	\centering
	\includegraphics[scale=0.33]{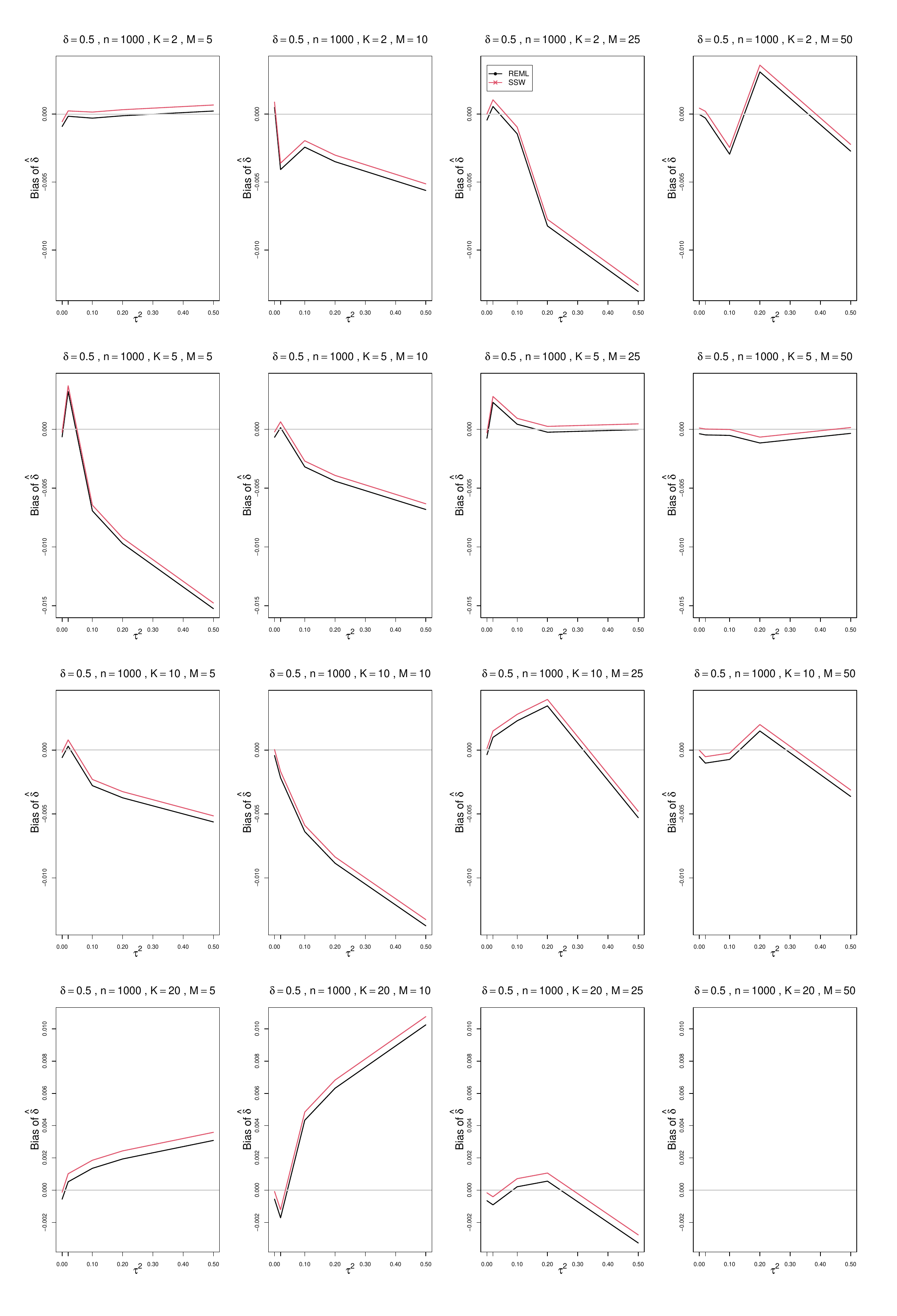}
	\caption{Bias  of estimators of overall effect $\delta$ (REML and SSW ) vs $\tau^2$, for $K$ = 2, 5, 10, and 20 studies per cluster and $M$ = 5, 10, 25, and 50 clusters; $\delta = 0.5$, and the sample size $n$ = 1000 in each study.  }
	\label{PlotBiasOfDelta_1000_05_HIER.pdf}
\end{figure}

\begin{figure}[ht]
	\centering
	\includegraphics[scale=0.33]{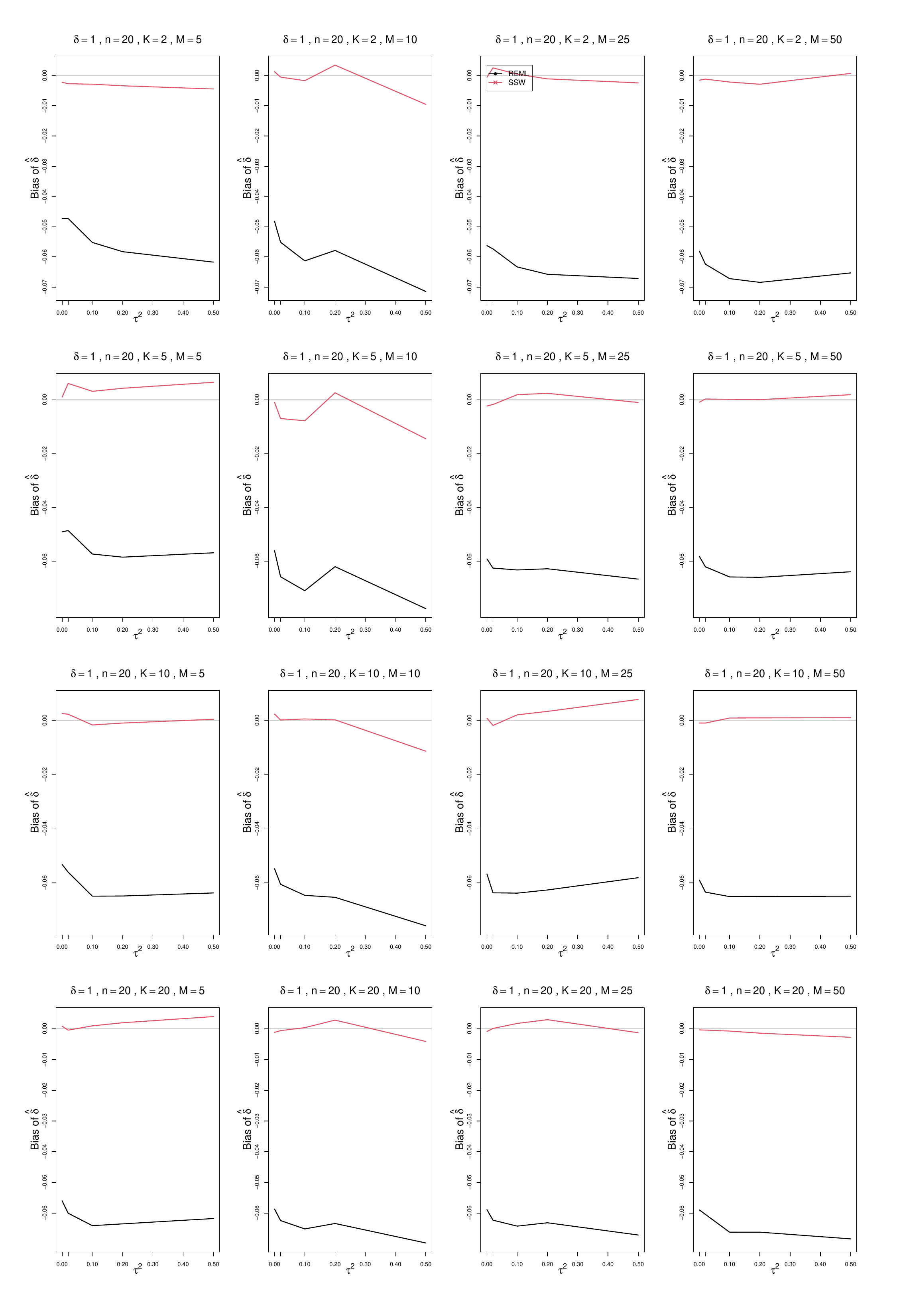}
	\caption{Bias  of estimators of overall effect $\delta$ (REML and SSW ) vs $\tau^2$, for $K$ = 2, 5, 10, and 20 studies per cluster and $M$ = 5, 10, 25, and 50 clusters; $\delta = 1$, and the sample size $n$ = 20 in each study.  }
	\label{PlotBiasOfDelta_20_1_HIER.pdf}
\end{figure}

\begin{figure}[ht]
	\centering
	\includegraphics[scale=0.33]{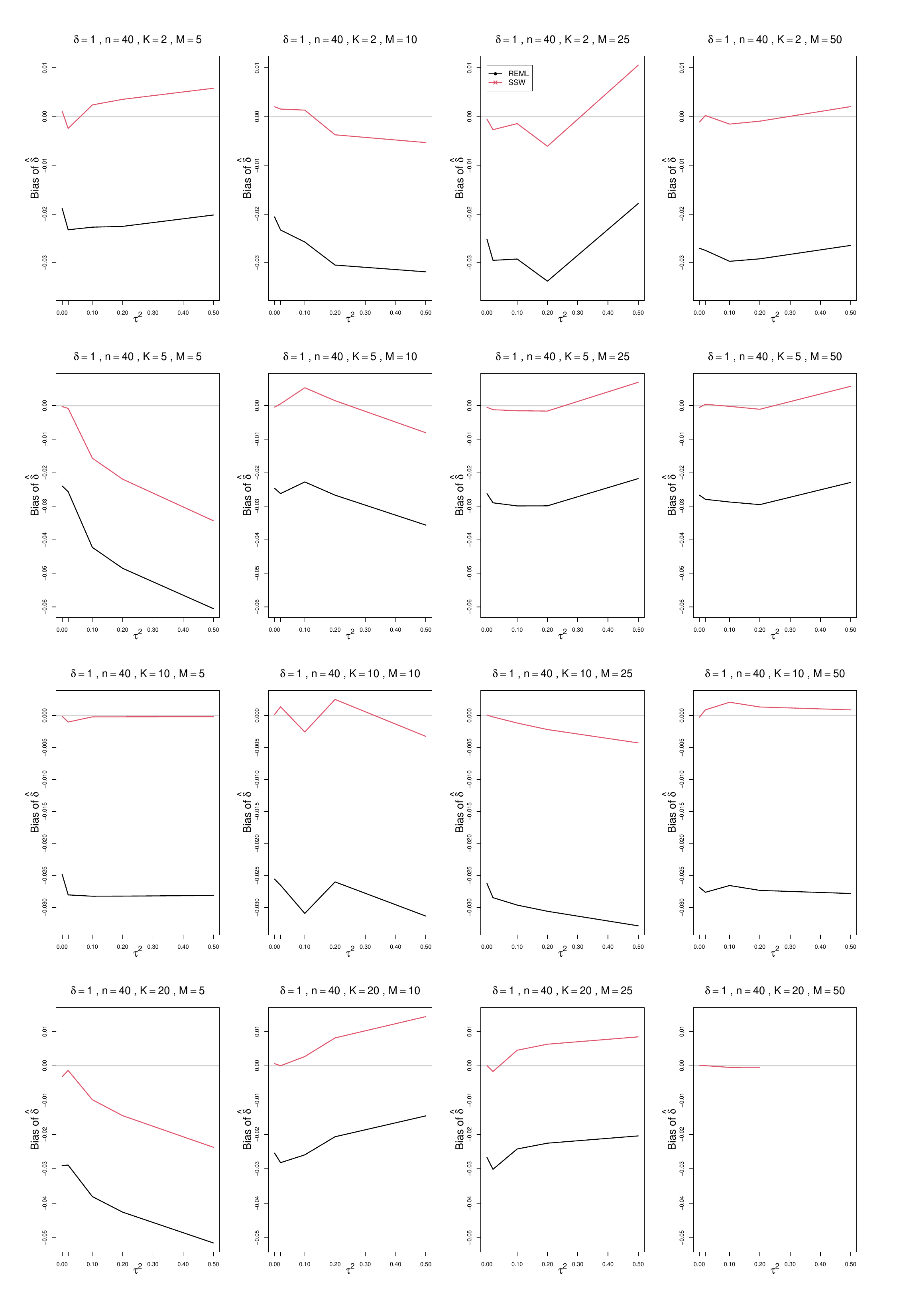}
	\caption{Bias  of estimators of overall effect $\delta$ (REML and SSW ) vs $\tau^2$, for $K$ = 2, 5, 10, and 20 studies per cluster and $M$ = 5, 10, 25, and 50 clusters; $\delta = 1$, and the sample size $n$ = 40 in each study.  }
	\label{PlotBiasOfDelta_40_1_HIER.pdf}
\end{figure}

\begin{figure}[ht]
	\centering
	\includegraphics[scale=0.33]{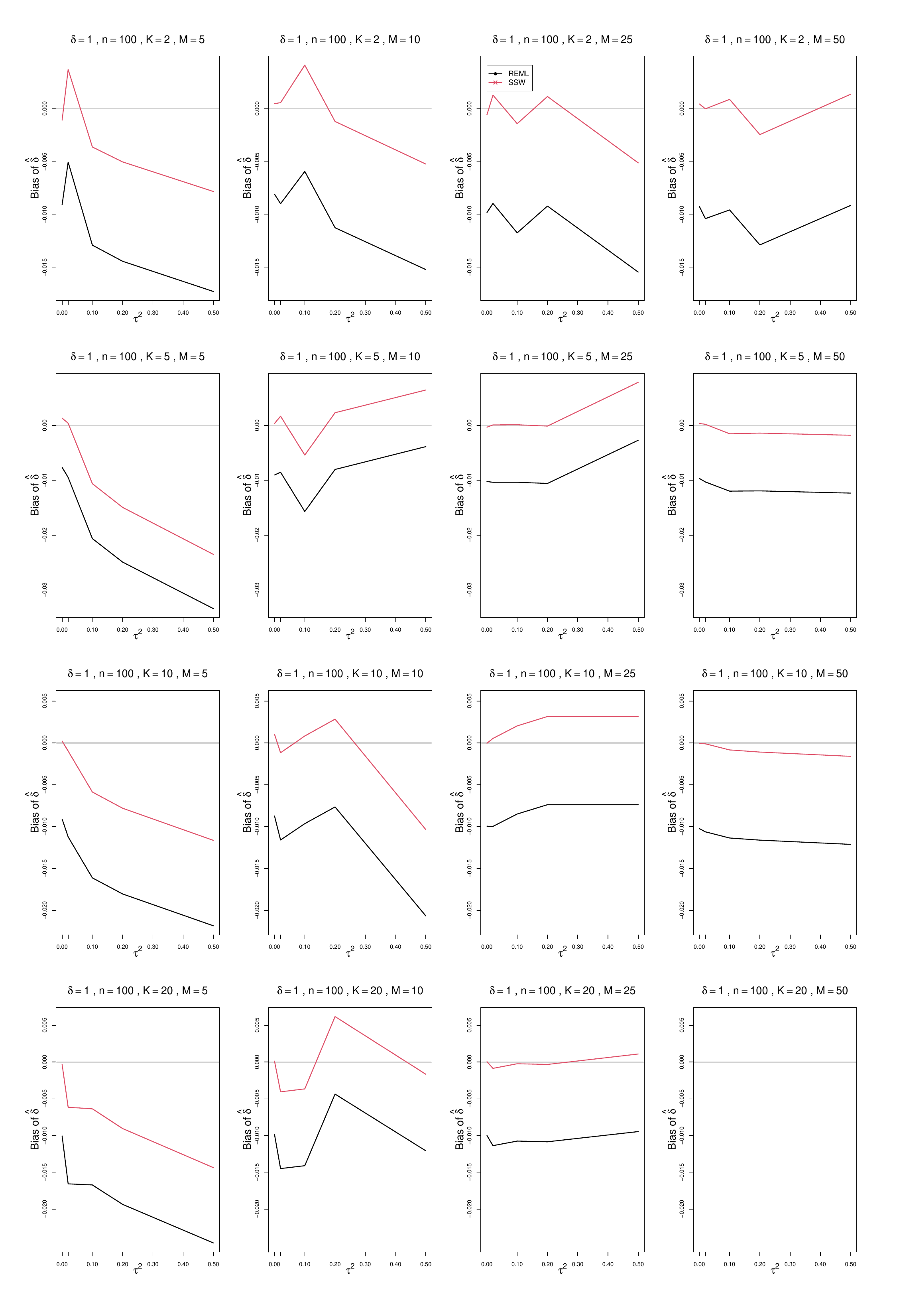}
	\caption{Bias  of estimators of overall effect $\delta$ (REML and SSW) vs $\tau^2$, for $K$ = 2, 5, 10, and 20 studies per cluster and $M$ = 5, 10, 25, and 50 clusters; $\delta = 1$, and the sample size $n$ = 100 in each study.  }
	\label{PlotBiasOfDelta_100_1_HIER.pdf}
\end{figure}

\begin{figure}[ht]
	\centering
	\includegraphics[scale=0.33]{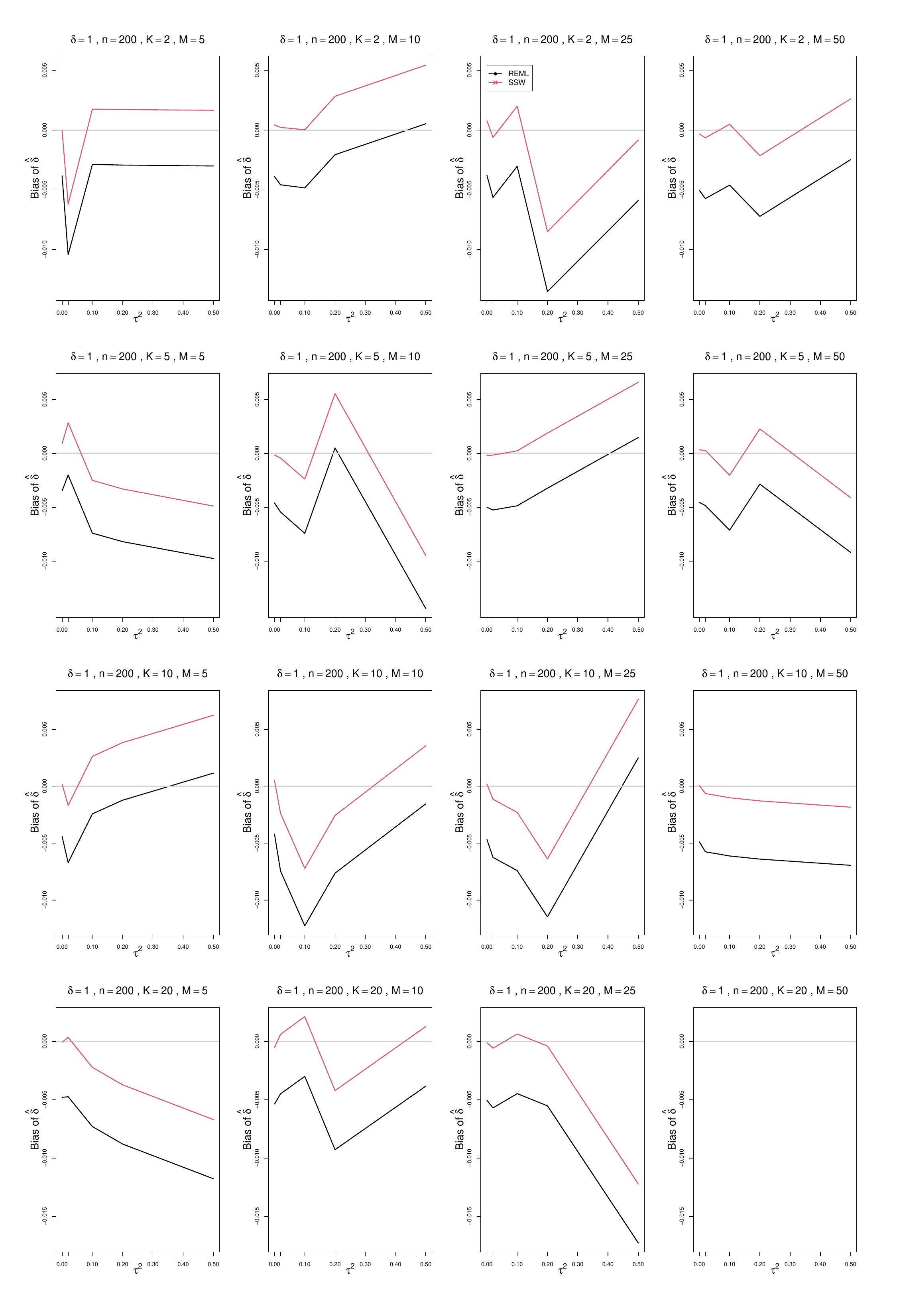}
	\caption{Bias  of estimators of overall effect $\delta$ (REML and SSW ) vs $\tau^2$, for $K$ = 2, 5, 10, and 20 studies per cluster and $M$ = 5, 10, 25, and 50 clusters; $\delta = 1$, and the sample size $n$ = 200 in each study.  }
	\label{PlotBiasOfDelta_200_1_HIER.pdf}
\end{figure}

\begin{figure}[ht]
	\centering
	\includegraphics[scale=0.33]{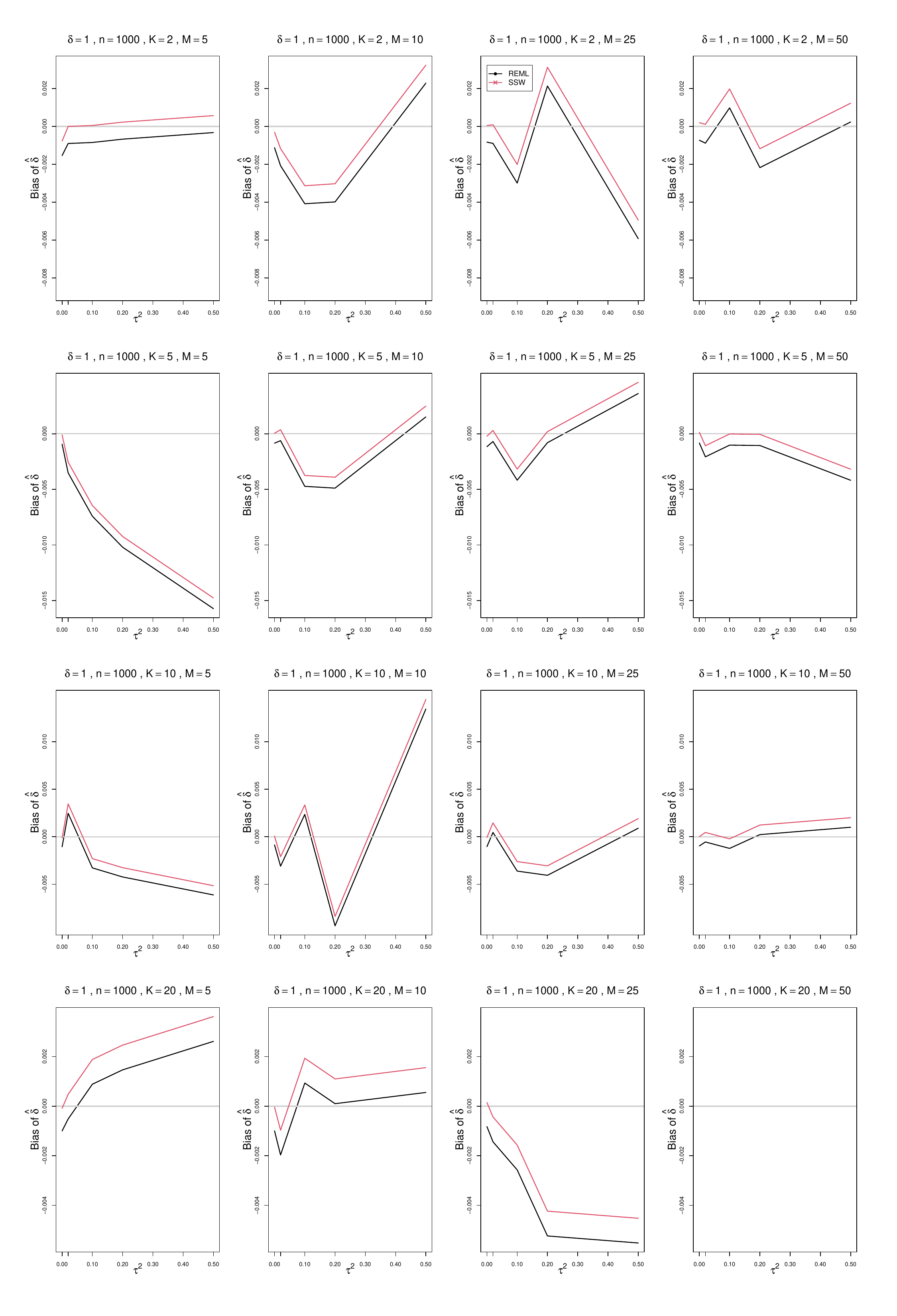}
	\caption{Bias  of estimators of overall effect $\delta$ (REML and SSW ) vs $\tau^2$, for $K$ = 2, 5, 10, and 20 studies per cluster and $M$ = 5, 10, 25, and 50 clusters; $\delta = 1$, and the sample size $n$ = 1000 in each study.  }
	\label{PlotBiasOfDelta_1000_1_HIER.pdf}
\end{figure}


\clearpage
\renewcommand{\thefigure}{H.\arabic{figure}}

\section*{Appendix H: Coverage of 95\% confidence intervals for the overall effect $\delta$}

Each figure corresponds to a value of the standardized mean difference ($\delta$  = 0, 0.2, 0.5, 1) and a value of the study sample size ($n$ = 20, 40, 100, 200, 1000).\\
The two variance components are held equal ($\tau^2 = \omega^2$).\\
For each combination of the number of studies in a cluster ($K$ = 2, 5, 10, 20) and the number of clusters ($M$ = 5, 10, 25, 50), a panel plots coverage of the overall effect $\delta$ at the 95\% nominal level versus $\tau^2$ (= 0, 0.02, 0.1, 0.2, 0.5).\\
The interval estimators of $\delta$ are centered at a point estimator of $\delta$. Their half-width equals the point estimator's standard error, calculated from REML or from moment-based estimators of variance components, combined with normal or $t$ critical values.
\begin{itemize}
\item REML N (REML, inverse-variance weights, normal critical values  {\it  rma.mv} in {\it metafor})
\item REML t (REML, inverse-variance weights, $t$ critical values  {\it  rma.mv} in {\it metafor})
\item $Q$ N (conditional moment-based method, effective-sample-size weights, normal critical values)
\item $Q$ t (conditional moment-based method, effective-sample-size weights,  $t$ critical values)
\end{itemize}

\clearpage
\setcounter{figure}{0}
\renewcommand{\thefigure}{H.\arabic{figure}}

\begin{figure}[ht]
	\centering
	\includegraphics[scale=0.33]{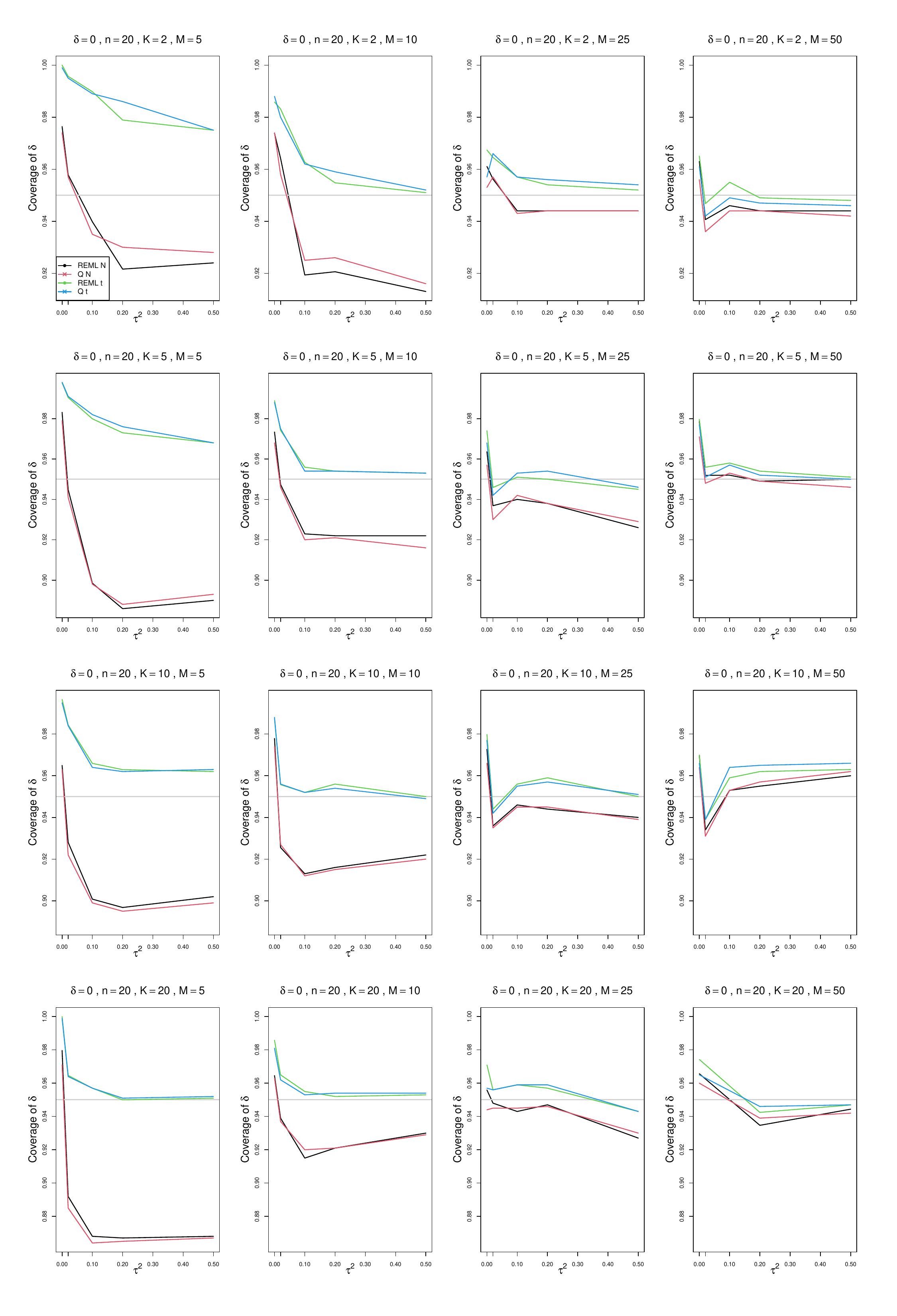}
	\caption{Coverage of 95\% confidence intervals for overall effect $\delta$ (REML N, REML t, $Q$ N, and $Q$ t) vs $\tau^2$, for $K$ = 2, 5, 10, and 20 studies per cluster and $M$ = 5, 10, 25, and 50 clusters; $\delta = 0$, and the sample size $n$ = 20 in each study.  }
	\label{PlotCoverageOfdelta_20_0_HIER.pdf}
\end{figure}

\begin{figure}[ht]
	\centering
	\includegraphics[scale=0.33]{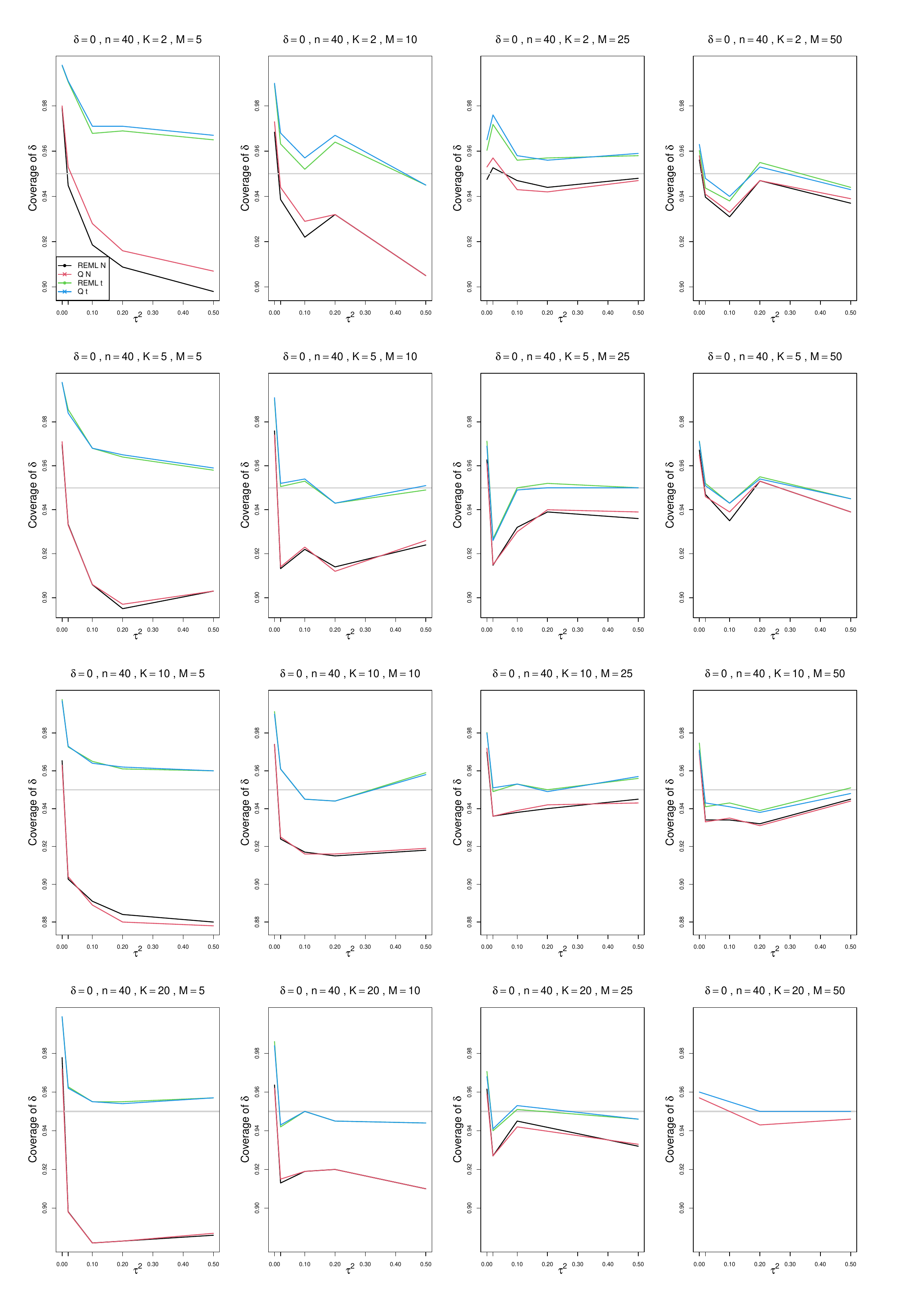}
	\caption{Coverage of 95\% confidence intervals for overall effect $\delta$ (REML N, REML t, $Q$ N and $Q$ t)  vs $\tau^2$, for $K$ = 2, 5, 10, and 20 studies per cluster and $M$ = 5, 10, 25, and 50 clusters; $\delta = 0$, and the sample size $n$ = 40 in each study.  }
	\label{PlotCoverageOfdelta_40_0_HIER.pdf}
\end{figure}

\begin{figure}[ht]
	\centering
	\includegraphics[scale=0.33]{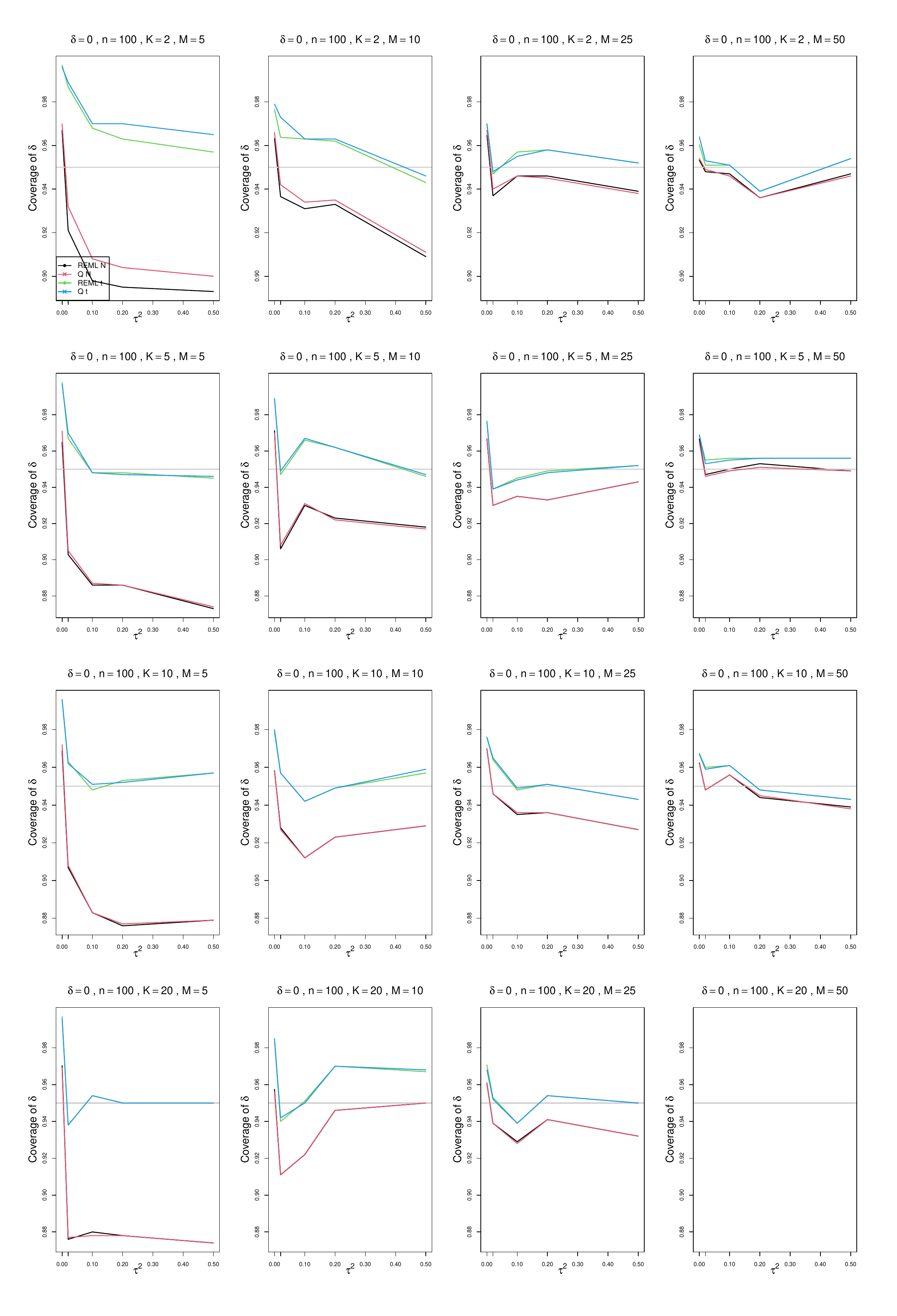}
	\caption{Coverage of 95\% confidence intervals for overall effect $\delta$ (REML N, REML t, $Q$ N and $Q$ t) vs $\tau^2$,  for $K$ = 2, 5, 10, and 20 studies per cluster and $M$ = 5, 10, 25, and 50 clusters; $\delta = 0$, and the sample size $n$ = 100 in each study.  }
	\label{PlotCoverageOfdelta_100_0_HIER.pdf}
\end{figure}

\begin{figure}[ht]
	\centering
	\includegraphics[scale=0.33]{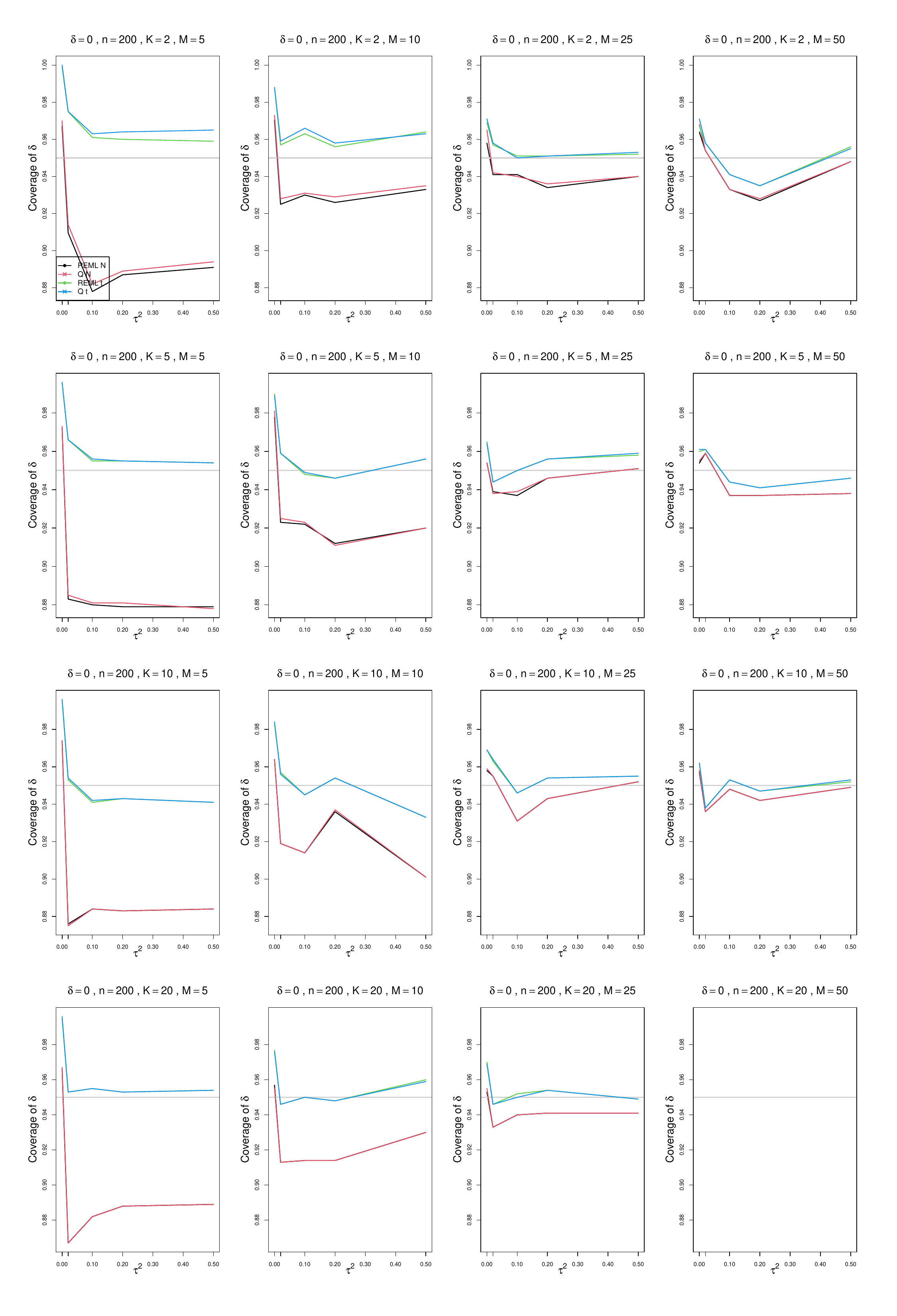}
	\caption{Coverage of 95\% confidence intervals for overall effect $\delta$ (REML N, REML t, $Q$ N and $Q$ t)   vs $\tau^2$,  for $K$ = 2, 5, 10, and 20 studies per cluster and $M$ = 5, 10, 25, and 50 clusters; $\delta = 0$, and the sample size $n$ = 200 in each study.  }
	\label{PlotCoverageOfdelta_200_0_HIER.pdf}
\end{figure}

\begin{figure}[ht]
	\centering
	\includegraphics[scale=0.33]{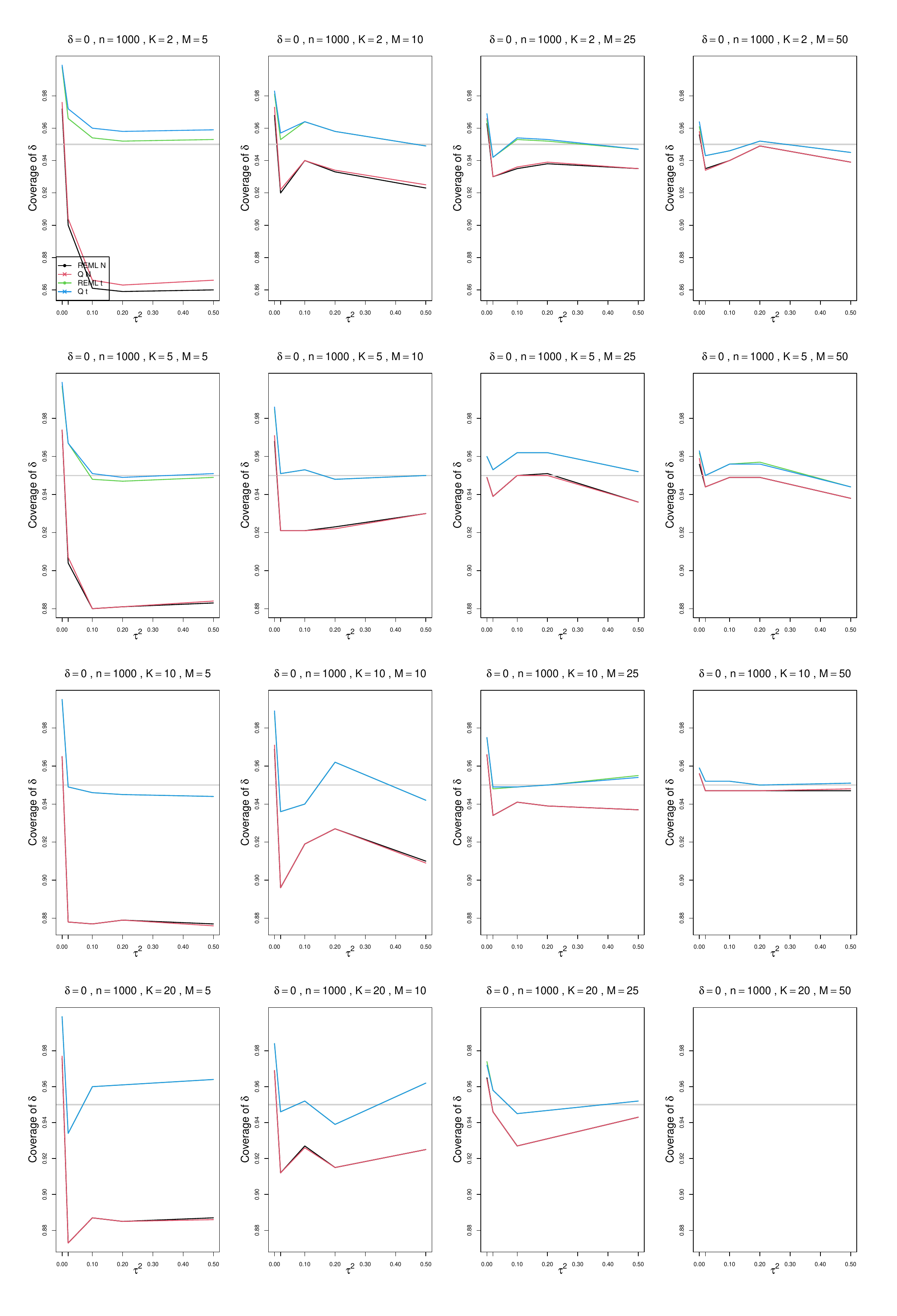}
	\caption{Coverage of 95\% confidence intervals for overall effect $\delta$ (REML N, REML t, $Q$ N and $Q$ t)   vs $\tau^2$,  for $K$ = 2, 5, 10, and 20 studies per cluster and $M$ = 5, 10, 25, and 50 clusters; $\delta = 0$, and the sample size $n$ = 1000 in each study.  }
	\label{PlotCoverageOfdelta_1000_0_HIER.pdf}
\end{figure}

\begin{figure}[ht]
	\centering
	\includegraphics[scale=0.33]{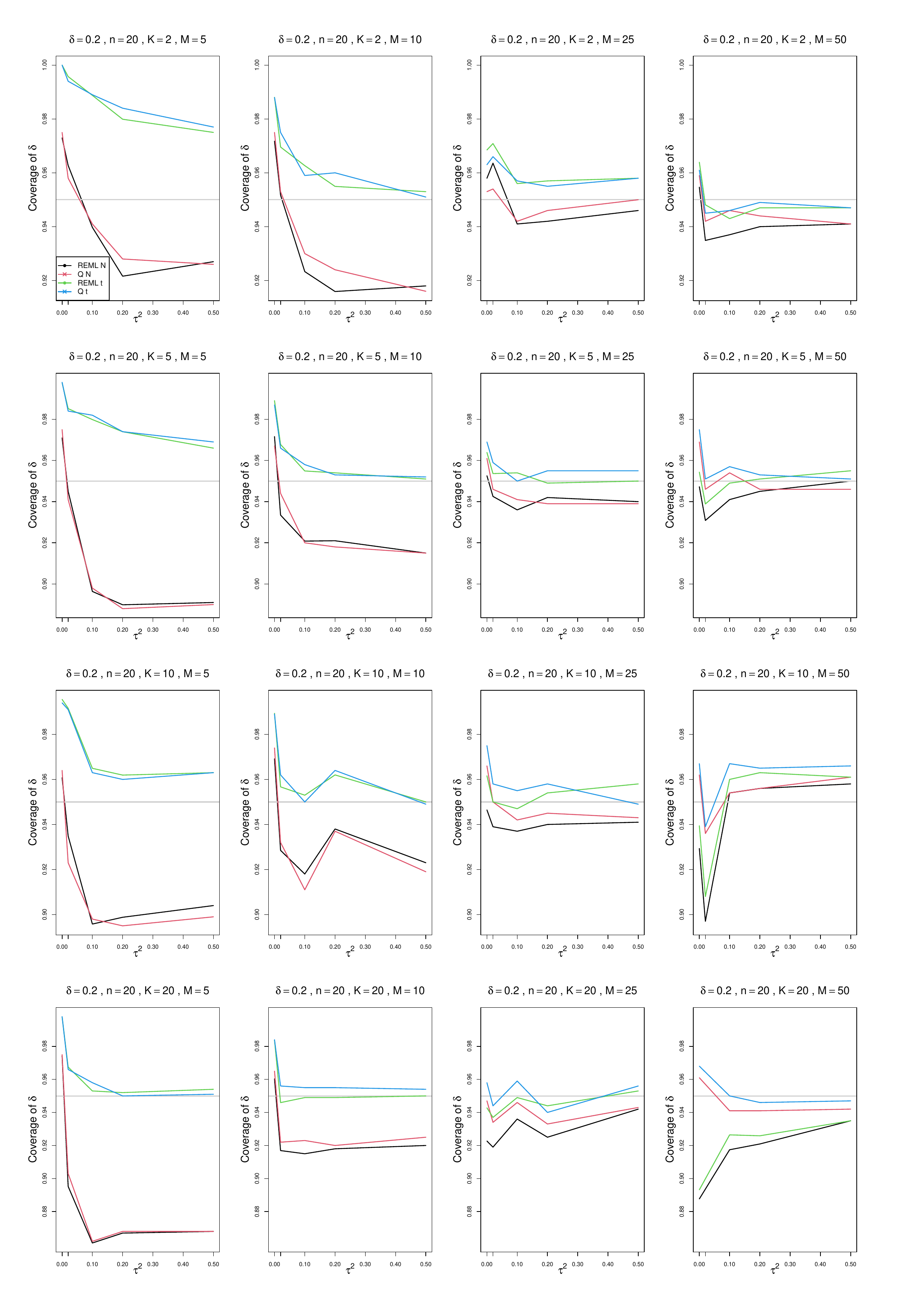}
	\caption{Coverage of 95\% confidence intervals for overall effect $\delta$ (REML N, REML t, $Q$ N and $Q$ t)   vs $\tau^2$,  for $K$ = 2, 5, 10, and 20 studies per cluster and $M$ = 5, 10, 25, and 50 clusters; $\delta = 0.2$, and the sample size $n$ = 20 in each study.  }
	\label{PlotCoverageOfdelta_20_02_HIER.pdf}
\end{figure}

\begin{figure}[ht]
	\centering
	\includegraphics[scale=0.33]{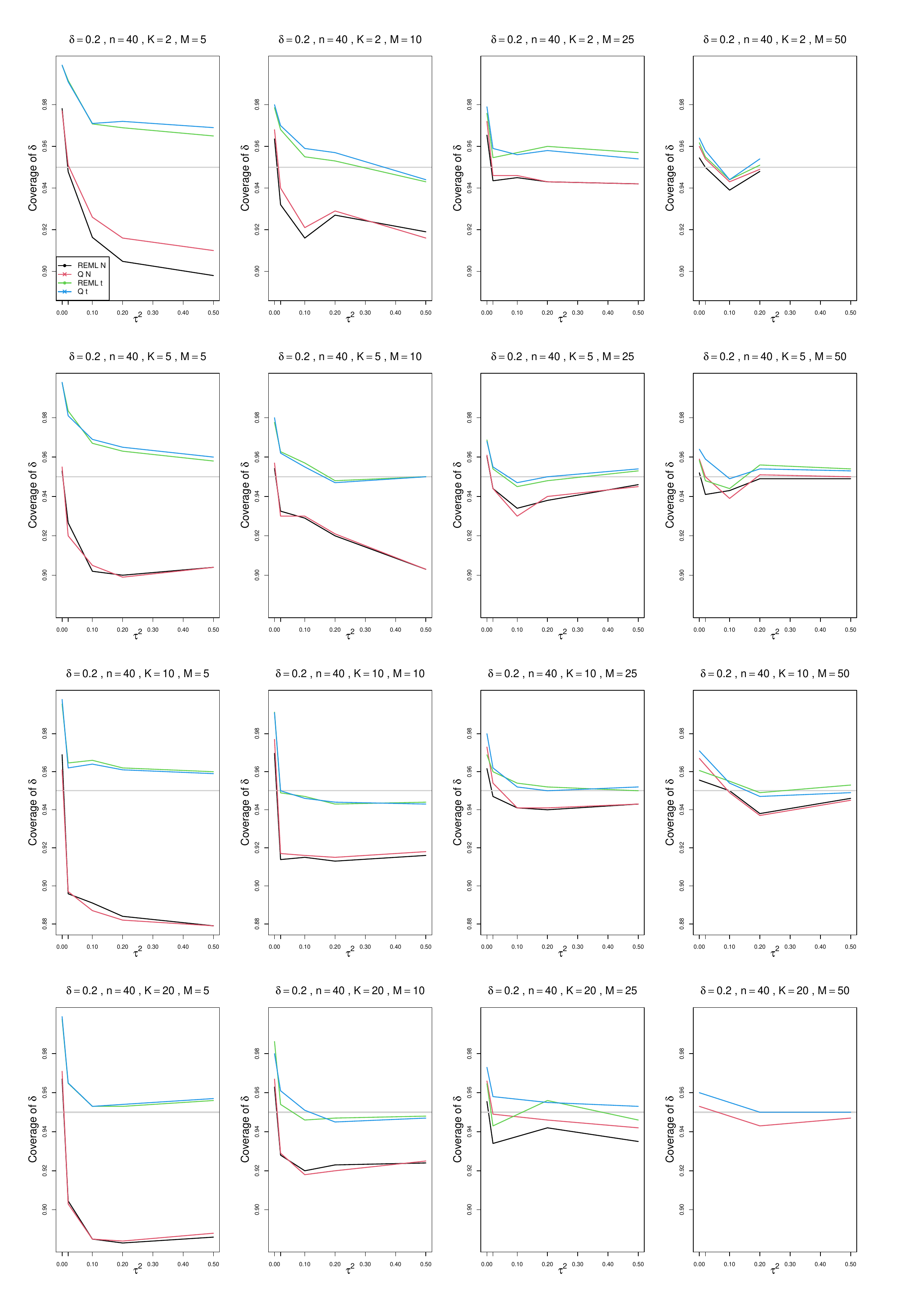}
	\caption{Coverage of 95\% confidence intervals for overall effect $\delta$ (REML N, REML t, $Q$ N and $Q$ t)   vs $\tau^2$, for $K$ = 2, 5, 10, and 20 studies per cluster and $M$ = 5, 10, 25, and 50 clusters; $\delta = 0.2$, and the sample size $n$ = 40 in each study.  }
	\label{PlotCoverageOfdelta_40_02_HIER.pdf}
\end{figure}

\begin{figure}[ht]
	\centering
	\includegraphics[scale=0.33]{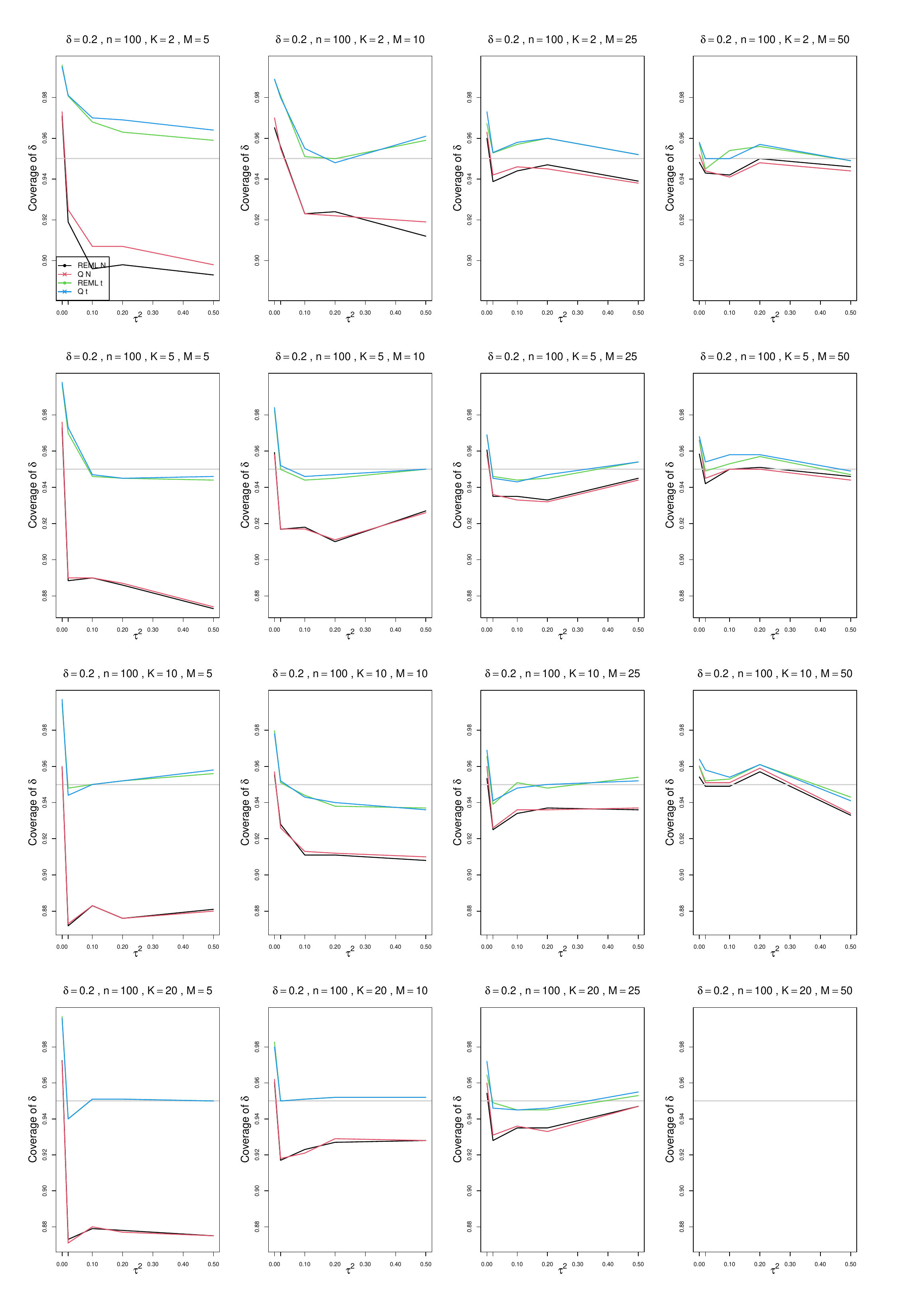}
	\caption{Coverage of 95\% confidence intervals for overall effect $\delta$ (REML N, REML t, $Q$ N and $Q$ t)  vs $\tau^2$, for $K$ = 2, 5, 10, and 20 studies per cluster and $M$ = 5, 10, 25, and 50 clusters; $\delta = 0.2$, and the sample size $n$ = 100 in each study.  }
	\label{PlotCoverageOfdelta_100_02_HIER.pdf}
\end{figure}

\begin{figure}[ht]
	\centering
	\includegraphics[scale=0.33]{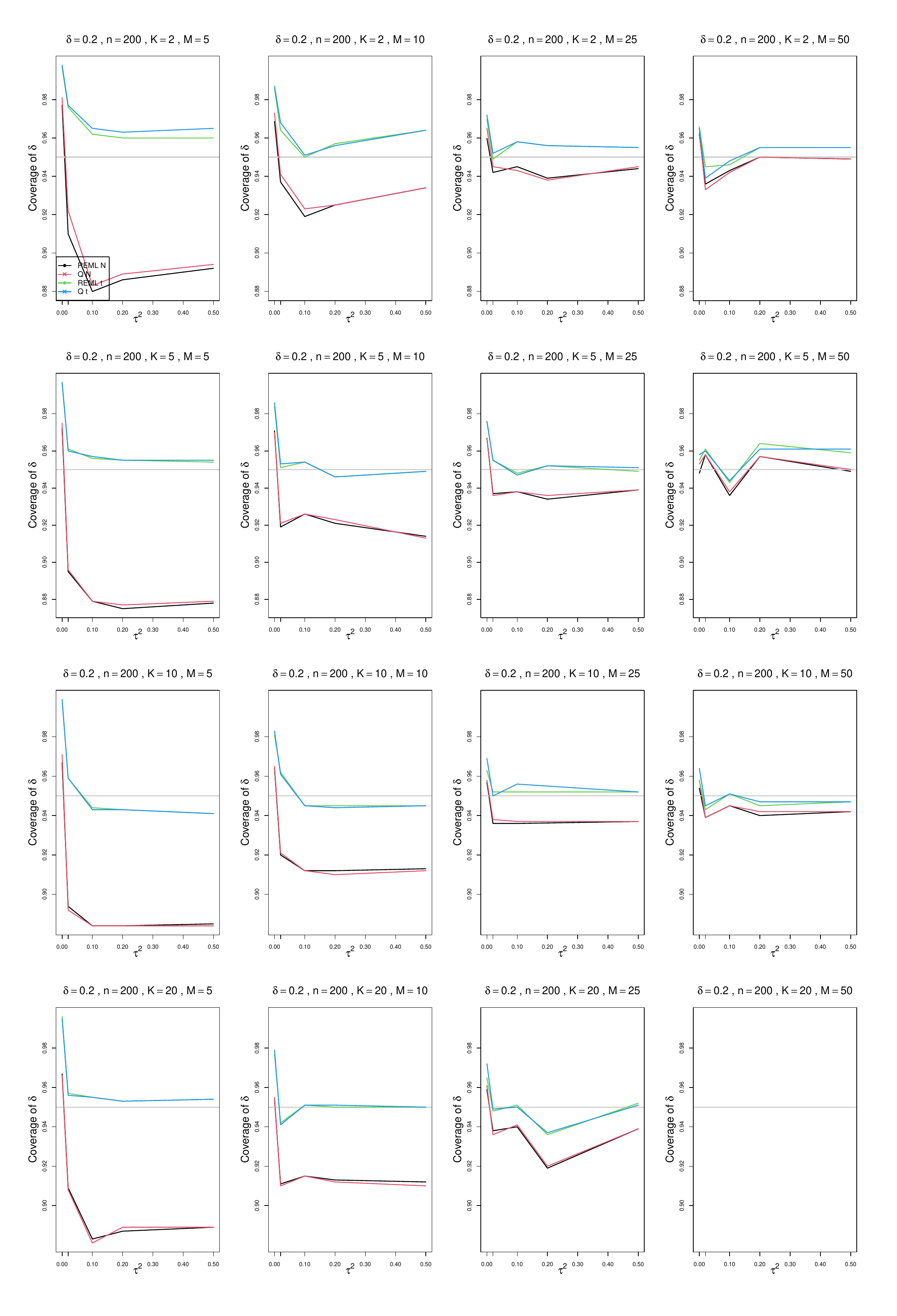}
	\caption{Coverage of 95\% confidence intervals for overall effect $\delta$ (REML N, REML t, $Q$ N and $Q$ t)   vs $\tau^2$, for $K$ = 2, 5, 10, and 20 studies per cluster and $M$ = 5, 10, 25, and 50 clusters; $\delta = 0.2$, and the sample size $n$ = 200 in each study.  }
	\label{PlotCoverageOfdelta_200_02_HIER.pdf}
\end{figure}

\begin{figure}[ht]
	\centering
	\includegraphics[scale=0.33]{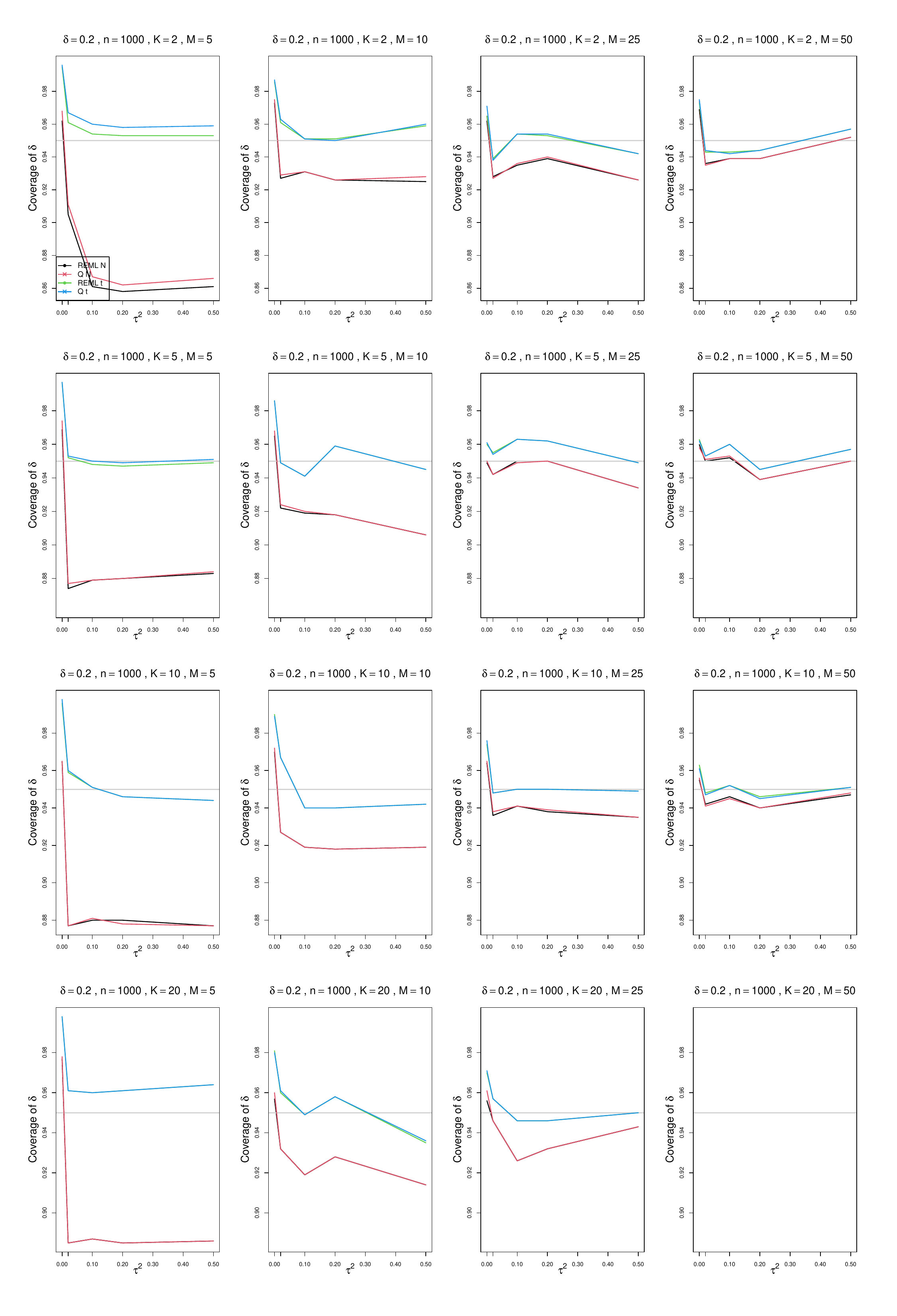}
	\caption{Coverage of 95\% confidence intervals for overall effect $\delta$ (REML N, REML t, $Q$ N and $Q$ t)   vs $\tau^2$, for $K$ = 2, 5, 10, and 20 studies per cluster and $M$ = 5, 10, 25, and 50 clusters; $\delta = 0.2$, and the sample size $n$ = 1000 in each study.  }
	\label{PlotCoverageOfdelta_1000_02_HIER.pdf}
\end{figure}

\begin{figure}[ht]
	\centering
	\includegraphics[scale=0.33]{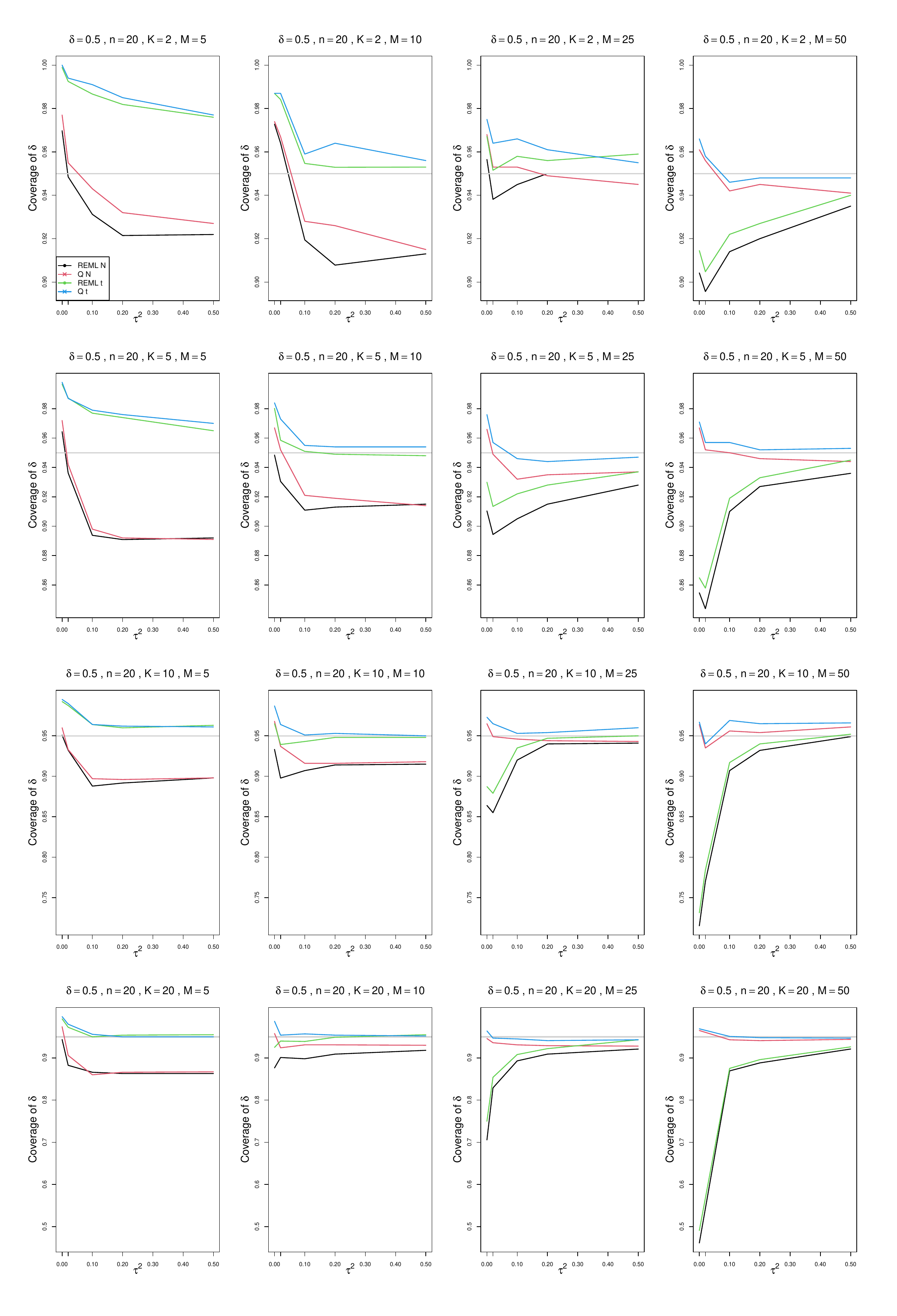}
	\caption{Coverage of 95\% confidence intervals for overall effect $\delta$ (REML N, REML t, $Q$ N and $Q$ t)  vs $\tau^2$, for $K$ = 2, 5, 10, and 20 studies per cluster and $M$ = 5, 10, 25, and 50 clusters; $\delta = 0.5$, and the sample size $n$ = 20 in each study.   }
	\label{PlotCoverageOfdelta_20_05_HIER.pdf}
\end{figure}

\begin{figure}[ht]
	\centering
	\includegraphics[scale=0.33]{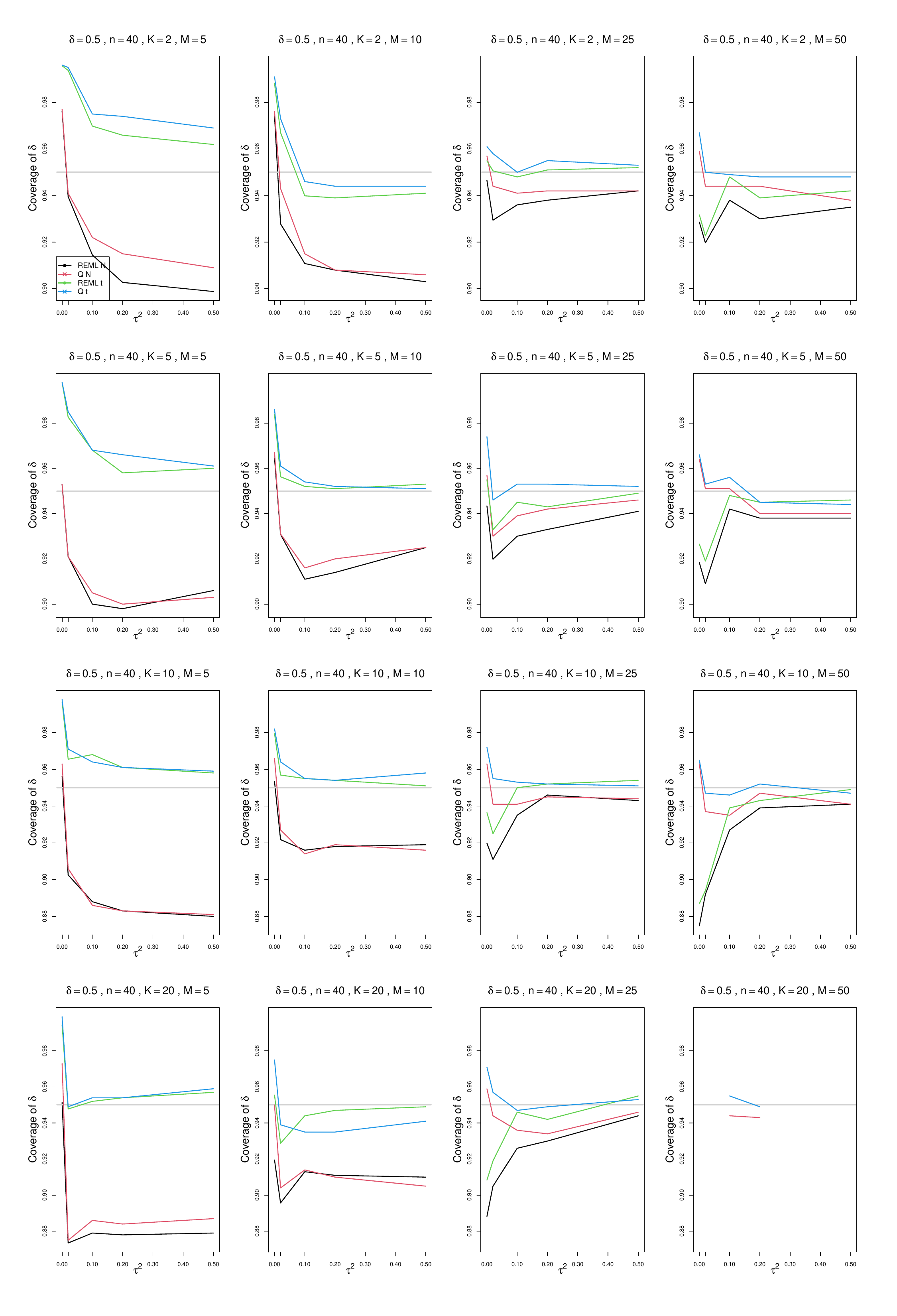}
	\caption{Coverage of 95\% confidence intervals for overall effect $\delta$ (REML N, REML t, $Q$ N and $Q$ t)   vs $\tau^2$, for $K$ = 2, 5, 10, and 20 studies per cluster and $M$ = 5, 10, 25, and 50 clusters; $\delta = 0.5$, and the sample size $n$ = 40 in each study. }
	\label{PlotCoverageOfdelta_40_05_HIER.pdf}
\end{figure}

\begin{figure}[ht]
	\centering
	\includegraphics[scale=0.33]{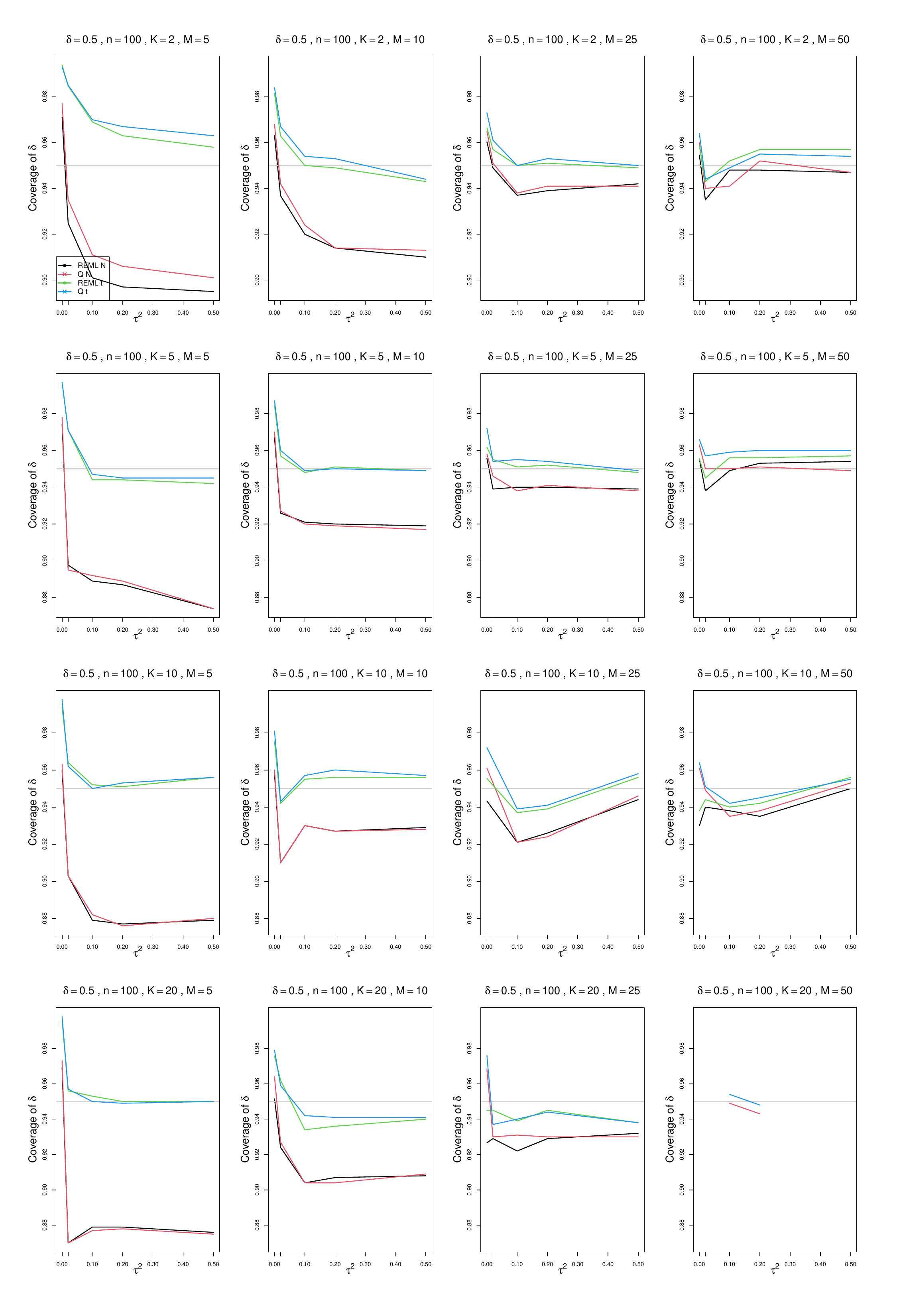}
	\caption{Coverage of 95\% confidence intervals for overall effect $\delta$ (REML N, REML t, $Q$ N and $Q$ t)   vs $\tau^2$, for $K$ = 2, 5, 10, and 20 studies per cluster and $M$ = 5, 10, 25, and 50 clusters; $\delta = 0.5$, and the sample size $n$ = 100 in each study. }
	\label{PlotCoverageOfdelta_100_05_HIER.pdf}
\end{figure}

\begin{figure}[ht]
	\centering
	\includegraphics[scale=0.33]{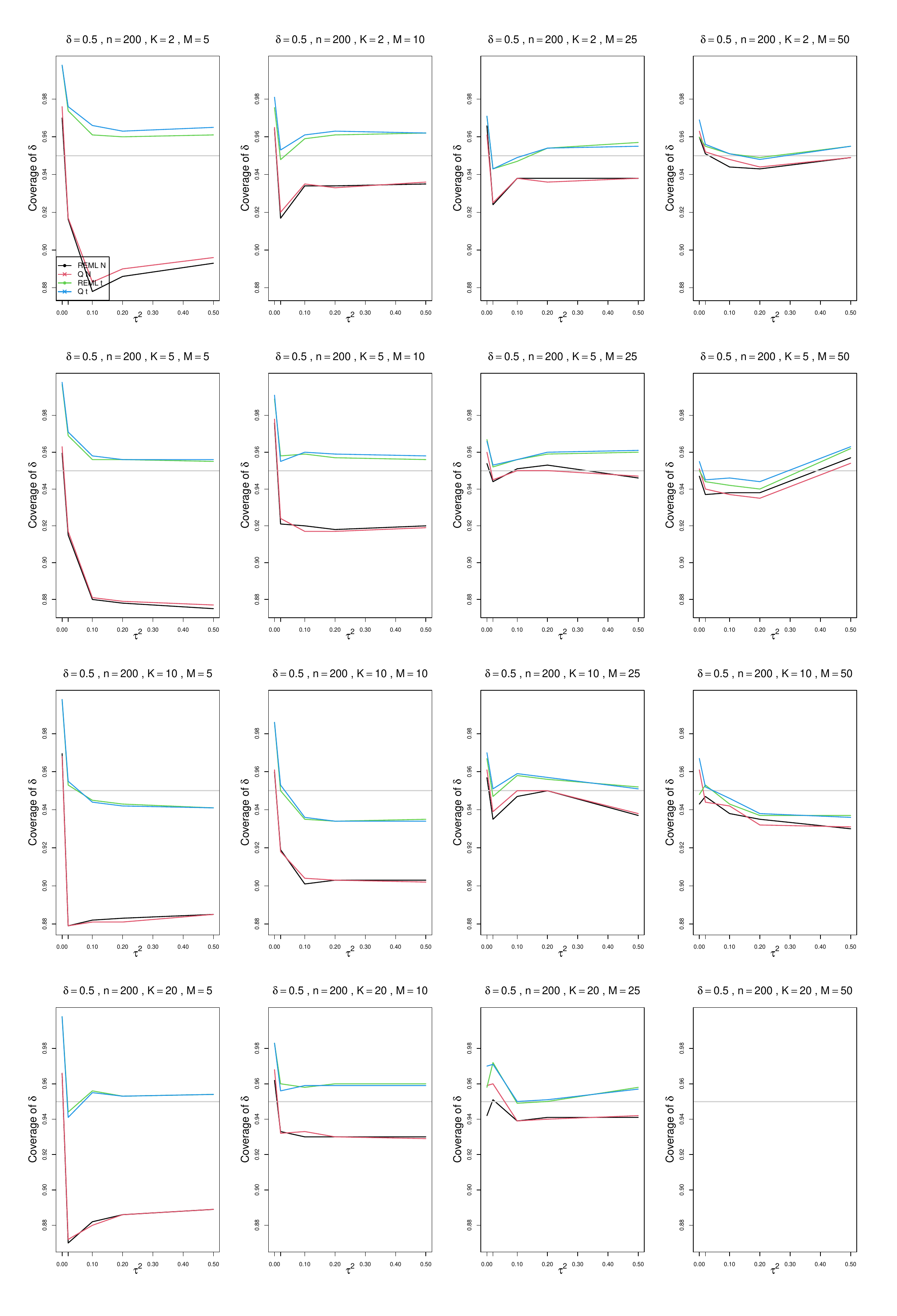}
	\caption{Coverage of 95\% confidence intervals for overall effect $\delta$ (REML N, REML t, $Q$ N and $Q$ t)   vs $\tau^2$, for $K$ = 2, 5, 10, and 20 studies per cluster and $M$ = 5, 10, 25, and 50 clusters; $\delta = 0.5$, and the sample size $n$ = 200 in each study.  }
	\label{PlotCoverageOfdelta_200_05_HIER.pdf}
\end{figure}

\begin{figure}[ht]
	\centering
	\includegraphics[scale=0.33]{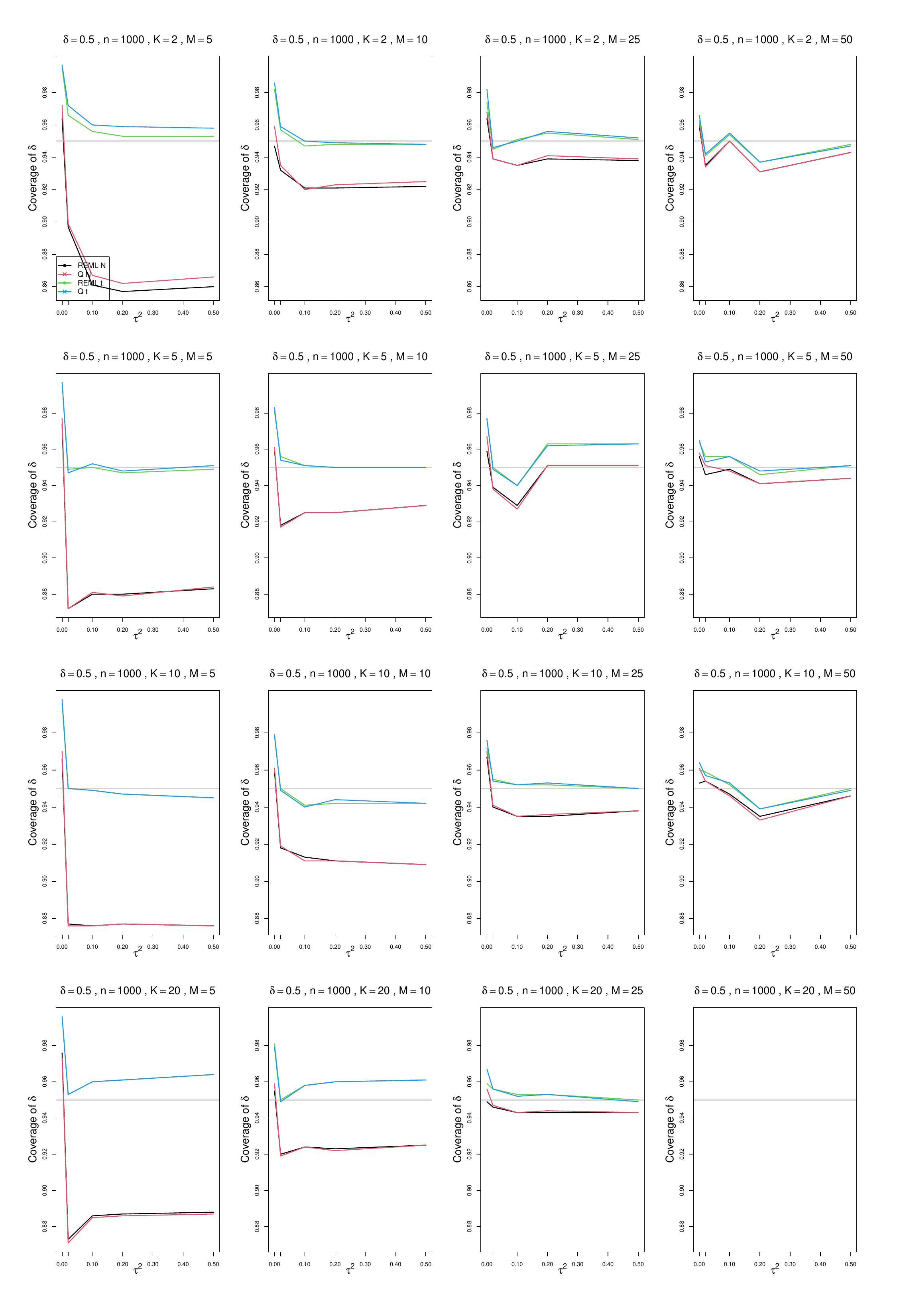}
	\caption{Coverage of 95\% confidence intervals for overall effect $\delta$ (REML N, REML t, $Q$ N and $Q$ t)   vs $\tau^2$, for $K$ = 2, 5, 10, and 20 studies per cluster and $M$ = 5, 10, 25, and 50 clusters; $\delta = 0.5$, and the sample size $n$ = 1000 in each study.  }
	\label{PlotCoverageOfdelta_1000_05_HIER.pdf}
\end{figure}

\begin{figure}[ht]
	\centering
	\includegraphics[scale=0.33]{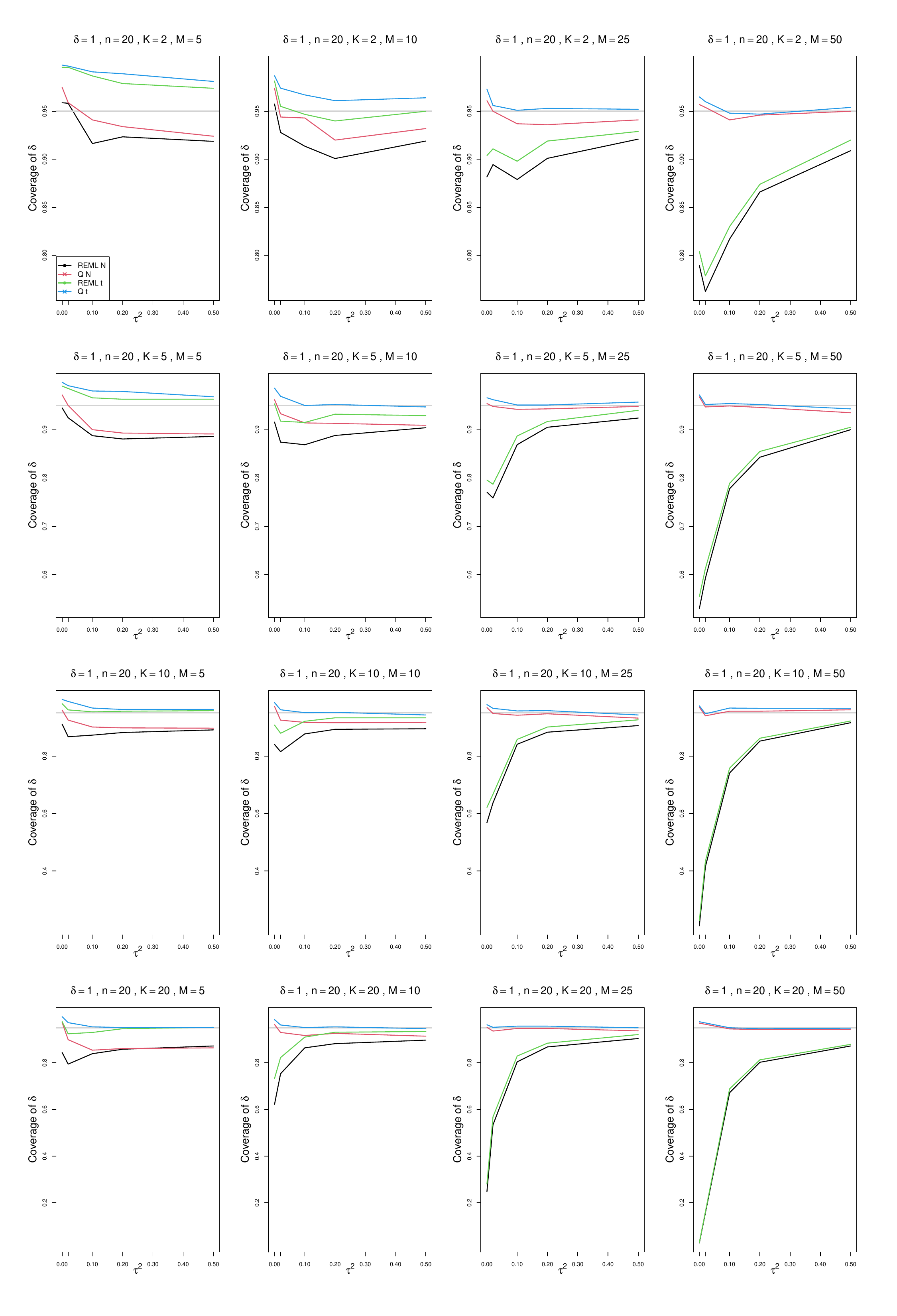}
	\caption{Coverage of 95\% confidence intervals for overall effect $\delta$ (REML N, REML t, $Q$ N and $Q$ t)   vs $\tau^2$, for $K$ = 2, 5, 10, and 20 studies per cluster and $M$ = 5, 10, 25, and 50 clusters; $\delta = 1$, and the sample size $n$ = 20 in each study.  }
	\label{PlotCoverageOfdelta_20_1_HIER.pdf}
\end{figure}

\begin{figure}[ht]
	\centering
	\includegraphics[scale=0.33]{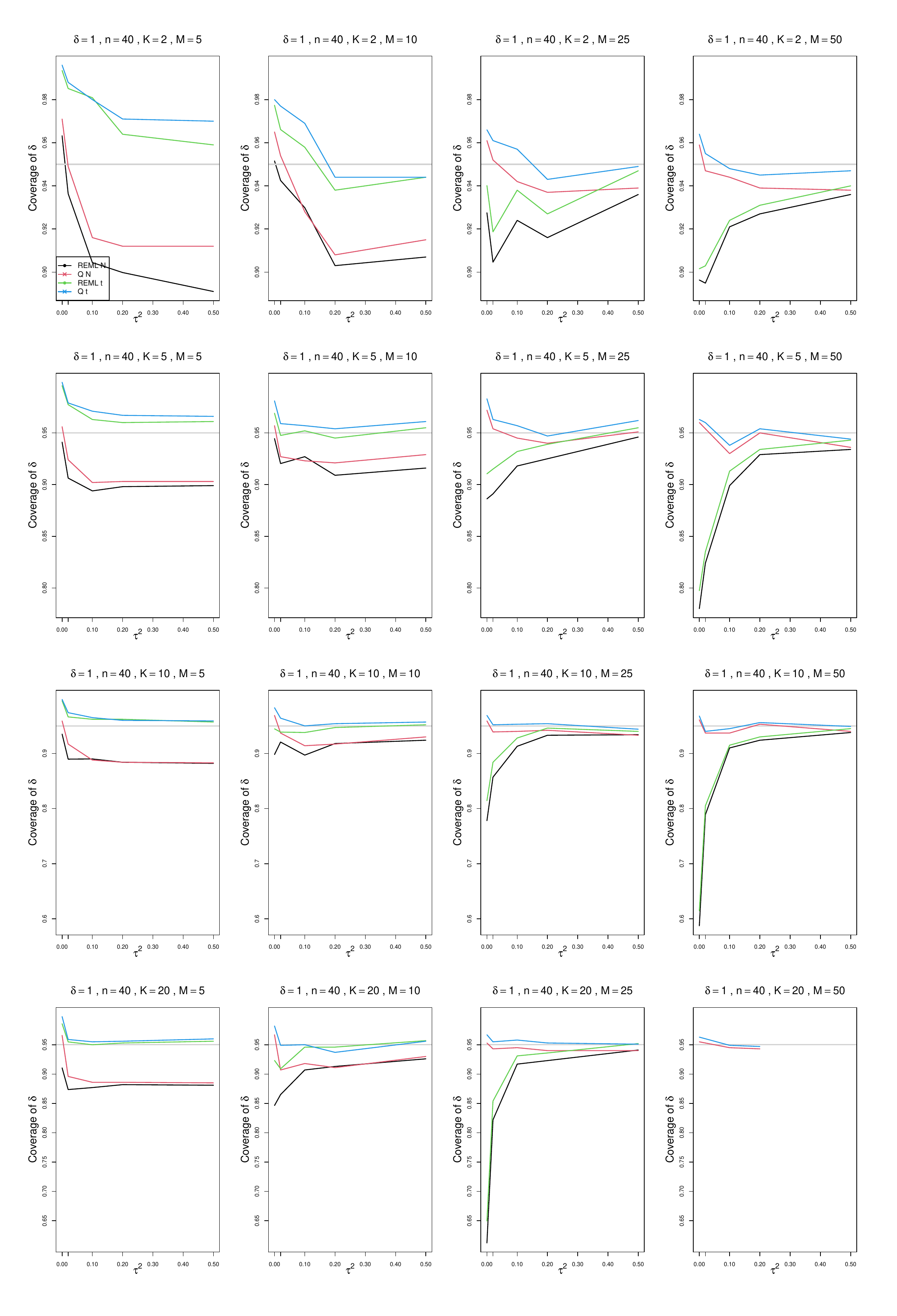}
	\caption{Coverage of 95\% confidence intervals for overall effect $\delta$ (REML N, REML t, $Q$ N and $Q$ t)   vs $\tau^2$, for $K$ = 2, 5, 10, and 20 studies per cluster and $M$ = 5, 10, 25, and 50 clusters; $\delta = 1$, and the sample size $n$ = 40 in each study.  }
	\label{PlotCoverageOfdelta_40_1_HIER.pdf}
\end{figure}

\begin{figure}[ht]
	\centering
	\includegraphics[scale=0.33]{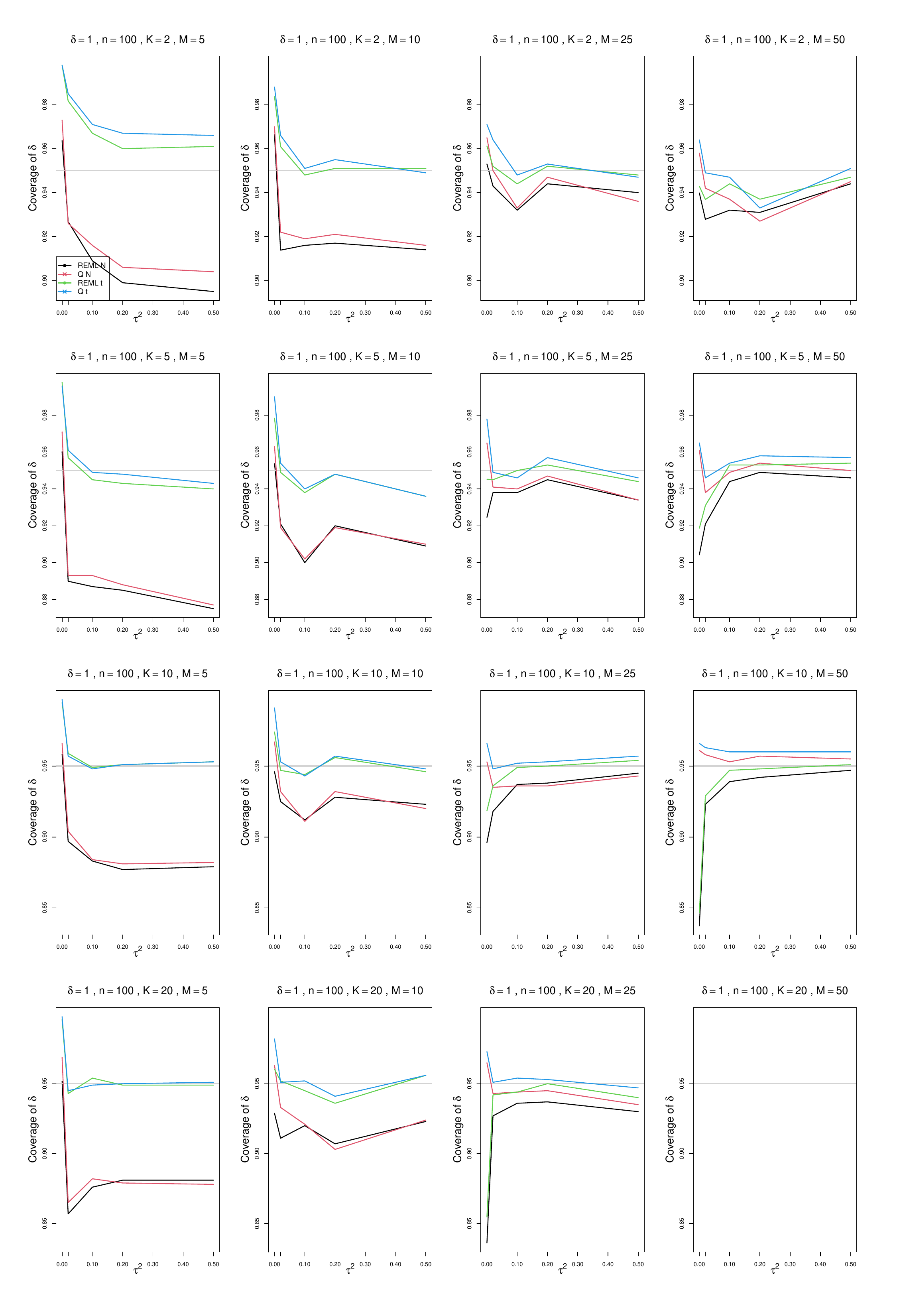}
	\caption{Coverage of 95\% confidence intervals for overall effect $\delta$ (REML N, REML t, $Q$ N and $Q$ t)   vs $\tau^2$,  for $K$ = 2, 5, 10, and 20 studies per cluster and $M$ = 5, 10, 25, and 50 clusters; $\delta = 1$, and the sample size $n$ = 100 in each study.  }
	\label{PlotCoverageOfdelta_100_1_HIER.pdf}
\end{figure}

\begin{figure}[ht]
	\centering
	\includegraphics[scale=0.33]{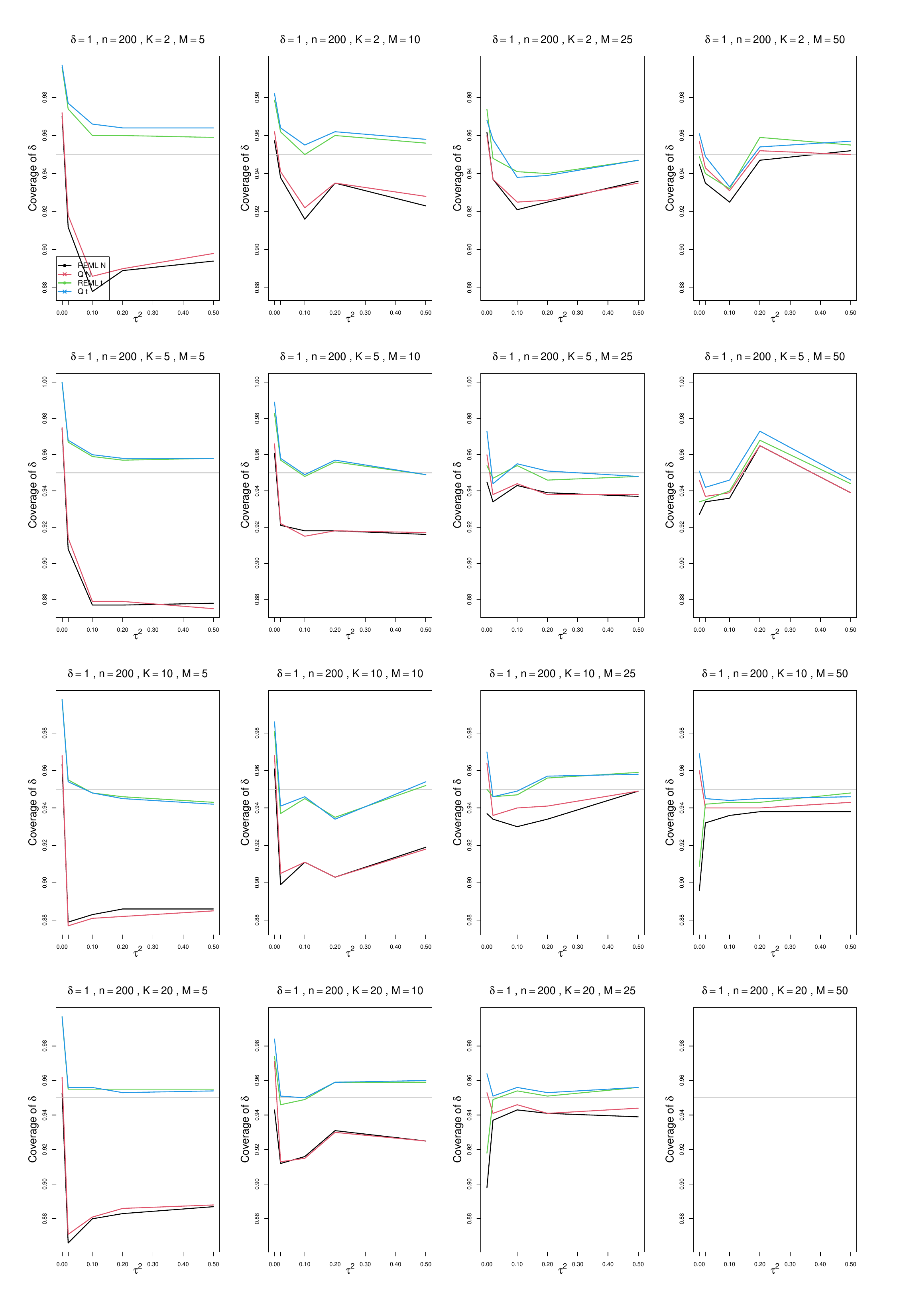}
	\caption{Coverage of 95\% confidence intervals for overall effect $\delta$ (REML N, REML t, $Q$ N and $Q$ t)  vs $\tau^2$, for $K$ = 2, 5, 10, and 20 studies per cluster and $M$ = 5, 10, 25, and 50 clusters; $\delta = 1$, and the sample size $n$ = 200 in each study.  }
	\label{PlotCoverageOfdelta_200_1_HIER.pdf}
\end{figure}

\begin{figure}[ht]
	\centering
	\includegraphics[scale=0.33]{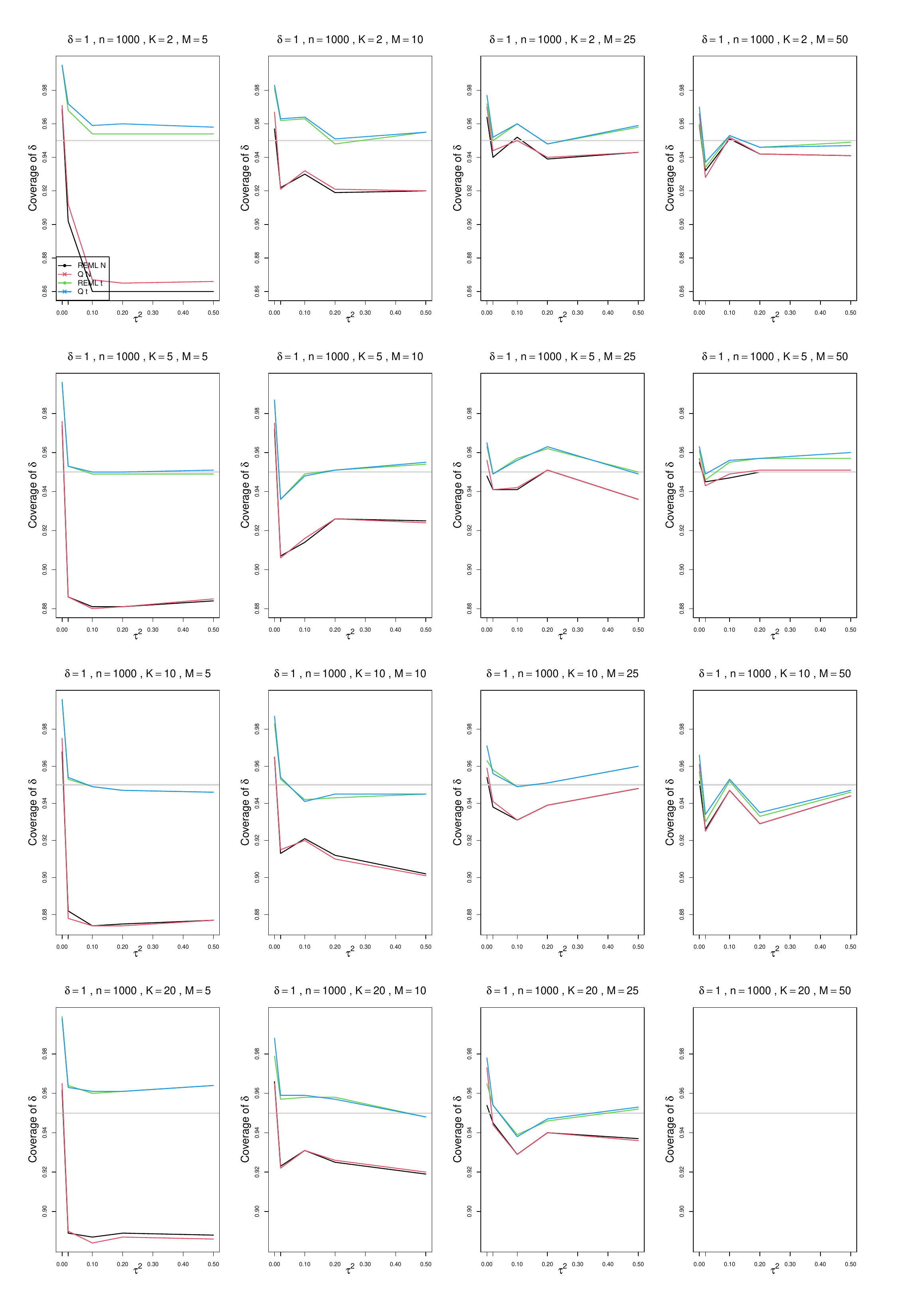}
	\caption{Coverage of 95\% confidence intervals for overall effect $\delta$ (REML N, REML t, $Q$ N and $Q$ t) vs $\tau^2$,  for $K$ = 2, 5, 10, and 20 studies per cluster and $M$ = 5, 10, 25, and 50 clusters; $\delta = 1$, and the sample size $n$ = 1000 in each study.   }
	\label{PlotCoverageOfdelta_1000_1_HIER.pdf}
\end{figure}

\clearpage

\section*{Appendix I: Additional simulations: bias in point estimators of  $\tau^2$, $\omega^2$ and $\delta$ and coverage of their 95\% confidence intervals when $\omega^2=4\tau^2$.}

 Each figure corresponds to a value  of between-study variance $\tau^2$ (= 0.02 or 0.1).
The ratio of two variance components is held constant  at $\omega^2=4\tau^2$.\\
For each value of  the standardized mean difference ($\delta$  = 0, 0.2, 0.5, 1), and a combination of the number of studies in a cluster ($K=5,\; 10$) and the number of clusters ($M$ = 10,\; 25), a panel plots bias or coverage at 95\% nominal level of a relevant point or interval estimator versus sample size $n$ (= 20, 40, 100, 200, 1000).\\
The point estimators of $\tau^2$ are
\begin{itemize}
\item REML method, inverse-variance weights, {\it  rma.mv} in {\it metafor}
\item $Q_A$ (conditional moment-based method, effective-sample-size weights)
\end{itemize}

The interval estimators of $\tau^2$ are
\begin{itemize}
\item PL (Profile-Likelihood method, inverse-variance weights,  {\it  rma.mv} in {\it metafor})
\item $Q_A$ (conditional moment-based method, effective-sample-size weights, Davies's approximation to the distribution)
\end{itemize}
The point estimators of $\omega^2$ are
\begin{itemize}
\item REML method, inverse-variance weights,  {\it  rma.mv} in {\it metafor})
\item $Q_F$ (conditional moment-based method, effective-sample-size weights)
\end{itemize}
The interval estimators of $\omega^2$ are
\begin{itemize}
\item PL (Profile-Likelihood method, inverse-variance weights,  {\it  rma.mv} in {\it metafor})
\item $Q_F$ (conditional moment-based method, effective-sample-size weights, Davies's approximation to the distribution)
\end{itemize}
The point estimators of $\delta$ are
\begin{itemize}
\item REML method, inverse-variance weights,  {\it  rma.mv} in {\it metafor})
\item SSW (conditional moment-based method, effective-sample-size weights)
\end{itemize}
The interval estimators of $\delta$ are centered at a point estimator of $\delta$. Their half-width equals the point estimator's standard error, calculated from REML or from moment-based estimators of variance components, combined with normal or $t$ critical values.\\
The interval estimators of $\delta$ are
\begin{itemize}
\item REML N (REML, inverse-variance weights, normal critical values  {\it  rma.mv} in {\it metafor})
\item REML t (REML, inverse-variance weights, $t$ critical values  {\it  rma.mv} in {\it metafor})
\item $Q$ N (conditional moment-based method, effective-sample-size weights, normal critical values)
\item $Q$ t (conditional moment-based method, effective-sample-size weights,  $t$ critical values)
\end{itemize}

\clearpage
\setcounter{figure}{0}
\renewcommand{\thefigure}{I.\arabic{figure}}

\begin{figure}[ht]
	\centering
	\includegraphics[scale=0.33]{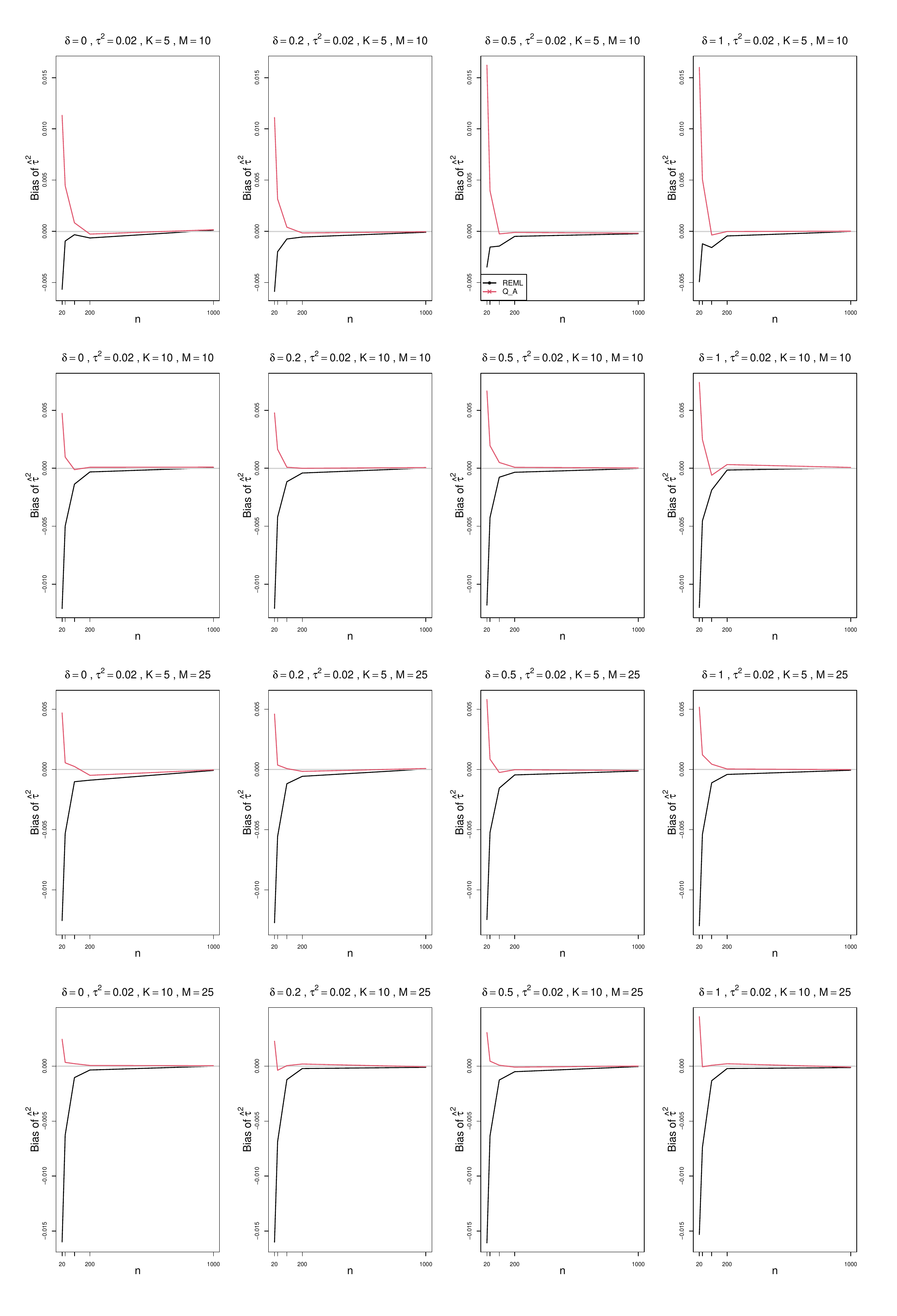}
	\caption{Bias of estimators of between-study variance $\tau^2=0.02$ of SMD (REML and $Q_A$) vs sample size $n$ = 20, 40, 100, 200, 1000 in each study, for $K$ =  5, 10 studies per cluster, $M$ =  10, 25 clusters and SMD $\delta = 0, 0.2, 0.5, 1$.   }
	\label{PlotBiasOfTau_x4_0.02_0_HIER.pdf}
\end{figure}

\begin{figure}[ht]
	\centering
	\includegraphics[scale=0.33]{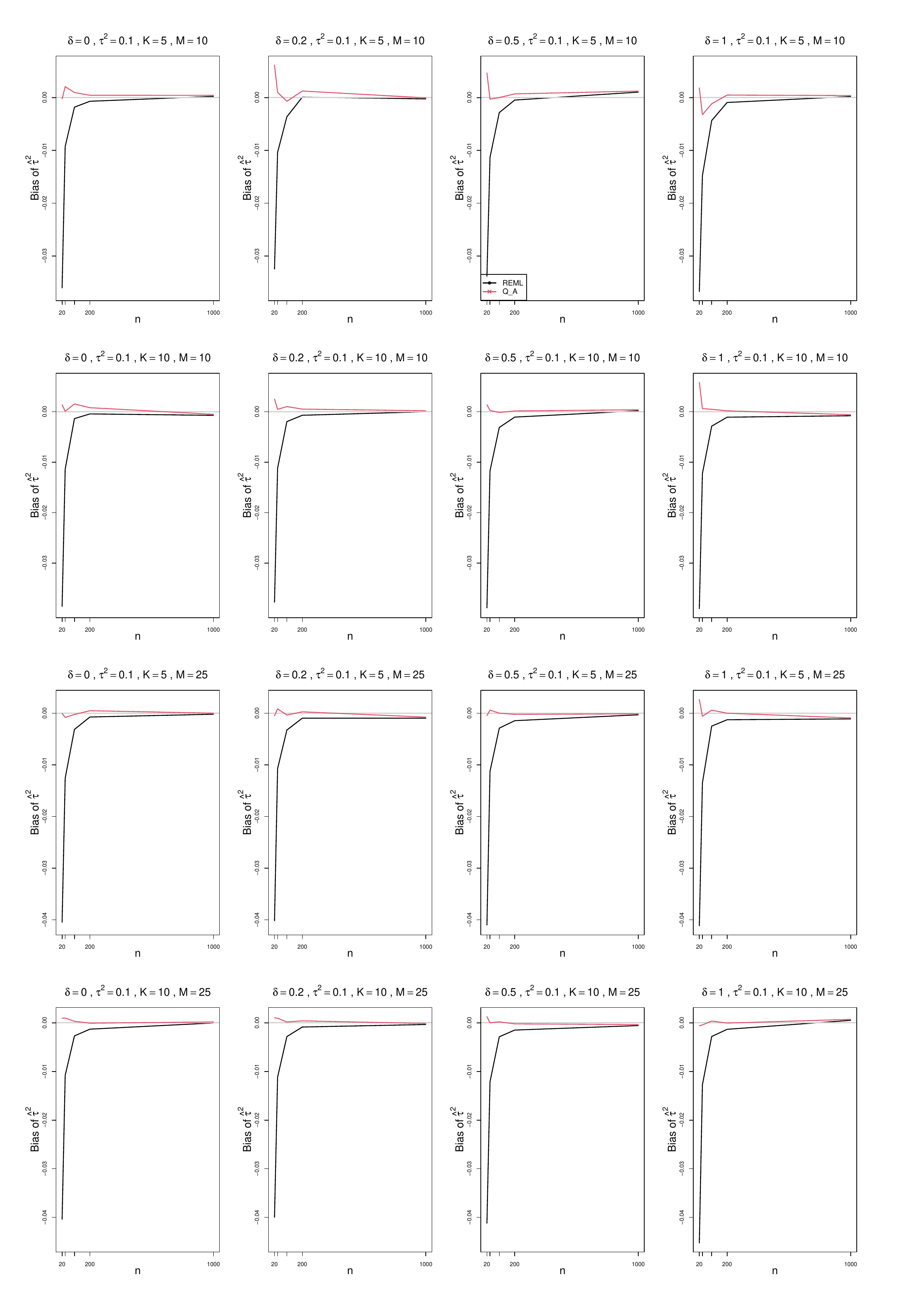}
	\caption{Bias of estimators of between-study variance $\tau^2=0.1$ of SMD (REML and $Q_A$) vs sample size $n$ = 20, 40, 100, 200, 1000 in each study, for $K$ =  5, 10 studies per cluster, $M$ =  10, 25 clusters and SMD $\delta = 0, 0.2, 0.5, 1$.   }
	\label{PlotBiasOfTau_x4_0.1_0_HIER.pdf}
\end{figure}

\begin{figure}[ht]
	\centering
	\includegraphics[scale=0.33]{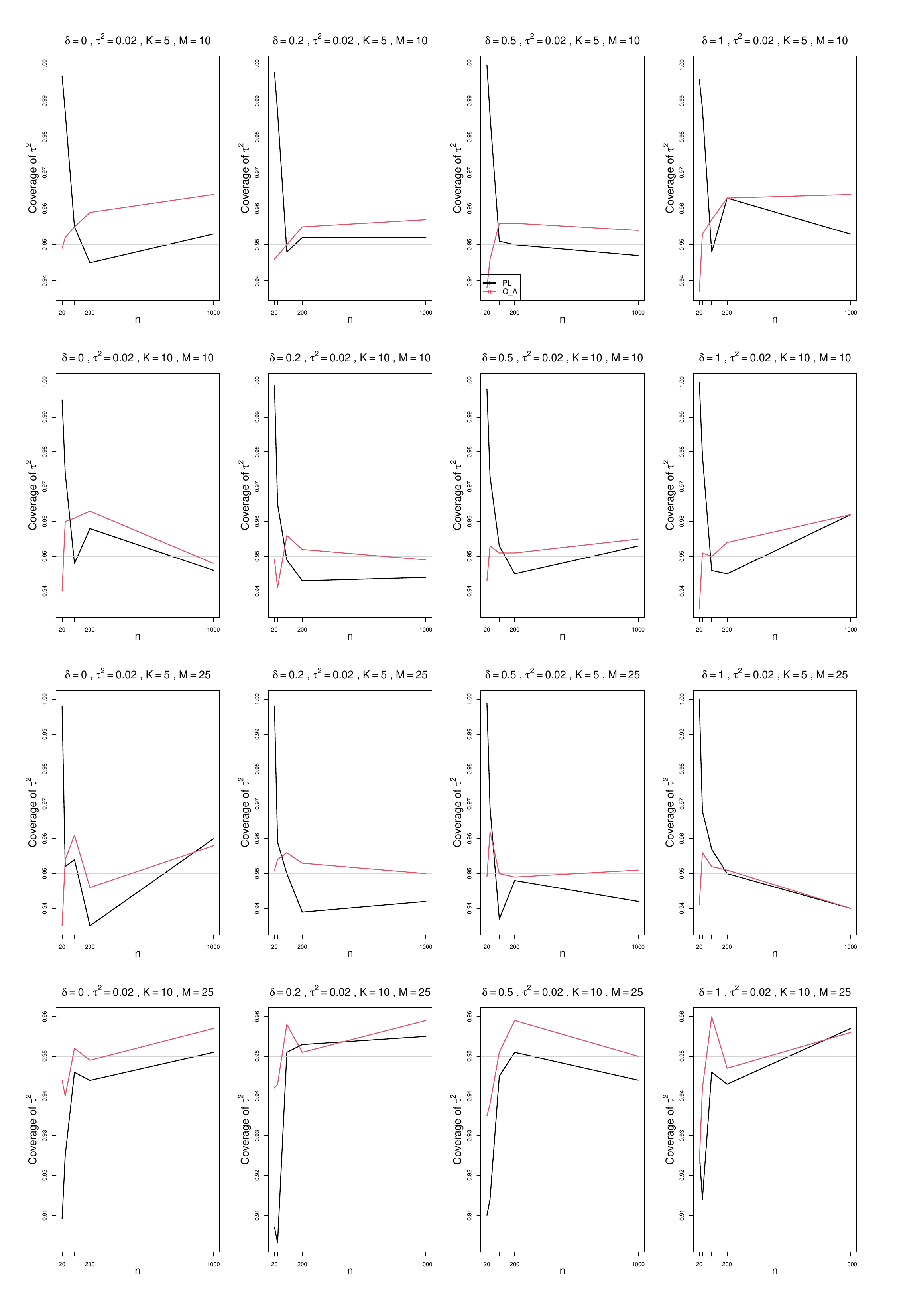}
	\caption{Coverage of 95\% confidence intervals for between-study variance $\tau^2=0.02$ of SMD (REML and $Q_A$) vs sample size $n$ = 20, 40, 100, 200, 1000 in each study, for $K$ =  5, 10 studies per cluster, $M$ =  10, 25 clusters and SMD $\delta = 0, 0.2, 0.5, 1$.   }
	\label{PlotCoverageOfTau_x4_0.02_HIER.pdf}
\end{figure}

\begin{figure}[ht]
	\centering
	\includegraphics[scale=0.33]{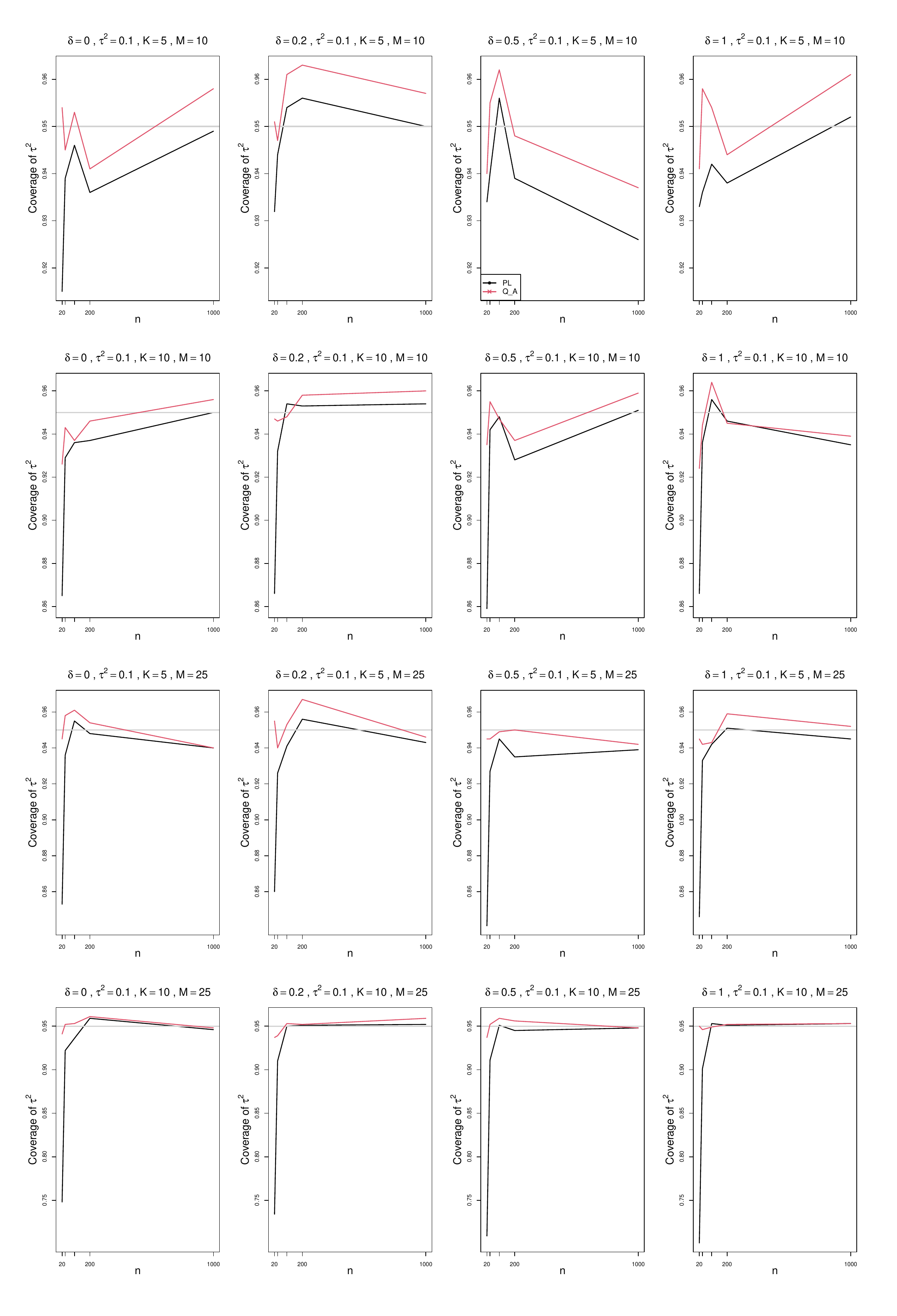}
	\caption{Coverage of 95\% confidence intervals for between-study variance $\tau^2=0.1$ of SMD (REML and $Q_A$) vs sample size $n$ = 20, 40, 100, 200, 1000 in each study, for $K$ =  5, 10 studies per cluster, $M$ =  10, 25 clusters and SMD $\delta = 0, 0.2, 0.5, 1$.   }
	\label{PlotCoverageOfTau_x4_0.1_HIER.pdf}
\end{figure}

\begin{figure}[ht]
	\centering
	\includegraphics[scale=0.33]{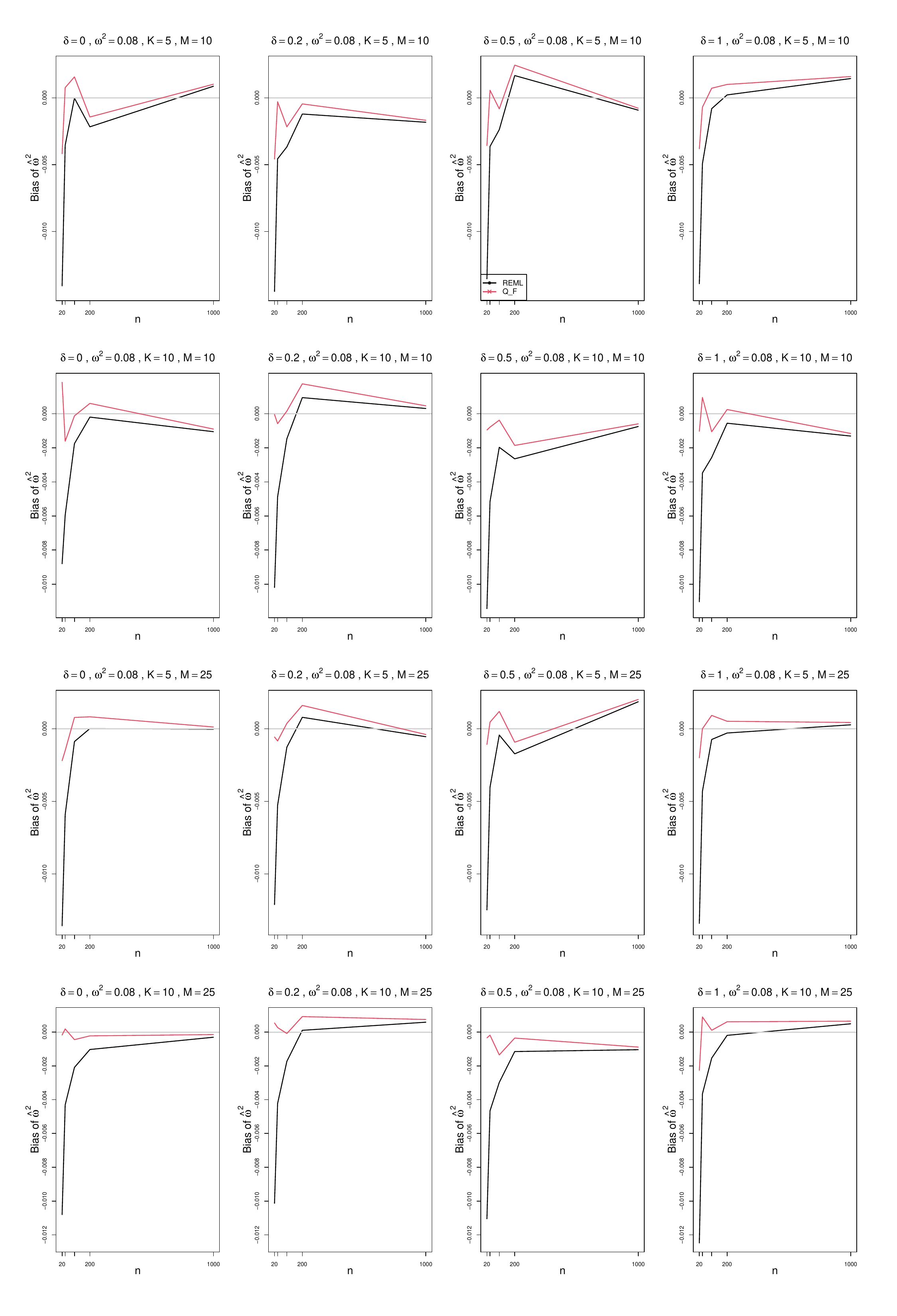}
	\caption{Bias of estimators of between-cluster variance $\omega^2=0.08$ of SMD (REML and $Q_F$) vs sample size $n$ = 20, 40, 100, 200, 1000 in each study, for $K$ =  5, 10 studies per cluster, $M$ =  10, 25 clusters and SMD $\delta = 0, 0.2, 0.5, 1$.   }
	\label{PlotBiasOfOmega_x4_0.02_0_HIER.pdf}
\end{figure}

\begin{figure}[ht]
	\centering
	\includegraphics[scale=0.33]{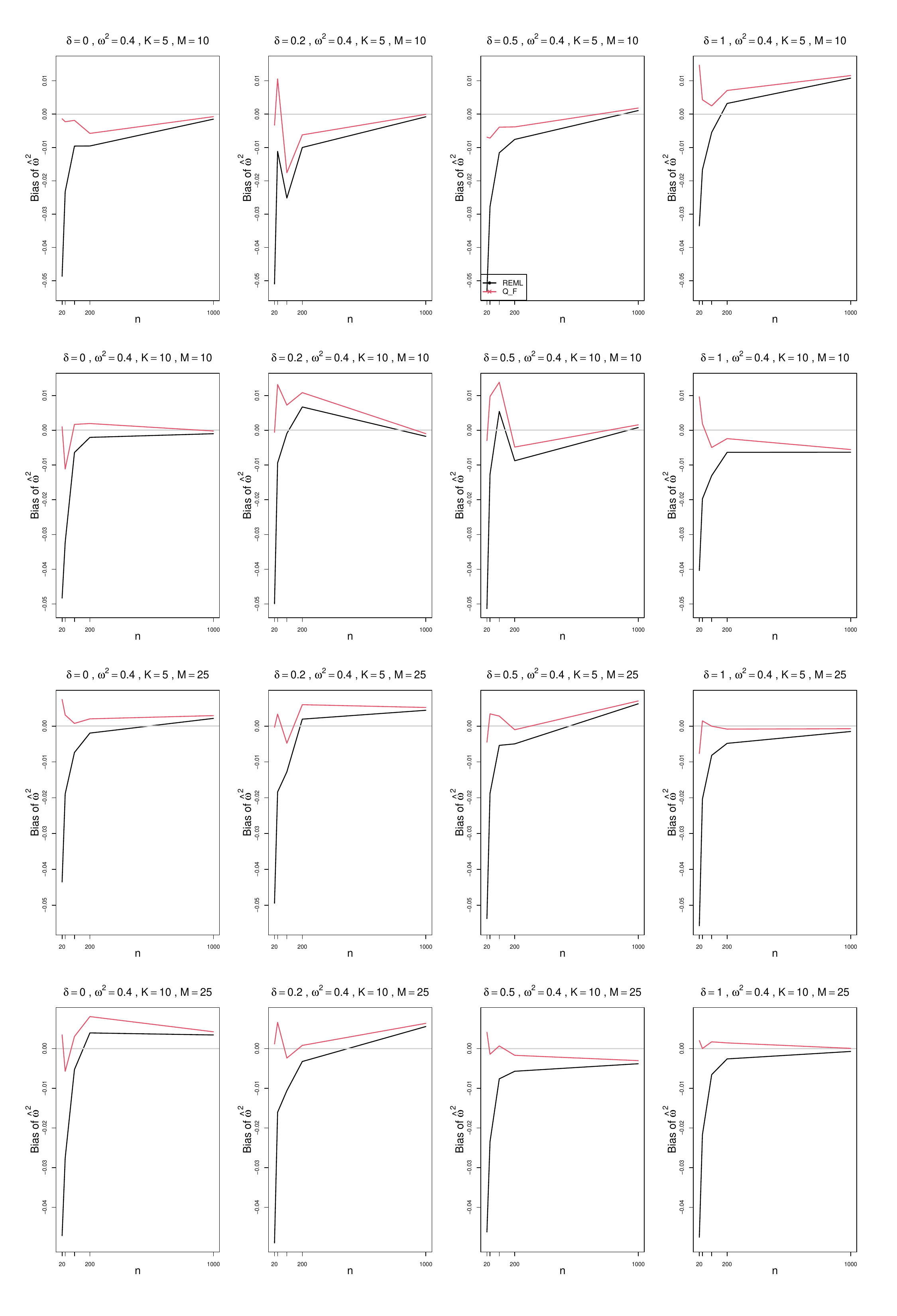}
	\caption{Bias of estimators of between-cluster variance $\omega^2=0.4$ of SMD (REML and $Q_F$) vs sample size $n$ = 20, 40, 100, 200, 1000 in each study, for $K$ =  5, 10 studies per cluster, $M$ =  10, 25 clusters and SMD $\delta = 0, 0.2, 0.5, 1$.   }
	\label{PlotBiasOfOmega_x4_0.1_0_HIER.pdf}
\end{figure}

\begin{figure}[ht]
	\centering
	\includegraphics[scale=0.33]{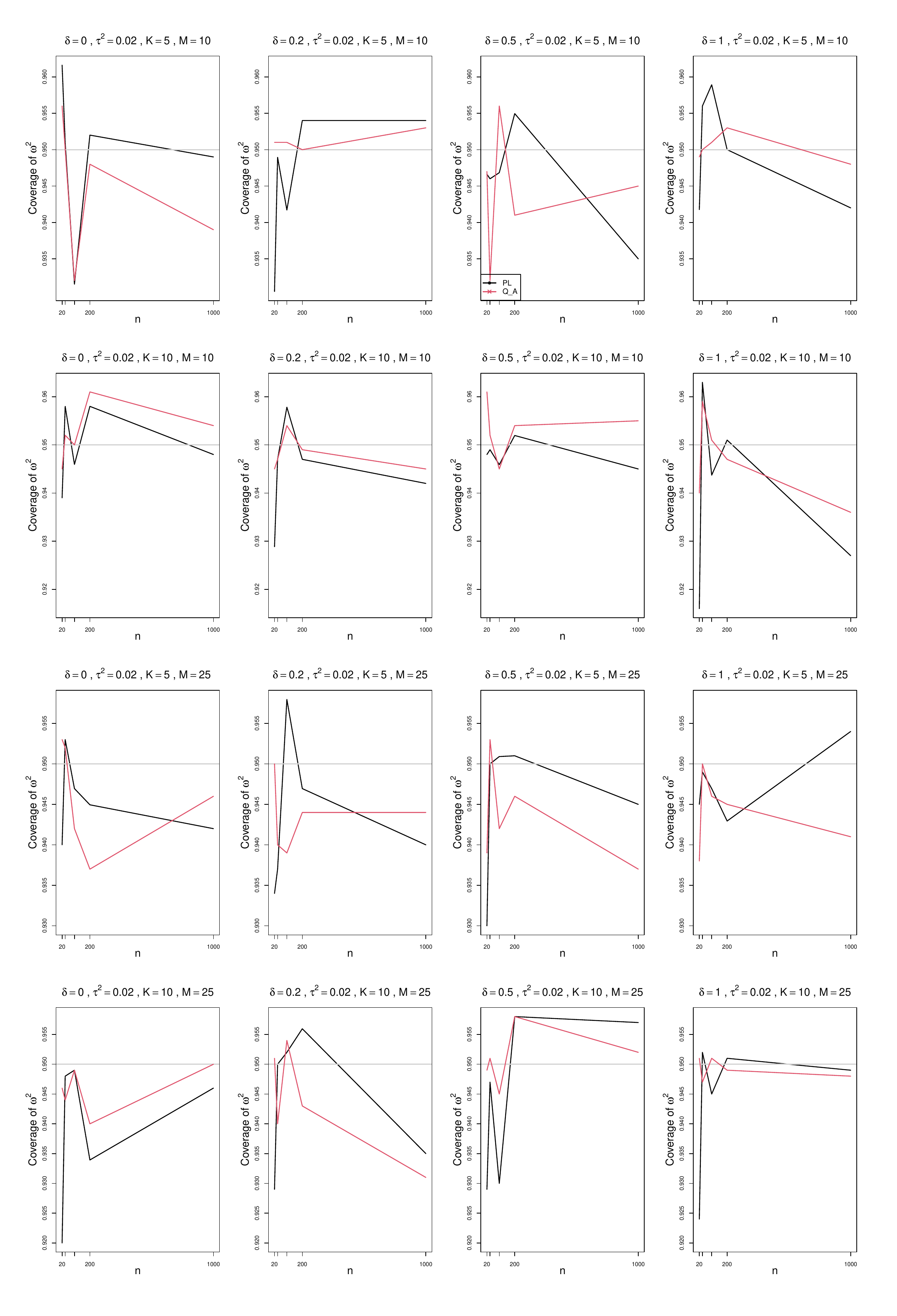}
	\caption{Coverage of 95\% confidence intervals for between-cluster variance $\omega^2=0.08$ of SMD (REML and $Q_F$) vs sample size $n$ = 20, 40, 100, 200, 1000 in each study, for $K$ =  5, 10 studies per cluster, $M$ =  10, 25 clusters and SMD $\delta = 0, 0.2, 0.5, 1$.   }
	\label{PPlotCoverageOfOmega_x4_0.02_HIER.pdf}
\end{figure}

\begin{figure}[ht]
	\centering
	\includegraphics[scale=0.33]{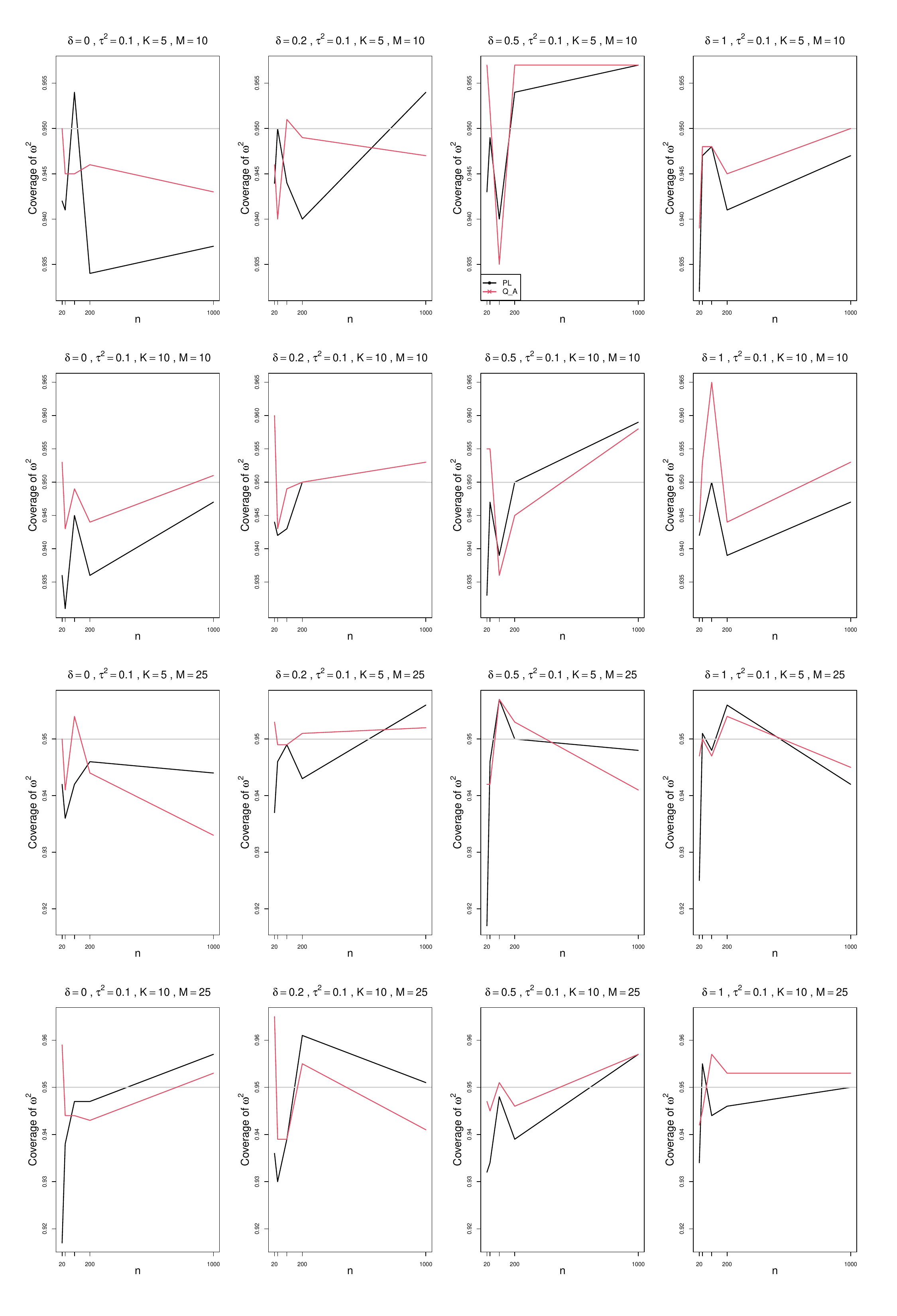}
	\caption{Coverage of 95\% confidence intervals for between-cluster variance $\omega^2=0.08$ of SMD (REML and $Q_F$) vs sample size $n$ = 20, 40, 100, 200, 1000 in each study, for $K$ =  5, 10 studies per cluster, $M$ =  10, 25 clusters and SMD $\delta = 0, 0.2, 0.5, 1$.   }
	\label{PlotCoverageOfOmega_x4_0.1_HIER.pdf}
\end{figure}

\begin{figure}[ht]
	\centering
	\includegraphics[scale=0.33]{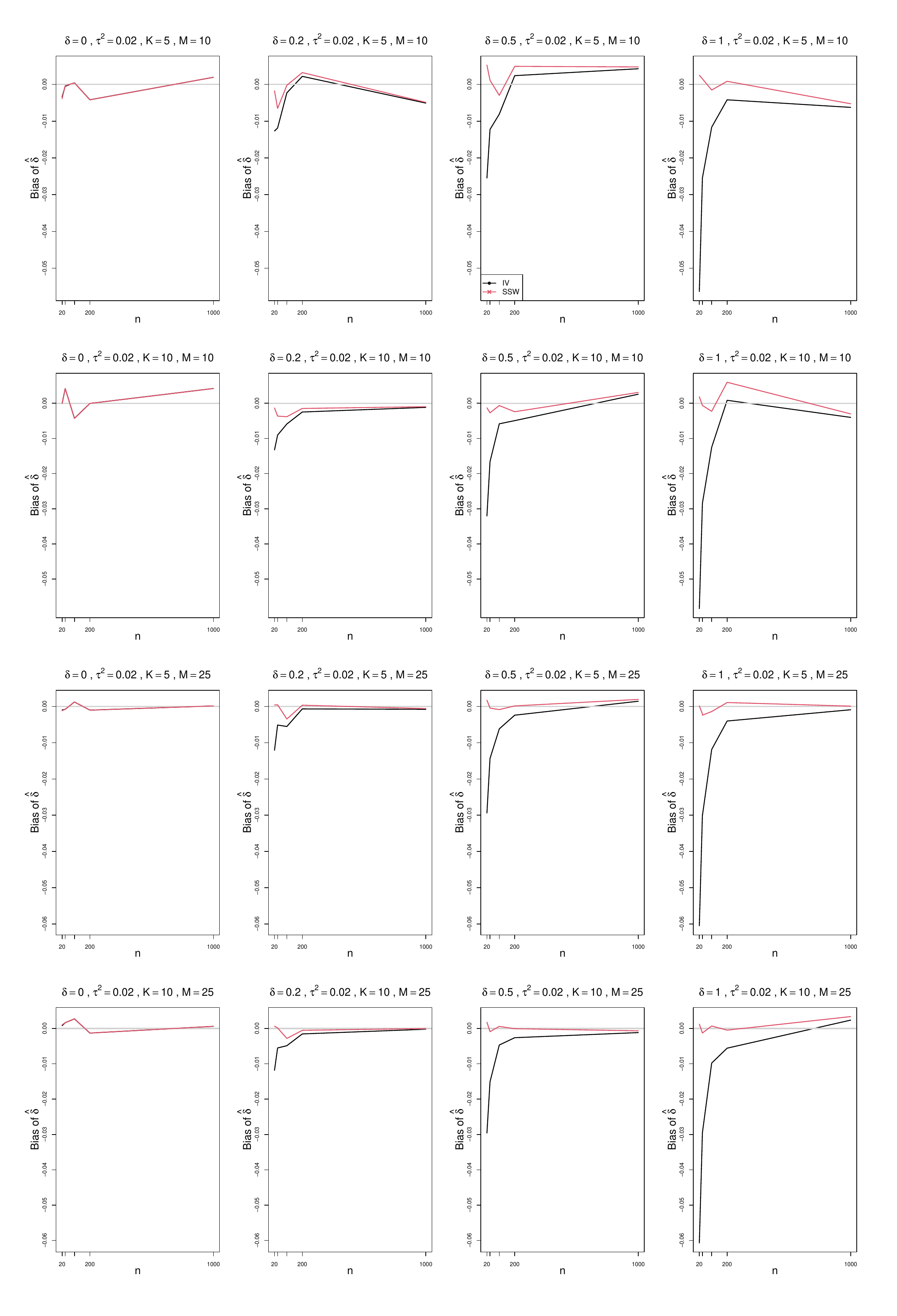}
	\caption{Bias of estimators of overall effect $\delta$  (REML and SSW ) vs sample size $n$ = 20, 40, 100, 200, 1000 in each study, for $\tau^2=0.02$, $\omega^2=0.08$, $K$ =  5, 10 studies per cluster, $M$ =  10, 25 clusters and SMD $\delta = 0, 0.2, 0.5, 1$.   }
	\label{PlotBiasOfDelta_x4_0.02_0_HIER.pdf}
\end{figure}

\begin{figure}[ht]
	\centering
	\includegraphics[scale=0.33]{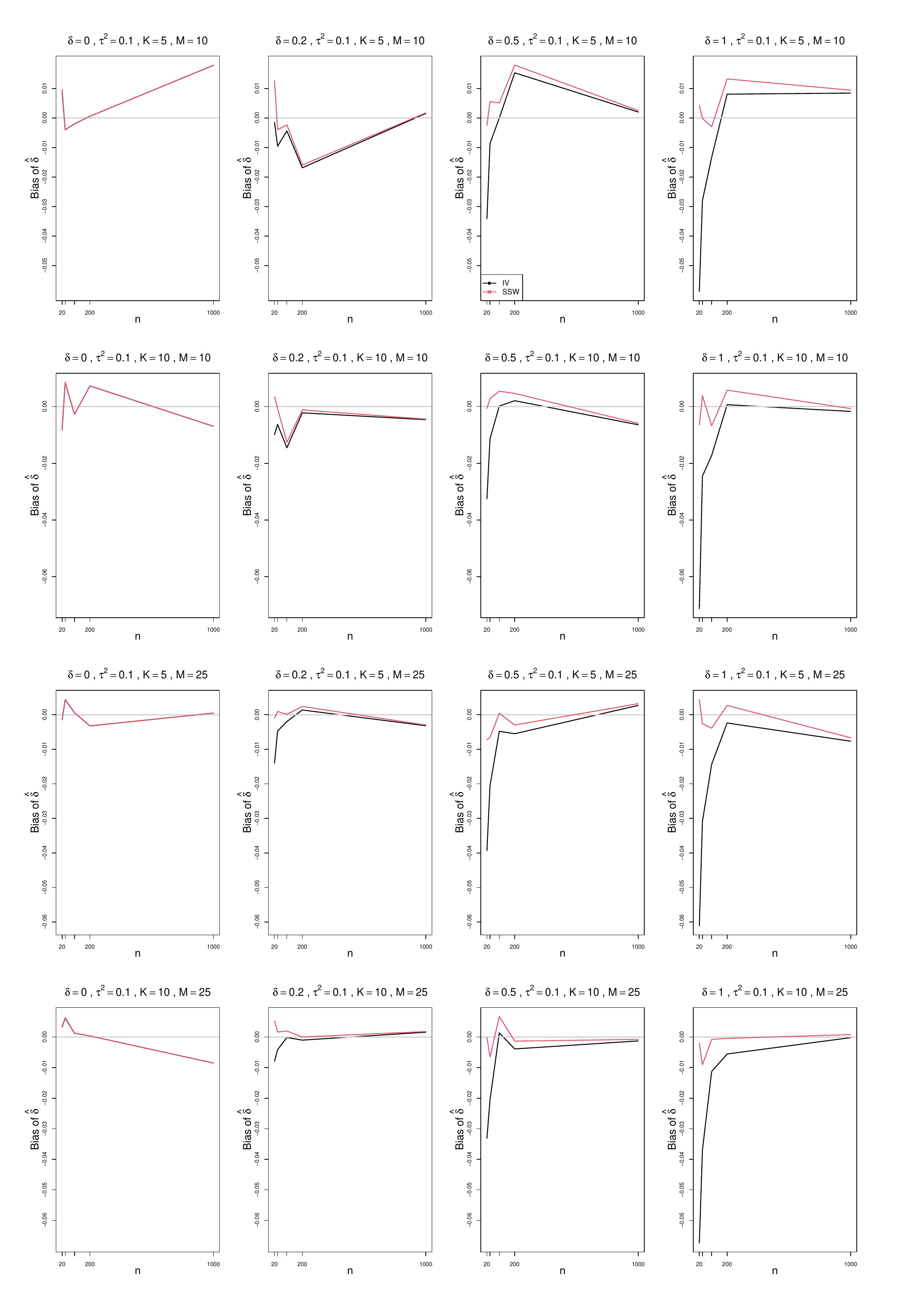}
	\caption{Bias of estimators of overall effect $\delta$  (REML and SSW ) vs sample size $n$ = 20, 40, 100, 200, 1000 in each study, for $\tau^2=0.1$, $\omega^2=0,4$,$K$ =  5, 10 studies per cluster, $M$ =  10, 25 clusters and SMD $\delta = 0, 0.2, 0.5, 1$.   }
	\label{PlotBiasOfDelta_x4_0.1_0_HIER.pdf}
\end{figure}

\begin{figure}[ht]
	\centering
	\includegraphics[scale=0.33]{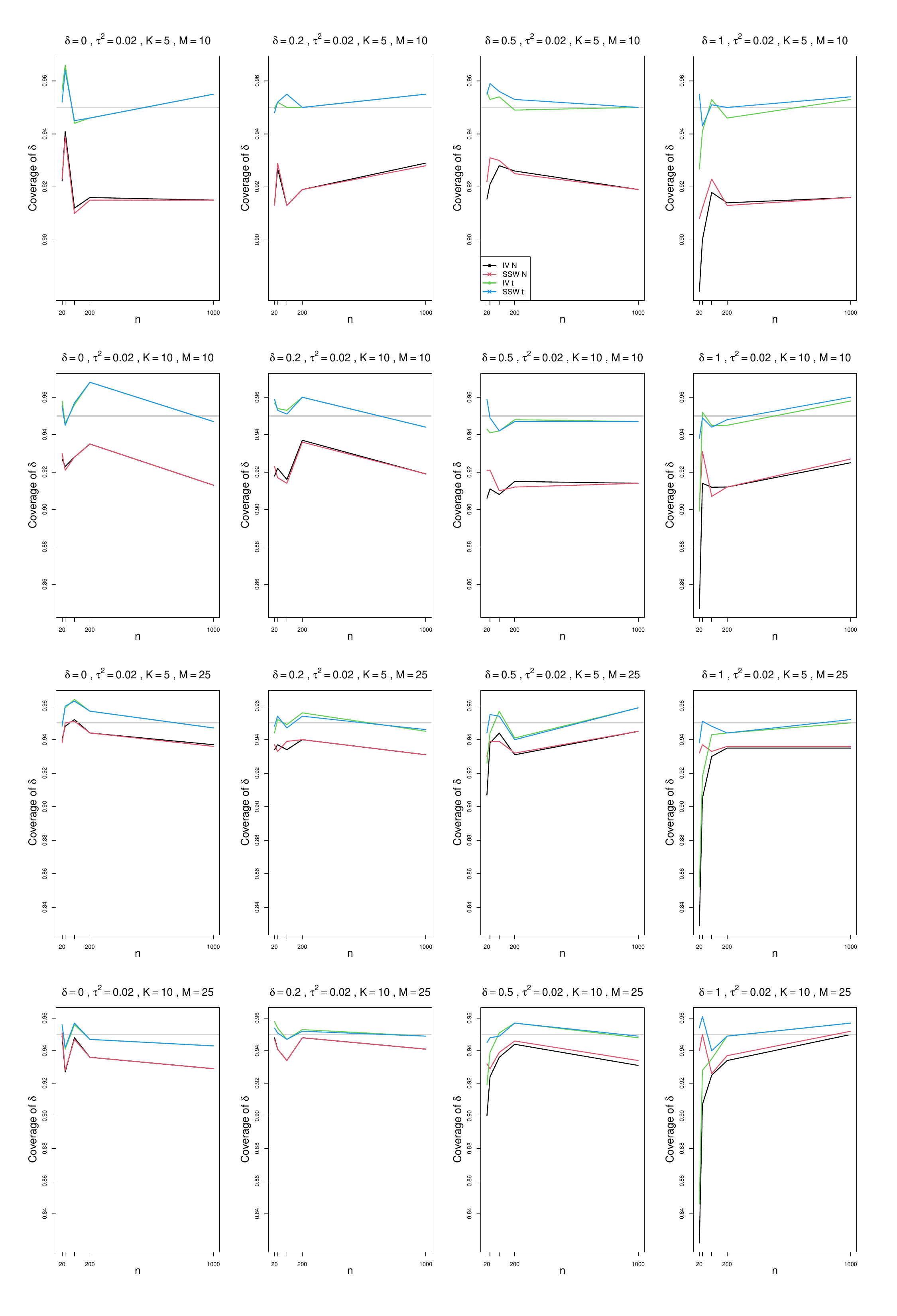}
	\caption{Coverage of 95\% confidence intervals for overall effect $\delta$ (REML N, REML t, $Q$ N, and $Q$ t)  vs sample size $n$ = 20, 40, 100, 200, 1000 in each study, for $\tau^2=0.02$, $\omega^2=0.08$, $K$ =  5, 10 studies per cluster, $M$ =  10, 25 clusters and SMD $\delta = 0, 0.2, 0.5, 1$.   }
	\label{PlotCoverageOfDelta_x4_0.02_HIER.pdf}
\end{figure}

\begin{figure}[ht]
	\centering
	\includegraphics[scale=0.33]{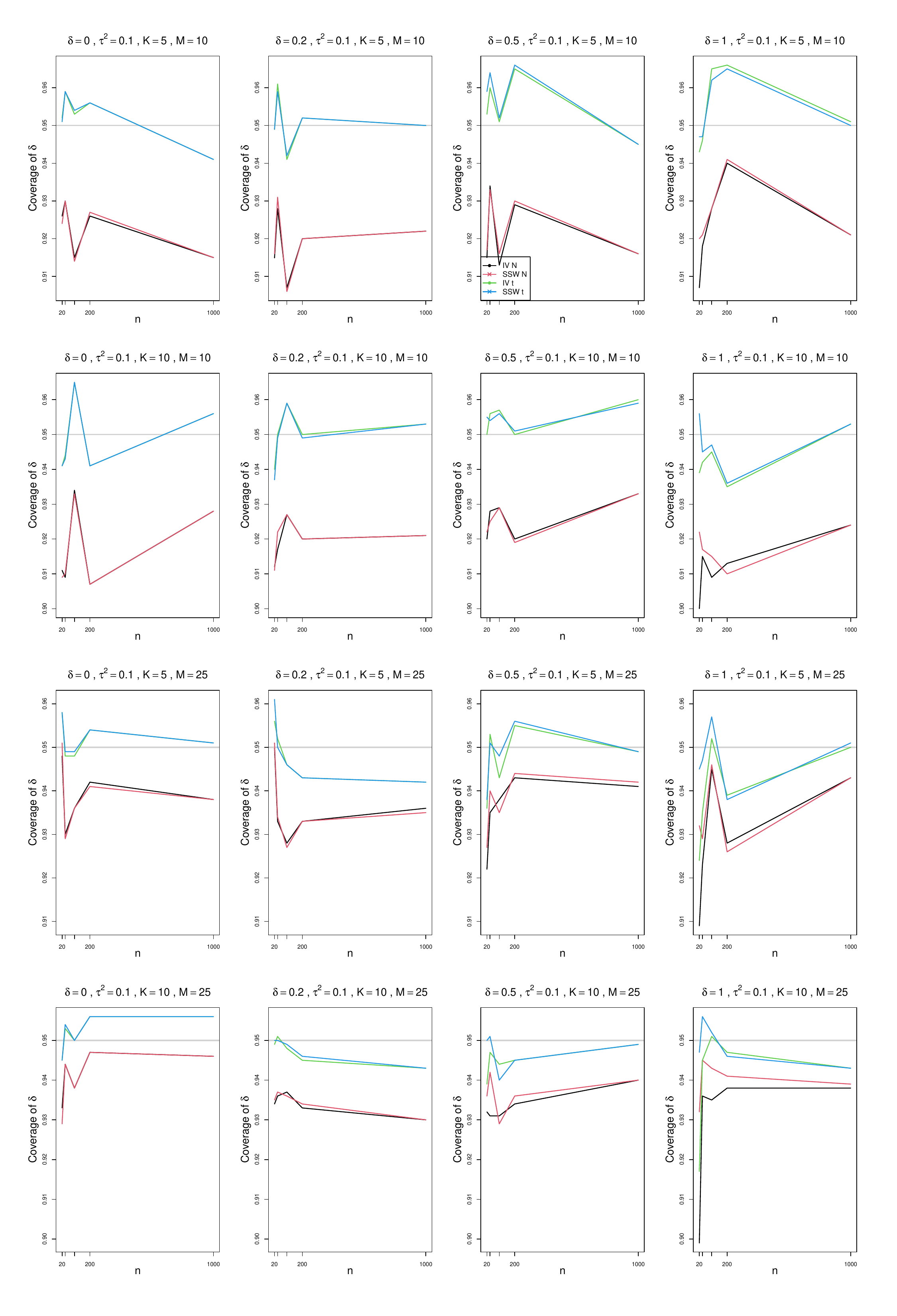}
	\caption{Coverage of 95\% confidence intervals for overall effect $\delta$ (REML N, REML t, $Q$ N, and $Q$ t)  vs sample size $n$ = 20, 40, 100, 200, 1000 in each study, for $\tau^2=0.1$, $\omega^2=0.4$, $K$ =  5, 10 studies per cluster, $M$ =  10, 25 clusters and SMD $\delta = 0, 0.2, 0.5, 1$.   }
	\label{PlotCoverageOfDelta_x4_0.1_HIER.pdf}
\end{figure}

\end{document}